\documentclass[aps,prd,showpacs,twocolumn,superscriptaddress,nofootinbib,preprintnumbers]{revtex4-1} 

\makeatletter
\def\p@subsection{}
\makeatother

\usepackage{graphicx,amsmath,amssymb,lipsum,bm}

\usepackage[usenames,dvipsnames]{xcolor}
\definecolor{xlinkcolor}{rgb}{0.7752941176470588, 0.22078431372549023, 0.2262745098039215}

\usepackage[colorlinks=true,citecolor=xlinkcolor,linkcolor=xlinkcolor,urlcolor=xlinkcolor, backref=false,pdfborder={0 0 0}]{hyperref}

\usepackage[normalem]{ulem}
\usepackage[utf8]{inputenc} 
\usepackage{mathtools}
\usepackage{aas_macros}
\usepackage{float}

\usepackage{ulem}
\usepackage{diagbox}

\usepackage[dvipsnames]{xcolor}
\usepackage{mathrsfs}

\newcommand{\be}{\begin{equation}}
\newcommand{\ee}{\end{equation}}
\newcommand{\beqa}{\begin{eqnarray}}
\newcommand{\eeqa}{\end{eqnarray}}

\newcommand\p{{\bm p}}
\renewcommand\k{{\bm k}}
\newcommand\q{\bm{q}}
\newcommand\x{\bm x}
\renewcommand\O{\Omega}

\newcommand\GG{\Gamma_3}
\newcommand\G{\mathcal{G}_2}

\renewcommand\a{\alpha}
\renewcommand\b{\beta}

\newcommand\F{\mathcal{F}}

\newcommand{\T}{\Theta}

\newcommand{\e}{\eta}

\def\d{\partial}
\newcommand{\bseq}{\begin{subequations}}
\newcommand{\eseq}{\end{subequations}}

\renewcommand{\ln}{\mathop{\rm ln}\nolimits}

\def\gsim{\raise0.3ex\hbox{$\;>$\kern-0.75em\raise-1.1ex\hbox{$\sim\;$}}}
\def\lsim{\raise0.3ex\hbox{$\;<$\kern-0.75em\raise-1.1ex\hbox{$\sim\;$}}}

\def\beqn#1{\begin{equation}\label{#1}}
\def\eeqn{\end{equation}}

\def\beqa#1{\begin{eqnarray}\label{#1}}
\def\eeqa{\end{eqnarray}}

\def\kmax{{k_\text{max}}}
\def\hMpc{h{\text{Mpc}}^{-1}}
\def\Mpch{h^{-1}{\text{Mpc}}}

\def\Z2{$\mathcal{Z_2}$}

\def\vpsi{{\boldsymbol{\psi}}}

\newcommand {\ignore}[1]{}

\begin{document}

\preprint{MIT-CTP/5762}

\title{Full-shape analysis with simulation-based priors: \\
cosmological parameters 
and the 
structure growth anomaly
}

\author{Mikhail M. Ivanov}
\email{ivanov99@mit.edu}
\affiliation{Center for Theoretical Physics, Massachusetts Institute of Technology, 
Cambridge, MA 02139, USA}

\author{Andrej Obuljen}
\affiliation{Department of Astrophysics, University of Zurich, Winterthurerstrasse 190, 8057 Zurich, Switzerland}

\author{Carolina Cuesta-Lazaro}
\email{cuestalz@mit.edu}
\affiliation{The NSF AI Institute for Artificial Intelligence and Fundamental Interactions, Cambridge, MA 02139, USA}
\affiliation{Department of Physics, Massachusetts Institute of Technology, Cambridge, MA 02139, USA}
\affiliation{Center for Astrophysics | Harvard \& Smithsonian, 60 Garden Street, MS-16, Cambridge, MA 02138, USA}

\author{Michael W. Toomey}
\email{mtoomey@mit.edu}
\affiliation{Center for Theoretical Physics, Massachusetts Institute of Technology, 
Cambridge, MA 02139, USA}

\begin{abstract} 
We explore 
full-shape analysis with simulation-based
priors, which is 
the simplest approach 
to galaxy clustering data analysis 
that combines effective field theory (EFT)
on large scales 
and numerical simulations
on small scales. The core ingredient 
of our approach is the 
prior density of EFT parameters
which we extract from a suite of 10500 
galaxy simulations based on the halo occupation distribution (HOD) model.
We measure the EFT parameters with 
the field-level forward model,
which enables us to cancel cosmic variance. 
On the theory side, 
we develop a new efficient 
approach to calculate field-level 
transfer functions using \textit{time-sliced perturbation theory}
and the logarithmic fast Fourier transform.
We find that the cosmology dependence of 
EFT parameters 
of galaxies is approximately degenerate
with the HOD parameters, 
and hence 
it can be ignored
for the purpose 
of prior generation. 
We use neural density estimation
to model the measured
distribution of EFT parameters.
Our distribution model is 
then used as a 
prior in a reanalysis of
the BOSS full-shape galaxy power spectrum data. 
Assuming the $\Lambda$CDM model, we find 
significant ($\approx 30\%$ and $\approx 60\%$)
improvements 
for the matter density fraction and the mass fluctuation amplitude, which are 
constrained to $\Omega_{m}=
0.315 \pm 0.010$ 
and $\sigma_8 = 0.671 \pm 0.027$. 
The value of the Hubble constant 
does not change, $H_0= 68.7\pm 1.1$~km/s/Mpc.
This
reaffirms earlier
reports of the structure growth 
tension from the BOSS data.
Finally, we use the measured
EFT parameters 
to constrain 
the galaxy-dark matter connection.
\end{abstract}

\maketitle

\section{Introduction}

The nature of dark matter, dark energy, 
and the origin of the Universe (cosmic inflation)
remain some of the greatest puzzles of fundamental physics. 
Ongoing and future galaxy surveys such as 
DESI~\cite{Aghamousa:2016zmz}, Euclid~\cite{Laureijs:2011gra}, LSST~\cite{LSST:2008ijt}, and Roman Space Telescope~\cite{Akeson:2019biv}
aim to deliver data that can shed light 
on these mysteries. 
However, extracting information from the 
maps of our Universe created by these surveys
will not be an easy task. It will require a detailed 
understanding of structure formation in
the non-linear regime.

Historically, there have been two leading approaches to this 
problem. The first one relies 
on cosmological perturbation theory, which proved to be 
extremely successful in the description
of cosmological fluctuations in the linear regime. 
The non-linear extension of cosmological 
perturbation theory relevant for 
structure formation
have been developed 
over the last fifty years~\cite{Zeldovich:1969sb}.
Most recently, this program culminated 
with the development of 
the effective field 
theory of large scale structure (EFT)~\cite{Baumann:2010tm,Carrasco:2012cv,Ivanov:2022mrd}. 
EFT provides a systematic and consistent 
program to describe 
galaxy clustering on scales larger than 
few Megaparsecs. The main practical advantages 
of EFT are its low computational cost
and flexibility. This made it a useful tool 
for tests of
the standard cosmological model\footnote{Based on the cosmological constant $\Lambda$ and cold dark matter (CDM), $\Lambda$CDM.}
and its various extensions with full-shape analyses of galaxy clustering data from BOSS and DESI, see e.g.~\cite{Ivanov:2019hqk,DAmico:2019fhj,Chen:2021wdi,Philcox:2021kcw,Chen:2024vuf,Maus:2024dzi}. 

Powerful as it is, EFT has important limitations.
First, it breaks down on small scales, which contain
significant information as measured by the number of Fourier modes. 
Second, even on large scales, the EFT predictions depend
on a large number of free (nuisance) parameters, which have to be 
marginalized over in order to obtain cosmological constraints. The number of free
parameters proliferates as one pushes to higher 
orders in perturbation theory, see e.g.~\cite{Baldauf:2014qfa,Baldauf:2015aha,Konstandin:2019bay}.
This leads to a significant degradation of constraining power. 
In particular, the constraints on single field inflation
degrade by orders of magnitude due to 
the degeneracy between EFT nuisance parameters
and the physical inflationary signal~\cite{Cabass:2022wjy,Cabass:2022ymb,Cabass:2022epm}. 
On the one hand, the EFT provides the most 
conservative framework that is absolutely 
agnostic about galaxy formation.
On the other hand, 
this approach ignores
years of intense theoretical 
and observational efforts that 
significantly improved our 
understanding of galaxy formation.
The rapid progress 
in this field is evidenced 
by the success of numerical simulations
to model galaxies, see e.g.
with the EAGLE, IllustrisTNG
and MillenniumTNG simulations~\cite{McAlpine:2015tma,Springel:2017tpz,Hernandez-Aguayo:2022xcl}.

This brings our discussion to the second approach to
model the cosmic web: numerical simulations. 
For pure dark matter clustering, numerical 
simulations provide a complete first principle
solution to the problem of 
cosmological structure formation. 
In particular, the N-body simulations 
show that the collapse of matter 
leads to the formation of dark matter halos.
However, the key challenge is to 
describe the observed luminous objects, such as 
galaxies, for which the 
first principle models may not be available. 
A number of approaches, ranging from 
theoretical to purely empirical, have been 
proposed, see~\cite{Wechsler:2018pic} for a review. 
In this work, we focus 
on the halo occupation distribution (HOD)
framework~\cite{Berlind:2001xk,Kravtsov:2003sg,Zheng:2004id,Hearin:2015jnf}.
The HOD models are based 
on a well established
fact that galaxies reside 
inside dark matter halos. 
The HOD then assigns 
galaxy positions 
based on the matter distribution 
within dark matter halos. 
From the EFT perspective, the HOD provides an ultraviolet (UV) complete 
galaxy clustering 
model, which is formally applicable 
even on small scales. 

Extracting
information from large-scale structure data
with simulations is part of a general 
``simulation-based inference'' (SBI) approach
of fitting data without explicit analytic models.
Various HOD-based SBI pipelines have been 
successfully tested 
in blind challenges~\cite{Beyond-2pt:2024mqz} and applied to actual data, see e.g.~\cite{Kobayashi:2021oud,Cuesta-Lazaro:2023gbv,Valogiannis:2023mxf,Hahn:2023kky,Hou:2024blc}.
This success raises the questions: 
How to improve EFT by  
incorporating 
galaxy formation information~\cite{Sullivan:2021sof,Ivanov:2024hgq,Cabass:2024wob}? 
And 
how to combine the benefits of both EFT 
and SBI~\cite{Obuljen:2022cjo,Modi:2023drt}?\footnote{In some sense, this question is addressed in phenomenological 
hybrid EFT models~\cite{Modi:2019qbt,Kokron:2021xgh,Pellejero-Ibanez:2022efv,Baradaran:2024jlh}. In our work, however, 
we apply EFT strictly within the range of its 
mathematical validity.}

In this work, we pursue possibly the simplest 
and least computationally expensive 
way to combine EFT and SBI: 
use simulation-based priors in EFT analysis~\cite{Sullivan:2021sof,Kokron:2021faa,Ivanov:2024hgq,Cabass:2024wob}. 
In particular, 
we use priors based on HOD models, as this approach 
offers great flexibility in galaxy modeling 
at a very modest computation cost (as compared e.g. to the 
hydrodynamic simulations).

A typical HOD model uses $O(10)$ parameters
to describe the clustering of galaxies 
in redshift space on all scales. 
However, the EFT power spectrum
model depends on roughly the same 
number of parameters already at next-to-leading (one-loop)
order. Formally, EFT expansion involves an infinite number
of parameters. Hence, if one assumes that the HOD is 
a true underlying model, the EFT parameters must be 
highly correlated. These correlations should 
bring additional information 
that can improve usual EFT full-shape analyses. 

A useful analogy here is provided by nuclear physics. 
The strong interaction at low energies is described
by chiral perturbation theory (ChPT) (see e.g.~\cite{Donoghue_Golowich_Holstein_2023,Donoghue:2017pgk} for reviews).
Being an EFT, ChPT at any given order 
formally depends on many unknown low energy constants. 
One can match these constants from UV complete models. 
A textbook example is given by matching ChPT 
to the linear sigma model, 
which produces tight 
constraints on combinations of ChPT parameters.  
More realistic examples
are provided by matching ChPT 
to results of numerical lattice simulations (see e.g.~\cite{Park:2021ypf,Abbott:2024vhj,Abbott:2023coj} and references therein)
or bootstrap calculations~\cite{He:2023lyy,He:2024nwd}, 
based on 
UV complete quantum chromodynamics (QCD).
In this context, an HOD model with some free parameters 
is equivalent to a QCD model whose free parameters are the gauge group, 
coupling constant, quark flavors and masses. 

Our approach follows the one outlined by 
us in ref.~\cite{Ivanov:2024hgq}.
There, we have produced
HOD-based priors for 
EFT parameters,
and applied them to the search of 
single field inflation primordial non-Gaussianity (PNG)
in the BOSS data. There are two 
simplifications that take place in this particular 
analysis. First, 
the underlying cosmological parameters  
of $\Lambda$CDM are kept fixed,
which is customary in PNG searches~\cite{Planck:2019kim}.
Second, the PNG constraints depend 
primarily on the bias parameters in real-space 
(i.e. in the galaxy rest frame)~\cite{MoradinezhadDizgah:2020whw,Cabass:2022wjy,Cabass:2022ymb}. 
Given these reasons, the HOD-based priors of ref.~\cite{Ivanov:2024hgq}
were produced for a fixed cosmology 
and only in real space. In this work 
we go beyond these simplifications. 
We derive priors on EFT parameters 
in redshift space, and also study its cosmology 
dependence (although eventually we find it to be negligible
for the purpose of EFT prior generation).
This allows us to explore the sensitivity of the full-shape 
analysis to
priors on redshift space counterterms. 
This is especially
relevant in the context of the structure growth tension, 
i.e. the discrepancy between the measurements of the mass
fluctuation amplitude parameter $\sigma_8$ (or the lensing parameter $S_8=(\Omega_m/0.3)^{0.5}\sigma_8$)
between cosmic microwave background and large-scale structure, e.g. 
in Baryon Oscillation Spectroscopic Survey (BOSS) data~\cite{BOSS:2016wmc}. 
Some evidence for low $\sigma_8$ 
has been reported 
by several independent analyses of galaxy clustering data~\cite{Ivanov:2019hqk,Philcox:2021kcw,Chen:2021wdi,Yuan:2022jqf,Chen:2022jzq,Lange:2023khv,Ivanov:2023qzb,Chen:2024vuf,Ibanez:2024uua}, 
see however~\cite{Kobayashi:2021oud,Paillas:2023cpk,Valogiannis:2023mxf,Sailer:2024coh,Chen:2024vvk}.

Our paper is structured as follows. 
We start with a summary of main results in Sec.~\ref{sec:main}.
Then we discuss the simulations that we use here in Sec.~\ref{sec:sims}.
The details of our field-level EFT technique and the theoretical 
calculations are presented in Sec.~\ref{sec:eft}.
Section~\ref{sec:priors}
presents our study of the cosmology
dependence of dark matter halos,
and the final distribution of EFT parameters 
from the HOD. There we also discuss the 
modeling of the EFT parameter density with 
machine learning tools. 
Section~\ref{sec:boss}
describes the re-analyses of the BOSS
galaxy clustering data with 
our HOD-based priors.
Sec.~\ref{sec:disc}
draws conclusions. 
Details of our theoretical calculations
and additional tests are reported 
in appendices.

\section{Main Results}
\label{sec:main}

The key goal of our paper is to 
produce a distribution of EFT parameters
from galaxy formation simulations
that can be used as a prior
in EFT-based full-shape analyses. 
We focus on the HOD-based models
here, noting that our approach is 
more general and in principle can be applied
to other types of simulations.

We produce 
a large sample of mock galaxy 
catalogs and extract their EFT parameters. 
We use 
neural density estimators
called
normalizing flows to obtain a 
model for the 
resulting 
distribution of EFT parameters
that can accurately capture intricate 
correlations between them. 
Finally, we apply this priors
in the actual analysis
of publicly available 
galaxy clustering data. 

In the process of 
producing simulation-based
priors, we have resolved a number 
of theoretical and practical 
problems, which are summarized in this Section,
together with our key discoveries and results.

\begin{figure*}
\centering
\includegraphics[width=0.99\textwidth]{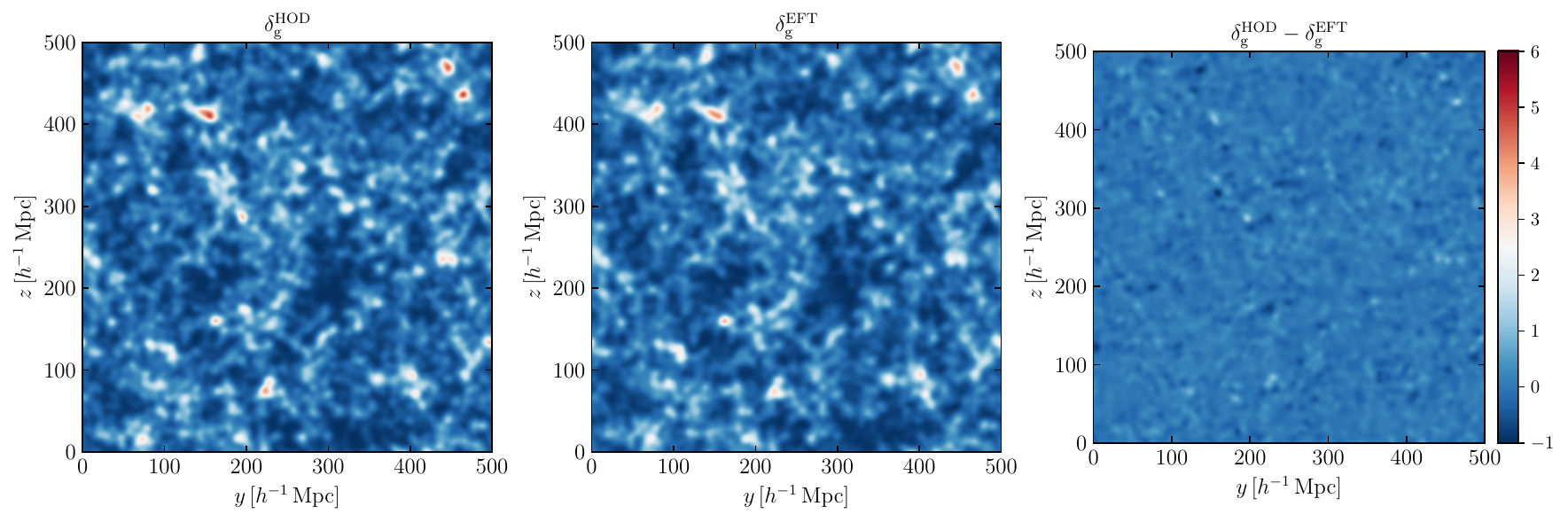}
\includegraphics[width=0.99\textwidth]
{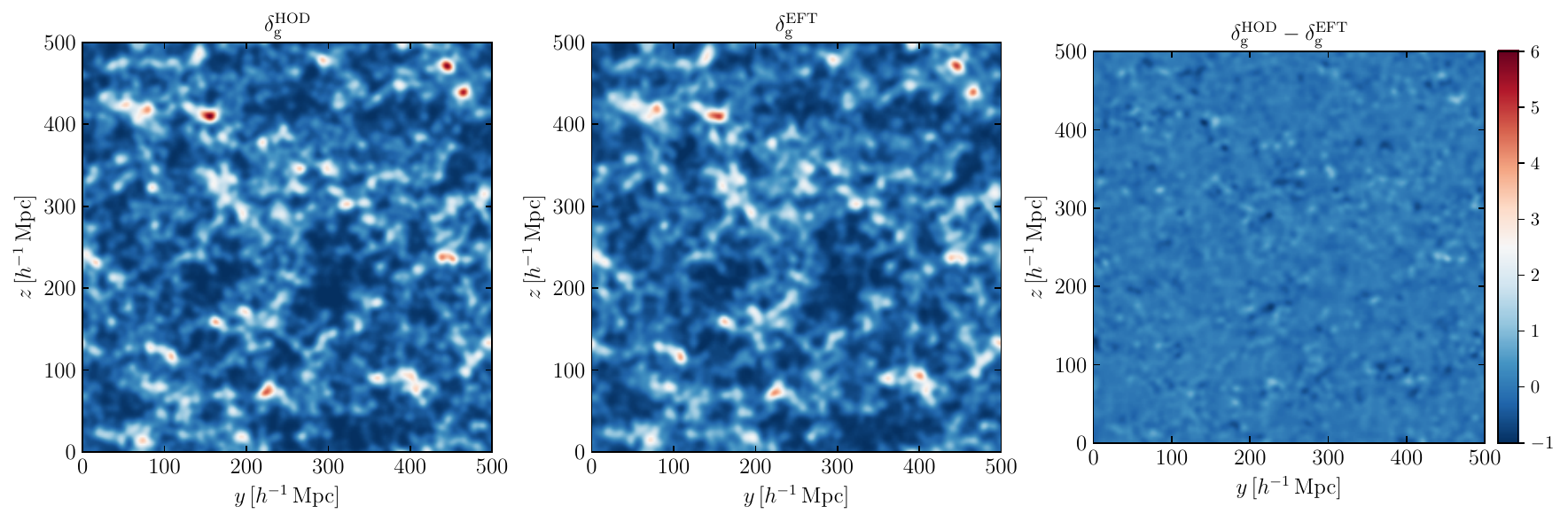}
   \caption{A typical HOD mock galaxy distribution in 
   real space (upper panel)
   and redshift space (lower panel)
   from our set (left), field-level EFT fit to it (center), 
   and the residuals (right). The overdensity field has been smoothed with a $R=4\,\Mpch$ 3D Gaussian filter, the depth of each panel is $\approx60\,\Mpch$, while the redshift space distortions are along the $z$-axis.
    } \label{fig:field}
\end{figure*}

\textbf{1. Redshift space field-level transfer functions.}
We use the field-level technique 
to measure EFT parameters from 
simulated galaxy catalogs. 
Specifically, we use the 
field-level EFT model of~\cite{Schmittfull:2018yuk,Schmittfull:2020trd} based on transfer functions 
and the shifted operators (see also refs.~\cite{Obuljen:2022cjo,Foreman:2024kzw} for the application to HI maps, 
and~\cite{Schmittfull:2014tca,Lazeyras:2017hxw,
Abidi:2018eyd,
Schmidt:2018bkr,
Elsner:2019rql,
Cabass:2019lqx,
Modi:2019qbt,
Schmidt:2020tao,
Schmidt:2020viy,
Lazeyras:2021dar,
Stadler:2023hea,
Nguyen:2024yth} for additional 
references on field-level EFT).
The field-level method allows for 
cosmic variance cancellation, 
and hence a precision measurement
of EFT parameters from 
computationally cheap small 
box simulation. To measure the EFT parameters, we need to fit the shape of the 
field-level transfer functions. 
Specifically, we are mostly 
interested 
in the transfer function of the 
matter density field $\beta_1$, which
contains the bulk of the information
on the EFT parameters. 
Schematically, 
the latter is defined as
\be 
\label{eq:field1}
\delta_g(\k)\Big|_{\rm EFT,~ field-lvl}=\beta_1(\k)\tilde{\delta}_1(\k)+...~\,,
\ee 
where 
$\tilde{\delta}_1$ is the shifted 
linear density field,\footnote{i.e. the linear density field shifted by the linear Zel'dovich displacement, see Section~\ref{sec:eft} for more detail.} 
$\delta_g$ is the forward model, 
while the above dots
stand for higher order operators. 
The model~\eqref{eq:field1}
is a simple extension
of the linear galaxy bias model $\delta_g = b_1\delta_1$. 
However, in eq.~\eqref{eq:field1} $b_1$
is promoted to be a scale-dependent 
transfer function
in order to capture higher order corrections (cubic operators and counterterms). 
The typical 
galaxy density snapshots in real space and
redshift space, as well as the best-fit field-level
EFT models and the residuals between them 
are shown in fig.~\ref{fig:field}. These are generated with the publicly available code \texttt{Hi-Fi mocks}\footnote{\url{https://github.com/andrejobuljen/Hi-Fi_mocks}}.

The one-loop perturbation theory
predictions for $\beta_1$
in real space
have been derived in ref.~\cite{Schmittfull:2018yuk}.
In this work,  for the first time, 
we calculate the one-loop model 
for the redshift space 
generalization of $\beta_1$ proposed in~\cite{Schmittfull:2020trd}. 
Although the initial 
model of~\cite{Schmittfull:2020trd}
is formulated in Lagrangian space, 
we carry out our calculation in Eulerian space, which is more
directly connected to 
physical observables.

\textbf{2. Efficient theoretical 
computation of field-level transfer functions.}
The redshift-space model for $\beta_1$
uses shifted operators that are defined
in Lagrangian space using 
the Zel'dovich displacements. 
Thus, the most natural setup for their 
calculation is 
Lagrangian perturbation theory~\cite{Schmittfull:2018yuk}. 
This approach, however, 
is relatively expensive
in terms of computational time 
and makes it hard to connect 
the field-level results with 
Eulerian EFT 
codes such as \texttt{CLASS-PT}~\cite{Chudaykin:2020aoj}.
The computation speed is
crucial for one of our main 
goals here: the application of EFT to 
a large ensemble of cosmological models.
To overcome these problems, we develop an 
equivalent, but more computationally 
efficient way to obtain all necessary 
one-loop corrections to $\beta_1$
in redshift space in $\sim 1$ second
per CPU 
per single cosmological model. 
Our approach is 
based on the fact that Lagrangian 
perturbation theory is equivalent 
to infrared (IR) resummed Eulerian perturbation theory.
The latter can be efficiently 
formulated with \textit{time-sliced perturbation theory}
(TSPT)~\cite{Blas:2015qsi,Blas:2016sfa,Ivanov:2018gjr,Vasudevan:2019ewf}.

TSPT is a formulation of EFT which makes 
manifest the IR safety of Eulerian correlation
functions and allows for an efficient 
description of the non-linear evolution
of baryon acoustic oscillations (BAO), known as IR resummation~\cite{Crocce:2007dt,Senatore:2014via,Baldauf:2015xfa}. 
We use TSPT to resumm the IR 
displacements in the Eulerian expansion
for the shifted correlation functions.
As a result, we obtain an expression
that accounts for the 
non-linear BAO and that can be easily 
evaluated with the standard tools of Eulerian
perturbation theory loop integrals,
such as the logarithmic Fast Fourier Transform (FFTLog)~\cite{Simonovic:2017mhp,Chudaykin:2020aoj}. 
As an example, we develop an external module 
for the \texttt{CLASS-PT} code
that evaluates the TSPT expressions
for shifted power spectra 
in redshift space. 
This model plays a central role 
in our exploration of the cosmology
dependence of EFT parameters.

\textbf{3. Cosmology dependence of EFT parameters for dark matter halos.} 
We extract EFT parameters 
of dark matter halos 
for 2000 cosmological $\Lambda$CDM  models simulated 
within the \texttt{Quijote} project~\cite{Villaescusa-Navarro:2019bje}.
These samples include, for the first time, 
redshift space EFT counterterms 
and redshift-space stochastic 
contributions. 
We have found that EFT parameters
depend on cosmology mostly through
the ``peak height'' parameter $\nu=\delta_c/\sigma_M(z)$, where $\delta_c\approx 1.686$ is the spherical collapse threshold
overdensity, and $\sigma^2_M$
is 
the mass variance 
at the Lagrangian size of the 
halo of mass $M$. 
This dependence can be understood 
as a consequence of 
the approximate universality of the 
halo mass function. 

As a result, the cosmology dependence 
is approximately degenerate 
with the halo mass. 
In particular,
simulation-based 
correlations between 
EFT parameters, e.g. functions 
$b_2(b_1)$
etc., are nearly 
cosmology-independent
as the variation of $\sigma^2_M$
that produces them can be 
mimicked with an appropriate 
variation of $M$.

\textbf{4. Cosmology dependence of EFT parameters for galaxies.}
In the HOD models, the EFT parameters
of galaxies 
are derived from those of halos
assuming an HOD. 
In particular, 
the galaxy bias parameters  are given by integrals over the halo mass function
weighted by the HOD~\cite{Seljak:2000gq,Cooray:2002dia}. 
We have found both
analytically and numerically
that the cosmology dependence encoded
by the HMF can be 
approximately absorbed 
by 
modest shifts of the basic HOD parameters~\cite{Zheng:2004id} such as the minimal halo mass to host a central galaxy.
Since 
the cosmology dependence of EFT parameters
of galaxies is approximately degenerate with the HOD parameters, 
it can be accounted for by 
choosing wide enough 
prior ranges.

In addition, we have found that the variation
of parameters of the
``decorated'' HOD models~\cite{Hearin:2015jnf}
produces 
a much wider spread of the EFT 
parameter distribution
than the variation of cosmology and the basic HOD parameters. 
This makes the 
distribution of EFT parameters
sampled from HOD models 
largely cosmology-independent. 

We have done additional
checks of the cosmology-independence
of the HOD-based EFT parameter
distribution. 
These include 
an explicit comparison
of our priors with
EFT samples
from two 
different cosmologies, and
tests of our 
priors on PT Challenge
simulations~\cite{Nishimichi:2020tvu} based on 
a cosmology
different from the one used
to generate our priors. 

Our analysis implies that 
our HOD-based priors 
can be used to analyze 
 extensions of $\Lambda$CDM,
which alter structure formation
primarily via the underlying 
linear matter power 
spectrum, 
e.g. spatial curvature, dynamical dark energy~\cite{Chudaykin:2020ghx},
and 
models with modified pre-recombination 
histories~\cite{Ivanov:2020ril,He:2023dbn,Camarena:2023cku,He:2023oke}.\footnote{Note that this approximation works well even in other cases, such as e.g. massive neutrinos~\cite{Chudaykin:2019ock,Ivanov:2019hqk}, ultralight axion~\cite{Rogers:2023ezo}, and 
light massive relics~\cite{Xu:2021rwg}, where 
it is the standard practice in EFT analyses to 
consider only the modification of the linear matter power 
spectrum 
and neglect 
dynamical nonlinear effects due to nontrivial 
clustering properties. It will be interesting to understand to  what extend our priors 
can be used in these models.}

\textbf{5. EFT parameters for HOD galaxies.} We have generated 
a sample of EFT parameters 
from 10,500 HOD models
that match the properties of 
luminous red galaxies (LRGs).
We assume flat wide priors on HOD 
parameters in our sampling 
procedure. Our sample is the largest 
and the most complete sample 
of bias and EFT parameters studied
to date. The redshift space
EFT counterterms, including the stochastic one, are 
analysed in detail
for the first time. 

An important observation is
a high degree of correlation
between the EFT parameters
from HOD models. This is expected
as the EFT by definition 
makes no assumptions about the 
small scale physics. Hence, strictly 
speaking, the EFT depends on an infinite 
tower of free parameters. Thus, within
any particular small-scale model, such as the HOD,  
the EFT parameters inevitable 
sample some low dimensional 
parameter space. Importantly, 
the correlations that we find 
are different from those
that appear when fitting 
the EFT parameters 
directly from the power spectrum, bispectrum etc.\footnote{For instance, the 
contributions from the stochastic $k^2\mu^2$ operator 
and the deterministic $k^4\mu^4 P_{11}$ term ($P_{11}$ is the linear matter power spectrum)  are degenerate 
at the galaxy power spectrum level~\cite{Ivanov:2019pdj,Chudaykin:2020hbf}, but completely 
independent at the field level.} 
Hence, 
field level simulation-based 
information efficiently
breaks parameter degeneracies,
leading to sizeable 
improvements of
parameter constraints.

 \begin{figure*}[ht!]
\centering
\includegraphics[width=0.7\textwidth]{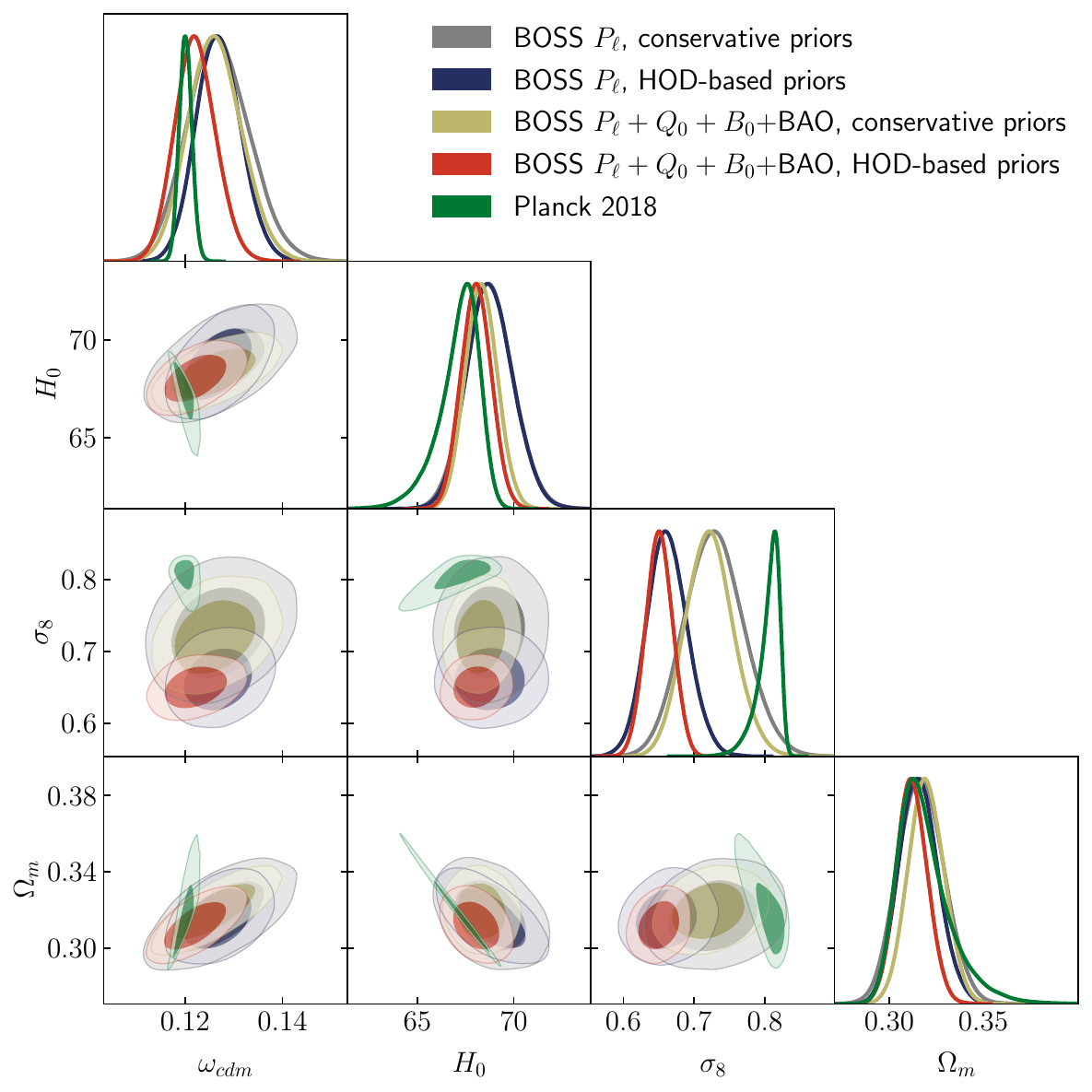}
   \caption{Cosmological parameters from the EFT-based full shape analysis of the BOSS power spectrum
   with conservative and informative simulation-based
   priors on EFT parameters. 
   Additionally, we show results
   from an extended data 
   vector that includes the real space
   power spectrum proxy $Q_0$, 
   BAO, 
   and the bispectrum monopole $B_0$. 
   For comparison, the \textit{Planck} 2018
   results for $\Lambda$CDM+$m_\nu$ are also shown. 
    } \label{fig:bossPk}
\end{figure*}

We have produced an 
accurate model for the 
distribution of the EFT parameters
with the 
normalizing flows. 
This distribution can be 
used as simulation-based priors (SBP) 
in actual full-shape 
analyses of LRG samples
from BOSS~\cite{BOSS:2016wmc} and DESI~\cite{Aghanim:2016sns}.

\textbf{6. Application to BOSS data.}
We have re-analyzed publicly available 
redshift
space galaxy clustering data 
from BOSS DR12~\cite{BOSS:2016wmc} with SBP.
Effectively, most of the EFT parameters 
are then determined by the priors
conditioned to information extracted 
from the data, such as the linear bias 
$b_1$ measured from (linear) redshift space distortions. 
In turn, tighter EFT parameters
reduce the freedom in fitting 
the small scale part of the power 
spectrum thereby allowing us to
extract more cosmological information.
As a result, we found significant improvements on $\sigma_8$
and $\omega_{cdm}$, by $60\%$
and $30\%$, respectively, 
from the power spectrum only analysis. 
This improvement 
is equivalent to doubling 
the survey volume. 

Our final parameter
estimates are presented in fig.~\ref{fig:bossPk}. 
There we show the posteriors 
of cosmological parameters 
from three BOSS analyses: the galaxy power
spectrum with conservative priors of~\cite{Philcox:2021kcw}, which is our benchmark, 
and SBP analyses of the galaxy power spectrum  
and of the full BOSS dataset from ref.~\cite{Philcox:2021kcw},
including the bispectrum monopole, BAO, and 
the real space power spectrum proxy $Q_0$.
The \textit{Planck} 2018 
constraints are displayed 
for comparison.

Our analysis 
confirms earlier
reports of the 
$\sigma_8$ tension in the 
BOSS data, see e.g.~\cite{Chen:2024vuf}. 
We find the tension nominally at the $\approx 5\sigma$ level
from the galaxy power spectrum alone,
which is the strongest evidence reported to date. 
Note that previous analyses such as~\cite{Chen:2024vuf} 
were performed 
with uninformative priors
on EFT parameters. Therefore, 
we additionally 
confirm 
that the BOSS 
$\sigma_8$ tension
is not driven by 
the choice of priors, see 
e.g.~\cite{Chen:2021wdi,Philcox:2021kcw,Chen:2024vuf} for earlier detailed
discussions.

\textbf{7. Constraining galaxy -- darm matter connection 
from EFT parameters.}
Our joint sample of 
EFT and HOD parameters ($\theta_{\rm EFT}$ and $\theta_{\rm HOD}$, respectively) can be used to 
build a conditional distribution
$p(\theta_{\rm HOD}|\theta_{\rm EFT})$, allowing one to translate
the measured values of EFT parameters
from data into HOD parameters.
This way one can get insights into
the 
galaxy -- dark matter connection
using large-scale
clustering information. 
We have 
implemented this approach 
in practice using appropriate 
normalizing flow models for
the conditional distribution $p(\theta_{\rm HOD}|\theta_{\rm EFT})$. 
We have obtained constraints 
on the threshold host 
halo masses from BOSS, 
which are consistent with 
SBI analyses based on HODs e.g.~\cite{Paillas:2023cpk}.
Notably, in agreement with these works, 
we have found 
evidence for environment-based assembly bias 
of central galaxies in the BOSS
data.

\textbf{8. Implications for 
the choice of priors in full-shape analyses.}
Our HOD-based priors are 
consistent with 
the conservative 
priors associated 
with the \texttt{CLASS-PT}~\cite{Chudaykin:2020aoj} and \texttt{velocileptor}~\cite{Chen:2020fxs,Chen:2020zjt}
codes. This confirms that 
these priors reflect well both
our current understanding of 
galaxy formation physics 
and EFT naturalness arguments. 
In contrast, 
the priors associated with the 
\texttt{PyBird} code~\cite{DAmico:2020kxu} 
appear to be overoptimistic,
which can bias
cosmological parameter estimates. 

An important observation
is that HOD models predict strong 
correlations between the EFT parameters.
It will be interesting to 
``orthogonolize'' the basis 
of EFT parameters to reflect the 
physics of HOD parametrizations, or other galaxy formation models, better. 
This may lead to new insights relevant for the
understanding 
of the physics of galaxy formation. 

\section{Simulations}
\label{sec:sims}

We discuss now the N-body simulations that we use in our work. 

\textbf{Quijote-LH}. The first type of simulations we use is a latin-hypercube 
\texttt{Quijote}
suite that samples 
2000 $\Lambda$CDM cosmologies~\cite{Villaescusa-Navarro:2019bje}.
Each cosmology 
is sampled from 
the following 
flat distributions, 
\be 
\begin{split}
& \Omega_m \in [0.1,0.5]\,,\quad 
\Omega_b \in [0.03,0.07]\,,\quad 
h\in [0.5,0.9]\,,\\
& n_s \in [0.8,1.2]\,,\quad 
 \sigma_8\in [0.6,1.0]\,.
\end{split}
\ee 
The neutrino mass
is set to zero. 
Note that these ranges 
significantly exceed
the actual constraints 
on these parameters
coming e.g. from \textit{Planck}~\cite{Ade:2015xua}. 
We use the high-resolution
\texttt{Quijote-LH} suite
where in simulation
was run with $1024^3$
particles. Each box
has a site length of 
1000 $\Mpch$. We use 
the snapshots 
at $z=0.5$. 
Note that even the high-resolution
\texttt{Quijote-LH} suite
is limited 
in terms of resolving low mass
halos. In particular, 
the Rockstar halo catalogs 
feature a saturation and a cutoff 
already for $\log_{10}(M_{\rm halo}/[M_{\odot}/h])<13$. 
The limited resolution is 
the main reasons why we do not use \texttt{Quijote-LH} 
for the HOD catalog
production. 

The generation of 
dark matter halo parameters for a fixed 
cosmology is done with 
the fiducial high-resolution \texttt{Quijote} simulation. 

\textbf{AbacusSummit small.}
Our main HOD samples
are based on the \texttt{AbacusSummit small}~\citep{Maksimova:2021ynf}, or covariance suite,
with box site length 
500~$\Mpch$. Each box
was run with $1728^3$
particles, which provides
an excellent resolution even for small halos. 
The fiducial cosmology 
of \texttt{AbacusSummit small} 
is the \textit{Planck} 2018
$\Lambda$CDM cosmology~\cite{Aghanim:2018eyx}
with a single massive neutrino 
of the minimal mass $M_\nu = 0.06$~eV. Our baseline redshift is $z=0.5$ as before.
We use halo catalogs computed from the full particle set 
using the CompaSO  
halo finder~\cite{Hadzhiyska:2021zbd}.

\textbf{AbacusSummit base.}
Additional
tests of the 
cosmology (in)dependence
of EFT parameters
are done with the \texttt{AbacusSummit base}
suite. Each box in the suite
has $6912^3$
particles and 2000~$\Mpch$ site length.
The suite covers 139 
different cosmological models. 
The cosmological models
that we use for our validation
tests are 
\texttt{abacus\_cosm000},
\texttt{abacus\_cosm003}, and 
\texttt{abacus\_cosm004}
from
\url{https://abacussummit.readthedocs.io/en/latest/cosmologies.html}.
The \texttt{cosm003}
and \texttt{cosm004}
cosmologies have
$\sigma_8=0.86$ and 
$\sigma_8=0.75$, respectively, 
which allow us to explore a large
range of the cosmological parameter
most relevant for structure formation.

\section{EFT at the field level}
\label{sec:eft}

In this section, we give a 
theoretical 
background on the field-level
EFT technique of refs.~\cite{Schmittfull:2018yuk,Schmittfull:2020trd}. 
The basis for the field level technique 
is the Eulerian bias model~\cite{Assassi:2014fva,Desjacques:2016bnm}, 
\be 
\label{eq:naive_eft}
\begin{split}
\delta^{\rm EFT}_g\Big|_{\rm n-pf}(\k) = b_1\delta + \frac{b_2}{2}\delta^2 +b_{\mathcal{G}_2}\mathcal{G}_2 \\
+b_{\Gamma_3}\Gamma_3 - b'_{\nabla^2\delta} \nabla^2 \delta
+\epsilon\,,
\end{split}
\ee 
where $\delta$ is the non-linear 
matter density field, 
$\mathcal{G}_2$ is the tidal operator, 
\be 
\label{eq:G2}
\begin{split}
\mathcal{G}_2(\k) & = \int_{\bm p} 
F_{\mathcal{G}_2}(\p,\k-\p)
\delta({\bm p})\delta(\k-{\bm p})\,,\\
 F_{\mathcal{G}_2}(\k_1,\k_2) & =\frac{({\bm k_1}\cdot \k_2)^2}{k_1^2 k_2^2}-1\,,
\end{split}
\ee 
$\int_{\k}\equiv \int \frac{d^3\k}{(2\pi)^3}$
and $\GG$ is the Galileon tidal operator, 
\be 
\begin{split}
F_{\GG}=\frac{4}{7}\left(1-\frac{(\k_1\cdot\k_2)^2}{k_1^2k_2^2}\right)
\left(\frac{((\k_1+\k_2)\cdot \k_3)^2}{(\k_1+\k_2)^2k_3^2}-1\right)\,.
\end{split}
\ee 
Here and in what follows we use the following
notation
for a general cubic operator $\mathcal{O}^{(3)} $
\be 
\label{eq:gamma3def}
\mathcal{O}^{(3)} =  \int_{\k_1}\int_{\k_2}\int_{\k_3}\left(\prod_{i=1}^3\delta(\k_i)\right)(2\pi)^3\delta_D^{(3)}(\k-\k_{123})F_{\mathcal{O}^{(3)}}\,.
\ee 
In our forward model we will replace $\delta$ in the above
integral expressions with the linear
matter over-density $\delta_1(\k,z)$, which is appropriate in perturbation theory. 
$b_1,b_2,b'_{\nabla^2\delta},b_{\Gamma_3}$
are Eulerian bias parameters. 
The field $\epsilon$ captures 
stochastic contributions to the observed
galaxy density. By definition, 
it does not correlate with any 
perturbative field. 

The model~\eqref{eq:naive_eft} requires 
a proper treatment of bulk flows (IR resummation).
At the field level, it means we have to 
keep the large-scale displacement
resummed. 
If $\delta_1(\k,z)$ is the initial 
density field, the associated
linear Zel'dovich 
displacement field 
in Fourier space is given by
\be 
 \bm{\psi}_1(\k,z)=\frac{i \bm{k} }{k^2}\delta_1(\k,z)\,.
\ee 
Note that the power spectrum
of $\delta_1$ by definition
is the linear matter 
power spectrum,
\be 
\langle \delta_1(\k)\delta_1(\k')\rangle =(2\pi)^3\delta_D^{(3)}(\k'+\k)P_{11}(k)\,.
\ee 
In what follows it will
be convenient to use the 
primed correlators with the 
Dirac delta function
stripped off, 
\be 
\langle \delta_1(\k)\delta_1(\k')\rangle'=P_{11}(k)\,.
\ee 

The field level model 
for galaxies is obtained by
writing down the 
perturbative bias
expansion, and shifting all terms
in it by the Zel'dovich displacement. 
The corresponding perturbative operators are dubbed ``shifted'' and denoted with tildas. In real space 
they are given by 
\be 
\tilde{\mathcal{O}}_{\rm real}(\k)
=\int d^3 \q~\mathcal{O}(\q)
e^{-i\k\cdot(\q+\vpsi(\q))}~\,.
\ee 
In redshift space, one has to add
an additional displacement 
along the line of sight $\hat{\bm z}$, 
\be 
\tilde{\mathcal{O}}(\k)
=\int d^3 \q~\mathcal{O}(\q)
e^{-i\k\cdot(\q+\vpsi(\q)+f\hat{\bm{z}}(\vpsi(\q)\cdot \hat{\bm{z}}))}~\,,
\ee 
where $f=d\ln D_+/d\ln a$ is the logarithmic growth factor, $D_+$ is the growth factor
and $a$ is the metric scale factor. 
Note that our model with the exponentiated 
Zel'dovich field is equivalent to 
IR resummed Eulerian or Lagrangian
EFT. 

Another important ingredient 
of the forward model is the 
orthogonalization of the 
relevant operators. 
Without orthogonalization, 
we have large operator mixing effects that 
make the measurements of bias 
parameters at the field level 
strongly cutoff-dependent. 
In order to reduce operator
mixing, following~\cite{Abidi:2018eyd,Schmittfull:2018yuk,Schmittfull:2020trd}
we use the orthogonalized operators
$\mathcal{O}^\perp_a$ that satisfy
\be 
\langle \mathcal{O}^\perp_m \mathcal{O}^\perp_n \rangle  = 0\quad \text{if}
\quad n\neq m\,.
\ee 
Once the set of operators is determined, they 
are orthogonalized using the Gram-Schmidt procedure. For instance, the first
three generic operators 
$\mathcal{O}_1,\mathcal{O}_2,\mathcal{O}_3$ are 
orthogonolized as
\be 
\begin{split}
& \mathcal{O}_1^\perp = \mathcal{O}_1 \,,\\
& \mathcal{O}_2^\perp =\mathcal{O}_2-\frac{\langle \mathcal{O}_2 \mathcal{O}_1 \rangle' }{\langle \mathcal{O}_1 \mathcal{O}_1 \rangle'}\mathcal{O}_1\,,\\
& \mathcal{O}_3^\perp =\mathcal{O}_3-\frac{\langle \mathcal{O}_3 \mathcal{O}_1 \rangle' }{\langle \mathcal{O}_1 \mathcal{O}_1 \rangle'}\mathcal{O}_1
-\frac{\langle \mathcal{O}_3 \mathcal{O}^\perp_2 \rangle' }{\langle \mathcal{O}^\perp_2 \mathcal{O}^\perp_2 \rangle'}\mathcal{O}^\perp_2
\,.
\end{split}
\ee 
The operators that we will use are those
that appear in the perturbative bias expansion. Explicitly, our 
field-level model will depend on the 
set of operators 
\[ 
\{1,\delta_1,\delta_2 \equiv \delta^2-\sigma^2_1,\mathcal{G}_2,\delta_3\equiv \delta_1^3\}\,,
\]
where $\mathcal{G}_2$ is defiend in \eqref{eq:G2},
and $\sigma_1^2$ is the 
mass variance, 
\be 
\sigma_1^2 = \int_\k P_{11}(k)\,.
\ee 
Note that the first operator in our set, 
the unity, after shifting, produces the 
Zel'dovich field which is relevant
for the redshift space model.

\subsection{Real space}

The real space density model is given by 
\be 
\label{eq:eft-field_real}
\begin{split}
\delta^{\rm EFT}_g (\k)=
&\beta_1(k)\tilde \delta_1(\k)
+\beta_2(k)(\tilde{\delta}_1^2)^\perp (\k) \\
& +\beta_{\mathcal{G}_2}(k)
\tilde{\mathcal{G}}_2^{\perp}(\k)
+\beta_3(k)
(\tilde{\delta}_1^3)^\perp (\k)\,,
\end{split}
\ee 
where the tidal operator is defined by eq.~\eqref{eq:G2} but 
using the linear density field, 
and 
$\beta_i(k)$
are transfer functions. Their shape
is fitted from the simulation snapshots $\delta_g^{\rm HOD}$, 
\be 
\label{eq:trfreal}
\beta_i(k)=\frac{\langle \mathcal{O}^*{}^\perp_i(\k)  \delta_g^{\rm HOD}(\k)\rangle'}{\langle
|\mathcal{O}^\perp_i (\k) |^2 
\rangle' }\,.
\ee 
In general, the transfer functions have 
a complicated scale dependence. 
Their low-$k$ limit, 
however, can be predicted 
in perturbation theory.  
Matching the two provides a 
practical way to extract the 
EFT parameters. 
The simplest way to do it is to replace
$\delta_g^{\rm HOD}$ with 
the usual EFT model density field 
$\delta_g^{\rm EFT}\big|_{\text{n-pf}}$
in eq.~\eqref{eq:trfreal}.
In contrast to eq.~\eqref{eq:eft-field_real}
with the transfer functions,  
$\delta_g^{\rm EFT}\big|_{\text{n-pf}}$
is the usual perturbative model
that is used in $n$-point function 
calculations. 
At the one-loop
order, one has
\be 
\label{eq:lowk_b1_real}
\begin{split}
& \beta_1(k) =  b_1 + (-b_1c_s+b'_{\nabla^2\delta})k^2
+\frac{b_2}{2}\frac{\langle \tilde \delta_1 \tilde \delta_2\rangle' }{\langle \tilde \delta_1\tilde \delta_1\rangle' }
-b_1
\frac{\langle \tilde\delta_1 \tilde{\mathcal{S}}_3\rangle'}{\langle \tilde\delta_1 \tilde\delta_1 \rangle' }
\\
& +\left(b_{\mathcal{G}_2}+\frac{2b_1}{7}\right)
\frac{\langle \tilde\delta_1 \tilde{\mathcal{G}}_2\rangle'}{\langle \tilde\delta_1 \tilde\delta_1 \rangle'}
+ \left( b_{\Gamma_3}+\frac{b_1}{6}+\frac{5}{2}b_{\mathcal{G}_2}\right)\frac{\langle \tilde\delta_1 \tilde \Gamma_3\rangle'}{\langle \tilde\delta_1 \tilde\delta_1 \rangle'}
\\
\end{split}
\ee 
where $b_1,b'_{\nabla^2\delta},b_2,b_{\mathcal{G}_2},b_{\Gamma_3}$
are the usual Eulerian bias parameters. 
$c_s$ is the dark matter speed of sound. In what follows we will use a definition 
\be 
b_{\nabla^2\delta} = -b_1c_s+b'_{\nabla^2\delta}~\,,
\ee 
since only this combination appears 
in the galaxy density counterterm. 
The extra operator that appears
in the r.h.s of eq.~\eqref{eq:lowk_b1_real}
is characterized by the kernel 
\be 
\begin{split}
F_{\mathcal{S}_3}=-\frac{3}{14}\left(1-\frac{(\k_1\cdot\k_2)^2}{k_1^2k_2^2}\right)
 \frac{((\k_1+\k_2)\cdot \k_3)}{|\k_1+\k_2|^2}\,,
\end{split}
\ee 
which is defined as in eq.~\eqref{eq:gamma3def}.
As far as other transfer functions are concerned, 
one can show that in the $k\to 0$ limit
\be 
\label{eq:lowk_betas}
\begin{split}
& \beta_2(k) = \frac{b_2}{2}+\mathcal{O}(k^2/k_{\rm NL}^2)\,,\\
& \beta_{\mathcal{G}_2}(k) = b_{\mathcal{G}_2}
+\frac{2}{7}b_1
+\mathcal{O}(k^2/k_{\rm NL}^2)\,,\\
& \beta_{3}(k) = \frac{b_3}{6}+\mathcal{O}(k^2/k_{\rm NL}^2)\,,
\end{split}
\ee 
where $k_{\rm NL}$
is the non-linear scale 
of perturbation theory, satisfying $k^3_{\rm NL}P_{11}(k_{\rm NL})\sim 1$.
The power spectrum of the error between the simulation 
and the perturbative model is defined as
\be 
P_{\rm err}(k) = 
\langle 
|\delta^{\rm EFT}(\k) - \delta^{\rm HOD}(\k)|^2
\rangle'\,.
\ee 
In perturbation theory, by definition\footnote{Note that $P_{\rm err}$ also absorbs 
the cutoff-dependent part of the 
auto-spectrum of $\delta^2$. This part, however, is negligibly small 
for our choice of the grid smoothing 
and we will ignore it in what follows. }
$P_{\rm err}$ should be 
equal to the power spectrum of
the stochastic field $\epsilon$. 
The EFT prediction
for it reads
\be 
\label{eq:Perr_real}
P_{\rm err} = \frac{1}{\bar n}\left(1+\alpha_0 + \alpha_1\left(\frac{k}{0.45~\hMpc}\right)^2\right)~\,,
\ee 
where $\bar n$ is the true number density of 
galaxies or halos. Note that 
here we use the convention
consistent with the EFT 
analysis pipeline based on the 
\texttt{CLASS-PT}
code. 


\subsection{Efficient calculation of the $\beta_1$
transfer function}

The 
shapes that appear in eq.~\eqref{eq:lowk_b1_real}
require loop calculations in Lagrangian space.
In order to calculate them
efficiently using the available tools for 
fast loop integration, 
it is convenient 
to work directly within the equivalent 
framework of IR resummed 
Eulerian perturbation theory. 
As a first step, we rewrite the model \eqref{eq:eft-field_real} as 
\be 
\delta_g^{\rm EFT}-
\sum_{\mathcal{O}_a={\tilde{\delta^2_1},\tilde{\mathcal{G}}_2,\tilde{\delta_1^3}}}\beta_a(k)\mathcal{O}^\perp_a=\beta_1(k)\tilde{\delta}_1\,.
\ee 
This representation is useful as 
the sum over non-linear operators 
in the l.h.s. above vanishes after 
correlating it with $\tilde \delta_1$.
Adding and subtracting $b_1\tilde{\delta}_1$ 
to the l.h.s. of the above equation, multiplying everything by 
$\tilde{\delta_1}(\k')$
and taking the expectation values we get 
\be 
\langle \delta_g^{\rm EFT}-b_1 \tilde{\delta_1}|\tilde\delta_1\rangle' +b_1 P_{\tilde 1 \tilde 1} = \beta_1(k) P_{\tilde 1 \tilde 1}\,,
\ee 
where $P_{\tilde 1 \tilde 1}(k) = \langle \tilde \delta_1(\k) \tilde \delta_1(\k)\rangle'$. 
Recall that primes denote stripping off 
the Dirac delta functions. 
The equation above is useful because 
its leftmost term starts with the second order in 
perturbation theory, i.e. at the linear level (equivalently $k\to 0$) we have $\beta_1(k)=b_1$. 
At the non-linear level, the leftmost term above 
is the cross-spectrum between $\tilde \delta_1$
and a new field $\Delta {\delta_g} $
defined as
\be 
\Delta {\delta_g} \equiv \delta_g^{\rm EFT} - b_1\tilde\delta_1\,.
\ee 

Both $P_{\tilde 1 \tilde 1}$
and $P_{\Delta g \tilde 1}\equiv \langle
\Delta\delta_g \tilde\delta_1
\rangle'$ can be computed in Eulerian perturbation theory. To this end,
we start with the Eulerian kernel expansion
for $\tilde\delta_1$,
\be 
\label{eq:d1_spt}
\tilde \delta_1 = \sum_{m=1}^3\left(
\prod_{n=1}^m
\int_{\k_n}\delta_1(\k_n)
\right)(2\pi)^3\delta_D^{(3)}(\k-\k_{1...n})\tilde K_n
\,,
\ee 
where we suppressed the explicit momentum dependence 
of $\tilde K_n$ above, and 
in real space
\be 
\label{eq:delta1_kern}
\begin{split}
&\tilde K_1 (\k) = 1\,,\\
& \tilde K_2(\k_1,\k_2) = 1+\frac{1}{2}\left(
\frac{(\k_2\cdot \k_1)}{k_1^2} + \frac{(\k_1\cdot \k_2)}{k_2^2} 
\right)\,,\\
&\tilde K_3 = \frac{1}{2} 
\frac{(\k_2\cdot \k)(\k_3\cdot \k)}{k_2^2k_3^2}~\,.
\end{split}
\ee 
$\tilde{K}_3$ above should be symmetrized w.r.t. 
individual momenta. 
Note that the kernels above 
have poles as one of their momentum arguments
becomes very soft. 
This is because by construction
they are produced by the Taylor 
expansion over the linear displacement. 
The auto spectrum of $\tilde{\delta_1}$ at one loop
order 
is given by
\be 
\label{eq:P11}
\begin{split}
& P_{\tilde{1}\tilde{1}} = P_{11}
+2\int_{\bm p} 
\tilde{K}_2^2(\k-\p,\p)P_{11}(|\k-\p|)P_{11}(p) \\
& + 6 \tilde{K}_1(\k)P_{11}(k)\int_\p 
\tilde{K}_3(\p,-\p,\k)P_{11}(p)\,,
\end{split}
\ee 
where $P_{11}$ is the linear matter 
power spectrum.

Using the kernels in eq.~\eqref{eq:delta1_kern}
we can obtain a series expansion for 
$\Delta{\delta_g} $ similar to 
eq.~\eqref{eq:d1_spt} but with the new kernels
\be 
\begin{split}
& \tilde F^b_1 = F^b_1 - b_1 \tilde K_1 = 0\,,\\
&\tilde F^b_2 = F^b_2 - b_1 \tilde K_2=\frac{b_2}{2}+\left(b_{\mathcal{G}_2}+\frac{2}{7}b_1\right)F_{\mathcal{G}_2}\,,\\
& \tilde F^b_3 = F^b_3 - b_1 \tilde K_3\,,
\end{split}
\ee 
where $F^b_n$ are usual Eulerian
bias kernels (i.e. the $Z_n$ kernels from~\cite{Ivanov:2019pdj} with $f\equiv 0$). 
Note that the 
Zel'dovich shifts
cancel in the 
new kernels 
$\tilde{F}^b_n$
by construction. 

Now one can 
calculate the cross-spectrum between 
$\tilde\delta_1$ and $\tilde{\delta_g}^{\rm EFT}$ in Eulerian
perturbation theory: 
\be 
\label{eq:P1g_rsd}
\begin{split}
& P_{\tilde 1 \Delta g}= 
 3 \tilde K_1(\k)P_{11}(k)\int_\p  \tilde F^b_3(\p,-\p,\k)P_{11}(p) \\
& +2\int_\p \tilde K_2(\k-\p,\p)\tilde F^b_2(\k-\p,\p)P_{11}(|\k-\p|)P_{11}(q) \,.
\end{split}
\ee 
Note that we need to supplement this expression with appropriate counterterms 
$P^{\rm ctr}_{\tilde 1 \Delta g}$, which will be discussed shortly. 
Loop
integrals in both
expressions \eqref{eq:P11} and 
\eqref{eq:P1g_rsd} can be evaluated with  
FFTLog.

As a last step, 
one can use the formalism of time-sliced
perturbation theory to 
perform IR resummation
directly at the level of the correlation
function $P_{\tilde{1}\Delta{g}}$
and $P_{\tilde{1}\tilde{1}}$. 
The details of this calculation are summarized 
in Appendix~\ref{sec:tspt}. The result of this calculation
is identical to the resummation of usual correlators
in Eulerian perturbation theory:
\be 
\begin{split}
P_{\tilde{1}X}=
& P_{\tilde{1}X}^{\rm tree}[e^{-k^2\Sigma^2}(1+k^2\Sigma^2)P_{ w} + P_{ nw}] \\
& +P_{\tilde{1}X}^{\rm 1-loop}[e^{-k^2\Sigma^2}P_{ w} + P_{ nw}]\,,
\end{split}
\ee 
where $X=\{\tilde{1},\Delta{g}\}$ and $P_w$
and $P_{nw}$ correspond to 
wiggled and de-wiggled parts 
of the linear power spectrum. 
The damping function is defined as
\be 
\Sigma^2 \equiv \frac{1}{6\pi^2}\int_0^{\Lambda_{\rm IR}}~dp P_{11}(p)[1-j_0(r_s p)+2j_2(r_s p)]\,,
\ee 
where $r_s$ is the position space BAO scale,
$j_\ell(x)$ are spherical Bessel functions, 
and $\Lambda_{\rm IR}$ is the IR separation
scale which we choose $\Lambda_{\rm IR}=0.2~\hMpc$ following~\cite{Blas:2016sfa}.

\subsection{Redshift-space model}

The redshift space model 
takes the following form, 
\be 
\label{eq:eft-field_rsd}
\begin{split}
&\delta^{\rm EFT}_g (\k,\hat{\bm z})= \delta_Z(\k,\hat{\bm z})-\frac{3}{7}\mu^2 f \tilde{\mathcal{G}_2}\\
&\beta_1(k,\mu)\tilde 
\delta_1(\k,\hat{\bm z})
+\beta_2(k,\mu)
\tilde{(\delta_1^2)}^\perp (\k,\hat{\bm z}) \\
& +\beta_{\mathcal{G}_2}(k,\mu)
\tilde{\mathcal{G}_2}^{\perp}(\k,\hat{\bm z})
+\beta_3(k,\mu)
\tilde{(\delta_1^3)}^\perp (\k,\hat{\bm z})\,,
\end{split}
\ee 
where $\mu$ is the cosine 
between the 
wavevector and the unit line-of-sight direction vector ${\bm z}$:
\be 
\mu = (\k\cdot \hat{\bm z})/k\,.
\ee 
One observes two new terms: 
the Zel'dovich field $\delta_Z$ and the term proportional to the tidal operator $\G$. 
These terms are needed to ensure that 
the low-$k$ limits of the transfer functions
do not depend on $\mu$
at the lowest order in perturbation theory.
On intermediate scales the transfer functions  explicitly 
depend on $\mu$.

\begin{figure*}
\centering
\includegraphics[width=0.99\textwidth]{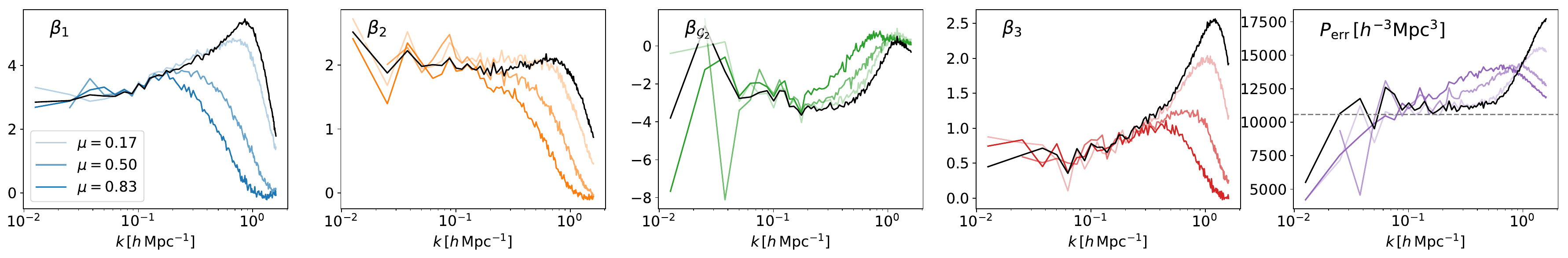}
   \caption{Typical forward model transfer functions 
   and noise power spectra
   for HOD galaxies 
   in real space (black lines)
   and in redshift space, for three $\mu$ bins.
   Note that we subtract 1 and 0.5 from the real space transfer functions for $\beta_1$ and $\beta_{\G}$
   to match the low-$k$
   limit of the redshift space transfer functions. Dashed line in the rightmost panel
   depicts the $\bar n^{-1}$
    Poisson prediction.
    } \label{fig:transfer}
\end{figure*}

Another important point is that the 
the error field in redshift space is
also 
$\mu$-dependent. Its power spectrum is given by
\be 
P_{\rm err}(k,\mu) = 
\langle 
|\delta^{\rm EFT}(\k,\hat{\bm z}) - \delta^{\rm HOD}(\k,\hat{\bm z})|^2
\rangle' \,.
\ee 
In EFT, the theoretical prediction for the error power spectrum at the first order in derivatives 
is given by~\cite{Perko:2016puo,Chudaykin:2020aoj} 
\be 
\label{eq:rsd_stoch}
P_{\rm err}(k,\mu) = \frac{1}{\bar n}\left(
\alpha_0+\alpha_1 \left(\frac{k}{k_{\rm S}}\right)^2
+\alpha_2\mu^2 \left(\frac{k}{k_{\rm S}}\right)^2
\right)\,,
 \ee 
where $k_S\sim k_{\rm NL}$, and $\alpha_{0,1,2}$ are EFT constants. In what follows we set $k_S=0.45~\hMpc$ following~\cite{Philcox:2021kcw}.
A typical set of transfer functions 
and $P_{\rm err}$ measurements from out fits
are shown in fig.~\ref{fig:transfer}.

\subsection{Redshift space transfer functions}

To measure EFT parameters, we need to understand the low-$k$ limit of the 
transfer functions. 
For $\beta_1(k,\mu)$, 
one can use the same strategy 
as in the real space case. 
We rewrite the model \eqref{eq:eft-field_rsd} as 
\be 
\begin{split}
& \beta_1(k,\mu)\tilde{\delta_1} = \\
&\delta_g^{\rm EFT}-\delta_Z+\frac{3}{7}f\mu^2 \tilde{\mathcal{G}}_2-
\sum_{\tilde{\mathcal{O}}_a={\tilde{\delta^2_1},\tilde{\mathcal{G}}_2,\tilde{\delta_1^3}}}\beta_a(k,\mu)\tilde{\mathcal{O}}^\perp_a \,.
\end{split}
\ee 
Then we add and subtract 
$(b_1-1)\tilde \delta_1$ from the r.h.s. above,
and correlate the resulting expression 
with $\tilde \delta_1$, which yields
\be
\label{eq:lowk_b1}
\begin{split}
& \beta_1(k,\mu) = b_1-1\\
& +\frac{1}{P_{\tilde 1\tilde 1}(k,\mu)} 
\langle \delta^{\rm EFT}_g -\delta_Z- (b_1-1)\tilde \delta_1  + \frac{3}{7}f\mu^2\tilde{\mathcal{G}}_2| \tilde{\delta_1} \rangle'\,.
\end{split}
\ee
We see that on large scales $\beta_1$
does not depend on $\mu$ and it is given 
by $b_1-1$. This simplicity 
was achieved by adding 
the Zel'dovich field to the model~\eqref{eq:eft-field_rsd}.
The second important observation is that 
the shape of $\beta_1$ in perturbation
theory is controlled by the cross 
spectrum of $\tilde \delta_1$
and 
\be 
\label{eq:deltag_rsd}
\Delta \delta_g = \delta^{\rm EFT}_g -\delta_Z- (b_1-1)\tilde \delta_1  + \frac{3}{7}f\mu^2\tilde{\mathcal{G}}_2
\,.
\ee 
This field by construction starts only at 
the quadratic order in $\delta_1$. 
Let us discuss now how to calculate
loop corrections in eq.~\eqref{eq:lowk_b1}.
We first calculate 
perturbative kernels 
for $\tilde\delta_1$
in redshift space:
\be 
\label{eq:delta1_kern_rsd}
\begin{split}
\tilde K_1 (\k) & = 1\,,\\
\tilde K_2(\k_1,\k_2) &  = 
\frac{\k\cdot \k_1}{2k_1^2} 
+ \frac{\k\cdot \k_2}{2k_2^2} 
\\
&+
\frac{(f\mu k )}{2}\left(\frac{k_{1z}}{k_1^2}+\frac{k_{2z}}{k_2^2}
\right)\,,\\
\tilde K_3 & = \frac{1}{2}\Bigg(
\frac{(\k_2\cdot \k )(\k_3\cdot \k)}{k_2^2k_3^2}+
f\frac{(\k_2\cdot\k)k_zk_{3z}}{k_2^2k_3^2}\\
&+
f\frac{(\k_3\cdot\k)k_zk_{2z}}{k_2^2k_3^2}
+ f^2\frac{k^2_zk_{2z}k_{3z}}{k_2^2k_3^2}\Bigg)~\,,
\end{split}
\ee 
where $k_{iz}=(\k_i \cdot\hat{\bm z})$,
and $\k\equiv \k_1+...\k_n$ for the n'th kernel. 
This allows us to compute the 
power spectrum 
$\tilde \delta_1 $
using the same expression~\eqref{eq:P11}
but with the 
above kernels. 
The perturbative kernels for the ``subtracted'' galaxy density ~\eqref{eq:deltag_rsd} 
are given by
\be 
\begin{split}
& \tilde Z_1 = Z_1 -F^Z_1 - (b_1-1)\tilde K_1 = 0\,,\\
&\tilde Z_2 = Z_2  -F^Z_2 - (b_1-1) \tilde K_2+\frac{3}{7}f\mu^2 F_{\mathcal{G}_2}\,,\\
& \tilde Z_3 = Z_3 - F^Z_3 - (b_1-1) \tilde K_3 \\
&~~~~~~~+\frac{3}{7}f\mu^2 F_{\mathcal{G}_2}(\k_1,\k_2)\left(\frac{(\k\cdot\k_3)}{k_3^2}+f\mu k\frac{k_{3z}}{k_3^2}\right) \,,
\end{split}
\ee 
where $Z_n$ are the usual 
redshift-space non-linear
kernles in Eulerian standard perturbation
theory~\cite{Bernardeau:2001qr,Ivanov:2019pdj}, and $F^Z_n$
are the Zel'dovich redshift space 
kernels in standard perturbation theory~\cite{Blas:2013bpa,Scoccimarro:1995if,Scoccimarro:1996se,Bernardeau:2001qr}. 
The cross-spectrum $P_{\tilde 1\Delta g}$
is given by 
\be 
\label{eq:P1g_rsd}
\begin{split}
& P_{\tilde 1\Delta g}= 
 3 \tilde K_1(\k)P_{11}(k)\int_\p  \tilde Z_3(\p,-\p,\k)P_{11}(p) \\
& +2\int_\p \tilde K_2(\k-\p,\p)\tilde Z_2(\k-\p,\p)P_{11}(|\k-\p|)P_{11}(q) \,.
\end{split}
\ee 
Note
again that we also need to add 
appropriate 
counterterms 
to renormalize the above loop
integrals. 
The loop integrals in eq.~\eqref{eq:P1g_rsd} 
are easy to compute with FFTLog
using the expansion over powers 
of $\mu^2$ developed in~\cite{Chudaykin:2020aoj}.

Finally, one can IR-resumm 
the infrared displacements 
using the redshift space version
of TSPT. The details are provided in 
Appendix~\ref{sec:tspt}. The final expression is given by
\be 
\label{eq:irres_rsd}
\begin{split}
P_{\tilde{1}X}=
& P_{\tilde{1}X}^{\rm tree}[e^{-\mathcal{S}(k,\mu)}(1+\mathcal{S}(k,\mu))P_{ w} + P_{nw}] \\
& +P_{\tilde{1}X}^{\rm 1-loop}[e^{-\mathcal{S}(k,\mu)}P_{ w} + P_{nw}]\,,
\end{split}
\ee 
where the redshift-space variant 
of the damping function is given by: 
\be 
\label{eq:damp_rsd}
\begin{split}
& \mathcal{S}(k,\mu) = k^2 \left(\Sigma^2(1+f\mu^2(2+f)) + \delta\Sigma^2f^2\mu^2(\mu^2-1) \right)\,,\\
&\delta \Sigma^2  = \int_0^{\Lambda_{IR}} \frac{dp}{2\pi^2} P_{11}(p) j_2(p r_s)\,.
\end{split}
\ee 
In practice, we have found that
the redshift-space corrections to the IR resummation formula for redshift-space 
one-loop spectra
are negligible 
for the purpose of fitting the transfer functions. Given that, our main results 
are obtained 
with approximate templates 
in which the loops are evaluated with the isotropic (real space)
exponential damping. 
If a better precision is required, 
it is straightforward to implement 
the templates with the full 
redshift-space corrections as in eq.~\eqref{eq:irres_rsd}.

As for the other transfer functions, 
one can obtain
that 
in the low-$k$
limit 
they are $\mu$-independent, 
and proportional 
to the real space ones, c.f.~eq.~\eqref{eq:lowk_betas},
\be 
\label{eq:lowk_betas2_rsd}
\begin{split}
& \beta_2(k,\mu) = \frac{b_2}{2}\,,\\
& \beta_{\mathcal{G}_2}(k,\mu) = b_{\mathcal{G}_2}
+\frac{2}{7}b_1
-\frac{1}{2}\,,\\
& \beta_{3}(k,\mu) = \frac{b_3}{6}\,.
\end{split}
\ee 
The corrections to these expressions
are both scale and $\mu$-dependent. 

\subsection{Counterterms}

In EFT, the leading order higher-derivative 
contributions 
to the galaxy density field 
in redshift space is given by~\cite{Senatore:2014vja,Perko:2016puo}
\be 
\label{eq:Loctr}
\delta_g^{\rm EFT}\Big|_{k^2}= (-b_1c_s+b'_{\nabla^2\delta}+f c_{s2}\mu^2 
+f^2 c_{s4}\mu^4 )k^2 \delta_1~\,,
\ee 
where $c_{s2},c_{s4}$ are redshift space
counterterms (Wilson coefficients). 
$c_{s4}$ above comes from the
renormalization of the velocity field. 
The equivalence principle
dictates that this counterterm 
be the same for galaxies and 
dark matter~\cite{Perko:2016puo}. $c_s$ is the real space 
dark matter counterterm. $b_{\nabla^2\delta}$
and $c_{s2}$ are higher-derivative bias 
and $\mu^2$ galaxy counterterms. 
These are expected to depend on the galaxy
population. Plugging this 
into $\Delta \delta_g$
and correlating 
with 
$\tilde{\delta_1}$
produces the following contribution 
to the transfer function $\beta_1$:
\be 
\label{eq:beta1k2}
\beta_1(k,\mu)\Big|_{k^2}
=(-b_1c_s+b'_{\nabla^2\delta}+f c_{s2}\mu^2 
+f^2 c_{s4}\mu^4 )k^2 ~\,.
\ee 
Let us discuss now how these counterterms are related to the 
actual physical observables. 
The leading order
counterterms in eq.~\eqref{eq:Loctr}
produce the following power 
spectrum contribution
\be 
\label{eq:Pctr_mu}
P^{\rm ctr}_{g} = 2(b_1+f\mu^2)(b_{\nabla^2\delta}+fc_{s2}\mu^2+f^2 c_{s4}\mu^4)P_{11}(k)\,.
\ee 
This can be re-arranged as:
\be 
\label{eq:Pctr_mu2}
P^{\rm ctr}_{g} = 2
\left(
\tilde{c}_0
+\tilde{c}_2f\mu^2 + 
\tilde{c}_4 f^2\mu^4
+ \tilde{c}_6 f^3 \mu^6
\right )k^2P_{11}\,,
\ee 
with 
\be 
\label{eq:ctilde}
\begin{split}
& \tilde{c}_0 = b_{\nabla^2\delta}b_1\,,\quad 
   \tilde{c}_2  = (b_1 c_{s2}+b_{\nabla^2\delta})\,,\\
  & \tilde{c}_4  = (b_1 c_{s4}+c_{s2})\,,\quad 
  \tilde{c}_6 = c_{s4}\,.
\end{split}
\ee 
The \texttt{CLASS-PT} code uses the following convention for the counterterms 
appearing in front of the 
redshift space multipoles, 
\be 
P^{\rm ctr}_\ell = - 2 c_{\ell} \frac{2\ell+1}{2}\int_{-1}^1 d\mu \mathcal{L}_\ell (\mu)f^{\ell/2}\mu^{\ell}\left( k^2 P_{11}\right)~\,.
\ee 
Comparing this with eq.~\eqref{eq:Pctr_mu2}
we get the following map between the field level coefficients 
and the counterterms used to fit the actual data:
\be 
\label{eq:classpt_conv_ctr}
\begin{split}
& c_0 = -\left(\tilde{c}_0+\frac{f}{3}\tilde{c}_2+\frac{f^2}{5}\tilde{c}_4+\frac{f^3}{7}\tilde{c}_6\right)\,,\\
& c_2 = -\left(\tilde{c}_2+\frac{6f}{7}\tilde{c}_4+\frac{5f^2}{7}\tilde{c}_6\right)\,,\\
& c_4 = -\left(\tilde{c}_4+\frac{15f}{11}\tilde{c}_6\right)\,,\\
& c_6 = - \tilde{c}_6\,,
\end{split}
\ee 
where $\tilde{c}_\ell$
are given in eq.~\eqref{eq:ctilde}.

Note that in redshift space the naive one-loop EFT model breaks down at relatively small
$k_{\rm max}\lesssim 0.1~\hMpc$ because 
the non-linear scale in redshift space
is lower than that in real space. 
The reach 
of the one-loop
model can be increased at the 
expense of introducing one 
additional
fitting parameter $b_4$ (also denoted as $\tilde c$)~\cite{Ivanov:2019pdj,
Chudaykin:2020hbf,
Ivanov:2021fbu,Taule:2023izt}.
It is a coefficient in front of 
the following higher order power 
spectrum counterterm
contribution
\be 
P_{\nabla^4_{\hat{\bm z}}\delta} =-b_4 k^4\mu^4 f^4 (b_1+f\mu^2)^2 P_{\rm 11}
\ee 
that can be used 
as a proxy
for two-loop effects. 
The presence of this term 
in the power spectrum model translates into 
an additional field level counterterm 
\be 
\delta_g^{\rm EFT}\Big|_{k^4}=-\frac{b_4}{2} k^4 \mu^4 f^4 (b_1+f\mu^2)\delta_1\,,
\ee 
which gives the following  correction 
to the transfer function:
\be 
\label{eq:beta1k4}
\beta_1(k,\mu)\Big|_{k^4}=
-\frac{b_4}{2} k^4 \mu^4 f^4 (b_1+f\mu^2)\,.
\ee

\subsection{Corrections for cutoff dependence}

The EFT parameters are scheme and cutoff dependent. 
The \texttt{CLASS-PT} code that we 
employ here 
uses dimensional 
regularization (dim. reg.) implemented via 
the FFTLog method. All converging 
loop
integrals are computed exactly, 
while the diverging ones are set 
to zero. 
For the one-loop power spectrum 
integrals that we encounter
in our calculation, the only 
divergent part 
is given by the mass variance
integral 
\be 
\label{eq:varL}
\sigma^2_\Lambda = \int_{|\k|\leq \Lambda}P_{11}(k)\,.
\ee 
One can show that the corresponding 
corrections to the transfer
functions are removed
at the field level 
after applying the 
orthogonalization~\cite{Abidi:2018eyd,Schmittfull:2018yuk}. 
Thus, we can ignore
corrections proportional 
to eq.~\eqref{eq:varL}
on both the power spectrum side
and the field level side. 
In practice, however, 
we measure EFT transfer functions
at a finite cutoff provided by
the grid resolution. Let us argue now, 
that even if this cutoff is formally present,
it does not require any additional adjustments
of the fitting procedure.

Let us discuss now the cutoff 
dependence of the transfer
functions that affects the matching 
of the converging loop
integrals.
On the field level side, 
our transfer function measurements 
are subject to grid resolution effects
that arise due to the clouds in cells (CIC)
interpolation. Although 
we apply the standard algorithm 
for the compensation of the CIC window, there is 
a residual bias  
that 
effectively
acts as a soft low pass filter 
whose characteristic scale 
is proportional to the Nyquist frequency
$k_{\rm Ny}=\pi N_{\rm mesh}/L_{\rm box}$~\cite{Hand:2017pqn}.
Thus, our 
EFT parameter measurements 
are taken, technically, 
at a finite cutoff $\Lambda \sim k_{\rm Ny}$.
To obtain their values at 
$\Lambda \to \infty$,
consistent with dim. reg., 
we need to add the missing 
UV pieces, i.e. to ``run''
the EFT constants up to infinity.

The first relevant transfer function
is $\beta_1$.
Let us start our discussion with the 
dark matter at the field level. 
In this case, the only 
relevant EFT parameters is the 
sound speed $c_s$, which renormalizes
the $k^2P_{11}$ contribution to the 
total power spectrum.
The physical coefficient 
in front of this term
is independent of the cutoff, 
i.e. 
\be 
\label{eq:cutoff1}
\begin{split}
& -2c_s(\Lambda) -\frac{61}{630\pi^2} \int_0^\Lambda dp~P_{11}(p)\\
& = -2c_s(\Lambda=\infty)-\frac{61}{630\pi^2} \int_0^\infty dp~P_{11}(p)~\,.
\end{split}
\ee 
where for the sake of the argument we approximated the low-pass 
filter as a sharp cutoff. 
$c_s(\Lambda=\infty)=c_s$ is the value from the fit of the power spectrum 
without any cutoff 
that reproduces the usual measurement
from the power spectrum with \texttt{CLASS-PT}. Eq.~\eqref{eq:cutoff1}
shows that if one were to fit  data 
with the one-loop integrals evaluated all the 
way to infinity, this would give 
$c_s(\Lambda=\infty)$
without the need to additionally 
``run'' it from $\Lambda$. 
This is true even if the actual 
fields have a cutoff, which 
is always the case in simulations.

The sound speed also enters
the $k^2$ correction 
to the $\beta_1(k)$
transfer function at the field
level. In order to fit it, 
we have to compute the shape 
of $\beta_1$ in 
perturbation theory. 
At the one loop 
order this calculation 
is identical to the usual 
power spectrum calculation~\cite{Schmidt:2018bkr}. 
It will give us the exact same 
expression 
for the cutoff-dependence 
of $c_s$ as in eq.~\eqref{eq:cutoff1}.
Hence, if one chooses
$\Lambda=\infty$ in the theory model 
for the transfer function,
this will produce the 
sound speed in the desired 
$\Lambda\to \infty$ limit. 
This argument is also true 
for bias tracers and in 
redshift space. 

All in all, we conclude 
that if one fits the $\beta_1$
transfer function
with perturbation theory shapes
computed without an explicit cutoff
in dim. reg., the results will be 
the same as if our fields did not 
have an explicit cutoff. This implies 
that we can ignore the cutoff altogether
in our PT calculations, and the 
EFT parameters extracted 
from $\beta_1$ should match those
of the $n$-point functions. 

As for the $\beta_2$
and $\beta_{\mathcal{G}_2}$
transfer functions, they must approach 
the dim. reg. values of 
$b_2/2$ and $b_{\mathcal{G}_2}$
on large scales thanks to the 
auto-spectrum of $\delta^2$, which 
will suppress the loop corrections 
on large scales~\cite{Schmittfull:2018yuk}. Hence, one can ignore
the cutoff dependence of these terms 
provided that the field-level cutoff
is reasonably small, such that the auto-spectrum of $\delta^2$
is saturated. This is the case for 
our baseline choice of the Nyquist frequency
$k_{\rm Ny}=1.6~\hMpc$
which we use both
for \texttt{Quijote}
and \texttt{Abacus}.

\subsection{Fitting procedure}

Let us discuss our fitting procedure 
to extract the EFT parameters
from the transfer functions 
and the error spectra. 
Let us start with the real 
space transfer functions 
$\beta_2,\beta_{\mathcal{G}_2},\beta_3$.
Based on eqs.~\eqref{eq:lowk_betas}
and~\eqref{eq:lowk_betas2_rsd}
we use the following 
template to fit them:
\be 
\label{eq:bias_fit}
\beta_{X} = a^X_0 + a^X_1 k^2 + a^X_2 k^4~\,,\quad 
X = \{\delta^2,\mathcal{G}_2,\delta^3\}\,.
\ee 
Then we take $a^X_0$
as a measurement of the 
corresponding bias parameter. 
Since by construction our
redshift space
transfer functions
do not bring 
additional information
on the nonlinear bias parameters,
we extract them from 
the real space measurements
only. It will 
be interesting to predict 
these transfer functions in
perturbation theory. 
We have found that the 
fit in eq.~\eqref{eq:bias_fit}
is stable up to $\kmax=0.4\hMpc$.
This is our baseline choice in
redshift space. 

As far as $b_1,b_{\Gamma_3}$
and $b_{\nabla^2\delta}$
are concerned, we fit them from 
the shape dependence of 
$\beta_1(k)$ in real space,
given by 
\be 
\beta_1(k) = b_1 +
\frac{P_{\tilde{1}\Delta g}(k)}{P_{\tilde{1}\tilde{1}}(k)}+b_{\nabla^2\delta}k^2~\,.
\ee 
$P_{\tilde{1}\Delta g}$ above depends
on $\{ b_1,b_{\mathcal{G}_2},b_2, b_{\Gamma_3} \}$.
In principle, 
one could fit these parameters from $\beta_1(k),\beta_2(k),\beta_{\G}(k)$ simultaneously, but in practice 
$b_2$ and $b_{\G}$ measurements are always dominated
by the corresponding transfer functions. 
For our main analysis, we fix $b_2$ and $b_{\G}$ 
to the transfer function best-fit values when
fitting $\b_1$, which gives us $\{b_1,b_{\Gamma_3},b_{\nabla^2\delta}\}$. 

The real space stochasticity EFT parameters are extracted 
by fitting $P_{\rm err}$ with the 
EFT prediction eq.~\eqref{eq:Perr_real}.
The redshift space $P_{\rm err}$ measurements give 
us the $\mu^2 k^2$ counterterm 
\eqref{eq:rsd_stoch}, which we fit from the 
$\mu^2$-dependence of the noise having fixed
$\alpha_0$ and $\alpha_1$ to their 
real space values, as required by the consistency 
of the EFT. Note that the non-linear corrections
are larger in redshift space due to virialized
motions of galaxies and dark matter particles, known as fingers-of-God~\cite{Jackson:2008yv}. 
For that reason, we use $\kmax=0.2~\hMpc$ 
in our fits to the redshift space transfer 
functions and $P_{\rm err}(k,\mu)$.
This choice is based on the observation
that the scaling of the stochastic noise
in redshift space starts deviating 
from the $k^2\mu^2$ behaviour for 
$k>0.2~\hMpc$.
We have also found that this choice minimized
the scale-dependence of $\alpha_2$, which 
otherwise becomes significant for 
$\kmax>0.2~\hMpc$. 
Details of this analysis
can be found in Appendix~\ref{sec:kmax}.
Note that our 
baseline value $\kmax=0.2~\hMpc$ for redshift-space 
is consistent
with the scale cut extracted from 
high fidelity simulations 
and the actual data~\cite{Nishimichi:2020tvu,Philcox:2021kcw}.

Finally, we fit the redshift space transfer 
function $\beta_1(k)$
in PT:
\be 
\beta_1(k,\mu) = b_1 -1 +
\frac{P_{\tilde{1}\Delta g}(k)}{P_{\tilde{1}\tilde{1}}(k)}+\beta_1\Big|_{k^2} + 
\beta_1\Big|_{k^4}\,,
\ee 
where $\b_1\Big|_{k^2,k^4}$
are the higher derivative counterterms 
defined in eq.~\eqref{eq:beta1k2}
and eq.~\eqref{eq:beta1k4}.
Then we convert the measured 
values of the redshift space counterterms 
to the counterterms used in \texttt{CLASS-PT}
with eq.~\eqref{eq:classpt_conv_ctr}.

Before we move on, let us note that
for our final HOD-based priors
we fit the universal counterterm $c_{s4}$
from dark matter
transfer functions in redshift space. 
In this case, we use $\kmax=0.15~\hMpc$
consistent with the EFT fits to 
redshift space dark matter 
from~\cite{Chudaykin:2020hbf,Taule:2023izt}. 
We have checked that raising it up 
to $\kmax=0.2~\hMpc$
does not have a significant impact 
on our results.

\section{Simulation-based priors}
\label{sec:priors}

\subsection{Cosmology dependence of EFT parameters for dark matter halos
and galaxies}

We start our investigation of the cosmology dependence of the EFT 
parameters with dark matter halos. 
This is a natural first step given that 
the cosmology-dependence of EFT parameters
for HOD galaxies is ``inherited'' 
from that of the underlying dark matter halos, up to corrections due to assembly bias. 
Indeed, for the linear
galaxy bias $b_{1,g}$ we have~\cite{Benson:1999mva,Desjacques:2016bnm}
\be 
\label{eq:linearbg}
b_{1,g}=\bar{n}_g^{-1}\int d\ln M  
\frac{d\bar{n}_h}{d\ln M}\langle N_c\rangle_M[1+\langle \mathcal{N}_s\rangle_M]b_1(M)\,,
\ee 
where $b_1(M)$
is the linear bias of halos, 
$\langle N_c\rangle_M,\langle \mathcal{N}_s =N_c N_s\rangle_M$
are expectation values for numbers of central and satellite 
galaxies in halos of mass $M$,
and $\bar n_{g}$
is the galaxy number density. We will
use $\langle N_g\rangle$ to denote
the total HOD, 
\be 
\langle N_g\rangle_M=\langle N_c\rangle_M[1+\langle \mathcal{N}_s\rangle_M]\,.
\ee 
The number density of halos of mass 
$M$, $d\bar n_{h}/d\ln M$,
or the halo mass function (HMF),
can be modeled analytically,
\be 
\label{eq:PSmf}
\begin{split}
& \frac{d \bar n_h}{d\ln M}=\frac{\bar\rho_m}{M}\left|\frac{d\ln\sigma_M}{d\ln M}\right|f\left(\frac{\delta_c}{\sigma_M}\right)\,,
\end{split}
\ee 
where the function $f(\nu)$ captures 
the fraction of the mass of the 
Universe locked up
in halos. Its simplest form is given by the Press-Schechter theory~\cite{Press:1973iz},
\be 
f_{\rm PS}(\nu) = \sqrt{\frac{2}{\pi}}\nu e^{-\nu^2/2}\,,
\ee 
where we introduced the ``peak height''
parameter 
\be 
\label{eq:peak_hei}
\nu\equiv \frac{\delta_c}{\sigma_M(z)}\,,
\ee 
where $\delta_c=1.686$ is the threshold overdensity,\footnote{In principle, $\delta_c$
depends on cosmology,
but this dependence is 
quite weak, $\lesssim 1\%$, see e.g.~\cite{Ivanov:2018lcg},
so this effect is negligible
for prior generation. } 
and $\sigma^2_M$ is 
the mass variance 
in spherical cells 
whose radius is equal to the Lagrangian
radius of the protohalo,
\be 
\label{eq:sigmaM}
\sigma^2_M(z) = \int_\k P_{11}(k,z)W^2_R(kR[M])\,,
\ee 
where $W_R(x)$ is the Fourier image of the position space spherically symmetric 
top-hat window function, and 
$R[M]=(3M/(4\pi \bar\rho_m)^{1/3}$
is the comoving Lagrangian radius,
with $\bar\rho_m$
standing for 
the background 
matter
density at redshift zero,
\be 
\bar\rho_m \equiv \rho_c \Omega_m \,,
\quad \rho_c=2.77\cdot 10^{11}h^2M_\odot \text{Mpc}^{-3}\,.
\ee 
Note that we measure distances and masses in units of $h^{-1}$Mpc and $h^{-1}M_\odot$, in which 
case 
the critical density above is constant 
for all \texttt{Quijote LH} catalogs.

While other parametrization exist 
for $f(\nu)$ in the literature~\cite{Sheth:1999mn,Tinker:2008ff}, 
the key property relevant for our 
discussion is that $f(\nu)$
depends only on the peak height,
i.e. it is universal to the changes of 
cosmology, redshift, and the halo mass. 
HMFs with this property are
called universal. Our arguments 
given below will apply to any 
universal HMF. 
Note that the universality of the HMF 
is a basic assumption
of the analytic halo models~\cite{Seljak:2000gq,Cooray:2002dia}.

Within the halo model, 
expressions analogous
to eq.~\eqref{eq:linearbg}
can be derived for other deterministic
EFT parameters as well. Namely, the  EFT parameters defined via\,,
\be 
\begin{split}
& \delta_h = \sum_{a=1} b_{\mathcal{O}_a}^h \mathcal{O}_a(\x)\,,\quad 
 \delta_g = \sum_{a=1} b_{\mathcal{O}_a}^g  \mathcal{O}_a(\x)\,,
\end{split}
\ee  
where $\mathcal{O}_a=\{\delta,\frac{1}{2}\delta^2,\G,...\}$, satisfy 
\be 
\label{eq:bg_via_bh}
b_{\mathcal{O}_a}^g= \bar n_g^{-1}
\int d\ln M 
\frac{d \bar n_h}{d\ln M}
\langle N_g\rangle_M 
b_{\mathcal{O}_a}^h(M)~\,.
\ee
The universality of the HMF suggests 
that
the EFT parameters of halos 
should depend on mass and 
cosmology via the peak height
parameter in eq.~\eqref{eq:peak_hei},
\be
\label{eq:bh_via_nu}
b_{\mathcal{O}_a}^h(M) = b_{\mathcal{O}_a}^h(\nu[M])\,.
\ee 
Analytic results 
for $b_1(\nu)$ and $b_2(\nu)$
following from the HMF are available
in the literature~\cite{Lazeyras:2015lgp,Desjacques:2016bnm}. 
Under the assumption 
of the HMF universality
the stochastic EFT parameters
are also expected to primarily 
depend on cosmology via 
the peak height, as follows
from halo exclusion models~\cite{Baldauf:2013hka}.
Eq.~\eqref{eq:bh_via_nu}
suggests that the cosmology-dependence
of the EFT parameters of halos
should be degenerate with the halo mass.
Let us investigate to what extent this is 
supported by direct 
measurements. 

\begin{figure*}[ht!]
\centering
\includegraphics[width=1.00\textwidth]{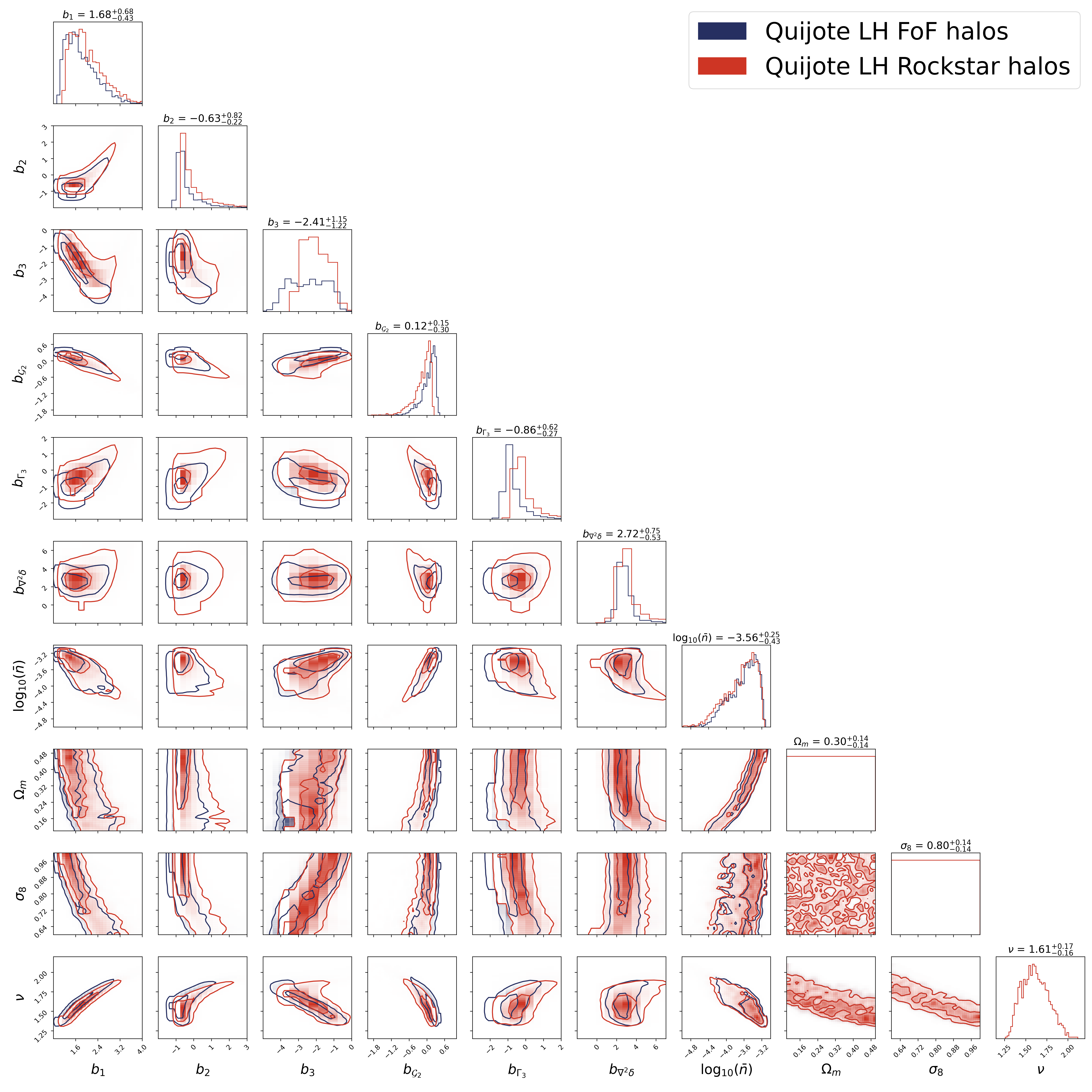}
   \caption{The joint distribution of EFT parameters and number densities of dark matter halos with masses in range $\log_{10} (M /[M_\odot/h]) =13-13.5$
   from 2000 \texttt{Quijote LH} 
   simulations at $z=0.5$.
   We show results 
   of both FoF
   and Rockstar halo finders. 
   For cosmological parameters,
   we only show the correlation
   with $\sigma_8$ and $\Omega_m$,
   and the halo peak heigh $\nu$.
   The correlation with other cosmological parameters
   varied in Quijote LH mocks, 
   ($n_s,h,\Omega_b$)
   is negligible. 
Density levels correspond to two-dimensional $1$-$\sigma$
  and $2$-$\sigma$ intervals (i.e. 39.3\% and 86.5\% of samples). 
    } \label{fig:dist_QH}
\end{figure*}

\begin{figure*}[ht!]
\centering
\includegraphics[width=1.00\textwidth]{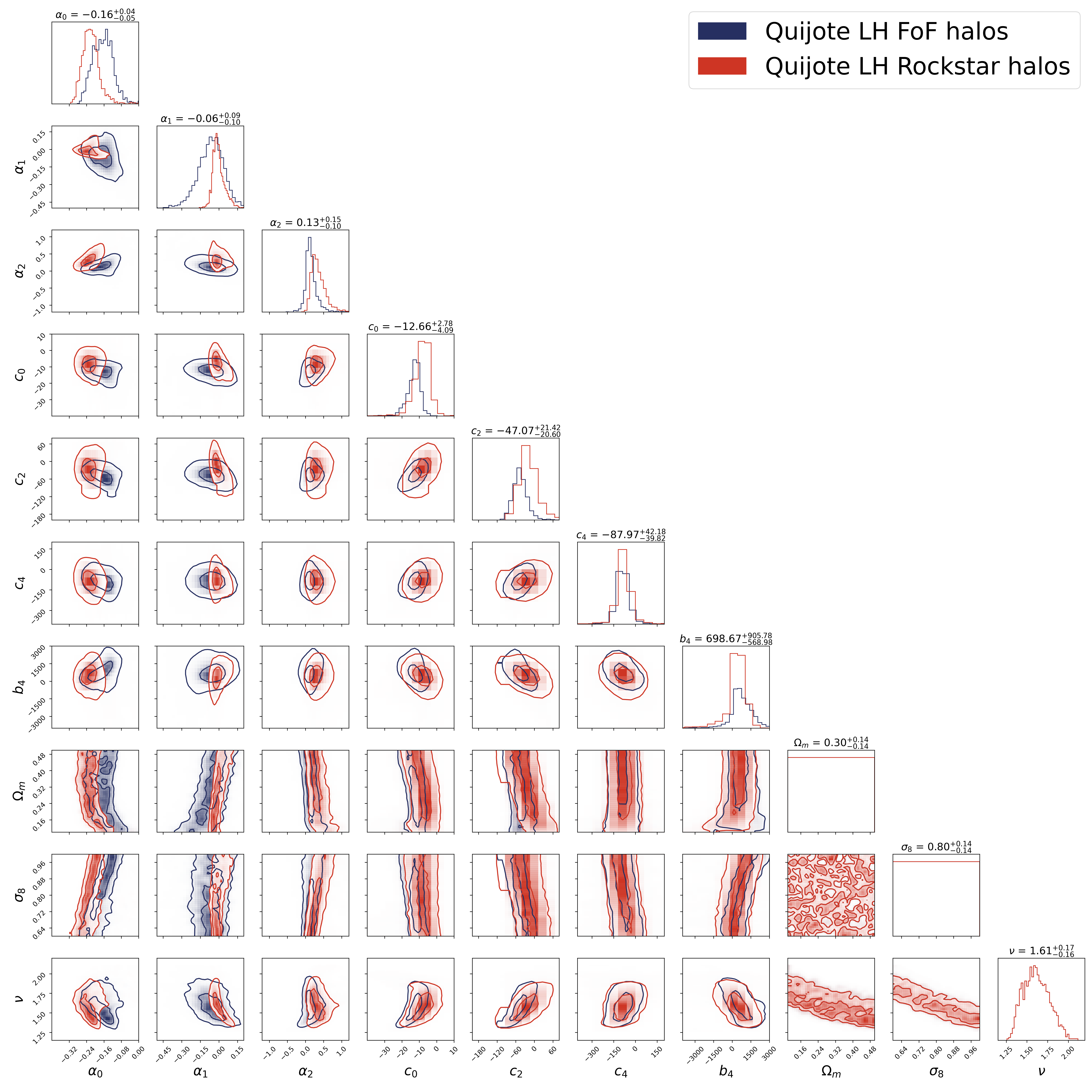}
   \caption{Same as fig.~\ref{fig:dist_QH}
   but for the stochastic 
   EFT parameters and redshift-space
   counterterms. The number density
   is omitted. 
    } \label{fig:dist_QH_ctr}
\end{figure*}

We perform EFT-based field-level fits 
of snapshots from \texttt{Quijote LH} catalogs that cover 2000
different cosmological models. 
Due to the limited mass resolution of \texttt{Quijote}, 
we focus on the
halo mass range  
$\log_{10} (M/[M_\odot/h]) =13-13.5$
for all catalogs at our baseline 
redshift $z=0.5$.
We produce two sets of EFT parameters
for friends-of-friends (FoF) and Rockstar
halo-finders. In principle, 
one should treat the halo finder
as a parameter of our UV model 
during the production of priors. 
We will comment on it in
detail shortly. 
Our results are shown in figs.~\ref{fig:dist_QH},~\ref{fig:dist_QH_ctr}.

First, we see that the bias
parameters and real space  
counterterms 
for FoF and Rockstar
are quite consistent. 
There are differences only at the level of 
the stochasticity parameters $\alpha_0$
and $\alpha_1$.
In addition, the RSD counterterms are
also somewhat different, which
is expected given differences in the 
velocity assignments between the two 
halo finders.

As far as cosmology dependence is concerned, 
we see a noticeable correlation
only with $(\Omega_m,\sigma_8)$
parameters.
Another important observation is that EFT 
parameter
samples with a fixed halo mass and 
varied cosmology-dependence that we study here
are quite similar 
to EFT 
parameter samples for a fixed cosmology
and varied halo masses, 
in line with the above expectations
based on the halo model. 
To illustrate this better, we measured
EFT parameters from FoF halo catalogs 
based on the \texttt{Quijote} fiducial 
simulation. The (overlapping) halo mass bins we consider are 
\be 
\begin{split}
& \log_{10} (M /[h^{-1}M_\odot])\in  [12,12.5]\,,
[12.5,13]\,,\\
& [13.2,13.8]\,, [13.4, 14]\,,  [13,13.5]\,, [13.5, 14]\,,  \\
&
[13.6, 14.2] \,,
[13.8 , 14.4]\,,
[14,14.5]\,.
\end{split}
\ee 
These results for the bias parameters $b_1,b_2,b_{\mathcal{G}_2},b_{\Gamma_3}$
and $b_{\nabla^2 \delta}$ 
and the corresponding peak
heights $\nu$ are shown
in fig.~\ref{fig:cosmo_vs_mhalo}. 
The picture is similar for
other EFT parameters, 
see Appendix~\ref{app:plots}.
In addition, we plot the 
EFT parameters for the same
halo mass bins from the baseline 
\texttt{Abacus}
simulation.\footnote{We have added two additional bins, $\log_{10} (M /[h^{-1}M_\odot])\in  [11,11.5]\,,
[11.5,12]$, which we could not extract from \texttt{Quijote} because of the limited resolution. } This allows us to 
bracket the uncertainty in the EFT 
parameters due to the use of the 
CompaSO halo finder~\cite{Hadzhiyska:2021zbd} (discussed shortly), which we find
to be comparable to that
of the FoF-Rockstar difference. 

Fig.~\ref{fig:cosmo_vs_mhalo} 
demonstrates that the variation of the halo mass
at a fixed cosmology acts like 
a variation of cosmology for a given halo mass. 
This confirms
the expectation 
that the cosmology dependence of the samples,
to a large extent, is degenerate with the
halo mass, as predicted
within the halo model. 
Comparing the \texttt{Quijote}
and \texttt{Abacus}
halo results we note 
that the exact dependence
of bias parameters on the peak height is halo-finder dependent. Nevertheless the functions such as $b_2(b_1)$ appear to be the virtually the same for 
the galaxy bias parameters. We do find some differences 
at the level of $b_{\nabla^2\delta}$
and other counterterms (see Appendix~\ref{app:plots} for more detail), but 
these differences are much smaller than the width of the
HOD distributions for these parameters.


In figs.~\ref{fig:dist_QH},~\ref{fig:dist_QH_ctr},  
we display the correlation between bias parameters 
and $\sigma_M(z=0.5)$ for $M=10^{13}M_\odot /h$.
First, we see that $\nu$ is 
naturally
highly 
correlated with $\Omega_m$ and $\sigma_8$.
Second, it is strongly correlated 
with the EFT parameters. In fact, it appears to be more correlated with the halo parameters than 
$\sigma_8$ and $\Omega_m$ themselves.
Note that the halo number
density is additionally correlated
with $\Omega_m$ through
the normalization in eq.~\eqref{eq:PSmf}.

All in all, 
our measurements confirm
the analytical argument that 
the cosmology-dependence of the EFT parameters should be 
approximately captured by
the peak height $\nu$
and hence is degenerate 
with the halo mass.
Hence, to the first
approximation, 
one can simply fix cosmology and vary the 
halo mass in order to generate the 
samples for EFT parameters.

Having established that
EFT parameters of halos
are approximately
cosmology-independent, 
let us now discuss
the 
implications
for the cosmology-dependence
of the HOD galaxies. 
Eq.~\eqref{eq:bg_via_bh}
can be rewritten as 
\be 
b_{\mathcal{O}_a}^g= 
\bar n_g^{-1}
\int d\nu \frac{d\bar n_h(\nu)}{d\nu}  
\langle N_g\rangle_{M[\nu]} 
b_{\mathcal{O}_a}^h(\nu)~\,.
\ee 
From this expression it is 
clear that the EFT parameters
of galaxies
can acquire 
cosmology-dependence 
only through the mapping
$\nu[M]$. 
We argue now
that for the standard HOD
functions such as~\cite{Zheng:2004id},
the bulk of 
this cosmology-dependence
can be absorbed by re-defining 
the HOD parameters themselves.

In the basic HOD models 
for galaxies, we integrate over
the halo mass distribution a
with the HOD weights
\be 
\label{eq:zheng_hod}
\begin{split}
& \langle N_c\rangle_M =\frac{1}{2}\left[1+\text{Erf}\left(\frac{\log_{10} M - \log_{10} M_{\rm cut}}{\sqrt{2}\sigma}\right)\right]~\,,\\
& \langle \mathcal{N}_s\rangle_M = \Theta_H(M-\kappa M_{\rm cut})\left(\frac{M-\kappa M_{\rm cut}}{M_1}\right)^\alpha\,,
\end{split}
\ee 
where $\Theta_H(x)$
is the Heaviside step function,
and 
$M_{\rm cut}$, $\sigma$, $M_1$, $\kappa$ and $\alpha$
are free parameters. 
The above HOD saturates
the relevant bias parameter 
integrals 
at $M_{\rm cut}$, which effectively 
acts as a cutoff that determines the
lower 
range of halos hosting
the relevant galaxies. 
As we vary this parameter, 
we effectively vary 
the mass of the host halo.
The latter, as we showed,
is equivalent to varying cosmology.
Hence, the variation of the halo mass
captures the bulk of 
cosmology dependence.

To make this argument more 
precise, we explicitly 
show in Appendix~\ref{app:HOD_analytic}
that the 
cosmology dependence of EFT 
parameters of HOD galaxies can be 
almost
entirely absorbed
by re-defining the HOD parameters.
We prove this analytically 
for a power-law cosmology
with the linear matter power spectrum
$P_{11}\propto k^{n}$, noting that
our conclusions in fact apply to
more realistic dependencies. 
Interestingly, 
we find that the satellite HOD in eq.~\eqref{eq:zheng_hod} 
introduces
a mild cosmology
dependence on the slope of the 
linear matter power spectrum $n$ when $M\sim \kappa M_{\rm cut}$. 
However, the dependence 
on the mass
fluctuation amplitude at the redshift 
of the sample $\sigma_8(z)$
can be absorbed by the HOD parameters exactly even for a realistic 
$\Lambda$CDM cosmology. 
For instance, 
a $\approx 15\%$ shift
in $\sigma_8$, similar to the difference between
the $\sigma_8$ of \texttt{Abacus}
and our best-fit from BOSS is equivalent 
to a shift of $\Delta \log_{10}M_{\rm cut}\approx 0.3$.
In what follows, we use sufficiently 
wide ranges
of HOD priors to account for such shifts.

In addition, variations of 
other HOD parameters beyond the basic ones,
e.g. assembly bias, 
will generate an extra scatter 
around the halo relations between the 
EFT parameters. As can be seen from comparing 
figs.~\ref{fig:dist_QH},~\ref{fig:dist_QH_ctr} for halos and Fig.~2
of~\cite{Ivanov:2024hgq},
this scatter is stronger than
the variation of cosmology itself, at least for the 
real space parameters. 
We will see shortly that this is also the case 
for the
redshift space counterterms.

Given these arguments, 
we will ignore the explicit 
cosmology dependence in what follows. 
We will generate EFT distributions 
by varying HOD parameters at 
a fixed cosmology. We choose this
cosmology to be the \texttt{Abacus} fiducial 
cosmology consistent with the 
\textit{Planck} 2018
best-fit $\Lambda$CDM model.

Before closing this section, let us note that
the statement about cosmology-independence
of EFT parameters is approximate. It relies
on several crucial approximations. 
First, we have assumed the universality 
of HMF. This approximation 
is accurate to $\lesssim 5\%$~\cite{Li:2024wco}.
Second, 
based the halo model arguments, 
we have assumed that the EFT parameters
of halos depend only on the peak height, which 
is consistent with our \texttt{Quijote}
measurements, but is still a simplification. 
Third, our proof of the degeneracy between
cosmology and the HOD parameters was carried out
for the power-law cosmology. 
Although our proof of the 
cosmology-independence is 
formally exact for the mass 
fluctuation amplitude $\sigma_8(z)$,
it remains approximate 
for the parameters controlling 
the effective slope
of the linear matter power 
spectrum.
Fourth, our proof relied on the standard HOD 
parametrization and did not include
velocity bias, assembly bias,
or other parameters of the ``decorated''
HOD 
extensions.  
In general, one may expect that departures
from these approximations generate 
a residual cosmology dependence 
of the EFT parameters.
However, our explicit tests
of the residual cosmology-dependence in
Appendix~\ref{sec:abacus_large}
suggest that 
it is negligible for the purpose 
of generating the ensemble of EFT parameters. 
Thus, our semi-analytic
arguments 
provide a basic 
explanation for the 
approximate cosmology-independence
suggested by simulations.

We note however that the residual 
dependence of EFT parameters 
may be important if one wants to 
systematically connect small and large scales 
via a conditional distribution
between the EFT and cosmological 
parameters~\cite{Modi:2023drt,Ivanov:2024hgq}. This task is beyond the scope 
of our paper, which primarily focuses on 
producing priors distributions for EFT parameters.

\begin{figure*}[ht!]
\centering
\includegraphics[width=0.99\textwidth]{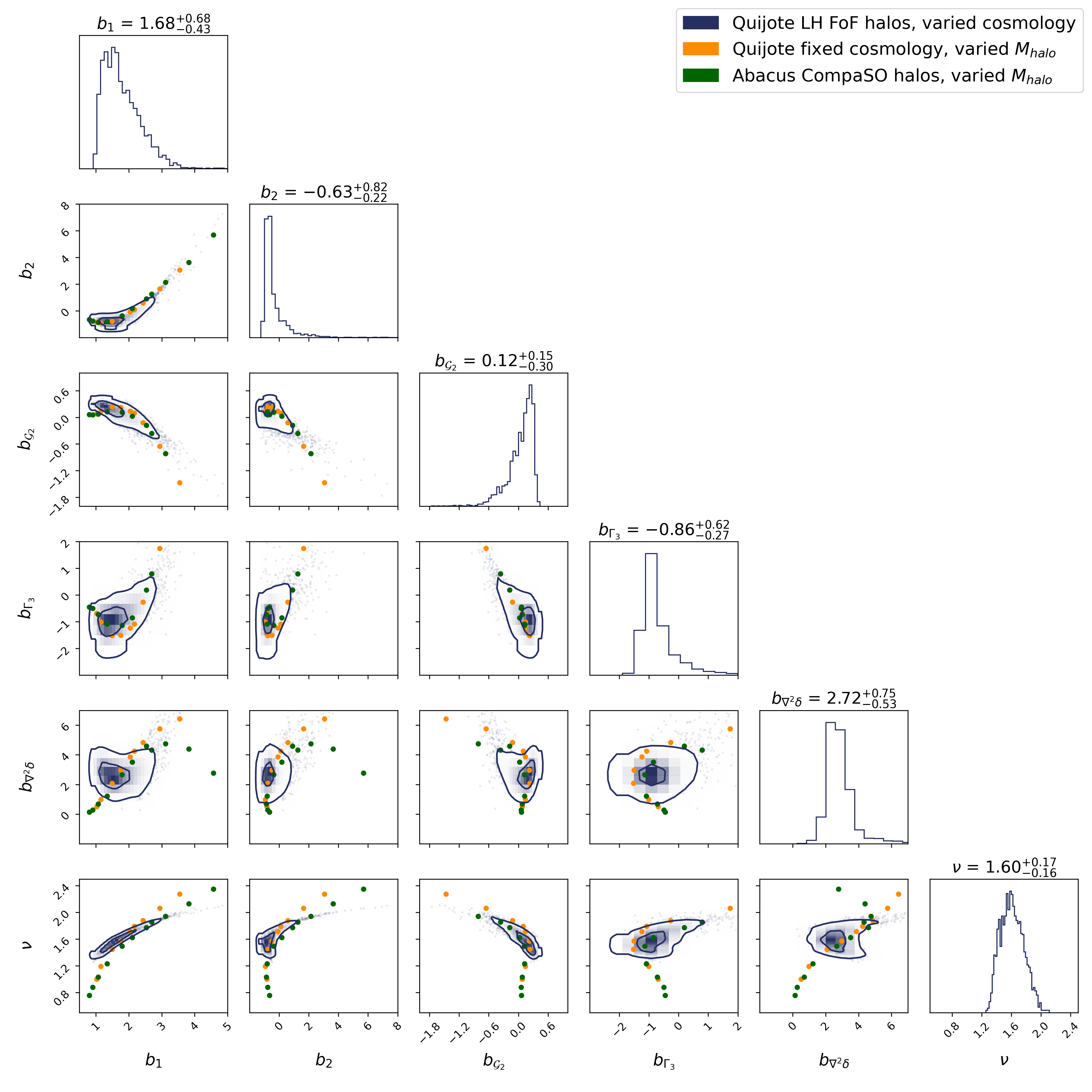}
   \caption{The distribution of EFT parameters from
   the FoF halo catalogs of the \texttt{Quijote} fiducial cosmology simulation 
   at $z=0.5$ versus the \texttt{Quijote} halo samples from Fig.~\ref{fig:dist_QH} with fixed halo mass, but varied cosmology. In addition, we also show
   results for \texttt{Abacus} halos
   at fixed cosmology. 
    } \label{fig:cosmo_vs_mhalo}
\end{figure*}

\subsection{HOD-based priors}

Having shown that the cosmology dependence 
of EFT parameters should have a negligible effect 
in the context of HOD models, we now 
generate HOD mock galaxy catalogs based 
on the \texttt{AbacusSummit} simulation (referred to as  \texttt{Abacus} for compactness in what follows) and extract their 
parameters at the field level.  We use the HOD models and code described in ~\citep{Yuan:2021izi}.

We generate 10500 HOD galaxy snapshots 
at $z=0.5$ using the fiducial cosmology of 
\texttt{Abacus}.
Each catalog is characterized by a set of HOD 
parameters which we randomly sample from
the following flat distribution:
\be 
\begin{split}
& 
\log_{10} M_{\rm cut}\in [12,14]\,,\quad 
\log_{10} M_1\in [13,15]\,,\\
& \log \sigma \in [-3.5,1.0]\,,\quad  \alpha \in [0.5,1.5]\,,\quad \\
& \alpha_c \in [0,1]\,,
\quad 
\alpha_s \in [0,2]\,,
\quad s\in [0,1]\,,
\quad  \kappa \in [0.0,1.5]\,,\\
& A_{\rm cen} \in [-1,1]\,,\quad 
A_{\rm sat} \in  [-1,1]\,,\\
& B_{\rm cen} \in [-1,1]\,,\quad 
B_{\rm sat} \in  [-1,1]\,.
\end{split}
\ee 
We use the spherical overdensity  CompaSO halo finder 
to identify dark matter halos from our
snapshots. 
We have found that
the CompaSO halo finder
produces results 
very similar to 
the other halo finders, see Appendix~\ref{app:plots}. 
In general, the 
variance between different  
halo finders is much smaller than the width of the HOD 
priors for galaxies.
This implies
that the choice of the halo
finder does not produce a significant 
uncertainty 
during prior
generation.

\begin{figure*}[ht!]
\centering
\includegraphics[width=0.99\textwidth]{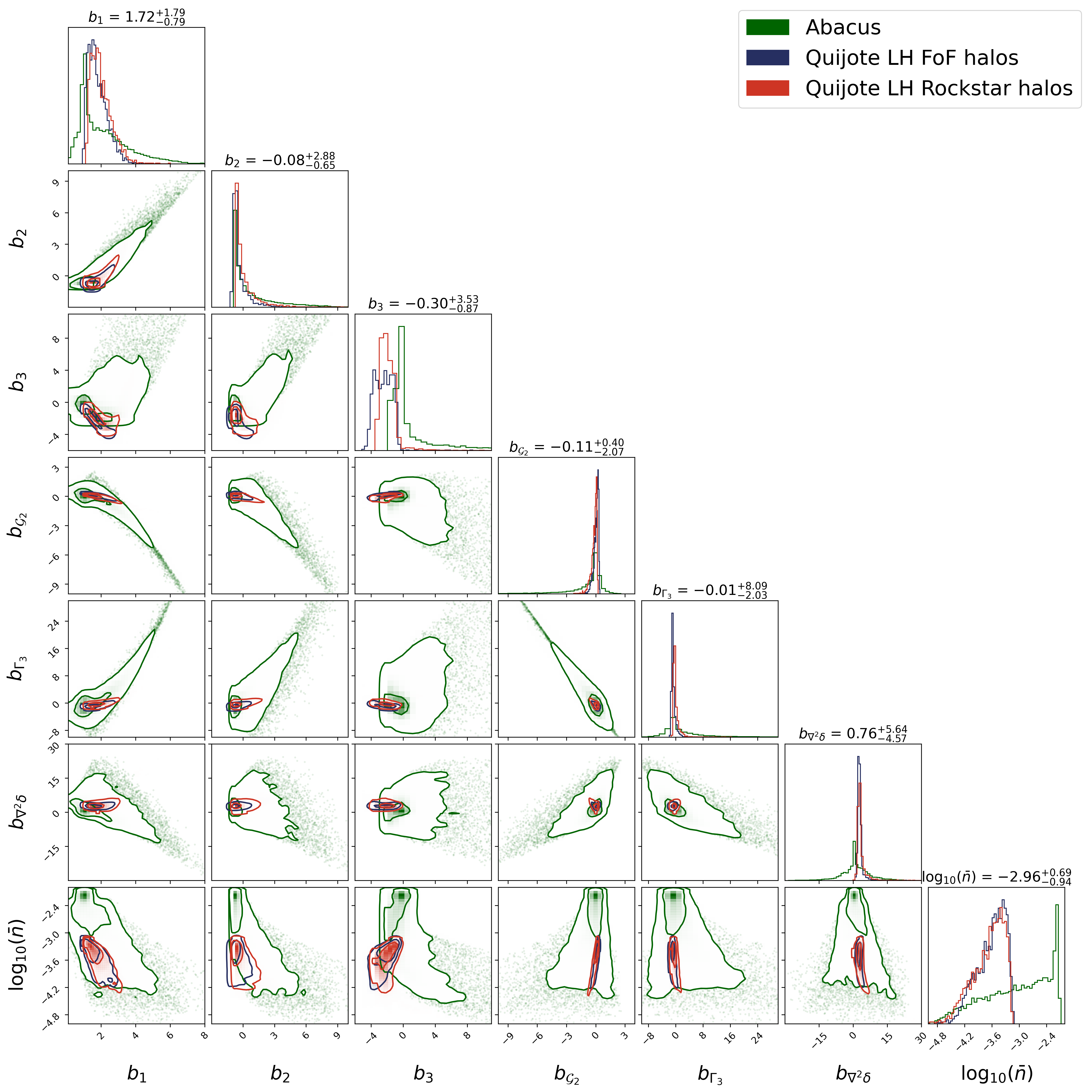}
   \caption{The distribution of EFT
   galaxy bias parameters and number densities of 10500 HOD galaxy models based on the Abacus fiducial cosmology simulation $z=0.5$ versus the halo samples from Fig.~\ref{fig:dist_QH}. 
    } \label{fig:dist_abacus_eft}
\end{figure*}

\begin{figure*}[ht!]
\centering
\includegraphics[width=0.99\textwidth]{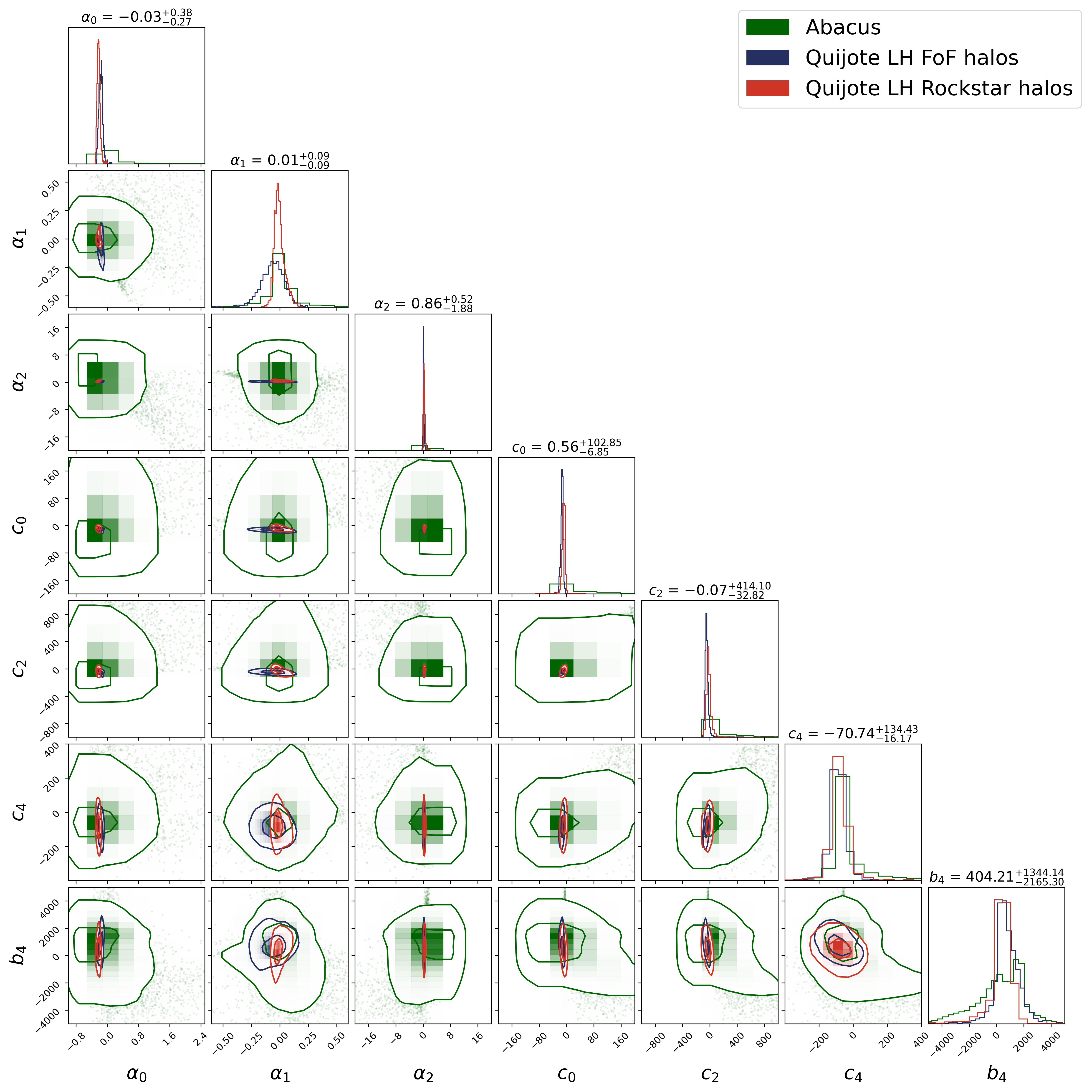}
   \caption{The distribution of EFT
  stochasticity 
  and counterterm parameters of 10500 HOD galaxy models based on the Abacus fiducial cosmology simulation $z=0.5$ versus the halo samples from Fig.~\ref{fig:dist_QH}. 
    } \label{fig:dist_abacus_eft_2}
\end{figure*}

The EFT parameters extracted 
from these samples are shown in 
figs.~\ref{fig:dist_abacus_eft},~\ref{fig:dist_abacus_eft_2},
along with the \texttt{Quijote} halo samples
from figs.~\ref{fig:dist_QH},~\ref{fig:dist_QH_ctr}.
The first relevant 
observation is that 
the HOD densities are much wider
than the halo ones. This confirms our
expectation that the variation of HOD 
parameters is much more important 
for distribution of EFT parameters 
than the variation of cosmology. 
Note that the tails of the halo distributions 
for certain parameters, 
see e.g. the upper part of the $b_{3}-b_1$ plane, ``leak'' outside the regions 
covered by the HOD densities. 
These ``leaks'' can be explained
by the fact that the HOD bias parameters are given
by the integrals over the halo biased weighted with 
the HOD and HMF, see eq.~\eqref{eq:bg_via_bh}. 
This weighting effectively shifts the 
bias parameters of galaxies away from the halo values~\cite{Ivanov:2024dgv}.

Also note that unlike our previous work~\cite{Ivanov:2024hgq}, here we do not
condition our samples on the number density
to match BOSS. 
We do so in order for our priors to be applicable to both BOSS and DESI. This also allows us 
to produce a wider, more conservative,
distribution of EFT parameters.
The 
histogram with the number densities 
of our samples can be seen in the lower panel of fig.~\ref{fig:dist_abacus_eft}. The total range covered is [$8.3\cdot 10^{-3},1.1\cdot 10^{-5}$] [$h$Mpc$^{-1}$]$^3$. The median of the distribution 
is $1.1\cdot 10^{-3}$ [$h$Mpc$^{-1}$]$^3$, 
while the
16th and 84th percentiles are 
$5.4\cdot 10^{-3}$ [$h$Mpc$^{-1}$]$^3$ and 
$1.3\cdot 10^{-4}$ [$h$Mpc$^{-1}$]$^3$, enclosing both 
the typical number densities of 
BOSS LRGs, 
$\bar n_g\approx 3.5 \cdot 10^{-4}$ [$h$Mpc$^{-1}$]$^3$~\cite{Reid:2015gra}
and DESI LRGs, $\bar n_g\approx 5\cdot 10^{-4}$ [$h$Mpc$^{-1}$]$^3$~\cite{DESI:2024mwx}.

As an additional test, we produced several 
HOD catalogs where we vary both 
HOD and cosmological parameters. To that 
end we select two cosmologies with 
larger and smaller values of $\sigma_8$
from the large \texttt{Abacus} boxes.
The analysis of large boxes 
is significantly more expensive than the baseline \texttt{Abacus} mocks, so we use the large boxes with 
varied cosmology only to test 
our baseline distribution. 
The details of this test are given in Appendix~\ref{sec:abacus_large}.
This analysis confirms 
that variations of cosmology are
negligible compared to the variations
of HOD parameters.

Before closing this section, let us discuss some interesting correlations between
the EFT parameters. 
While the dependencies
of non-linear bias
parameters on the linear bias, e.g. $b_2(b_1)$, 
trace the halo ones very well,
for stochasticity 
parameters 
we see some interesting 
correlations that 
appear specific to galaxies. 

For halos, 
the stochastic counterterm 
$\alpha_0$ is quite close
to zero, see fig.~\ref{fig:dist_QH_ctr},
with $\sim 10\%$ deviations consistent
with the halo exclusion estimates~\cite{Casas-Miranda:2001dwz,Baldauf:2013hka}. 
For galaxies, however, $\alpha_0$
can be as large as $\mathcal{O}(10)$. 
This is because galaxies trace dark matter
halos, and hence their $P_{\rm err}$
is close to that
of the halos on large scales~\cite{Baldauf:2013hka},
which can be much larger than $\bar n_g^{-1}$.
Indeed, in agreement with~\cite{Baldauf:2013hka},
we found that large $\alpha_0$
are specific to HODs with large $M_{\rm cut}$
and low $M_1$, which correspond to 
samples with a large number of satellites
living in massive halos. In this case 
it is natural to have the hierarchy
$P_{\rm err} \sim \bar{n}^{-1}_h \gg \bar{n}^{-1}_g$, 
resulting in large $\alpha_0$.

As far as the stochastic counterterm $\alpha_1$
is concerned, its behavior
is also consistent with~\cite{Baldauf:2013hka},
and it can be linked with the 
exclusion effects. 
In particular, $\alpha_1$ is typically
positive for samples with a low 
satellite fraction. In this case
the scale-dependence associated with $\alpha_1$ 
describes the increase of 
$P_{\rm err}\simeq \bar{n}_{h}^{-1}$ on large scales up to $\bar{n}_{g}^{-1}$
on small scales.
$\alpha_1$ can also be negative for samples with a large satellite 
fraction. 
This is because on large scales 
$P_{\rm err}\sim \bar{n}^{-1}_h$ (the twiddle accounts for corrections due to exclusion),
while on small scales 
$P_{\rm err}\sim \bar{n}^{-1}_g \ll \bar{n}^{-1}_h$,
implying the 
reduction of stochasticity as one increases $k$.

Quite interestingly, we have found 
that in redshift space, the 
restoration of the $\bar{n}_g^{-1}$
behavior of the noise power spectrum 
on small scales 
happens faster for the modes along the line 
of sight.
The velocity field ``assists'' to recover
the high$-k$
behavior quicker 
by means of the 
redshift-space 
stochastic 
counterterm $\alpha_2$. 
Specifically, samples with small/large satellite 
fraction tend to have positive/negative $\alpha_2$,
needed to reach the
expected $\bar{n}_g^{-1}$
small scale behavior. 
In particular, one can see that
the negative tail of  $\alpha_2$ in fig.~\ref{fig:dist_abacus_eft_2} is strongly 
correlated with the super-Poisson
stochasticity, $\alpha_0 \gg 1$.
It will be interesting to develop
a better physical understanding of
the peculiar behavior 
of $\alpha_2$ in the context of the halo model
along the lines of~\cite{Baldauf:2013hka,Maus:2024dzi}.
We additionally study 
a 
representative extreme outlier from this regime in 
Appendix~\ref{app:plots}.

\subsection{Including information from 
dark matter}

\begin{figure*}[ht!]
\centering
\includegraphics[width=0.7\textwidth]{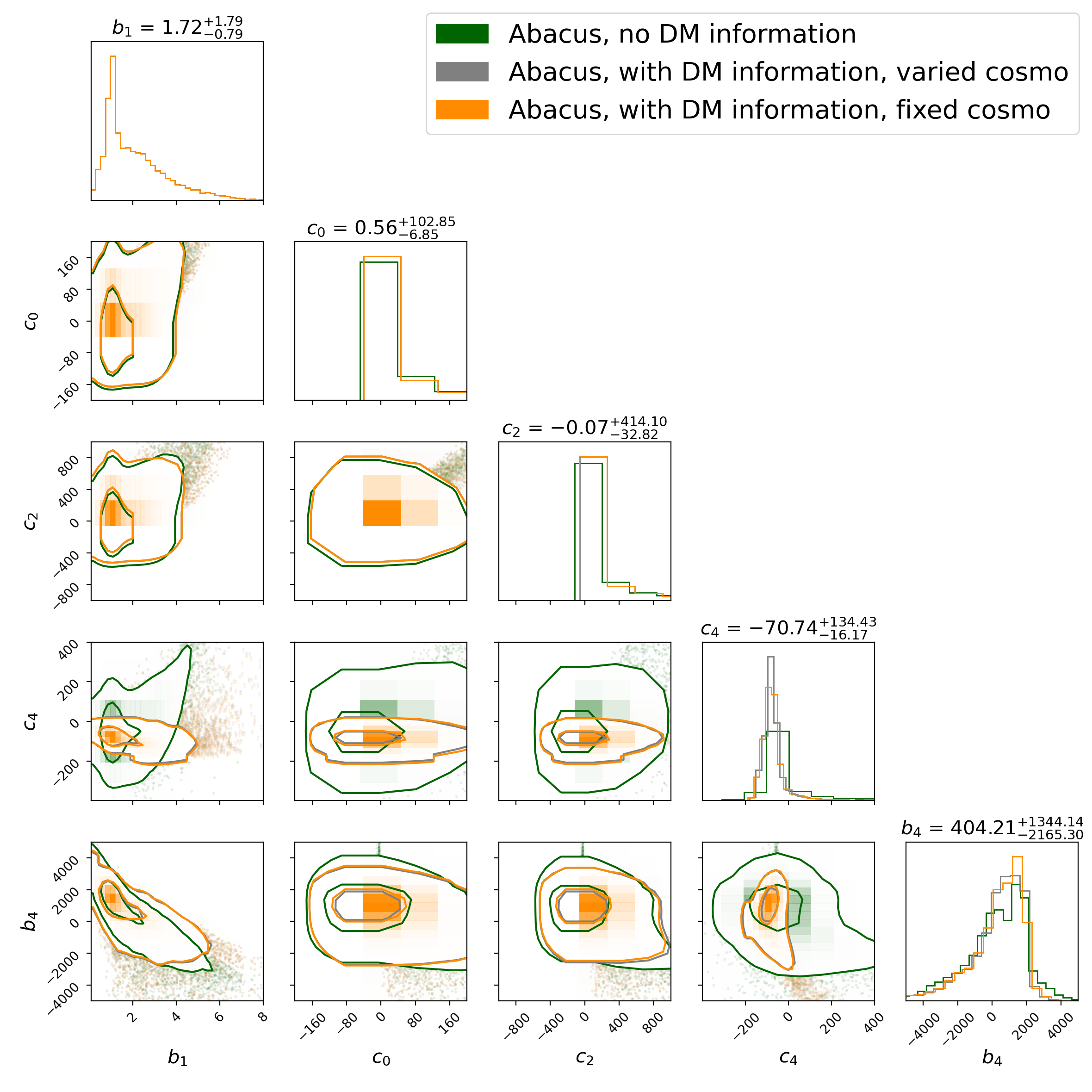}
   \caption{The distribution of $b_1$
   and EFT counterterms for redshift space multipoles without and with the 
   dark matter information in the form of the 
   $c_{s4}$ redshift space counterterm measurement.
    } \label{fig:counterterms}
\end{figure*}

The equivalence principle dictates that the redshift space counterterm $c_{s4}$ 
be the same for galaxies 
and dark matter. This information helps 
us reduce the 
number of free parameters in our fits to HOD transfer functions. We have measured the counterterm 
from the dark matter snapshot of 
\texttt{Abacus}, and 
used this measurement as a prior 
in fits to galaxies. This allowed us to 
significantly improve the precision of 
the $c_4$ field-level measurement and 
reduce the scatter 
in the $c_2 - b_4$ plane. We show this change in 
the counterterm measurements in fig.~\ref{fig:counterterms},
where for convenience we also display $b_1$.
As expected, we see that the DM prior on 
$c_{s4}$ has mostly affected the $c_4$
and $b_4$ counterterms that are sensitive 
to the $\mu\approx 1$ modes.

Since we are using 
the 
$c_{s4}$ prior on dark matter 
from the \texttt{Abacus} cosmology, one may be concerned about the 
propagation of the 
cosmology dependence in this prior.
To address this, we measure the standard deviation of the
$c_{s4}$ parameters from \texttt{Quijote}
LH mocks, and apply it as a standard deviation
to the \texttt{Abacus} dark matter $c_{s4}$
prior in our fit of EFT parameters
from the \texttt{Abacus} HODs. We find virtually no
difference in the EFT parameter measurements 
in this case (the means and the standard deviations 
are affected by less than $1\%$), see the gray contours in fig.~\ref{fig:counterterms}. This implies
that the cosmology-dependence of $c_{s4}$
can be neglected in practical 
applications.

The new counterterms and the real space 
parameters presented in figs.~\ref{fig:dist_abacus_eft},
~\ref{fig:dist_abacus_eft_2}
constitute our final EFT parameter 
distribution, which we will use as SBP in our full-shape analysis.

\subsection{Density estimation}

To model the marginal density distribution of the EFT parameters, we use the same method as in our ref.~\cite{Ivanov:2024hgq}.
Briefly, we use
Masked Autoregressive 
normalizing flows \citep{papamakarios2018maskedautoregressiveflowdensity} 
to approximate the density 
of the EFT parameters. 
$10\%$ of our EFT samples 
are used in validation, while the rest
is used for training. 
The flow is trained with 30 000 steps
and batch size 128, using the 
Adam optimizer with learning rate $3\cdot 10^{-4}$.
We use \texttt{nflows}\footnote{\url{https://github.com/bayesiains/nflows}} library, and \texttt{PyTorch}~\cite{paszke2019pytorch}
for training and evaluation.
All other aspects of our normalizing flow implementation
and training routine are the same 
as in our ref.~\cite{Ivanov:2024hgq}.
To extract optimal HOD parameters from our EFT 
chains from BOSS data we have also buit 
a conditional distribution 
$p(\theta_{\rm HOD}|\theta_{\rm EFT})$,
following~\cite{Ivanov:2024hgq}.
In principle, one can
use the normalizing flows to produce 
a conditional model for the distribution
of the EFT parameters given the HOD ones, $p(\theta_{\rm EFT}|\theta_{\rm HOD})$,
which can provide insights into 
physics behind the EFT parameters. 
We leave this for future work. 

Note that we do not use
$b_3$ in our priors as 
this parameter does not appear 
in the one-loop power spectrum
and three-level bispectrum
calculations which we use
in our fitting models. 

Note that the samples that we generated 
depend on the number density $\bar n$ 
which is a measurable 
parameter. One possible approach is to
condition the distribution 
to the number density. 
Fig.~\ref{fig:dist_abacus_eft}, however, 
shows that the EFT parameters for our HOD samples 
are consistent with a broad range of number density values, which suggests that the use of the 
conditional model will not 
significantly shrink the EFT prior
density. We have validated this explicitly
by producing a conditional model, and generating 
samples with a fixed number density 
$\bar n\approx 2\cdot 10^{-4}~[\hMpc]^3$ similar to that of the BOSS CMASS sample. We obtained a distribution 
of the EFT parameters 
which matches the original one 
(where we effectively marginalize over $\bar n$)
very closely. Thus, the approach of conditioning 
on the number density does not lead to 
a significant narrowing of the EFT distribution
in the context of the BOSS survey, 
but it may be more useful 
for other surveys such as DESI.

\subsection{Effects of Realistic Surveys}

There are two effects relevant for the 
actual survey that we would like to 
discuss: the effective redshift
and fiber collision 
corrections. 

Since the actual galaxies 
are observed on a past lightcone, 
their observed power spectrum 
is weighted with the selection function
$\bar n(z)$:
\be 
P^{\rm obs}_\ell(k) =
\frac{\int d^3r \bar n^2(r)P_\ell(k,z(r))}{\int d^3r \bar n^2(r)}~\,.
\ee 
The above integral is approximated
as $P_\ell(k,z_{\rm eff})$,
where $z_{\rm eff}$
is defined as 
\be 
z_{\rm eff} =\frac{\int d^3r \bar n^2(r)z(r)}{\int d^3r \bar n^2(r)}\,.
\ee 
The rationale behind this approximation is that 
the Taylor expansion
of the power spectrum 
around the effective redshift 
then starts only at a second order,
\be 
P^{\rm obs}_\ell(k) =
P_\ell(k,z_{\rm eff})
+
\frac{\int d^3r \bar n^2(r)(z-z_{\rm eff})^2P''_\ell(k,z_{\rm eff})}{2\int d^3r \bar n^2(r)}\,,
\ee 
where we ignored higher order terms. It is customary
to ignore the rightmost term above in cosmological 
analyses. In this approximation, all 
nuisance parameters
can be treated as constants
taken at the effective redshift.

A comment on the 
redshift dependence of our
priors is in order. 
We have calibrated our 
sample of EFT parameters
at $z=0.5$, which is very close, but not exactly the same
as the effective redshifts 
of the BOSS data that we will use, $z_{\rm eff}=0.38,~0.61$~\cite{BOSS:2016wmc}.
We argue now that the effects of redshift 
evolution are largely irrelevant for 
the HOD-based priors. 
First, 
as we have established, 
the EFT parameters for dark matter halos
depend on cosmology primarily through 
a single parameter, $\sigma_M$. 
Hence, a change 
of $\sigma_M$ due to redshift 
can be fully compensated by a change of the halo mass $M$.
Hence, as long as our prior
on the HOD parameter $M_{\rm cut}$ related to $M$ is wide enough, its variation
can effectively absorb the redshift
mismatch. Second, our full HOD model
with extra effects such as 
the assembly bias, 
assumes that the distribution of galaxies
is determined only by properties
of dark matter at a given redshift, 
i.e.\ there is no explicit dependence
on the past evolution. Together with the 
(approximate) universality of the halo mass function, this suggests 
that the HOD galaxies at different redshifts
should be self-similar, which implies the redshift-independence 
of the EFT priors.

The second important effect 
is the collision of optical fibers. 
At the power spectrum, it 
can be modeled 
with the effective window 
approach of~\cite{Hahn:2016kiy}.
It was pointed out in
ref.~\cite{Ivanov:2019pdj,Ivanov:2021zmi,Chudaykin:2022nru}
that on small scales the effect of this
effective window 
can be fully absorbed 
by  the stochastic counterterms, which 
makes the implementation of the full effective window model unnecessary.\footnote{The uncorrelated fiber collision contribution produces a ``non-local'' $k^{-1}$ correction sizeable for $k<0.02~\hMpc$.  This contribution can be easily included in an analysis pipeline at no extra cost because it does not depend on the galaxy power spectrum.}
Indeed, 
for $k>0.02~\hMpc$
the fiber collision effective 
window leads to 
to the following typical shifts of the 
constant shot noise contribution: 
\be 
\begin{split}
&\text{z=0.61}: ~~~\Delta \alpha_0^{\rm fc.}=-0.04 \,,\\
&\text{z=0.38}: ~~~\Delta \alpha_0^{\rm fc.}=-0.03\,,
\end{split}
\ee 
while the $k^2\mu^2$
counterterm shifts by 
$\Delta \alpha^{\rm fc.}_2 = 0.06$.
The shift in $\alpha_1$
is too small to be robustly detected. 
The above numbers have been
obtained for the best-fit 
BOSS power spectra
from~\cite{Philcox:2021kcw},
and they are consistent with 
results of~\cite{Hahn:2016kiy}.
To implement the fiber collision
shifts at the level of the priors consistently,
one has to compute the galaxy power
spectrum 
for each point in the sample,
apply the effective window, 
and extract the shifted 
stochastic counterterms 
$\alpha_i$'s. 
While it is straightforward 
to implement this, 
the overall effect of fiber collisions 
is negligibly small to affect 
our results. 
A typical shift 
of $\Delta \alpha_i$ 
is a tiny
fraction
of the statistical error $\sigma_{\alpha_i}$ on these 
parameters from the data even when using the SBP, e.g.
$\Delta\alpha_0\simeq 0.16\sigma_{\alpha_0}$
for BOSS NGCz1. Given the smallness
of the fiber collision corrections, 
we will ignore them in what follows.

\subsection{Validation on PT Challenge simulations}

We validate our simulation-based
priors on the PT Challenge (PTC) simulation
data~\cite{Nishimichi:2020tvu}. 
First, the PTC suite  
provides high fidelity 
mock power spectrum measurements
with $\sim 0.1\%$
statistical errorbars, 
which are ideal for 
precision tests of the theory model
and the priors. 
Second, their 
underlying cosmology is different 
from \texttt{Abacus},
and hence it can be used 
to test our assumption
that the cosmology dependence
of the EFT parameters 
is negligible for the 
purpose of generating the SBP.

We fit the galaxy power
spectrum monopole and 
quadrupole of the PTC mocks
at $z=0.61$. We choose this dataset
primarily because it was used in ref.~\cite{Nishimichi:2020tvu}. 
Our power spectrum likelihood is 
Gaussian, with the linear theory covariance computed using the total 
simulation volume $566~[\Mpch]^3$
as in~\cite{Nishimichi:2020tvu}.
The hexadecapole can be easily added, 
but its role is negligible 
within $\Lambda$CDM. While the 
real-space proxy, and bispectrum
multipoles of the same mock can be 
added to the datavector as well, 
we prefer not to do it here in 
order to clearly access the 
improvements of the $P_0+P_2$ 
analysis thanks for SBP. 
Indeed, the bispectrum and $Q_0$
mostly help by breaking 
degeneracies between the 
cosmological and EFT parameters 
which are present at the power 
spectrum level. In our setup, the same
degeneracies are expected to be 
broken by SBP. The comparison
between our analysis and the traditional analysis
of the 
$P_\ell+Q_0+B_\ell$ data plus 
uninformative priors 
will show us how much information
there is in the large-scale power
spectrum once we do not have to 
pay the price of nuisance parameter
marginalization. In addition, 
the comparison between the EFT parameter
posteriors in both cases will serve
us as a consistency test. 

We compare our results with those 
obtained with conservative priors~\cite{Chudaykin:2020aoj,Philcox:2021kcw}, 
\be
\begin{split}
& b_1\in \text{flat}[0,4]\,, \quad b_2\sim \mathcal{N}(0,1^2)\,, 
\\ 
& b_{\mathcal{G}_2}\sim \mathcal{N}(0,1^2) \,,
\quad
b_{\Gamma_3}\sim \mathcal{N}\left(\frac{23}{42}(b_1-1),1^2\right)\,,\\
\end{split}
\label{eq:ptpb_priors_bias1}
\ee
\be
\begin{split}
& \frac{c_0}{[\text{Mpc}/h]^2} \sim \mathcal{N}(0,30^2)\,,\quad 
\frac{c_2}{[\text{Mpc}/h]^2} \sim \mathcal{N}(30,30^2)\,,\\ 
& \frac{c_4}{[\text{Mpc}/h]^2} \sim \mathcal{N}(0,30^2)\,,\quad 
\frac{b_4}{[\text{Mpc}/h]^4} \sim \mathcal{N}(500,500^2)\,,\\
\end{split}
\label{eq:ptpb_priors_bias2}
\ee
\be
\begin{split}
& \alpha_{0} \sim \mathcal{N}(0,1^2)\,,\quad \a_{1}
\sim \mathcal{N}(0,1^2)\,,\\ 
& \a_2\sim \mathcal{N}(0,1^2)\,.
\end{split}
\label{eq:ptpb_priors_bias3}
\ee

We fit the full set of nuisance 
parameters plus three 
cosmological parameters 
of the base $\Lambda$CDM model: the physical 
density of dark matter $\omega_{cdm}$,
the Hubble constant $H_0$ and the mass fluctuation
amplitude $\sigma_8$.
The rest of the cosmological
parameters are set to their
true values. 
Since the 
PT challenge is ongoing, 
we will report here only 
the deviations of
the cosmological parameters
from their true values.
In addition, we will 
blind the values 
of $b_1$, 
$b_2$ and $b_{\G}$. 
We will show their deviations w.r.t.
the values estimated from
the PTC bispectrum multipoles~\cite{Ivanov:2023qzb} (consistent with the tree-level and one-loop monopole analyses~\cite{Ivanov:2021kcd,Philcox:2022frc}). 
Note that unlike the 
cosmological parameters, 
these are not the true values, 
but rather their estimates,
which are subject to 
their own biases and uncertainties. 
We will not blind 
the values of other parameters
as we believe their 
knowledge 
will not provide a significant
advantage to potential 
challenge participants. 
We use our datavector up to $\kmax=0.14~\hMpc$
validated 
in~\cite{Nishimichi:2020tvu}. 
This is lower than the baseline $\kmax$
for our transfer function 
fits and the BOSS data because 
of the outstandingly small statistical error of PTC data, for which the two loop
corrections become important at lower momenta 
than in the actual data.

\begin{figure*}[ht!]
\centering
\includegraphics[width=0.99\textwidth]{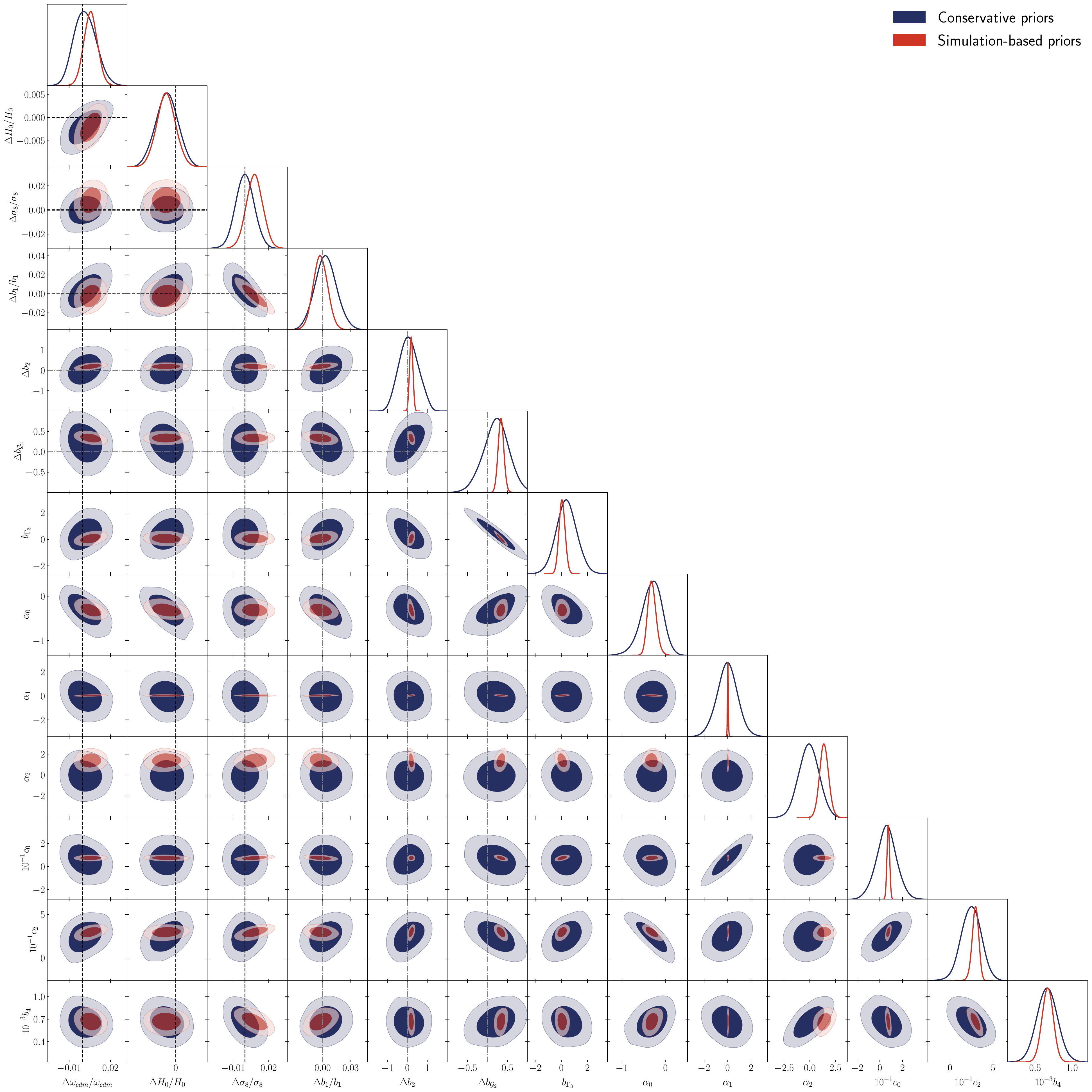}
   \caption{Cosmological and EFT 
   parameters extracted from the PT Challenge simulation with the usual conservative priors 
   and the SBP derived in this work. For the cosmological parameters
   we show deviations from the true values,
   which are depicted by solid dashed lines. 
   We show residuals of bias parameters $b_1,b_2,b_{\G}$ 
   w.r.t. the PTC bispectrum
   measurements from~\cite{Ivanov:2023qzb}; 
   the corresponding values are marked by 
   grey
   dotted-dashed lines. 
    } \label{fig:ptc}
\end{figure*}

\begin{table*}
\begin{tabular}{|l|c|c|c|c|}
 \hline
 \multicolumn{5}{|c|}{PTC $P_0+P_2$ data with conservative priors} \\
\hline
Param & best-fit & mean$\pm\sigma$ & 95\% lower & 95\% upper \\ \hline
$\frac{\Delta \omega_{cdm}}{\omega_{cdm}}$  &$0.00065$ & $0.0016^{+0.0075}_{-0.0085}$ & $-0.0129$ & $0.0178$ \\
$\Delta H_0/H_0 $ &$-0.0021$ & $-0.0020^{+ 0.0024}_{-0.0024} $ & $-0.0068$ & $0.0028$ \\
$\Delta \sigma_8/\sigma_8$ &$-0.0015$ & $0.0000^{+ 0.0078}_{-0.0078}$ & $-0.0152$ & $0.0178$ \\
$\Delta b_1/b_1 $ &$0.0041$ & $0.003^{+0.012}_{-0.012}$ & $-0.0187$ & $0.0267$ \\
$\Delta b_{2 }$ &$0.045$ & $0.07^{+0.48}_{-0.48} $ & $-0.83$ & $1.02$ \\
$\Delta b_{\mathcal{G}_2}$ &$-0.24$ & $0.22^{+0.32}_{-0.32} $ & $-0.42$ & $0.84$ \\
$b_{\GG}$ &$0.3243$ & $0.3964_{-0.79}^{+0.74}$ & $-1.136$ & $1.975$ \\
$\a_0$ &$-0.2642$ & $-0.3097_{-0.2}^{+0.26}$ & $-0.8006$ & $0.1372$ \\
$\a_{1 }$ &$-0.20$ & $-0.02_{-0.83}^{+0.87}$ & $-1.8$ & $1.687$ \\
$\a_{2 }$ &$0.3759$ & $-0.09155_{-1}^{+1}$ & $-2.066$ & $1.891$ \\
$10^{-1}c_{{0} }$ &$0.5635$ & $0.5742_{-0.86}^{+0.86}$ & $-1.212$ & $2.36$ \\
$10^{-1}c_{{2} }$ &$2.385$ & $2.443_{-1.2}^{+1.2}$ & $0.2061$ & $4.778$ \\
$10^{-3}b_{4 }$ &$0.7123$ & $0.6634_{-0.14}^{+0.14}$ & $0.3846$ & $0.941$ \\
\hline
 \end{tabular} 
 \begin{tabular}{|l|c|c|c|c|}
 \hline
  \multicolumn{5}{|c|}{PTC $P_0+P_2$ data with simulation-based priors} \\
\hline
Param & best-fit & mean$\pm\sigma$ & 95\% lower & 95\% upper \\ \hline
$\frac{\Delta \omega_{cdm}}{\omega_{cdm}}$ &$0.0043$ & $0.0056\pm 0.0050$ & $-0.0041$ & $0.015$ \\
$\Delta H_0/H_0$ &$-0.024$ & $-0.0021\pm 0.0021$ & $-0.0057$ & $0.0024$ \\
$\Delta \sigma_8/\sigma_8$ &$0.0077$ & $0.0080 \pm 0.0069 $ & $-0.006$ & $0.022$ \\
$\Delta b_1/b_1$ &$-0.0014$ & $-0.0024\pm 0.0078$ & $-0.018$ & $0.013$ \\
$ \Delta b_{2 }$ &$0.19$ & $0.193\pm 0.081$ & $0.03$ & $0.35$ \\
$\Delta b_{\mathcal{G}_2}$ &$0.31$ & $0.340\pm 0.072$ & $0.2$ & $0.48$ \\
$b_{\GG }$ &$0.1264$ & $0.0586_{-0.25}^{+0.24}$ & $-0.4202$ & $0.5359$ \\
$\a_0$ &$-0.3419$ & $-0.3088_{-0.11}^{+0.088}$ & $-0.5005$ & $-0.1093$ \\
$\a_{1 }$ &$0.02429$ & $0.02933_{-0.04}^{+0.035}$ & $-0.04861$ & $0.1071$ \\
$\a_{2 }$ &$1.427$ & $1.401_{-0.48}^{+0.44}$ & $0.4841$ & $2.336$ \\
$10^{-1}c_{{0} }$ &$0.7494$ & $0.7454_{-0.12}^{+0.12}$ & $0.5118$ & $0.9788$ \\
$10^{-1}c_{{2} }$ &$3.102$ & $2.946_{-0.36}^{+0.42}$ & $2.164$ & $3.73$ \\
$10^{-3}b_{4 }$ &$0.656$ & $0.6693_{-0.078}^{+0.08}$ & $0.5118$ & $0.8247$ \\
\hline
 \end{tabular} 
 \caption{Best-fits and 1d marginalized limits for cosmological and EFT parameters 
 from PT Challenge simulation data. }
 \label{eq:tab1}
 \end{table*}

Our results 
for the full-shape analysis 
of PTC data with the 
conservative and simulation-based
priors are displayed in fig.~\ref{fig:ptc} 
and in table~\ref{eq:tab1}. 
First, we have recovered the true input 
cosmological parameters without bias 
in the SBP analysis. 
The bias parameters $b_1,b_2$
are also recovered without bias, 
but $b_{\G}$ is found in a $\approx 4\sigma$
bias w.r.t.\ the bispectrum 
measurement. Given that the
maximum of the $b_{\G}$ posterior in the
conservative analysis is also shifted w.r.t. the bispectrum value, 
it is suggestive that this tension
is a result of a statistical 
fluctuation. 
It could also be due to systematic effects, such as a bias
in the bispectrum
estimate 
or 
the use of the Gaussian covariance 
matrix in our power spectrum likelihood. 
Alternatively, this tension could arise
due to the residual cosmology dependence
of the EFT parameters which we ignored
while  generating our priors. 
The shift of $b_{\G}$, however, does not exceed the statistical 
precision with which we measure this 
parameter from BOSS, and therefore 
we do not expect it to affect 
our final constraints 
at a statistically significant level. 

The second important observation
is that the posteriors for EFT 
parameters are consistent in both analysis.  
Moreover, as anticipated, they 
shrink dramatically
with the SBP. As for the 
cosmological parameters, constraints on 
$\omega_{cdm},H_0$
and $\sigma_8$ improve by 
$60\%,14\%$ and $13\%$,
respectively. The improvements 
for $\omega_{cdm}$ are quite 
significant because SBP help 
break its degeneracies with $\alpha_0,\alpha_1$
and $b_{\Gamma_3}$. The improvements 
on $\omega_{cdm}$ and $\sigma_8$
are
more significant than those coming 
from the addition of external datasets such
as the bispectrum and the real space power spectrum proxy $Q_0$~\cite{Ivanov:2021fbu}.
This clearly illustrates the power
of the SBP. 

Finally, let us note that we have clearly detected both redshift-space counterterms $b_4$
and $\a_2$. They are degenerate at the 
power spectrum level, but the degeneracy 
is lifted with the SBP. While $b_4$
is detected even with conservative priors, 
a statistically significant detection of 
$\a_2$ with PT Challenge data is possible only with the SBP.

All in all, we conclude that 
our HOD-based priors successfully 
pass the validation on the PT Challenge 
data.

\section{Applications to BOSS}
\label{sec:boss}

In this section we redo the EFT-based
full-shape analysis of the BOSS galaxy clustering data using the SBP. 
Our dataset and likelihood
have been extensively 
used in previous
full-shape analysis
by~\cite{Philcox:2021kcw,Ivanov:2023qzb},
which we refer to for the technical details. 
We analyze BOSS DR12 
galaxy clustering data, which are split across two patches of the sky, north galactic cap (NGC) and 
south galactic cap (SGC),
and two non-overlapping redshift bins.
In this work, we use the z1-z3 split used in most of the previous BOSS analysis starting with~\cite{Beutler:2016arn}. 
The two 
redshift bins z1 and z3 in this case have 
effective redshifts $0.38$ and $0.61$, respectively. 
This choice is done mainly 
to ease the comparison 
with previous works 
such as~\cite{Philcox:2021kcw,Ivanov:2023qzb},
which used a similar split. 
We use the publicly available 
power spectrum and bispectrum measurements 
obtained with the window-free estimators~\cite{Philcox:2020vbm,Philcox:2021ukg}. Our scale cuts are the same 
as in~\cite{Philcox:2021kcw,Ivanov:2023qzb}.
In particular, we use $k_{\rm max}=0.2~\hMpc$
for the power spectrum multipoles, 
which is consistent with the scale
cut used to fit the EFT 
parameters from redshift space
transfer functions.

We fix the cosmological parameters 
to the \textit{Planck} best-fit values~\cite{Aghanim:2018eyx}
except $\omega_{cdm},H_0$
and $\ln(10^{10}A_s)$ ($A_s$ is the amplitude of primordial scalar perturbations), which we vary in our Markov Chain Monte Carlo (MCMC) chains. For each BOSS data chunk we use a separate set of EFT 
parameters. We compare our results with the 
analysis based on conservative priors 
carried out in~\cite{Philcox:2021kcw}
on the same data. 

\subsection{Results for BOSS NGCz1 sample}

\begin{figure*}[ht!]
\centering
\includegraphics[width=0.99\textwidth]{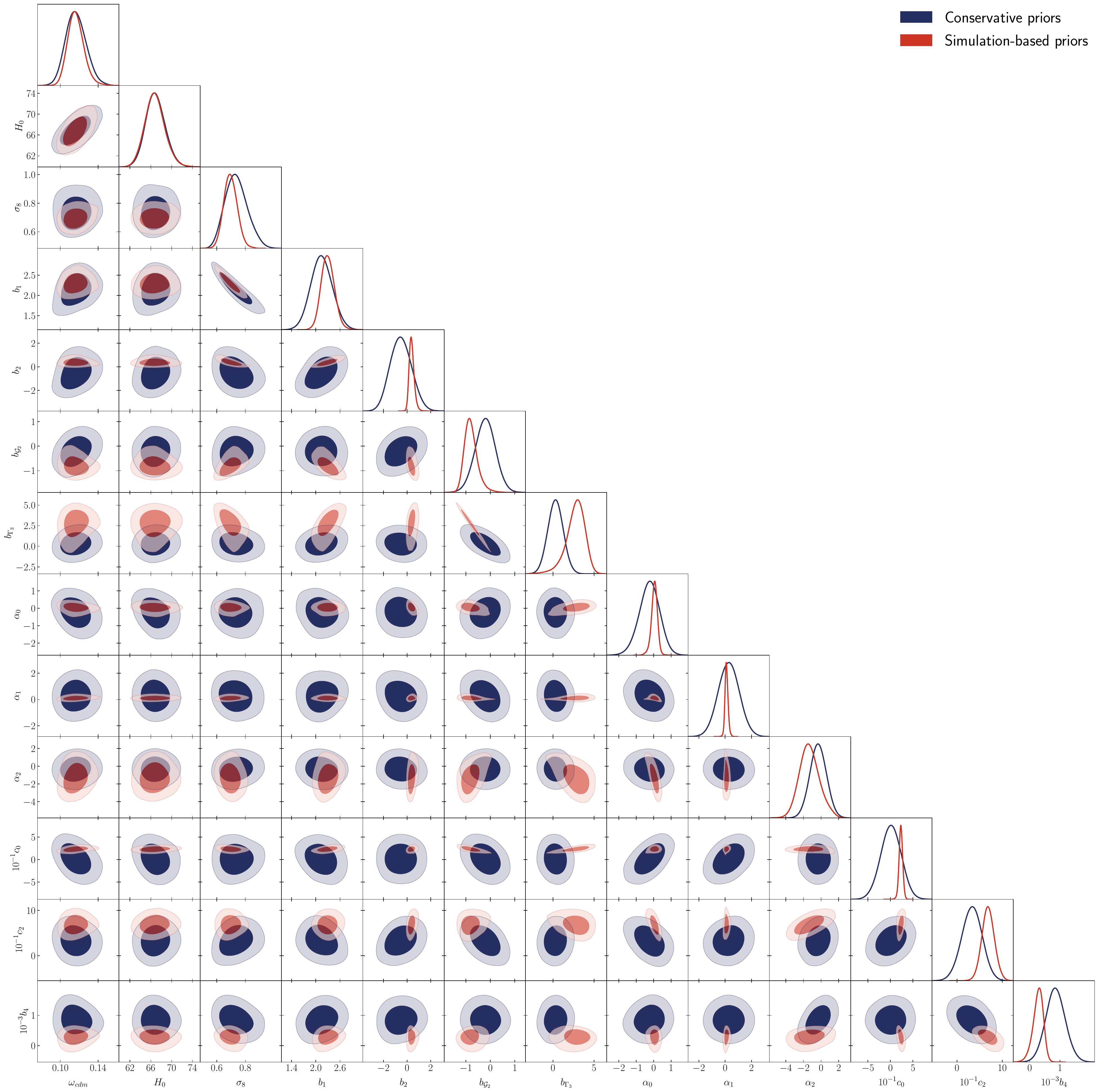}
   \caption{Cosmological and EFT 
   parameters extracted from the BOSS NGCz1 data with the usual conservative priors 
   and the SBP derived in this work.
    } \label{fig:ngcz1}
\end{figure*}

To start off, we re-analyze
the BOSS NGCz1 data
from~\cite{Philcox:2021kcw}
with $z_{\rm eff}=0.38$.
We use the galaxy 
power spectrum monopole 
$P_0$ and quadrupole $P_2$ data 
from the likelihood described in~\cite{Philcox:2021kcw}.
The corner plot and 1d marginalized
posteriors for the SBP
and the conservative prior analyses 
are shown in fig.~\ref{fig:ngcz1}.
The best-fit values and 
1d marginalized limits are given in table~\ref{eq:tab2}. 
The first important observation 
is that the SBP results are consistent
with the conservative ones: both posteriors
largely overlap. 
The second observation is that 
the SBP affect the posteriors
of EFT parameters in this analysis
in a very non-uniform way.
EFT parameters $b_2,\a_0,\a_1,c_0$
are much better constrained with SBP, 
$b_1,b_{\G},c_2,b_4$ show moderate improvements, 
while $b_{\Gamma_3}$ and $\a_2$
are less constrained now. 
In the latter case, 
the priors are not informative enough, and at the same time 
the power spectrum 
data alone cannot break the remaining degeneracies 
$b_4-\a_2$ and $b_{\G}-b_{\GG}$, which results 
in worse constraints. 
Note that our constraint of $b_1$
is driven by linear redshift-space distortions 
in the data and not by the priors 
themselves. The priors only help break 
degeneracies between $b_1$, $\sigma_8$
and other nuisance parameters
that appear in the non-linear regime at large wavenumbers.

The posteriors of cosmological 
parameters $\omega_{cdm}$ and $\sigma_8$ 
narrow quite noticeably 
as a result of using the SBP. 
We find no improvement 
on $H_0$. 
The improvement in the case of $\omega_{cdm}$
is $\approx 30\%$.
The $\sigma_8$ constraint improves by more than $60\%$.
This is achieved by breaking the degeneracy between
$\sigma_8$ and the redshift space counterterms.
The strongest residual degeneracy that $\sigma_8$
has after applying the SBP is with $b_{\GG},b_2$
and $b_{\G}$.

\begin{table*}
\begin{tabular}{|l|c|c|c|c|}
 \hline
 \multicolumn{5}{|c|}{BOSS NGCz1 $P_0+P_2$ with conservative priors} \\
\hline
Param & best-fit & mean$\pm\sigma$ & 95\% lower & 95\% upper \\ \hline
$\omega{}_{cdm }$ &$0.1122$ & $0.1167_{-0.012}^{+0.01}$ & $0.09525$ & $0.1385$ \\
$H_0$ &$66.15$ & $66.92_{-2}^{+1.8}$ & $63.17$ & $70.76$ \\
$\ln(10^{10}A_{s })$ &$3.031$ & $2.862_{-0.25}^{+0.22}$ & $2.399$ & $3.34$ \\
$b_{1 }$ &$1.941$ & $2.139_{-0.28}^{+0.26}$ & $1.624$ & $2.669$ \\
$b_{2 }$ &$-1.102$ & $-0.513_{-0.98}^{+0.87}$ & $-2.262$ & $1.284$ \\
$b_{\G}$ &$-0.3471$ & $-0.2069_{-0.42}^{+0.43}$ & $-1.039$ & $0.628$ \\
$b_{\GG }$ &$0.1284$ & $0.3103_{-0.96}^{+0.95}$ & $-1.561$ & $2.183$ \\
$\a_0$ &$-0.2216$ & $-0.2489_{-0.57}^{+0.62}$ & $-1.444$ & $0.9123$ \\
$\a_1$ &$0.4947$ & $0.2724_{-0.82}^{+0.81}$ & $-1.309$ & $1.88$ \\
$\a_2$ &$-0.4268$ & $-0.3283_{-0.96}^{+0.96}$ & $-2.2$ & $1.565$ \\
$10^{-1}c_{{0} }$ &$0.9671$ & $0.1047_{-2.3}^{+2.3}$ & $-4.496$ & $4.637$ \\
$10^{-1}c_{{2} }$ &$4.127$ & $3.306_{-2.2}^{+2.3}$ & $-1.187$ & $7.731$ \\
$10^{-3}b_{4 }$ &$0.7137$ & $0.852_{-0.33}^{+0.32}$ & $0.2073$ & $1.506$ \\
\hline
$\Omega{}_{m }$ &$0.3076$ & $0.3104_{-0.021}^{+0.018}$ & $0.2726$ & $0.3496$ \\
$\sigma_8$ &$0.7802$ & $0.7383_{-0.087}^{+0.07}$ & $0.5912$ & $0.8929$ \\
\hline
 \end{tabular} 
 \begin{tabular}{|l|c|c|c|c|}
 \hline
  \multicolumn{5}{|c|}{BOSS NGCz1 $P_0+P_2$ with simulation-based priors} \\
\hline
Param & best-fit & mean$\pm\sigma$ & 95\% lower & 95\% upper \\ \hline
$\omega{}_{cdm }$ &$0.1193$ & $0.1165_{-0.0094}^{+0.0077}$ & $0.09956$ & $0.134$ \\
$H_0$ &$67.43$ & $66.77_{-2}^{+1.8}$ & $63$ & $70.63$ \\
$ln10^{10}A_{s }$ &$2.783$ & $2.752_{-0.16}^{+0.15}$ & $2.445$ & $3.06$ \\
$b_{1 }$ &$2.216$ & $2.298_{-0.18}^{+0.15}$ & $1.974$ & $2.629$ \\
$b_{2 }$ &$0.2628$ & $0.3964_{-0.25}^{+0.18}$ & $-0.02381$ & $0.8499$ \\
$b_{\G  }$ &$-0.7999$ & $-0.7897_{-0.31}^{+0.21}$ & $-1.313$ & $-0.207$ \\
$b_{\Gamma_3}$ &$2.77$ & $2.72_{-0.9}^{+1.3}$ & $0.2394$ & $4.921$ \\
$\a_0$ &$0.09192$ & $0.04499_{-0.16}^{+0.19}$ & $-0.3444$ & $0.4149$ \\
$\a_{1 }$ &$0.1218$ & $0.1043_{-0.13}^{+0.11}$ & $-0.1361$ & $0.3694$ \\
$\a_{2 }$ &$-1.854$ & $-1.32_{-1.2}^{+1.1}$ & $-3.506$ & $1.09$ \\
$10^{-1}c_{{0} }$ &$2.167$ & $2.309_{-0.42}^{+0.43}$ & $1.451$ & $3.187$ \\
$10^{-1}c_{{2} }$ &$6.098$ & $6.878_{-1.5}^{+1.4}$ & $4.066$ & $9.757$ \\
$10^{-3}b_{4 }$ &$0.3993$ & $0.2749_{-0.15}^{+0.2}$ & $-0.09049$ & $0.6056$ \\
\hline
$\Omega{}_{m }$ &$0.3116$ & $0.3117_{-0.019}^{+0.015}$ & $0.2779$ & $0.3469$ \\
$\sigma_8$ &$0.7197$ & $0.6965_{-0.049}^{+0.046}$ & $0.6036$ & $0.7898$ \\
\hline
 \end{tabular} 
 \caption{Best-fits and 1d marginalized limits for cosmological and EFT parameters 
 from BOSS NGCz1 data. The lower two rows show derived cosmological parameters. }
 \label{eq:tab2}
 \end{table*}

\subsection{On the consistent choice of priors}

A comment is in order on
the prior choice. Overall, the priors associated 
with the \texttt{CLASS-PT} and \texttt{velocileptor}~\cite{Chen:2020fxs,Chen:2020zjt}
codes are consistent with the 
HOD-based priors and with the 
posteriors from the BOSS data, see fig.~\ref{fig:ngcz1}.
One can also compare these priors choices 
with the one associated with the \texttt{PyBird}
code~\cite{DAmico:2019fhj}. 
While we defer a detailed comparison
to a separate paper, let us make a few
comments which may be relevant 
for the interpretation of 
results obtained with this code. 
\texttt{PyBird} analyses assume $b_4=0$
and a prior on $\alpha_2$ (proportional to $c_{\epsilon,quad}$ of~\cite{DAmico:2019fhj})
\be 
\label{eq:a2Guido}
\alpha_2 \sim \mathcal{N}(0,0.5^2)~\,,
\ee 
for their LOWZ sample\footnote{Note that 
the values of number densities assumed  in~\cite{DAmico:2019fhj} do not match the ones measured from data.} similar to our z1. 
Besides,~\cite{DAmico:2019fhj} 
do not account for the effects of exclusion
of dark matter halos
and galaxies~\cite{Casas-Miranda:2001dwz,Baldauf:2013hka} (i.e. non-Poissonian sampling) in their prior
on the constant stochasticity parameter $c_{\epsilon,1}=\alpha_0$, 
for which they assume 
\[
\alpha_0 \sim \mathcal{N}(0,0.1^2)~\,,
\]
based on the fiber collision corrections only. 
Both our HOD samples and the actual measurements 
from the data strongly rule out 
this assumption.

The small
number $0.5$ in~\eqref{eq:a2Guido} is 
also 
challenging to 
justify from the theoretical perspective. 
In addition, such an overoptimistic 
prior on $\alpha_2$ is not supported 
by data and simulations. In particular, 
this is evidenced 
from figures 25 and 26 of~\cite{DAmico:2019fhj},
which display a tension between 
the actual posteriors for 
$c_{\epsilon,quad}$ and the prior for
this parameter. 
Arguments for non-zero $b_4$ 
based on the relevance
of the two-loop corrections
were given 
in~\cite{Ivanov:2019pdj,Chudaykin:2020hbf,Taule:2023izt}. 
The large value of the $k^2P_{11}$ quadrupole
counterterm from data 
suggests a sizeable $k^4P_{11}$
correction as well. 

In addition to these arguments, 
our HOD-based priors and posteriors presented in 
fig.~\ref{fig:ngcz1}
suggest that the prior from eq.~\eqref{eq:a2Guido}
and the choice $b_4=0$
are inconsistent with a large 
sample of BOSS-like HOD models. 
This sample includes
HODs for the 
PT Challenge mocks.
We caution against 
overoptimistic 
priors as they can 
bias parameter estimation.

\begin{table*}
\begin{tabular}{|l|c|c|c|c|}
 \hline
 \multicolumn{5}{|c|}{BOSS $P$ with conservative priors} \\
\hline
Param & best-fit & mean$\pm\sigma$ & 95\% lower & 95\% upper \\ \hline
$\omega_{cdm }$ &$0.1252$ & $0.1268_{-0.0068}^{+0.0062}$ & $0.114$ & $0.1398$ \\
$H_0$ &$68.64$ & $68.76_{-1.3}^{+1.2}$ & $66.29$ & $71.27$ \\
$\ln(10^{10}A_{s })$ &$2.805$ & $2.75_{-0.13}^{+0.12}$ & $2.504$ & $2.994$ \\
$b^{(1)}_{1 }$ &$2.236$ & $2.319_{-0.17}^{+0.16}$ & $1.997$ & $2.648$ \\
$b^{(1)}_{2 }$ &$-1.291$ & $-1.1_{-1}^{+0.94}$ & $-3.032$ & $0.8797$ \\
$b^{(1)}_{{\G} }$ &$-0.05125$ & $0.01557_{-0.46}^{+0.47}$ & $-0.9242$ & $0.9429$ \\
$b^{(2)}_{1 }$ &$2.386$ & $2.462_{-0.17}^{+0.15}$ & $2.143$ & $2.79$ \\
$b^{(2)}_{2 }$ &$-0.5497$ & $0.02503_{-0.97}^{+0.95}$ & $-1.876$ & $1.949$ \\
$b^{(2)}_{{\G} }$ &$-0.3622$ & $-0.1452_{-0.47}^{+0.5}$ & $-1.117$ & $0.8138$ \\
$b^{(3)}_{1 }$ &$2.162$ & $2.218_{-0.15}^{+0.14}$ & $1.932$ & $2.513$ \\
$b^{(3)}_{2 }$ &$-0.7661$ & $-0.2975_{-0.9}^{+0.8}$ & $-1.975$ & $1.431$ \\
$b^{(3)}_{{\G} }$ &$-0.44$ & $-0.1732_{-0.4}^{+0.39}$ & $-0.9581$ & $0.6304$ \\
$b^{(4)}_{1 }$ &$2.201$ & $2.243_{-0.15}^{+0.14}$ & $1.958$ & $2.54$ \\
$b^{(4)}_{2 }$ &$-0.2233$ & $-0.07458_{-0.92}^{+0.84}$ & $-1.805$ & $1.69$ \\
$b^{(4)}_{{\G} }$ &$0.04152$ & $0.2367_{-0.42}^{+0.41}$ & $-0.5914$ & $1.076$ \\
\hline 
$\Omega_{m }$ &$0.3146$ & $0.317_{-0.013}^{+0.012}$ & $0.2928$ & $0.3419$ \\
$\sigma_8$ &$0.7422$ & $0.7288_{-0.045}^{+0.04}$ & $0.6458$ & $0.8141$ \\
$S_8$ &$0.7726$ & $0.749^{+0.048}_{-0.048}$ & $0.6593$ & $0.8459$ \\
\hline
 \end{tabular} 
 \begin{tabular}{|l|c|c|c|c|}
 \hline
  \multicolumn{5}{|c|}{BOSS $P$ with simulation-based priors} \\
\hline
Param & best-fit & mean$\pm\sigma$ & 95\% lower & 95\% upper \\ \hline
$\omega{}_{cdm }$ &$0.1276$ & $0.1269_{-0.0046}^{+0.0045}$ & $0.1181$ & $0.1363$ \\
$H_0$ &$67.61$ & $68.79_{-1.3}^{+1.1}$ & $66.47$ & $71.22$ \\
$\ln(10^{10}A_{s })$ &$2.544$ & $2.533_{-0.084}^{+0.092}$ & $2.354$ & $2.712$ \\
$b^{(1)}_{1 }$ &$2.675$ & $2.674_{-0.12}^{+0.11}$ & $2.443$ & $2.9$ \\
$b^{(1)}_{2 }$ &$0.6511$ & $0.8556_{-0.31}^{+0.21}$ & $0.3716$ & $1.405$ \\
$b^{(1)}_{{\G} }$ &$-0.9653$ & $-0.8646_{-0.32}^{+0.23}$ & $-1.383$ & $-0.2882$ \\
$b^{(2)}_{1 }$ &$2.758$ & $2.774_{-0.12}^{+0.12}$ & $2.531$ & $3.016$ \\
$b^{(2)}_{2 }$ &$0.835$ & $0.9431_{-0.32}^{+0.21}$ & $0.4362$ & $1.516$ \\
$b^{(2)}_{{\G} }$ &$-1.056$ & $-1.078_{-0.26}^{+0.19}$ & $-1.537$ & $-0.5721$ \\
$b^{(3)}_{1 }$ &$2.453$ & $2.47_{-0.11}^{+0.099}$ & $2.262$ & $2.681$ \\
$b^{(3)}_{2 }$ &$0.5009$ & $0.5357_{-0.21}^{+0.15}$ & $0.1841$ & $0.9123$ \\
$b^{(3)}_{{\G} }$ &$-0.9306$ & $-0.9451_{-0.24}^{+0.16}$ & $-1.344$ & $-0.5018$ \\
$b^{(4)}_{1 }$ &$2.484$ & $2.468_{-0.11}^{+0.1}$ & $2.257$ & $2.684$ \\
$b^{(4)}_{2 }$ &$0.5287$ & $0.5722_{-0.22}^{+0.16}$ & $0.1896$ & $0.9783$ \\
$b^{(4)}_{{\G} }$ &$-1.058$ & $-0.9933_{-0.28}^{+0.19}$ & $-1.452$ & $-0.4821$ \\
\hline 
$\Omega{}_{m }$ &$0.3282$ & $0.3155_{-0.011}^{+0.0095}$ & $0.2962$ & $0.3364$ \\
$\sigma_8$ &$0.6661$ & $0.6632_{-0.027}^{+0.027}$ & $0.6086$ & $0.7212$ \\
$S_8$ &$0.7114$ & $0.679^{+0.031}_{-0.035} $ & $0.6186$ & $0.7466$ \\
\hline
 \end{tabular} 
 \caption{Best-fits and 1d marginalized limits for cosmological and EFT parameters 
 from BOSS data. For EFT parameters, we display the linear and quadratic bias parameters only. The lower two rows show derived cosmological parameters. The bias parameters' superscripts $(1),(2),(3),(4)$ correspond to NGCz3, SGCz3, NGCz1, and SGCz1 samples, respectively.  }
 \label{eq:tab3}
 \end{table*}

\subsection{Results for full BOSS DR12}

We reanalyze now the power spectrum 
multipoles
from the 
complete BOSS DR12 data.\footnote{We also include the hexadecapole $P_4$ here using the methodology of~\cite{Chudaykin:2020ghx}.} This includes 
four data slices: NGCz1, SGCz1, NGCz3
and SGCz3. The results are
presented in fig.~\ref{fig:bossPk}
and table~\ref{eq:tab3}.
For comparison, we also show
the \textit{Planck} 2018
results for the $\Lambda$CDM+$m_\nu$
model which is appropriate
for comparison with 
the BOSS EFT full-shape likelihood~\cite{Ivanov:2019pdj,Ivanov:2019hqk}.
We see that the SBPs improve $\omega_{cdm}$
and $\sigma_8$ limits by $\approx 30\%$
and $\approx 60\%$, respectively. 
As before, the constraint on $H_0$
does not improve noticeably.
Our final results nominally suggest 
a $\approx 5\sigma$ tension
with \textit{Planck}
for $\sigma_8$, in agreement
with the recent analysis
of the BOSS data~\cite{Chen:2024vuf}
that used an extended dataset, 
consisting 
of the post-reconstructed BAO~\cite{Philcox:2020vvt}, 
real-space power spectrum proxy $Q_0$~\cite{Ivanov:2021fbu},
the bispectrum mulipoles~\cite{Ivanov:2023qzb},
and the galaxy-CMB lensing 
cross correlations~\cite{Chen:2022jzq}. 

It is interesting to compare 
our results with those of~\cite{Chen:2024vuf}.
This analysis assumed
conservative priors 
on EFT parameters as in eqs.~(\ref{eq:ptpb_priors_bias1}-\ref{eq:ptpb_priors_bias3}).
Their final results 
for $\Omega_{m}$
and $\sigma_8$ are 
consistent with ours 
both in terms of errorbars
and the mean values. The improvement 
on cosmological parameters from adding 
more data while keeping 
the conservative priors
are similar to the results that we 
get from the power spectrum alone, 
but with the better priors. 
This confirms the argument
that the marginalization
over EFT parameters leads to a strong
degradation of constraining power~\cite{Wadekar:2020hax,Cabass:2022epm}. 

Our results also suggest that the 
$\sigma_8$
anomaly 
was present at a significant level 
already in the galaxy power spectrum,
but its signal was absorbed 
into the EFT parameters. 
Once the freedom in the 
variation of EFT parameters is 
reduced, the evidence 
for the anomaly in the 
power spectrum increases. 
This effect was first pointed out 
in the analysis of~\cite{Philcox:2021kcw},
which found that fitting the BOSS
galaxy power spectrum with the 
\textit{Planck} cosmology
requires values of bias
parameters that are not consistent 
with dark matter halo expectations.
We will comment on 
this point in detail
shortly.

A comment on the primordial spectral tilt is in order. 
We have fixed this parameter to the \textit{Planck}
best-fit $\Lambda$CDM value $n_s=0.9649$
in our main analysis. To understand the impact of the SBP 
on measuring this parameters, we have re-done our main $P_\ell$ 
analysis with free $n_s$ and report 
the results in Appendix~\ref{sec:free_ns}. 
We have found that the SBP somewhat reduce the preference
of the BOSS data for low $n_s$ values, and alleviate the 
discrepancy 
with \textit{Planck} from $\approx 1.1\sigma$ to 
$\approx 0.9\sigma$. The best-fit value of $n_s$
from the SBP analysis matches the \textit{Planck}
value quite closely, suggesting that the remaining $0.9\sigma$
shift above is a likely projection effect. 
The SBP improve the limit on $n_s$ (and other cosmological parameters
except $\sigma_8$) only marginally, by $\lesssim 10\%$. 
The structure growth parameter measurement, however, 
improves drastically, just like in our baseline analysis, 
from $\sigma_8 =0.704^{+0.044}_{-0.049}$
to $\sigma_8 =0.660^{+0.026}_{-0.03}$.

At this point one may wonder how much the addition
of other data, such as the BAO, 
real-space proxy $Q_0$,
the bispectrum
can help improve cosmological
constraints. Using the full dataset of~\cite{Philcox:2021kcw}, we found a very significant
improvement in
cosmological constraints. In particular, 
for the same choices as in the baseline $P_\ell$+BAO+$Q_0$+$B_0$ analysis of~\cite{Philcox:2021kcw}, 
we find 
\be 
\begin{split}
    & \sigma_8=0.6511~(0.6508)_{-0.018}^{+0.019}\,,\\
    & \omega_{cdm}=0.1221~(0.1211)_{-0.0045}^{+0.004}\,,\\
    & H_0=68.05~(67.82)_{-0.8}^{+0.79}\,,\\
    &\O_m=0.312~(0.3119)_{-0.0088}^{+0.0078}\,.
\end{split}
\ee 
where we quote best-fit values 
in the parentheses. 
The posteriors are
also shown in fig.~\ref{fig:bossPk}.
These compare favorably with the 
$P_\ell$+BAO+$Q_0$+$B_0$ results
obtained under the 
conservative priors from~\cite{Philcox:2021kcw}: 
\be 
\begin{split}
    & \sigma_8=0.7221~(0.7248)_{-0.037}^{+0.032}\,,\\
    & \omega_{cdm}=0.1262~(0.1242)_{-0.0059}^{+0.0053}\,,\\
    & H_0=68.32~(68.09)_{-0.86}^{+0.83}\,,\\
    &\O_m=0.3197~(0.3176)_{-0.01}^{+0.0095}\,.
\end{split}
\ee 
In particular, we still see 
a significant ($\approx 80\%$) improvement on
$\sigma_8$, 
along with a $\approx 30\%$
improvement on $\omega_{cdm}$.
We have found that most 
of the additional improvements
w.r.t. the $P_\ell$ SBP analysis
is produced by the bispectrum.
We believe that these 
extra
improvements are due to 
the fact that the degeneracy
directions between the 
EFT parameters in the bispectrum
are different from those implied
by the priors, and hence 
combining the two 
allows one to gain 
more information from the degeneracy breaking.

Nominally, the growth of structure in 
the  $P_\ell$+BAO+$Q_0$+$B_0$ 
analysis is constrained to 
better than $3\%$.
However, we caution these limits are preliminary.
At our new level of precision, we may have to 
re-calibrate the scale cuts and analyses 
choices for the bispectrum and $Q_0$
made in~\cite{Philcox:2021kcw}. This 
requires a detailed study 
which we defer to a separate publication. 

 \begin{figure*}[ht!]
\centering
\includegraphics[width=0.8\textwidth]{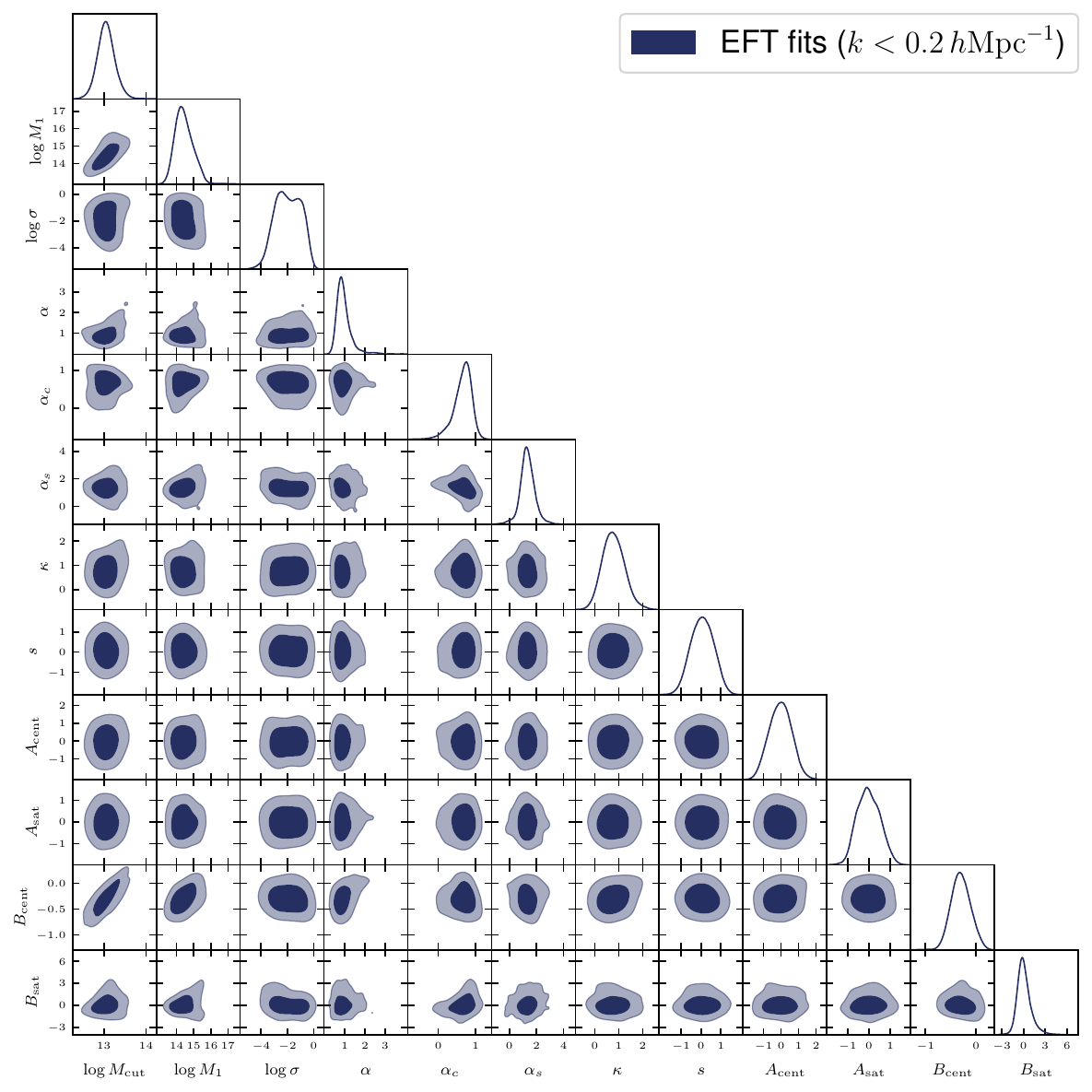}
   \caption{Constraints on HOD parameters of the 
   BOSS NGCz3 galaxies
   from the 
   EFT parameters of the  
   galaxy power spectrum. 
    } \label{fig:hods_boss}
\end{figure*}

\subsection{On the origin of the structure
growth anomaly in the context of the simulation-based priors}

In this section we report the 
results of our investigation of the 
origin of the low $\sigma_8$
value in our analysis. 
We limit ourselves here with the main
conclusions of our study,
and refer an interested reader  
to Appendix~\ref{app:lowsigma8}
for technical details.

The first relevant observation
that one can make is that the posteriors
for $b_1$ are dominated by 
the likelihood.
Although our $b_1$
prior is peaked at low $b_1$,
it does not ``pull'' the posterior values
of this parameter to lower values. 
Indeed, our posteriors
for $b_1$ are much narrower than the 
width of the prior, and the shape 
of the $b_1$ prior is quite flat 
in the range $b_1\in [2,3$], relevant 
for the BOSS data.

The second relevant observation 
is that the $\approx 2\sigma$ shift of the mean value of $\sigma_8$ due to SBP is driven
by the high-$k$ end of the 
data. To verify this we re-analyzed
the BOSS data with the conservative priors
and SBP with a momentum cut $\kmax=0.1~\hMpc$,
and found highly consistent results, $\sigma_8 = 0.696_{-0.055}^{+0.044}$ (conserv. priors)
and $\sigma_8 = 0.677^{+0.033}_{-0.039}$ (SBP), 
with the difference between the means at the level 
of $\sim 0.5\sigma$.

Investigating the high-$k$ part of the data,
we found that it can be fitted with a high 
value of $\sigma_8\approx 0.74$, but 
this requires highly unlikely values
of the EFT parameters. Specifically, 
as can be appreciated from table~\ref{eq:tab3},
the high $\sigma_8$ fit requires
$b_2^{\rm NGCz3}\approx -1$
and $b_1^{\rm NGCz3}\approx 2.2$,
which are impossible to reproduce
within the HOD framework, i.e. we have no samples with these values.\footnote{The impossibility of these values is dictated by the shape of the HOD for Luminous Red Galaxies~\cite{Ivanov:2024dgv}.} The tension 
further extends to other EFT parameters,
such as the stochastic counterterms 
$\alpha_0$ and $\alpha_2$. The high $\sigma_8$
fit for NGCz3 requires $\alpha_0\approx 0.4$
and $\alpha_2\approx -2$, which correspond
to HODs with unlikely large satellite fractions.  (The NGCz1 parameters 
quoted
in table~\ref{eq:tab2} exhibit a similar
behavior.) 
All in all, 
the likelihood of EFT parameters required
to fit the data at $\sigma_8=0.74$ 
appears to be 
quite low from the HOD modeling viewpoint.

Once the EFT parameters are restricted
by the HOD-based priors, it becomes impossible
to maintain a good fit to data at 
large $\sigma_8$ values. The lower 
$\sigma_8$ values we obtain appear
to be ``a compromise'' between the HOD 
priors
and the data likelihood. This can be estimated
on the basis of the $\chi^2$
statistics. If we consider only the 
likelihood of the galaxy $P_{0,2}$ 
data, 
the raw $\chi^2_{\text{like}}$ of 
the best fit at $\sigma_8 =0.67$ 
from the SBP analysis is notably worse
than that of the conservative analysis with 
$\sigma_8 = 0.74$: 
\be 
\label{eq:chi2}
\Delta \chi^2_{\text{like}}=\chi^2_{\text{like}}\Big|_{\rm SBP}-
\chi^2_{\text{like}}\Big|_{\rm cons.}\approx 8\,.
\ee 
However, if we compute the 
``effective $\chi^2$'' of the prior
as $\chi^2_{\text{eff, prior}}=-2\ln \mathcal{L}_{\rm EFT}$ (with $\mathcal{L}_{\rm EFT}$ being the likelihood of the EFT parameters from our sample estimated 
by  the normalizing flow), we find a very significant improvement over the 
conservative analysis,
\be  
\Delta \chi^2_{\text{eff, prior}}=\chi^2_{\text{eff, prior}}\Big|_{\rm SBP}-
\chi^2_{\text{eff, prior}}\Big|_{\rm cons.}\approx -391\,.
\ee 
This large improvement\footnote{As a point of 
comparison, on can use $\Delta \chi^2$ as a function of confidence levels
for 10 parameters (as in our $P_{0,2}$
analysis), $\Delta \chi^2=11.5~(68\%)$, 
$18.4~(95\%)$,  
as follows from the $\chi^2$ distribution. 
}  
mostly comes from the very unlikely values
of the $\{b_1,b_2\}$
pair ($290$ units of
the effective $\chi^2$). The rest is
distributed between
the RSD counterterms ($33$ units)
and the real space EFT parameters
($68$).

As a final note, it is worth
pointing out that 6.05 out of 8 units of the $\chi^2$ difference 
between the SBP and conservative
best-fits in eq.~\eqref{eq:chi2}
stem from the power spectrum 
monopole $P_0$
data, which is not sensitive
to $\sigma_8$ in linear theory. 
Taken together with 
the difference in the inferred values 
of $b_1\sigma_8$ between the 
conservative and SBP analyses (see table~\ref{eq:tab3}),
the dependence
of our results on the 
high-$k$ end, this
suggest that the shifts 
and improvements in $\sigma_8$
constraints 
that we have obtained 
have resulted from non-linear
one-loop corrections. 
Indeed, in principle, the 
one-loop corrections break the 
degeneracy between $b_1$
and $\sigma_8$ (see e.g.~\cite{Chudaykin:2020hbf}),
but this mechanism is inefficient 
at the power spectrum level because
of the degeneracy with the 
bias parameters. 
The degeneracy with the non-linear
bias parameters also affects the 
extraction of $b_1\sigma_8$
from the monopole data. 
The HOD priors provide 
non-perturbative information
on the relations between the bias parameters, thereby breaking 
this degeneracy, and allowing 
to extract the extra
$\sigma_8$ information
from the amplitude 
of the one-loop power spectrum monopole
contributions.

\subsection{Learning galaxy-halo 
connection
from EFT parameters on large scales}

One interesting application of our
joint sample of EFT and HOD parameters
is that we can build a model for the 
mapping $\{\theta_{\rm EFT}\}\to \{\theta_{\rm HOD} \}$ and use it to convert the
MCMC samples for EFT parameters 
from data into HOD samples. 
This approach allows us to 
constrain galaxy properties
from large scales, and get
insights into the optimal HOD 
parameters for a given galaxy population
even without carrying out a full
HOD-based SBI on the data. 
This approach also provides
a consistency test between 
large and small scale galaxy clustering
analyses. 

As an concrete example, we have 
built a model 
for the distribution
of HOD parameters conditioned 
on the EFT parameters following 
the methodology of~\cite{Ivanov:2024hgq}, and 
generated HOD samples 
from the MCMC chains 
for BOSS NGCz3 galaxies. 
This sample was previously analyzed 
with various HOD-based SBI techniques, 
e.g.~\cite{Paillas:2023cpk,Valogiannis:2023mxf}.
The resulting corner plot is shown in fig.~\ref{fig:hods_boss}. 
The posteriors of HOD parameters are quite wide,
which is expected given that they 
are constrained using the large scale 
modes only. Nevertheless, our 
HOD samples for threshold host halo masses $M_{\rm cut}$
and $M_{1}$, 
and some other parameters, 
are narrower than the priors.
Besides, we also see a marginal evidence
in favour of a negative
assembly bias parameter for 
centrals $B_{\rm sat}$. 

Our results are consistent with 
those of the HOD-based analysis 
utilizing the density split statistic
~\cite{Paillas:2023cpk}. 
We note, however, that 
in contrast to~\cite{Paillas:2023cpk}
we use only large scales ($k_{\rm max}=0.2~\hMpc$), 
our HOD model is more flexible (i.e. 
we have additional assembly bias 
parameters $A_{\rm cent/sat}$)
and our priors for HOD parameters are wider
than those considered in~\cite{Paillas:2023cpk}.
These factors resulted in the reduction
of constraining power as compared to~\cite{Paillas:2023cpk}.

\section{Discussion and Outlook}
\label{sec:disc}

We have generated the largest to date sample 
of EFT parameters for galaxies 
from HOD-based simulations. We suggest to use this sample 
as a prior in EFT-based full shape analyses. Our main 
results are listed in Sec.~\ref{sec:main}. Let us now
compare our results with the literature, 
discuss the main lessons learned, and outline the
directions for future improvement.

It has been known for a long time that the EFT-based
full-shape constraints can improve dramatically with 
better priors on EFT parameters.
For example, the constraint 
on the amplitude of primordial fluctuations $A_s$
(strongly correlated with $\sigma_8$ in full-shape analysis),
improves by more than a factor of 2 at $\kmax=0.2~\hMpc$ once the EFT parameters are fixed~\cite{Wadekar:2020hax}.
In this work we found a similar, though slightly 
less impressive improvement on $\sigma_8$. 
Therefore, our results are not so surprising. 
The other parameters, such as $\omega_{cdm}$
and $H_0$ do not respond to SBP 
as strongly
as suggested by~\cite{Wadekar:2020hax} 
because the eventual distribution
of the EFT parameters is not narrow enough
to break degeneracies relevant for these parameters. (Note
that~\cite{Wadekar:2020hax} assumed the perfect knowledge of all
EFT parameters, which does not take place for HOD models.)

Naively, our results appear in conflict with those of
ref.~\cite{Kobayashi:2021oud}, which analyzed 
the same data sample as us with the HOD-based emulator, 
but found very minor improvements 
at $\kmax\simeq 0.2~\hMpc$ over the usual EFT analysis
with conservative priors. We argue, however, that there is no
contradiction between our work and ref.~\cite{Kobayashi:2021oud} because many technical aspects our analyses are quite different.  
First, \cite{Kobayashi:2021oud} used 
an HOD-based emulator for the power
spectrum only, while here we calibrate our 
priors from the entire galaxy field, including 
higher-order correlations. This 
way our priors include clustering information
beyond the two point function. 
Second,
\cite{Kobayashi:2021oud} 
marginalizes over the additional constant shot noise contribution
similar to our $\alpha_0$. 
This parameter appears to be 
correlated with $\ln (10^{10}A_s)$, which 
introduces additional uncertainties 
in the $\sigma_8$ limit. 
In our approach we assume that $\alpha_0$ is fully determined by 
HOD models, which provides extra information.
Third, we use 
an extended HOD model
with assembly bias parameters. 
It will be interesting to compare our approach with 
that of ref.~\cite{Kobayashi:2021oud}
with similar analysis settings. 

In terms of the final error bars, our results 
appear to be similar with those of 
simulation-based analyses of refs.~\cite{Lange:2021zre,Paillas:2023cpk,Valogiannis:2023mxf}.
For instance, \cite{Lange:2021zre} finds
a 5$\%$ constraint on $f\sigma_8$
from the BOSS LOWZ sample, which is similar to our
$\approx 6\%$ constraint from the analogous 
NGCz1 data chunk. We also find $\sim 3\%$
constraint on $\sigma_8$
from the BOSS galaxy power spectrum, which is 
similar to the $2.3\%$ limit from~\cite{Valogiannis:2023mxf}.
\cite{Paillas:2023cpk} applied their pipeline to the BOSS CMASS data only, but the final constraints appear to be 
similar to ours in terms of the $\sigma_8$
errorbar.
Note however, that our datasets are very different. 
While~\cite{Valogiannis:2023mxf} uses wavelet scattering 
transforms, our main measurement is based purely on the two-point function. 
This suggests an interesting hypothesis 
that the small-scale beyond the two point function
information in HOD-based SBI actually mostly constrains
the HOD parameters. The better knowledge of HOD parameters
then leads to a better cosmological parameter estimation
from large scales. This picture suggests that the
bulk of the cosmological information
is stored in the large-scale power spectrum.
It will be interesting to test this hypothesis in future. 

The posteriors of our SBP-based analyses of BOSS
are consistent with the traditional ones obtained 
with conservative priors. The priors mostly change 
the widths of the cosmological parameter contours, 
while their positions
remain mostly unaffected. 
In particular, our analyses confirm reports 
of the $\sigma_8$ tension
from the earlier large-scale EFT analyses 
of the BOSS data with conservative priors. 
This additionally 
confirms that the preference for the low $\sigma_8$
in the BOSS data is not 
driven by priors. It will be interesting to see 
if the significant tension continues to be present in the DESI data. 

We note that our priors were derived for 
the HOD designed for luminous red galaxies. 
Other galaxy types, such as 
emission line galaxies (ELGs), 
require a different form of the HOD, and hence
our priors may not be applicable to them. 
In addition, the HOD models do not include
physical effects such as the baryonic feedback 
and the dependence on the past evolution. 
Therefore, it will be important to extend 
our priors to new galaxy types and include additional
effects beyond the HOD models. 
This suggests that
the EFT-based analysis 
of public data on ELGs and quasars from
eBOSS~\cite{deMattia:2020fkb,eBOSS:2020uxp,Ivanov:2021zmi,Chudaykin:2022nru}
with 
simulation priors 
should be among the first 
extensions of our work. 
In addition, it will be interesting to 
test our priors against other galaxy formation approaches,
such as hydrodynamical simulations and
abundance matching.

Our sample of EFT and HOD parameters
can be used to build conditional models, 
which can help understand the physics 
behind EFT parameters~\cite{Ivanov:2024hgq}.
These conditional models can also be used 
to translate results of the SBI to 
EFT-based analyses and vice versa, which can be a useful 
tool to test the consistency of large-scale 
and small-scale parameter estimation pipelines. 
We leave this and other research directions 
for future work.

\section*{Acknowledgments}

We thank 
Ryan Abbott, 
Kazuyuki Akitsu, 
Stephen Chen, 
Will Detmold,
Cora Dvorkin, 
Yosuke Kobayashi, 
Siddharth Mishra-Sharma,
Chirag Modi, 
Azadeh Moradinezhad,
Oliver Philcox,
Marko Simonovi\'c,
Fabian Schmidt, 
Uros Seljak, 
Jamie Sullivan, 
Masahiro Takada, 
Georgios Valogiannis,
and Matias Zaldarriaga
for useful discussions.
This work is supported by the National Science Foundation under Cooperative Agreement PHY-2019786 (The NSF AI Institute for Artificial Intelligence and Fundamental Interactions, \url{http://iaifi.org/}). This material is based upon work supported by the U.S. Department of Energy, Office of Science, Office of High Energy Physics of U.S. Department of Energy under grant Contract Number  DE-SC0012567. AO acknowledges financial support from the Swiss National Science Foundation (grant no CRSII5{\_}193826). MWT  acknowledges financial support from the Simons Foundation (Grant Number 929255).

\section*{Data availability}

Our work is based on the publicly available 
\texttt{Abacus} simulation~\cite{Maksimova:2021ynf}
\url{https://abacussummit.readthedocs.io/en/latest/}
and \texttt{Quijote} simulation data~\cite{Villaescusa-Navarro:2019bje} \url{https://quijote-simulations.readthedocs.io/en/latest/}.

\appendix

\section{Time Sliced Perturbation Theory for Shifted Correlators}
\label{sec:tspt}

In this section we carry our IR resummation for the correlators of the 
shifted fields using 
time-sliced perturbation theory~\cite{Blas:2015qsi,Blas:2016sfa,Ivanov:2018gjr,Vasudevan:2019ewf}. 

We start our review of TSPT with the integral equations 
for Eulerian perturbation theory 
of the pressureless perfect fluid in Fourier space~\cite{Bernardeau:2001qr}, 
\be
\label{eq:eomsreal}
\begin{split}
& \d_\e\delta_\k -\T_\k=\int [dp]^2\delta^{(3)}(\k-\p_{12})
\alpha(\p_1,\p_2)\T_{\p_1}\delta_{\q_2}\,,\\
& \d_\e \Theta_\k +\frac{1}{2}\Theta_\k - \frac{3}{2}\delta_\k  \\
& \quad \quad \quad = \int [dp]^2\delta^{(3)}(\k-\q_{12})\beta(\p_1,\p_2)\T_{\p_1}\T_{\p_2}\,,
\end{split}
\ee
where $[dp]=d^3p/(2\pi)^3$, $\delta^{(3)}=(2\pi)^3\delta^{(3)}_D$,
and the non-linear kernels
\be
\label{alphabetareal}
\begin{split}
\alpha(\k_1,\k_2)\equiv\frac{(\k_1+\k_2)\cdot \k_1}{k_1^2}\,, \\
\b(\k_1,\k_2)\equiv\frac{(\k_1+\k_2)^2(\k_1\cdot \k_2)}{2k_1^2k_2^2}\,. 
\end{split}
\ee
The density field $\delta$ can always be expressed in terms of $\T$
as 
\be
\label{eq:psi1}
\begin{split}
&\delta_\k=\delta[\T;\e,\k]\equiv\\
& \sum_{n=1}^\infty\frac{1}{n!}
\int [dp]^n K_n^{(r)}(\p_1,...,\p_n)\,\delta^{(3)}(\k-\p_{1...n})
\prod_{j=1}^n\T(\e,\p_j)  \,,
\end{split}
\ee
with $K_1^{(r)}=1$.
One can use representation \eqref{eq:psi1} 
to eliminate the density field from Eq.~\eqref{eq:eomsreal}. 
The final equation for the velocity divergence
only reads:
\be
 \label{eq:In}
 \begin{split}
&\d_\e\T(\e,\k)= \mathcal{I}[\T]\equiv\\
&\sum_{n=1}^\infty\frac{1}{n!}
\int [dq]^n I^{(r)}_n(\p_1,...,\p_n)\,\delta^{(3)}(\k-\p_{1...n})
\prod_{j=1}^n\T(\e,\p_j)  \,,
 \end{split}
\ee
with the growing mode condition
$I^{(r)}_1\equiv 1$. 
The kernels $K^{(r)}_n$ and $I^{(r)}_n$  
can be calculated
recursively~\cite{Ivanov:2018gjr}.

The central object of 
TSPT is the generating 
functional for 
the cosmological fields velocity divergence and density fields $\T$ and $\delta$:
\be
\label{eq:ztfp}
\begin{split}
& Z[J,J_\delta;\e]=\int [\mathcal{D}\T]\;{\mathcal P}[\T;\e]\;\exp\left\{J\T+J_{\delta}\delta\right\}\,,\\
&J\T+J_{\delta}\delta\equiv \int [dk] \left(\T_{\k} J(-\k)+\delta[\T;\e,\k]J_{\delta}(-\k)\right)\,.
\end{split}
\ee
Functional derivatives with respect to the  sources $J$ and
$J_\delta$ produce equal-time
correlation functions of $\T$ and $\delta$.
${\mathcal P}$ above is the 
probability density functional that 
satisfies the Liouiville equation,
\be
\label{Liouville}
\frac{\d}{\d \e} \mathcal{P}[\T;\e] + \int [dk]\frac{\delta}{\delta
  \Theta(\k)}(\mathcal{I}[\T;\e]\mathcal{P}[\T;\e])=0\,. 
\ee
In perturbation theory one can expand
$\mathcal{P}[\T;\e]$ as
\be
\label{eq:statweight}
\begin{split}
& \mathcal{P}[\T;\e]=\mathcal{N}^{-1}\exp\left\{-W\right\} \,,\\
& W\equiv \sum_{n=1}^{\infty}\frac{1}{n!}\int
[dk]^n\; 
\Gamma_n^{(r)\,tot}(\e;\k_{1},...,\k_{n})\; \prod^n_{j=1} \T_{\k_j}\,,
\end{split}
\ee
where ${\cal N}$ is a normalization constant. Substituting this
representation into (\ref{Liouville}) and 
using
Eq.~\eqref{eq:In} we obtain the following chain of equations on  
the vertices,
\begin{equation}
\begin{split}
  & \d_{\e}\Gamma_n^{(r)\,tot}(\eta;\k_1,...,\k_n) \\
  & +\sum_{m=1}^n \frac{1}{m!(n-m)!} \sum_{\sigma} I^{(r)}_m(\eta;\k_{1},...,\k_{m}) \\
  &\times \Gamma_{n-m+1}^{(r)\,tot}(\eta; \sum_{l=1}^m \k_{\sigma(l)},\k_{\sigma(m+1)},...,\k_{\sigma(n)}) 
    \\
   &=\delta^{(3)}_D\left(\k_{1...n}\right) \int [dp] I^{(r)}_{n+1}(\eta; \p, \k_1,...,\k_n)\;.
  \label{eq:liouvillegamma}
\end{split}
\end{equation}
The solution to this equation is decomposed into 
two pieces, 
\be 
\label{eq:splitCn}
\Gamma_n^{(r)\,tot}=\Gamma^{(r)}_n+C^{(r)}_n\,,
\ee
where $\Gamma^{(r)}_n$ 
is the homogeneous solution,
while $C^{(r)}_n$ 
are ``counterterms'' that stem 
from the inhomogeneous equations. 
The latter are irrelevant for IR 
resummation so we will ignore them
in what follows. 

For the Gaussian initial the vertices $\Gamma^{(r)}_n$ can be computed 
exactly. In particular, their 
time-dependence factorizes,
\be
\Gamma^{(r)}_n= \delta^{(3)}(\k_{1...n})\frac{\bar \Gamma'^{(r)}_n}{g^2(\e)}\,,
\ee
where $g(\e)\equiv D_+(\e)$
plays a role
of the coupling constant in TSPT. 
The key observation that makes 
TSPT a useful representation 
is that IR singularities 
of Eulerian perturbation theory
at low momenta are contained
in $\bar \Gamma'^{(r)}_n$. The second 
important fact is that only these vertices
depend on the initial power spectrum,
in contrast to $K^{(r)}_n$ and $C_n$. 
This allows one to identify
enhanced diagrams and resumm them
to all orders in perturbation
theory. For the $\T$ auto spectrum, 
this leads to the following 
one-loop IR resummed 
expression:
\be 
\label{eq:P1Lres}
\begin{split}
P^{\rm 1-loop, IR}_{YY}(k)=
P^{\rm tree}_{YY}
[e^{-\hat{\mathcal{S}}}(1+\mathcal{S})P_{11}]
+P^{\rm 1-loop}_{YY}
[e^{-\hat{\mathcal{S}}}P_{11}]\,,
\end{split}
\ee 
where $\{YY\}=\{\T\T\}$, 
and the operator $\hat{\mathcal{S}}$
is formally defined 
as
\be 
\label{eq:Shat}
\hat{\mathcal{S}}P_{11}=\int_{|\p|\leq \Lambda_{IR}} P_{11}\frac{(\k\cdot\p)^2}{p^4}(1-\cosh(\p\nabla_{k'}))P_{11}(k')\Big|_{k'=k}
\ee 
This formula 
requires some explanation. 
The original 
works on TSPT 
assumed a decomposition
of the linear matter power
spectrum 
into the wiggly and 
smooth parts to separate 
the BAO from the rest 
of the spectrum. While the latter
makes IR resummation
conceptually simple, 
this separation is,
strictly speaking, 
is only an approximation. 
But the TSPT formulas 
can be used even without 
performing the actual 
split. In this case, the split
should be considered a formal 
tool for power counting.
Using the notations of 
ref.~\cite{Blas:2016sfa},
from 
the formal point of view, 
one can always ``pull'' $P_{nw}$
through the operator 
$e^{-\hat{\mathcal{S}}}$
back to $P_w$ because it does not
change the IR power counting.
Analogously, the relevant 
expressions for the IR enhancements
in vertices $\Gamma_n^{(r)}$
can be rewritten in terms of the 
finite differences of the total linear power spectra. 
Therefore, the wiggly-smooth 
split is not a cornerstone 
part of the TSPT formalism. 

The integral that appears
in eq.~\eqref{eq:Shat}
can be transformed 
to position space, 
in which case the final 
IR resummed expression
will be identical 
to that of Lagrangian
EFT~\cite{Blas:2016sfa}. 
That said, we will continue 
with the wiggly-smooth split
in what follows because while being approximate, it
offers a very significant
computational advantage. 
Ref.~\cite{Chudaykin:2020aoj}
suggested that the error 
associated with this split,
while being formally small, 
can be absorbed into
the nuisance parameters.\footnote{This is confirmed by 
comparison with numerical simulations
such as PT Challenge.}
Hence, in order to match 
the \texttt{CLASS-PT} values of EFT 
parameters, we have to use 
the wiggly-smooth decomposition
for consistency. 

IR resummation of the density 
field is very similar to that
of the velocity divergence.
Once we establish that the kernels 
eq.~\eqref{eq:psi1}
do not have IR singularities, 
it follows 
straightforwardly that they can only come 
from vertices of the velocity
field. Their resummation
yields the same formula
as \eqref{eq:P1Lres} but with 
$\{YY\}=\{\delta\delta\}$. 

The bias tracers 
can be described in TSPT 
in full analogy 
with the matter density field 
$\delta$. To that end one rewrites 
the bias expansion as 
\be
\label{eq:psi2}
\begin{split}
&\delta_g(\k)=\delta_g[\T;\e,\k]\equiv\\
& \sum_{n=1}^\infty\frac{1}{n!}
\int [dq]^n M_n^{(r)}(\p_1,...,\p_n)\,\delta^{(3)}(\k-\p_{1...n})
\prod_{j=1}^n\T(\e,\p_j)  \,,
\end{split}
\ee
and finds that the Eulierian
bias kernels $M_n^{(r)}$
do not contain IR singularities. 
This is the consequence 
of the fact that the bias 
expansion must satisfy 
the equivalence principle. 
As a result of that, 
only the diagrams that contain 
the vertices $\bar \Gamma_n$
have to be resummed. 
This produces the same result as 
\eqref{eq:P1Lres} but with 
$\{YY\}=\{gg\}$. 

The formal generating functional
for correlators of the shifted fields is given by
\be
\label{eq:ztfp2}
\begin{split}
& Z[J_{\tilde{1}},J_{\Delta\delta_g};\e]=\int [\mathcal{D}\T]\;{\mathcal P}[\T;\e]\;\exp\left\{J_{\tilde{1}}\tilde{\delta}_1
+J_{\Delta\delta_g}\Delta\delta_g\right\}\,.
\end{split}
\ee
From the above discussion it is clear that 
in order to show that the algorithm 
\eqref{eq:P1Lres}
applies to the shifted operator 
field, it is sufficient 
to demonstrate that the kernels
that appear in an analog of 
eq.~\eqref{eq:psi2}
for shifted operators
are IR safe. 
Plugging the ansatz
\be
\label{eq:delta1TSPT}
\begin{split}
&\tilde\delta_1(\k)=\tilde\delta_1[\T;\e,\k]\equiv\\
& \sum_{n=1}^\infty\frac{1}{n!}
\int [dq]^n L_n^{(r)}(\p_1,...,\p_n)\,\delta^{(3)}(\k-\p_{1...n})
\prod_{j=1}^n\T(\e,\p_j)  \,,
\end{split}
\ee
we find a sequence of IR safe expressions
\be 
\begin{split}
& L^{(r)}_1 = \tilde K_1=1\,,\\
& L^{(r)}_2 = 2\left(
\tilde{K}_2
- G_2\right)=\frac{8}{7}\left(1-\frac{(\k_1\cdot\k_2)^2}{k_1^2 k_2^2} \right)\,,
\end{split}
\ee 
etc., where 
$\tilde{K}_2$ are Eulerian
kernels 
for $\tilde{\delta}_1$
and $G_2$ is the velocity 
divergence kernel in 
standard perturbation theory~\cite{Bernardeau:2001qr}.

As for the $\Delta \delta_g=\delta_g - b_1\tilde{\delta}_1$
field, it is a difference 
of two terms whose 
series expansions~\eqref{eq:delta1TSPT}
and \eqref{eq:psi2}
are IR safe, and hence TSPT
$\Delta \delta_g$ kernels $\Delta M_n^{(r)}$
are IR safe as well. 
This implies 
that eq.~\eqref{eq:P1Lres}
describes the end result of IR resummation for both $\langle \tilde{\delta}_1 \tilde{\delta}_1\rangle$
and $\langle \Delta \delta_g \tilde{\delta}_1\rangle$ correlators that 
appear in the transfer function
calculations. 

Let us discuss now redshift space distortions
in TSPT. The key idea is to 
consider the redshift space coordinate 
mapping ($v_z=\hat{{\bm z}}\cdot {\bm v}$, with $v$ being the matter velocity
field ${\bm v}$),
\be 
\label{eq:rsdmap}
{\bm s}={\bm x} + \hat{{\bm z}}v_z/(aH)~\,,
\ee 
as a flow in a second time $\mathcal{T}$
that ranges from 0 to $1/(aH)$,
\be 
{\bm s}={\bm x} + \hat{{\bm z}}v_z\mathcal{T}~\,.
\ee
The initial conditions in this fictitious 
flow
correspond to real space fields.
The fields at $\mathcal{T}=1/(aH)$
are redshift space fields $\T^{(s)}$ and $\delta^{(s)}$. 
The flow then satisfies the usual free
hydrodynamical equations,
\be
\label{eq:finaleoms}
\begin{split}
& \d_\mathcal{F}\delta^{(s)}_\k -\frac{k_z^2}{k^2}\T^{(s)}_\k=\int [dp]^2\delta^{(3)}(\k-\p_{12})
\alpha^{(s)}(\p_1,\p_2)\T^{(s)}_{\p_1}\delta^{(s)}_{\p_2}\,,\\
& \d_\mathcal{F} \Theta^{(s)}_\k  =  \int [dq]^2\delta^{(3)}(\k-\p_{12})\beta^{(s)}(\p_1,\p_2)\T^{(s)}_{\p_1}\T^{(s)}_{\p_2}\,,
\end{split}
\ee
where $\mathcal{F}=f\mathcal{T}aH$ and
\be 
\label{eq:kernab}
\begin{split}
& \a^{(s)}(\p_1,\p_2)\equiv
\frac{p_{1,z}(p_{1,z}+p_{2,z})}{p_1^2}\,,\\
& \b^{(s)}(\p_1,\p_2)\equiv \frac{(\p_1+\p_2)^2p_{1z}p_{2z}}{2p^2_1p_2^2}\,.
\end{split}
\ee
The above flow induces the transformation
of the TSPT generating function
similar to the transformation
produced by the usual 
cosmological evolution. 
The new (redshift space) generating
functional will now have 
vertices
\be
\label{eq:TSPTeq1}
\begin{split}
& \d_\mathcal{F} \Gamma^{(s)\,tot}_n(\F;\k_1,...,\k_n)
\\
&+\sum_{m=1}^n \frac{1}{(n-m)!m!}\sum_{\sigma}
I^{(s)}_m(\k_{\sigma(1)},...,\k_{\sigma(m)}) \\
&\Gamma^{(s)\,tot}_{n-m+1}\Big(\mathcal{F};\sum_{l=1}^m\k_{\sigma(l)},\k_{\sigma(m+1)},...,\k_{\sigma(n)}\Big)\\
&=\delta^{(3)}\left(\sum_{i=1}^n \k_i \right)\int [dp]
I^{(s)}_{n+1}(\mathcal{F};\p,\k_1,...,\k_n)\,, 
\end{split} 
\ee
where the r.h.s. stems from 
the fictitious Eulerian dynamics 
equation~\eqref{eq:finaleoms} for the redshift space velocity divergence $\T^{(s)}$:
\be 
\begin{split}
& \d_\F \T^{(s)}_\k=\\
& \sum_{n=1}^\infty \frac{1}{n!}\int [dp]^n\delta^{(3)}(\k-\p_{1...n}) I^{(s)}_n(\p_1,...,\p_n)\T^{(s)}_{\p_1}...\T^{(s)}_{\p_n}\,.
\end{split} 
\ee
This new fictitious dynamics 
generates new IR divergences,
which can be traced back to the fact 
that the redshift-space mapping
depends on the velocity w.r.t. the observer's frame, 
which violates the equivalence
principle. This introduces
additional IR singularities
in vertices 
$\Gamma^{(s)}_n$ that 
are defined just like in eq.~\eqref{eq:splitCn}.
The redshift space density kernels $\mathcal{M}_n^{(s)}$ are IR safe. 
The IR enhancements 
from the vertices
can be IR resummed
in the diagrammatic expansion, 
yielding the result 
\be 
\label{eq:P1LresRSD}
\begin{split}
P^{\rm 1-loop, IR}_{YY}(k)&=
P^{\rm tree}_{YY}
[e^{-\hat{\mathcal{S}}^{(s)}}(1+\mathcal{S}^{(s)})P_{11}]\\
&+P^{\rm 1-loop}_{YY}
[e^{-\hat{\mathcal{S}}^{(s)}}P_{11}]\,,
\end{split}
\ee 
where $\{YY\}=\{\T\T\}$, 
and the operator $\hat{\mathcal{S}}^{(s)}$
is given by
\be 
\label{eq:Shat}
\begin{split}
&\hat{\mathcal{S}}^{(s)}P_{11}=P^{ab}P^{cd}k^ak^c\times\\
&\int_{p\leq \Lambda_{IR}}[dp] P_{11}(p)\frac{p^bp^d}{p^4}(1-\cosh(\p\nabla_{k'}))P_{11}(k')\Big|_{k'=k}~\,,
\end{split}
\ee 
where $P_{ab}=\delta_{ab}+f\hat{z}_a\hat{z}_b$.
As before, if we use the wiggly-smooth split, it produces a simple
expression,
which can be further simplified
using method of~\cite{Ivanov:2018gjr},
leading to the damping factor~\eqref{eq:damp_rsd}.

IR resummation of the matter 
and galaxy density 
fields in redshift space
takes the same form as 
eq.~\eqref{eq:P1LresRSD}
by virtue of the fact that 
the their redshift space TSPT 
kernels are IR safe.  

As far as the shifted operators 
are concerned, it is convenient 
to treat our forward model as 
a redshift space mapping of a particular 
Eulerian biased tracer in real space. 
The bias expansion for this tracer depends on the line-of-sight
just like in the case of selection effects~\cite{Desjacques:2018pfv}
or the Lyman alpha forest~\cite{Ivanov:2023yla,Ivanov:2024jtl}.
Then it is easy to show that the TSPT kernels 
of the shifted operators 
$L_n^{(s)}$
and $\Delta M_n^{(r)}$
are IR safe 
since the shifted fields do not break 
the equivalence principle.
This 
leads to eq.~\eqref{eq:irres_rsd}
for their correlators, 
analogous 
to the galaxy density 
field in redshift space. 

Alternatively, instead of doing the 
redshift space mapping 
with the full 
velocity field~\eqref{eq:rsdmap},
one can consider a mapping 
with the Zel'dovich velocity,
as we do in our redshift space
forward model. 
This can be interpreted
as a fictitious flow 
\be 
{\bm s}={\bm x} + \hat{{\bm z}}v^{\rm ZA}_z\mathcal{T}~\,,
\ee
where $v^{\rm ZA}$ is the velocity 
in the Zel'dovich approximation. 
The TSPT equations for this 
fictitious Zel'dovich 
flow
will be the same as~\eqref{eq:finaleoms}.
This implies that our Zel'dovich 
flow will automatically capture 
IR singularities
in the fictitious evolution. 
This is not surprising as
the IR enhancements 
precisely appear from 
the Zel'dovich 
displacements. 
Then one can compute the flow 
of the kernels of the shifted 
operators $L_n^{(s)}$
and $\Delta M_n^{(r)}$,
find that they are IR safe, 
and carry out the IR resummation 
in the usual way.

\section{On the accuracy 
of the perturbative description 
for the galaxy density field
in redshift space}
\label{sec:kmax}

This section pursues two main goals. 
On the practical side, 
we would like to determine 
$\kmax$ for our fits 
of $P_{\rm err}$
in redshift space, which 
should yield the stochastic counterterms 
$\alpha_{0,1,2}$. From the conceptual side,
we would like to understand 
what is the reach of the 
perturbative field for galaxies
and redshift space,
and whether there is any sign
of the breakdown of EFT
for galaxies 
on large scales, similar 
to the reports of the 
flattening of the stochastic
power spectrum for the dark matter
density field in ref.~\cite{Baldauf:2015zga}.

\begin{figure*}
\centering
\includegraphics[width=0.47\textwidth]{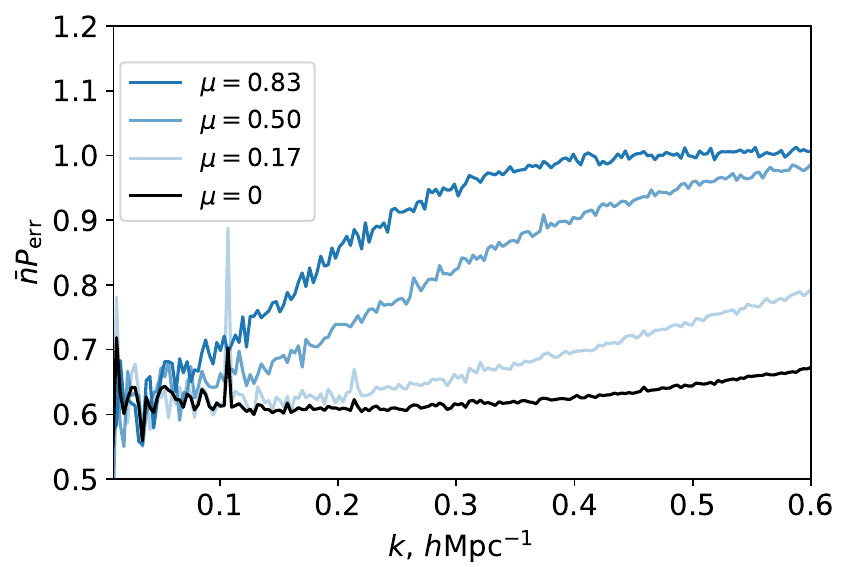}
\includegraphics[width=0.47\textwidth]{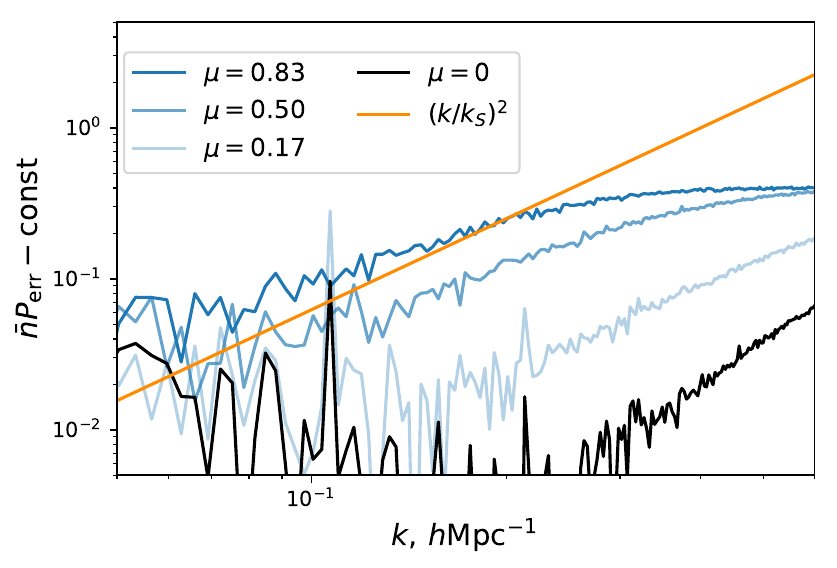}
   \caption{The noise power spectrum in redshift space (left panel),
   and its behavior after subtracting the constant white noise contribution (right panel). 
    } \label{fig:perr_0}
\end{figure*}

\begin{figure*}
\centering
\includegraphics[width=0.47\textwidth]{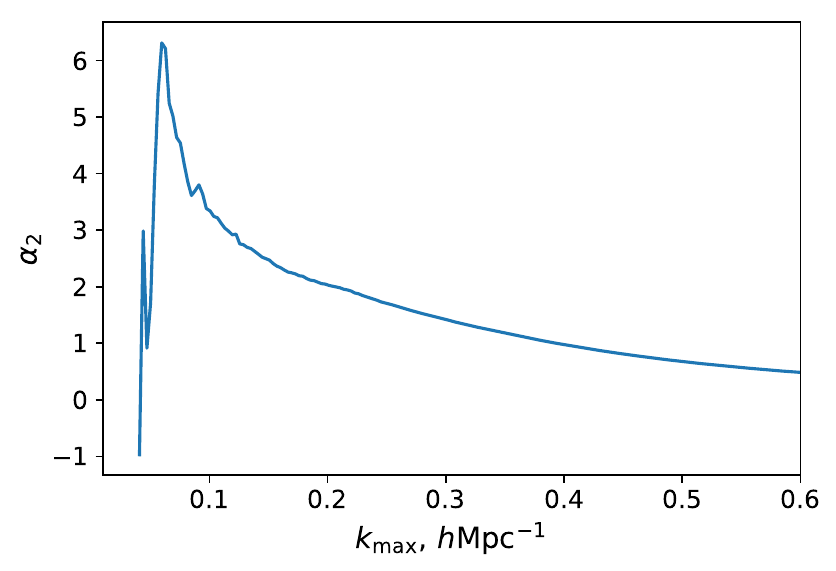} 
\includegraphics[width=0.49\textwidth]{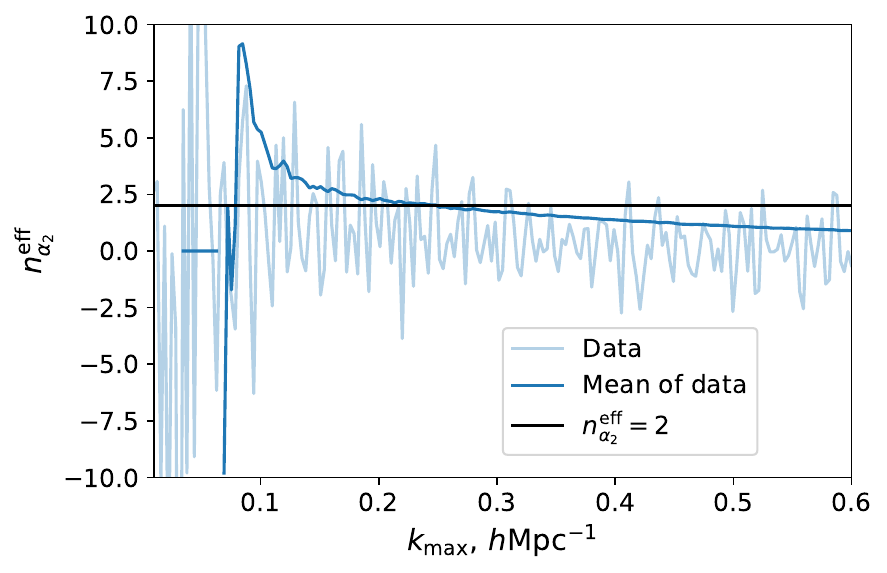}
   \caption{Left panel: the fit of $\alpha_2$
   from the error power spectrum as a function of $\kmax$. Right panel: the effective slope
   of the redshift space part 
   of the noise power spectrum $n_{\alpha_2}^{\rm eff}$
   as a function of $\kmax$
    measured from the data, 
   along with the cumulative mean up to 
   $\kmax$, and the EFT  
   prediction $n_{\alpha_2}^{\rm eff}=2$.
    } \label{fig:alpha2}
\end{figure*}

\begin{figure*}
\centering
\includegraphics[width=0.60\textwidth]{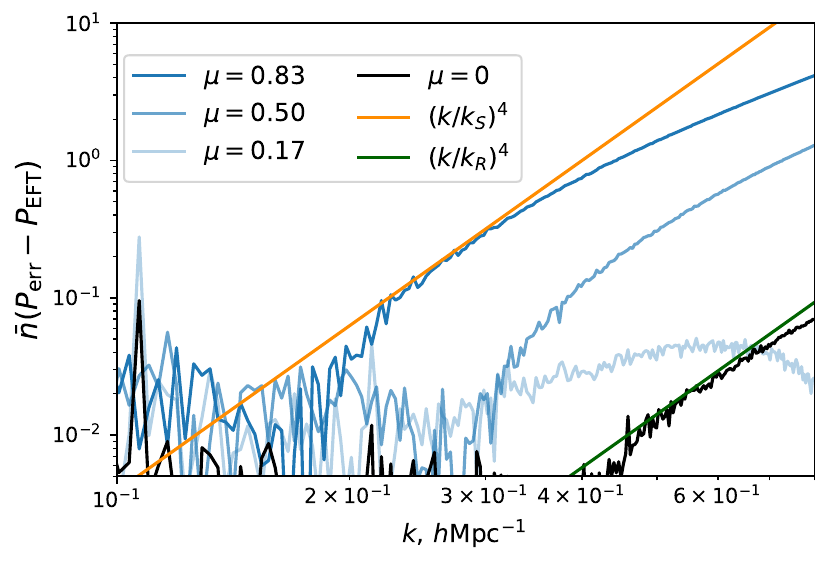}
   \caption{The residuals between the noise power spectrum of the BOSS-like HOD simulation and the next-to-leading order
   EFT model from eq.~\eqref{eq:rsd_stoch}.
   Straight lines show the predicted slopes 
   of the residuals for the next-to-next-to leading EFT corrections.
    } \label{fig:NNLO}
\end{figure*}

To answer these questions, 
we used HOD mocks 
produced for the large \texttt{Abacus}
boxes, which helps us reduce
the residual statistical 
scatter in our field-level 
calculations. 
For concreteness, we focus on
one particular HOD  model
similar to that of the BOSS CMASS
sample. The results are similar 
for other HOD models from our sample.

The noise power spectrum
is shown in fig.~\ref{fig:perr_0}.
As expected in theory, 
it is given by the white noise 
constant power on very large scales. 
On quasilinear scales
the noise depends both on 
the 
wavevector norm $k$
and its cosine with the line-of-sight $\mu$.
Its power spectrum 
has the theoretically 
expected scaling $k^2\mu^2$
up to $k\approx 0.2~\hMpc$, 
see the right panel of 
fig.~\ref{fig:perr_0}.
For $k>0.2~\hMpc$ the noise becomes 
more shallow, suggesting that
higher order corrections may become important. 
This is the first piece
of evidence that the choice of 
$\kmax=0.2~\hMpc$ is an adequate one. 

At a second step, we study the dependence
of our $\alpha_2$ measurements on $\kmax$.
We find that $\alpha_2$ exhibits 
significant scale dependence, see the left panel of 
fig.~\ref{fig:alpha2}. On large scales the main 
source of the scale dependence is
the statistical scatter, 
while on small scales it is
produced by higher order EFT 
corrections. To understand better where
the scale dependence due to the higher order corrections actually 
starts, we consider the redshift-space part of the noise:
\be 
P_{\rm err}-\frac{1}{\bar n}\left(\alpha_0+\alpha_1 \frac{k^2}{[0.45\hMpc]^2}\right)=P^{\rm FoG}_{\rm err}(k,\mu)\,,
\ee 
and calculate numerically the slope 
$n^{\rm eff}_{\alpha_2}$
of this spectrum in $k$ space, 
\be 
n^{\rm eff}_{\alpha_2}\equiv \frac{d\ln P^{\rm FoG}_{\rm err}(k,\mu)}{d\ln k}~\,.
\ee 
The numerical slope from the data 
is displayed in the right panel of 
fig.~\ref{fig:alpha2}. The $n^{\rm eff}_{\alpha_2}(k_{\rm max})$
curve becomes much smoother if 
we consider cumulative 
static averages of 
$n^{\rm eff}_{\alpha_2}$ 
up to $\kmax$, shown 
as the ``mean of data.''
We see that the mean of 
$n^{\rm eff}_{\alpha_2}$
crosses the theoretically expected 
value $2$ around $\kmax\approx 0.2~\hMpc$.
This is the second piece of evidence 
that confirms out baseline choice. 

Finally, fig.~\ref{fig:NNLO}
shows the residual between
$P_{\rm err}$
and the EFT model 
with the values of $\alpha_2$
and $\alpha_0,\alpha_1$
extracted at our baseline 
scale cuts $0.2~\hMpc$
and $0.4~\hMpc$,
respectively. 
First, we see that on large scales
the EFT model perfectly predicts
the measured noise power spectrum. 
The residuals between the 
theory and the data 
scale as $k^4$, 
which is precisely the 
prediction 
for the higher order corrections.
We find that the relevant EFT cutoff 
scales for the noise power spectrum
are $k_S\approx 0.45~\hMpc$
in redshift space and 
$k_R\approx 1.45~\hMpc$
in real space.

Fig.~\ref{fig:NNLO} shows 
that the noise power spectrum 
can be well described by the 
EFT derivative expansion 
even for wavenumbers 
larger than $0.2~\hMpc$ ($0.4~\hMpc$)
in redshift (real) space
provided that the appropriate 
higher order corrections are taken into 
account. 
Thus, unlike the dark matter case~\cite{Baldauf:2015zga}, 
we do not see 
evidence for the shallower shape 
of the noise power spectrum 
on the quasi-linear scales 
in the case of galaxies. 

We point out, however, 
that around the EFT cutoff,  
$k\approx 0.4~\hMpc$, 
the derivative expansion for the 
noise power spectrum breaks down.
Modeling the non-trivial scale-dependence of the stochastic noise at this scale 
may pose a serious challenge
to the phenomenological
``hybrid EFT'' approaches e.g.~\cite{Ibanez:2024uua}.

\section{Analytic proof of the approximate cosmology-independence of the HOD priors}
\label{app:HOD_analytic}

Let us present an analytic proof
that the cosmology-dependence 
of the HOD-based priors can be absorbed 
into the HOD parameters. 
Our proof will rely on four key assumptions:
\begin{enumerate}
    \item the HMF is universal (i.e. depends only on the peak height $\nu$),
    \item  the EFT parameters of halos depends only the peak height, 
    \item the standard real-space HOD parametrization~\eqref{eq:zheng_hod}
    \item power-law cosmology with $P_{11}\propto k^n$.  
\end{enumerate}

First, let us derive the relationship
between the EFT parameters of halos
and HOD galaxies.
We start by writing down the perturbed number density of halos as
\be 
n_h(\x)=\frac{d\bar n_h}{d\ln M}(1+\delta_h(\x))\,,
\ee 
where $\frac{d\bar n_h}{dM}$ is the halo mass function. Given an average 
HOD $\langle N_g\rangle $, 
we can obtain a similar expression for galaxies
\be
\begin{split}
n_g(\x)=\int dM \frac{d\bar n_h}{dM} \langle N_g \rangle_M (1+\delta_h(\x))\,,
\end{split}
\ee
with $\bar{n}_g = \int dM \frac{d\bar n_h}{dM} \langle N_g\rangle_M$, 
from which it immediately follows that 
\be 
\delta_g = \frac{n_g(\x)}{\bar n_g}-1
=\frac{1}{\bar{n}_g}\int dM \frac{d\bar n_h}{dM} \langle N_g\rangle_M \delta_h(\x)\,,
\ee
implying 
\be
\label{eq:eft_par_g_int}
b_{\mathcal{O}}^g = \frac{1}{\bar n_g}\int dM \frac{d\bar n_h}{dM} \langle N_g\rangle b_{\mathcal{O}}^h(M)\,.
\ee
Following the analytic halo models, 
let us assume that the HMF depends only on the peak height  $\nu = \frac{\delta_c}{\sigma_M(z)}$.
This relation defines the map between $M$ and $\nu$. Changing the variables in
the mass integral~\eqref{eq:eft_par_g_int}
we obtain
\be 
\label{eq:bg_via_nu_2}
b_{\mathcal{O}_a}^g = \frac{1}{\bar n_g} \int d\nu \frac{d\bar n_h}{d\nu}(\nu) \langle N_g\rangle_{M(\nu)} 
b_{\mathcal{O}_a}^h(\nu)\,,
\ee
where we took into account 
that the halo bias parameters
depend on $M$  via $\nu$, see
eq.~\eqref{eq:bh_via_nu}.
Note that $\bar \rho$ in the HMF is canceled by the denominator $\bar n_g$.

It is easy to show now that for the standard HOD, the expression above is approximately cosmology-independent. 
To that end, let us assume a power-law universe, 
\be 
P_{11}(k)=\frac{2\pi^2}{k^3_{\rm NL}} \frac{k^n}{k_{\rm NL}^{n}}\,.
\ee 
The amplitude 
of the linear matter power spectrum 
scales with redshift as $P_{11}\propto D^2_+(z)$,
implying 
\be 
k_{\rm NL}= k_{\rm NL}(z=0)D^{-\frac{2}{3+n}}_+(z)\,.
\ee 
The filtered mass variance in this cosmology 
is given by
\be
\begin{split}
\sigma_M^2(z) & = \frac{9\cdot 2^{-n}(1+n)\Gamma(n-1)}{n-3}\frac{\sin\left(\frac{\pi n}{2}\right)}{(k_{\rm NL}R)^{n+3}}\,,\\
& =\left({\tilde{M}_{\rm NL}}/{M}\right)^{\frac{n+3}{3}}\,,
\end{split}
\ee 
where the
cosmology
dependence is captured by 
\be 
\tilde{M}_{\rm NL}\equiv 
\left(\frac{9\cdot 2^{-n}(1+n)\Gamma(n-1)\sin\left(\frac{\pi n}{2}\right)}{n-3}\right)^{\frac{3}{n+3}}\frac{4\pi \Omega_m \rho_c}{3 k^3_{\rm NL}}\,,
\ee 
and we used $4\pi R^3\rho_c \Omega_m/3=M$. 
The peak height is given by 
\be
\label{eq:map}
\begin{split}
\nu = \delta_c \left(\frac{M}{\tilde{M}_{\rm NL}}\right)^{\frac{n+3}{6}}\equiv \left(\frac{M}{M_{\rm NL}}\right)^{\frac{n+3}{6}}\,,\quad M=M_{\rm NL}\nu^{\frac{6}{n+3}}\,.
\end{split}
\ee
We can use the above formulas 
to change the variables from $M$
to $\nu$ in the integral   
\eqref{eq:bg_via_nu_2},
so that the galaxy bias parameters
become universal. 
Using the standard HOD parametrization~\eqref{eq:zheng_hod}, we obtain the following re-mapping for the centrals:
\be
\begin{split}
\langle N_c \rangle & = \frac{1}{2}\left[1+\text{Erf}\left(\frac{\ln(M/M_{\rm cut})}{\ln10 \sqrt{2}\sigma}\right)\right] \\
& = \frac{1}{2}\left[1+\text{Erf}\left(\frac{\ln(\nu/\nu_{\rm cut})}{\ln10 \sqrt{2}\sigma'}\right)\right]\,,
\end{split}
\ee
where we defined 
\be 
\begin{split}
& \nu_{\rm cut} \equiv \frac{\delta_c}{\sigma_{M_{\rm cut}}}=
\left(\frac{M_{\rm cut}}{M_{\rm NL}}\right)^{\frac{3+n}{6}}\,,\\
&\sigma'=\frac{6}{n+3}\sigma\,.
\end{split}
\ee
This shows that the bias parameters
of the centrals are universal:
they depend
on cosmology and redshift only
via combinations $\sigma'$
and $\nu_{\rm cut}$, and hence a change 
in cosmology can be fully 
compensated by an adjustment 
of the HOD parameters. In this 
sense the cosmology-dependence
is degenerate with the HOD parameters. 
In particular, since $M_{\rm NL}\propto \sigma_8^{\frac{6}{n+3}}$, the change 
in $\sigma_8$ is equivalent to 
a rescaling of $\nu_{\rm cut}$ as
\be 
\label{eq:nucutabs}
\nu_{\rm cut}\to \nu'_{\rm cut} = \nu_{\rm cut}\frac{\sigma_8}{\sigma'_8}\,.
\ee 
In fact, the rescaling above 
removes the $\sigma_8(z)$ dependence 
\textit{exactly} even in 
a $\Lambda$CDM cosmology, 
as it simply follows from the definition
of the peak height parameter.

As for the satellites, 
using the mapping \eqref{eq:map}
we find 
\be
\label{eq:satNnu}
\langle N_s \rangle = \langle N_c \rangle \left(\frac{\nu^{\frac{6}{n+3}}-\kappa \nu^{\frac{6}{n+3}}_{\rm cut}}{M_1/M_{\rm NL}}\right)^{\alpha}\,.
\ee
The first relevant
observation is that one can 
completely absorb the dependence on 
$M_{\rm NL}$, which captures
the sensitivity to the mass 
fluctuation amplitude $\sigma_8(z)$
at the redshift of interest,
\be 
M_1' = M_1/M_{\rm NL}, 
\ee 
while the rest of it is absorbed into
$\nu_{\rm cut}$ through~\eqref{eq:nucutabs}. 
Thus, we have shown
that the dependence on
the amplitude 
$\sigma_8(z)$
can be absorbed 
into the HOD parameters
\textit{exactly}.

In general, however, it is impossible to fully
adsorb the dependence of the satellite HOD on the power spectrum slope $n$ 
by $M_1$, $\alpha$,
and $\kappa$.
One way out is to 
introduce an additional HOD parameter 
$\beta$ in \eqref{eq:satNnu},
\be 
\begin{split}
\frac{6}{n+3}\to \beta\,,
\end{split}
\ee
However, let us 
recall that for the satellites 
the physically interesting regime
is $M \gg \kappa M_{\rm cut}$. 
In this case the cosmology dependence can be absorbed into $M_1$ and $\alpha$
completely,
\be
\begin{split}
\langle N_s \rangle 
& \approx \langle N_c \rangle 
\left(\frac{\nu^{\frac{6\alpha}{n+3}}}{(M_1/M_{\rm NL})^\alpha}\right)\equiv \frac{\nu^{\alpha'}}{\nu^{\alpha'}_1}\,.
\end{split}
\ee

Let us note that our discussion
suggests that the breaking of the cosmology universality 
seems to be an artifact of the 
usual HOD
parametrization. From the 
approximate universality of the HMF, it is more natural 
to \textit{define}
the HOD models in terms
of the peak height, 
in which case the 
cosmology-independence
will be manifest. 
Therefore, the residual
cosmology dependence 
of the EFT parameters
of the HOD galaxies
may simply be 
an unphysical artifact
of the standard 
phenomenological
parametrization.

Let us estimate now 
a typical shift of the HOD
parameters needed to absorb
the cosmology dependence. 
Let us fix $M_{\rm cut}$ and consider two  
cosmologies with different $\sigma_8$. 
Let us compute now how much do we need to shift 
$M_{\rm cut}$ to compensate for the change in $\sigma_8$.
We have
\be 
M_{\rm cut} = M_{\rm NL}\nu^{\frac{6}{n+3}}_{\rm cut}
\ee 
Now we want to find $\nu'^{\frac{6}{n'+3}}_{\rm cut}$
that gives the same $b_{\mathcal{O}_a}^g$. 
For that we need 
to have the same HOD as a function of $\nu$
in both cosmologies.
For the centrals we have 
\be 
\frac{n'+3}{6}\ln(\nu/\nu'_{\rm cut})=\frac{n+3}{6}\ln(\nu/\nu_{\rm cut})\,,
\ee 
i.e. $\nu_{\rm cut}=\nu'_{\rm cut}$
since we assume that the effective tilt does not change, $n'=n$.
 We get
\be 
M'_{\rm cut}=\frac{M'_{\rm NL}}{M_{\rm NL}}M_{\rm cut}=\left(\frac{\sigma'_8}{\sigma_8}\right)^{\frac{6}{n+3}}M_{\rm cut}\,.
\ee
Since we are sampling $\log M_{\rm cut}$, we find 
\be 
\log M'_{\rm cut}-\log M_{\rm cut}=\frac{6}{n+3}\log \left(\frac{\sigma'_8}{\sigma_8}\right)\,.
\ee 
For $\sigma'_8=0.7$, $\sigma_8 = 0.8$, and $n=-2$ we get 
\be 
\log M'_{\rm cut}-\log M_{\rm cut}=-0.34\,,
\ee 
which is a very moderate shift that is much smaller than the width
of our prior on $\log M_{\rm cut}$.

\begin{figure*}[ht!]
\centering
\includegraphics[width=0.99\textwidth]{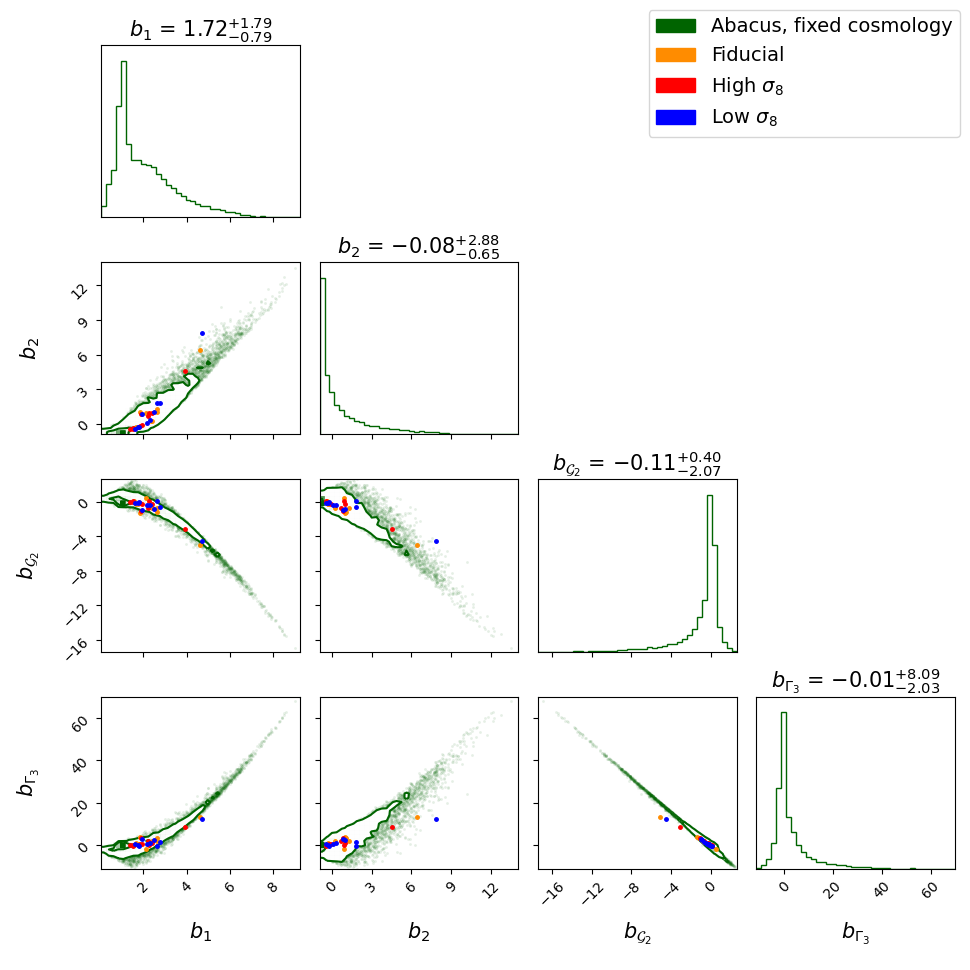}
   \caption{The baseline 
   distribution of EFT parameters 
   for a fixed cosmology (our baseline distribution),
   along with a sample of 10 HOD models 
   from three large \texttt{Abacus} boxes: the fiducial
   one (\texttt{cosm000}), which we 
   use as a control sample, 
   the one with low $\sigma_8=0.75$  (\texttt{cosm004}),
   and the one with high $\sigma_8=0.86$  (\texttt{cosm003}). The HOD parameters
   are the same for each cosmology. 
    } \label{fig:cosmo_abacus}
\end{figure*}

\section{Additional test of cosmology independence of HOD priors}
\label{sec:abacus_large}

In this section 
we study how much the HOD-based EFT parameters
change under the variation of the underlying 
cosmology. To that end we select three cosmological models 
from the large \texttt{Abacus} boxes:
the fiducial one, the one with a higher $\sigma_8=0.86$,
and the one with a lower $\sigma_8=0.75$. 
The fiducial large box has the same cosmology
as our baseline \texttt{Abacus} mocks, 
and hence it will serve us
as a control sample. We chose the boxes
with very different $\sigma_8$
because it is very closely related 
to $\sigma_M$, which dominates
the response of EFT parameters to 
cosmology.

As a next step, we make 10 samples of HOD 
parameters from our priors
and produce 10 galaxy catalogs 
for each cosmology. Then we measure the 
corresponding EFT parameters. Our results for 
the bias parameters $b_1,b_2.b_{\G}$
and $b_{\GG}$ shown in fig.~\ref{fig:cosmo_abacus}.
The picture is similar for other parameters. 

First, we see that all samples lie within the 
baseline distribution spanned by variations of 
HOD parameters at a fixed cosmology. 
Second, we see that for a fixed HOD model, 
the variation of cosmology has a relative small effect.
In particular, the typical spread of EFT parameters 
due to cosmology variation is smaller 
than the spread due to the HOD variation. 
Also note that our test confirms the basic 
intuition gained in our study of dark matter halos:
lowering/raising $\sigma_8$ for a fixed HOD 
is equivalent to raising/lowering the host halo mass parameters,
respectively. 

We have one outlier HOD model in each cosmology sample,
with $b_1\approx 4$. We see that the 
spread is somewhat larger for this parameter,
although the low and high $\sigma_8$
points are still quite close to the 
fiducial one. This suggests that 
the variation of cosmology
might be important at the tails of
the distribution. However, the corresponding 
values of the EFT parameters, e.g. $b_1\simeq 5$, $b_2\simeq 9$, 
are far away 
from the ones we probe with BOSS, $b\simeq 2.5,~b_2\simeq 2$,
which suggests that this should not have 
a significant effect on our 
main results. 

All in all, we conclude that the variation
of cosmology is a weaker effect than the 
variation of HOD for the purposes 
of the prior generation.

\begin{figure*}[ht!]
\centering
\includegraphics[width=0.99\textwidth]{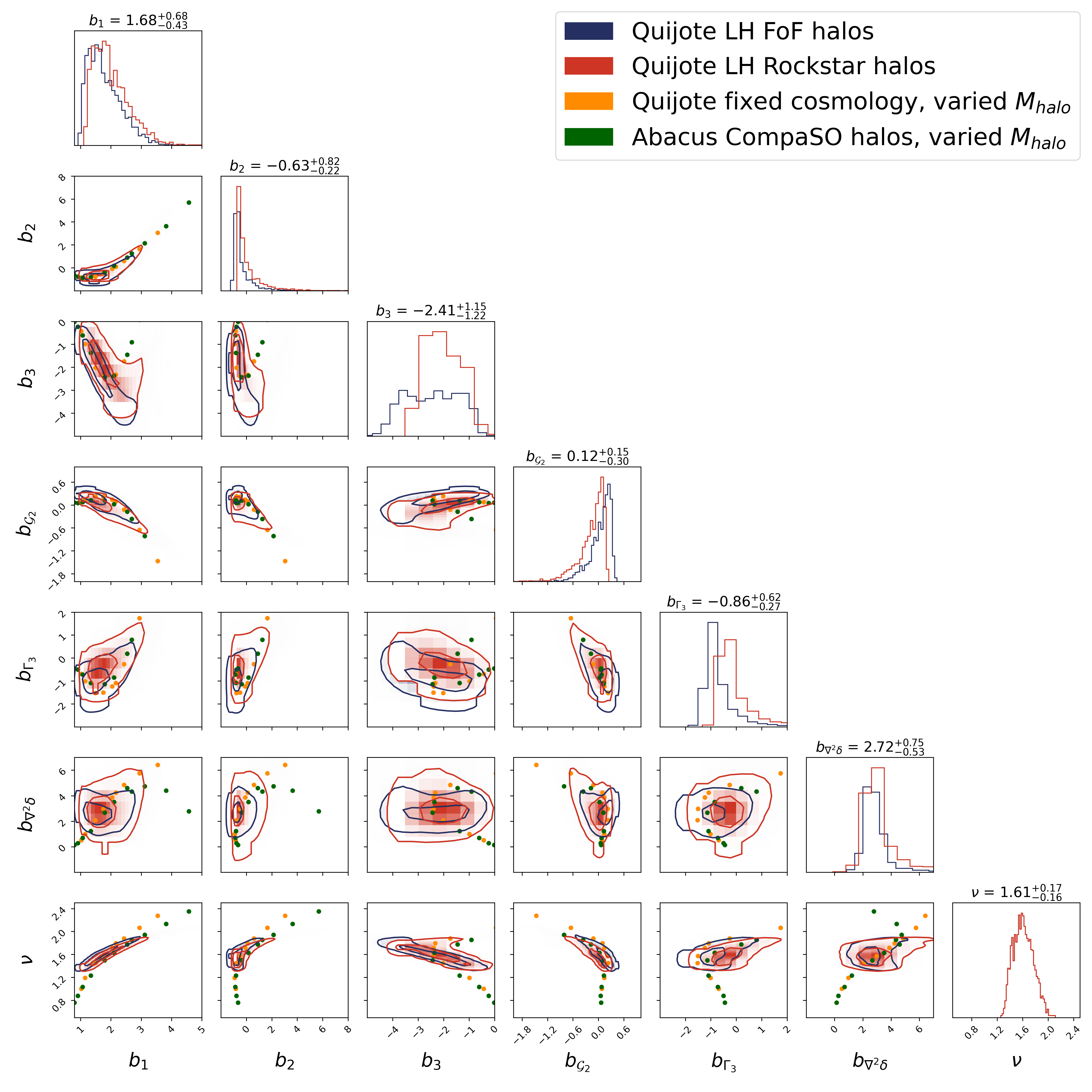}
   \caption{Same as fig.~\ref{fig:cosmo_vs_mhalo}
   but with the Rockstar halos from \texttt{Quijote LH}
   simulations. 
    } \label{fig:cosmo_vs_mhalo_app_bias}
\end{figure*}

\begin{figure*}[ht!]
\centering
\includegraphics[width=0.99\textwidth]{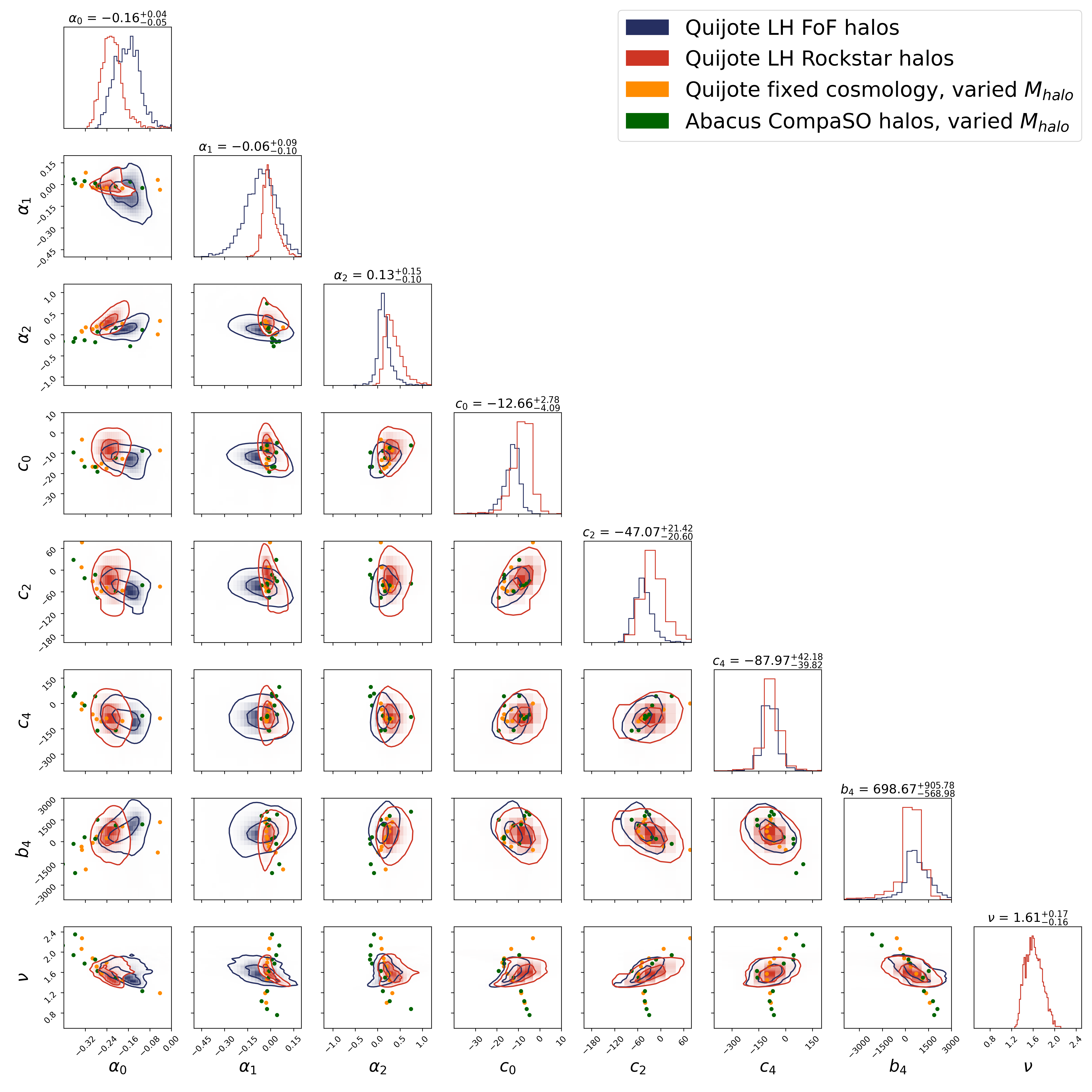}
   \caption{Same as fig.~\ref{fig:cosmo_vs_mhalo_app_bias}
   but for the stochasticity parameters and 
   redshift-space EFT counterterms. 
    } \label{fig:cosmo_vs_mhalo_app_ctr}
\end{figure*}

\begin{figure*}
\centering
\includegraphics[width=0.99\textwidth]{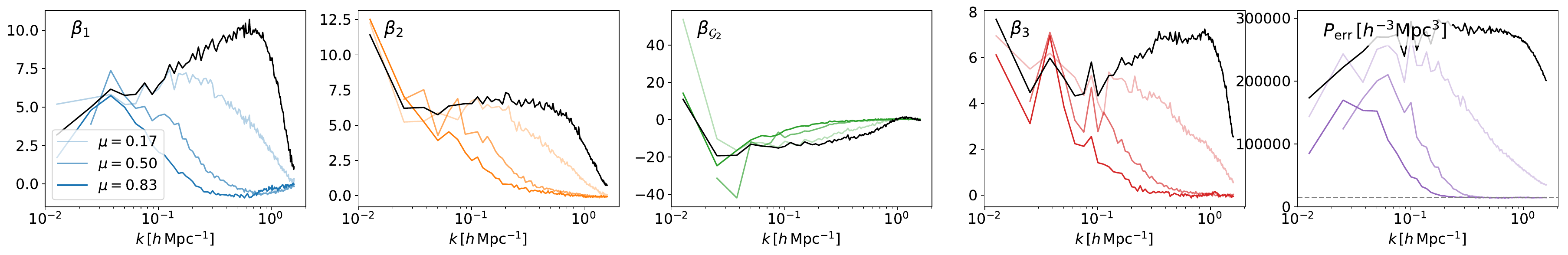}
   \caption{Same as fig.~\ref{fig:transfer} but
   for an outlier HOD sample
   with extreme values of EFT parameters. 
    } \label{fig:transfer_outlier}
\end{figure*}

\section{Additional plots and tests}
\label{app:plots}

Fig.~\ref{fig:cosmo_vs_mhalo_app_bias},
\ref{fig:cosmo_vs_mhalo_app_ctr}
display the EFT parameters 
of dark matter halos from: (a) FoF and (b) Rockstar
catalogs of the \texttt{Quijote LH} simulations, (c)
FoF catalogs of the fiducial 
\texttt{Quijote} simulation and (d) CompaSO 
catalogs 
the fiducial \texttt{Abacus} simulation.
Note that the distribution of the galaxy bias parameters
is very similar across all the halo finders. 
However, CompaSO and Rockstar produce slightly 
more consistent 
results for $\alpha_1$ and $c_2$.

In fig.~\ref{fig:transfer_outlier} 
we display field-level transfer functions
and the noise power spectra
for an HOD sample with extreme values
of EFT parameters. This sample 
has $\log_{10}M_{\rm cut}\approx 14$
and $\log_{10}M_1\approx 13$ ($\kappa\approx 0.5$)
and $\alpha_s\approx 2$ at the edges of the 
priors of the HOD parameters that we sample.
Physically, this sample has very few centrals, 
but a gigantic number
of satellites, resulting in the satellite fraction
$\approx 93\%$. 
The number density of the satellites in this sample 
is a factor of $\mathcal{O}(10)$ greater than 
the number density of the centrals. 
Thus, we have a highly biased tracer
with low number density. 
Since the number density of satellites 
is an order of magnitude 
greater than the number density of centrals, 
we obtain $P_{\rm err}\sim \frac{1}{\bar n_c}\sim  \mathcal{O}(10)\frac{1}{\bar n_g}$ from the arguments of~\cite{Baldauf:2013hka}
and in full agreement with the measurements. 
We point out, however, that the real space transfer 
functions are smooth up to $k_{\rm max}\approx 0.5~\hMpc$, so this sample does not
require any modification of the 
fitting procedure in real space.

In addition, the sample also features
a very large satellite velocity bias. 
Together with an extremely large 
satellite fraction this generates  
a very strong fingers-of-God suppression,
which can be seen both in the transfer 
function of $\beta_1(k,\mu)$
and the error power spectrum 
$P_{\rm err}$ in redshift space. 
In particular, we see that 
the fingers-of-God for this sample
lead to $\sim 100\%$ corrections
already $k\sim 0.2~\hMpc$, and thus 
become non-perturbative around our fiducial 
$k_{\rm max}$. To obtain 
more robust results, in principle, 
one should use $k_{\rm max}=0.1~\hMpc$
for this sample. 
However, such extreme catalogs
constitute only $0.3\%$
of our HOD sample, and hence modify
only the very tail of the distribution,
which we do not expect to model well
with the normalizing flows anyway.

All in all, while the EFT parameters
of this sample appear quite extreme, 
it has a physically sound behavior.
While this sample does not appear to 
match any observationally
relevant galaxy population, 
it represents a physically
plausible scenario.

\section{The origin
of low $\sigma_8$ preference 
in the BOSS analysis}
\label{app:lowsigma8}

In this appendix we investigate 
the origin of the preference for 
low $\sigma_8$
in our SBP analysis and also get 
insights into the extra 
information brought by the 
SBP.

\begin{figure*}[ht!]
\centering
\includegraphics[width=0.7\textwidth]{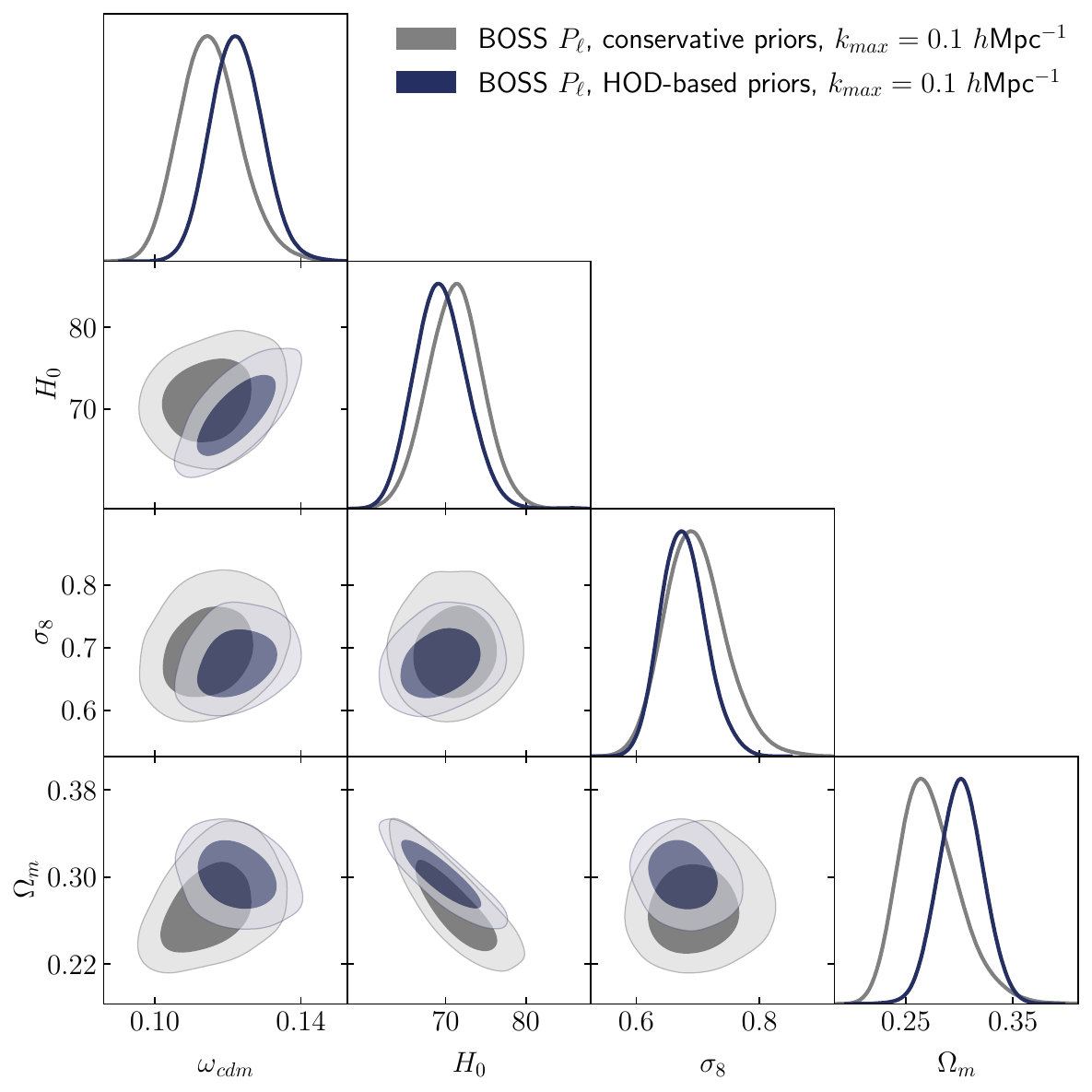}
   \caption{Cosmological parameters from the EFT-based full shape analysis of the BOSS power spectrum
   with conservative and informative simulation-based
   priors on EFT parameters for the data cut
   $\kmax=0.1~\hMpc$.
    } \label{fig:bossPk_010}
\end{figure*}

As a first step, we locate the 
k range of data that drives the 
constraints. To this end we 
re-analyze the baseline BOSS 
power spectrum dataset $P_\ell$
restricted to modes $k<0.1~\hMpc$.
We carry out both the conservative 
and HOD-informed analyses. 
The 1D marginalized 
constrains from these analyses 
are presented in table~\ref{eq:tab4} (only the linear and quadratic bias parameters are shown), 
while the corner plot for cosmological
parameters is displayed in fig.~\ref{fig:bossPk_010}.

 \begin{figure*}[ht!]
\centering
\includegraphics[width=0.7\textwidth]{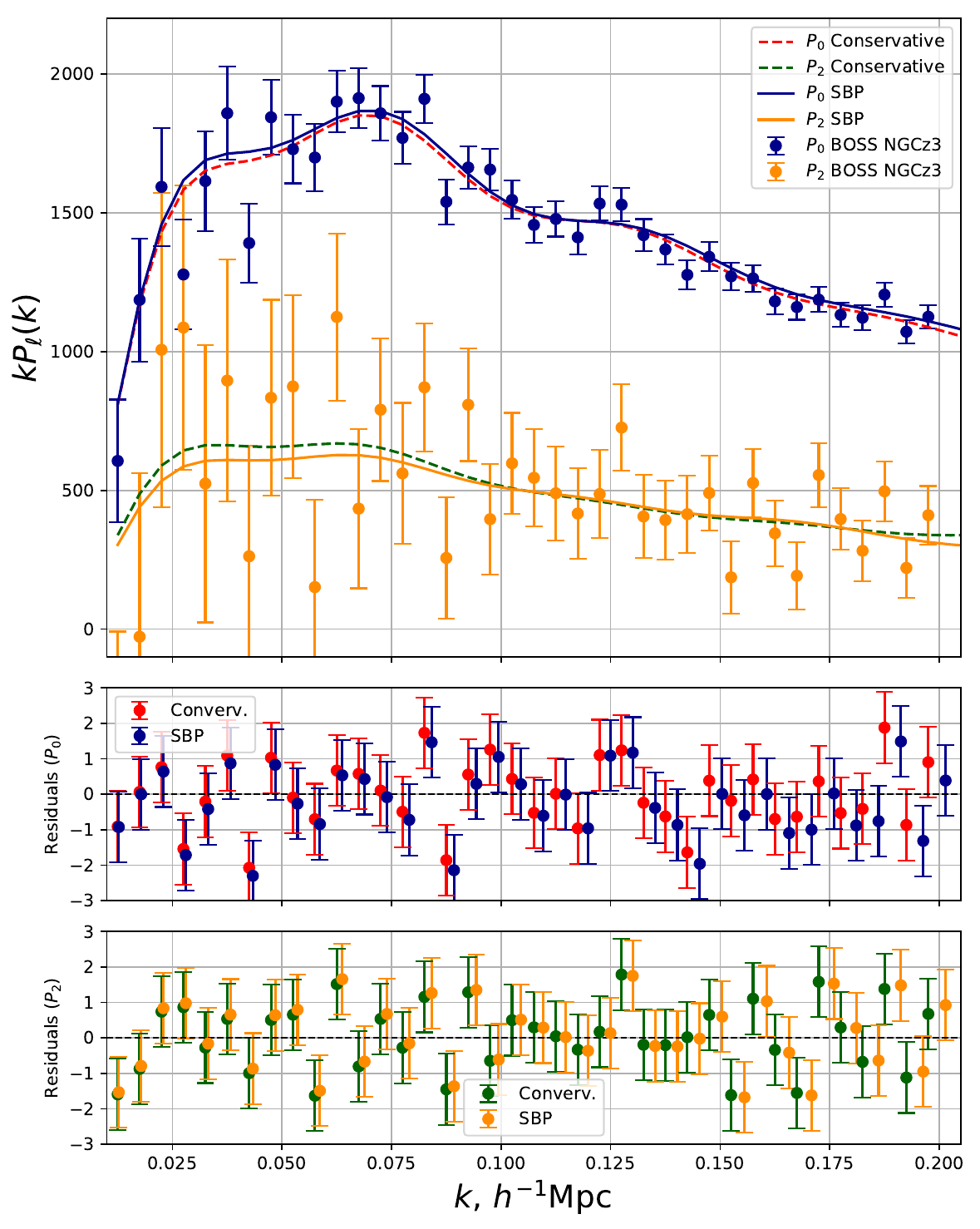}
   \caption{\textit{Upper panel:}  BOSS NGCz3 power spectrum monopole 
   and quadrupole data $(z=0.61)$ along the best-fit models 
   from the analyses with conservative and simulation-based priors. 
   Below are the residuals between the models and the data for 
   the monopole (\textit{central panel})
   and the quadrupole (\textit{lower panel})
   data. 
    } \label{fig:res}
\end{figure*}

 \begin{figure*}[ht!]
\centering
\includegraphics[width=0.99\textwidth]{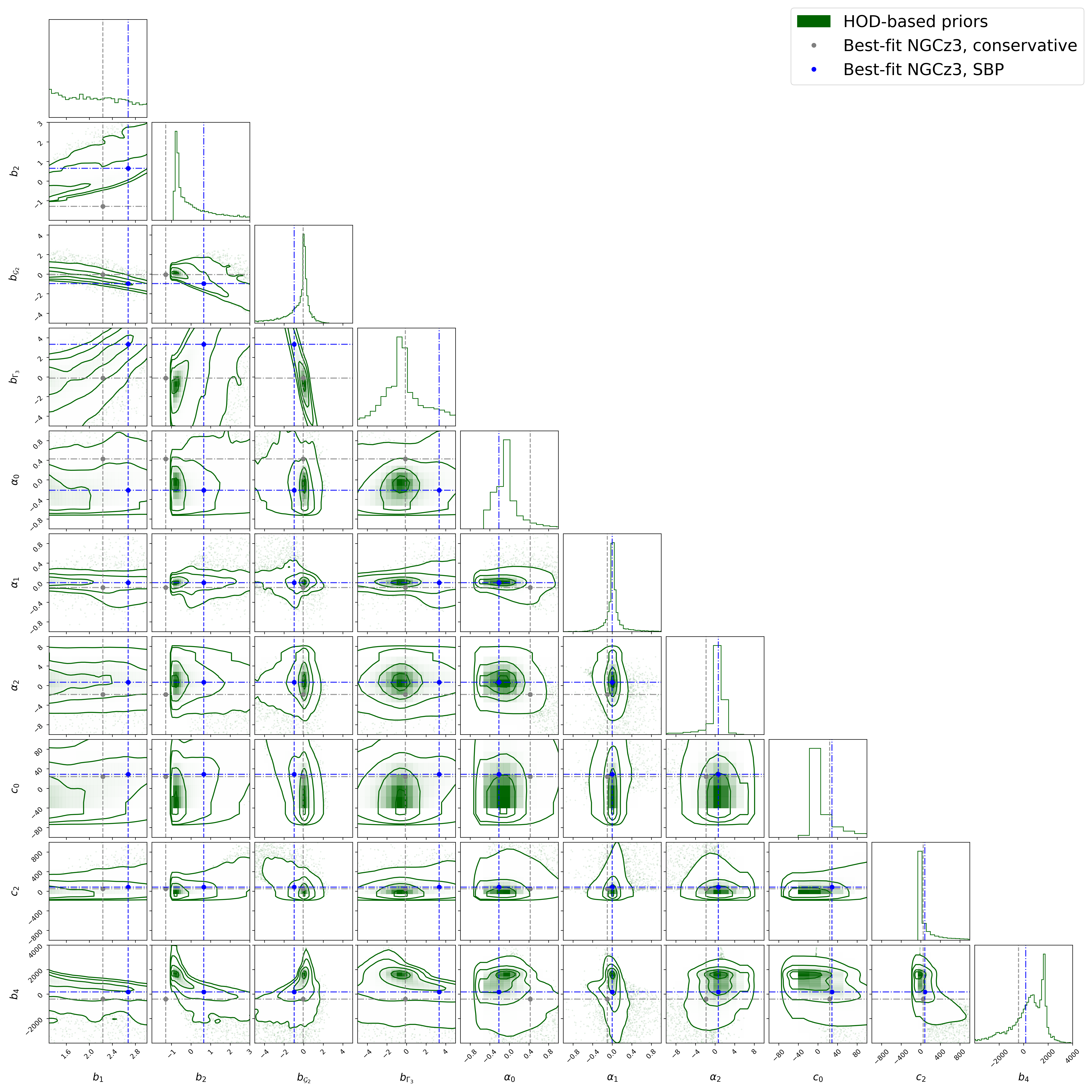}
   \caption{Best-fit values of EFT parameters from the BOSS NGCz3 
   sample in the conservative (gray dots)
   and simulation-based prior analysis (blue dots). 
   The green density on the background depicts the HOD-based
   priors. 
    } \label{fig:bf_vs_priors}
\end{figure*}

We see that the results are highly
consistent: the posteriors of 
$H_0$ and $\sigma_8$ largely overlap
and have comparable widths. 
The $\omega_{cdm}(\Omega_m)$ constraint
is visibly worse in the 
conservative case. This is expected given
that this parameter is measured
from the shape of the matter power
spectrum monopole at the high $k$
end of the allowed data range~\cite{Chudaykin:2019ock,Ivanov:2019pdj,Philcox:2020xbv}. The shape information
is strongly degenerate with the nuisance parameters for
$k<0.1~\hMpc$ which results in poor
constraints in the conservative analysis.
This also explains the improvement
that we get in the SBP analysis thanks
for better priors on nuisance parameters.

An important point here 
is that in both analyses the $\sigma_8$
posteriors are much wider than those
of our baseline analysis at $\kmax=0.2~\hMpc$,
which implies that
our main constraints are driven
by the momentum range $0.1<k/(\hMpc)<0.2$. 
Fig.~\ref{fig:bossPk_010}
also implies that the shift 
of $\sigma_8$ between the 
conservative analysis
and the SBP analysis at $\kmax=0.2~\hMpc$ in fig.~\ref{fig:bossPk} stems from the same large
wavenumber end of the data.

To understand further what drives this 
difference, let us plot take a look at the 
best-fitting curves from the conservative 
and SBP analyses. We consider
the largest data sample NGCz3, which
dominates the final constraints. 
The best-fit curves, data, and their
residuals are presented in fig.~\ref{fig:res}.
For better visibility, 
we offset the residual
points of the 
SBI analysis along the $k$ axis. 
We see that most of the difference 
between the two best-fits 
is generated at the level of $P_0$
data for $0.1<k/(\hMpc)<0.2$. 
This suggests that the main source
of information on $\sigma_8$ in SBP 
is not the quadrupole, but actually 
the monopole. To further verify this, 
we calculate the $\chi^2$
difference between the SBP and
conservative best-fits. 
We obtain
\be 
\chi^2_{\rm like} \Big|_{\rm SBP}=84.96\,,\quad \chi^2_{\rm like} \Big|_{\rm cons}=76.97\,,
\ee 
where ``like'' means that we only included
the data likelihood in our calculation (i.e. ignored the priors). At face value, 
the SBP fit is notably worse. This can be 
visually confirmed by looking at the bottom
panel of fig.~\ref{fig:res}, where we see somewhat larger residuals in the SBP analysis for the monopole around $k\simeq 0.17~\hMpc$.
Note that 
6.05 out of 8 total units of 
$\chi^2$ difference 
stem from the 
monopole part of the likelihood.

To understand 
the impact of the prior, 
we display the 
best-fitting EFT parameters in both analyses
on the background of the 
HOD-based priors in fig.~\ref{fig:bf_vs_priors}.
As before, we use the NGCz3 sample
in our discussion, noting that 
the picture is qualitatively the same
for other samples of the BOSS data. 
The first important observation
is that the best-fitting 
value of $b_2$ from the conservative analysis
is incompatible 
with the HOD models. 
The second important observation
is that the conservative 
analysis prefers values
of stochastic parameters
that correspond to a rather 
exotic HOD model with 
$\alpha_0\sim 0.4$
and $\alpha_2\sim -2$,
which corresponds to HOD samples
with a relatively large satellite fraction.

To get a quantitative 
measure of the role of the priors, 
let us introduce the effective 
$\chi^2$ statistic of the prior, 
$\chi^2_{\rm eff, prior}=-2\ln \mathcal{L}_{\rm EFT}$, 
where $\mathcal{L}_{\rm EFT}$
is the likelihood of the 
EFT parameters from our HOD
samples estimated by
the normalizing flows. In this case
we find that the best-fit EFT parameters needed to 
fit the BOSS data for a  cosmology with 
$\sigma_8=0.74$ 
have a very low probability 
as compared to the best-fitting 
parameters of the SBP analysis. 
Namely, we find 
\be  
\Delta \chi^2_{\text{eff, prior}}=\chi^2_{\text{eff, prior}}\Big|_{\rm SBP}-
\chi^2_{\text{eff, prior}}\Big|_{\rm cons.}\approx -391\,,
\ee 
which can be roughly interpreted as 
a $\sim 36\sigma$ tension w.r.t.
the best-fit values from the 
SBP analysis, as follows from the 
$\chi^2$ distribution for 10 parameters. 
Most of this difference 
is generated by the unlikely 
combination of $\{b_1,b_2\}$. 
Indeed, if we keep all other parameter
fixed, but replace $\{b_1,b_2\}$
with the SBP best-fit values ($b_1\simeq 2.6$, $b_2\simeq 0.6$),
we obtain 
\be  
\Delta \chi^2_{\text{eff, prior}}=\chi^2_{\text{eff, prior}}\Big|_{\rm SBP}-
\chi^2_{\text{eff, prior}}\Big|_{\text{cons.}+\{b_1,b_2\}}\approx -101\,,
\ee 
i.e. alleviating the tension down to 
$9\sigma$. The rest of this tension
can be attributed to the 
unlikely combination
of remaining real-space 
parameters (68 units of $\chi^2$, 
or $\approx 6\sigma$)
and RSD counterterms (33 units of $\chi^2$, or $\approx 3.\sigma$).

It was already pointed out
in~\cite{Philcox:2021kcw}
that fitting the BOSS power
spectrum data in a cosmology with 
\textit{Planck}-like $\sigma_8$
requires uncommon combinations
of EFT parameters that appear 
in tension with the halo-based
predictions. 
Once the bispectrum data was included
in~\cite{Philcox:2021kcw}, the optimal values of 
$\sigma_8$ went down as the bispectrum 
of BOSS by itself ruled out 
the unlikely combination of $\{b_1,b_2\}$
preferred by the power spectrum
with the conservative priors. 
In our work, we
keep seeing the drift towards low 
$\sigma_8$ as we impose stronger
restrictions on the EFT parameters. 
These restrictions come
from the field-level EFT priors, 
which effectively bring in 
higher order information, e.g. 
measurements of $b_2,b_{\G}$ etc.  
If we treat EFT parameters not a nuisances, 
but as physical parameters that can be 
predicted by the HOD models, 
the most likely fit to the galaxy power
spectrum data is the one with $\sigma_8\approx 0.66$.

\begin{table*}
\begin{tabular}{|l|c|c|c|c|}
 \hline
 \multicolumn{5}{|c|}{BOSS $P_\ell$ with conservative priors} \\
\hline
Param & best-fit & mean$\pm\sigma$ & 95\% lower & 95\% upper \\ \hline
$\omega{}_{cdm }$ &$0.1162$ & $0.1152_{-0.0092}^{+0.0076}$ & $0.09881$ & $0.1321$ \\
$H_0$ &$72.76$ & $71.19_{-3.4}^{+3.4}$ & $64.38$ & $78$ \\
$\ln(10^{10}A_{s })$ &$2.791$ & $2.74_{-0.16}^{+0.15}$ & $2.43$ & $3.045$ \\
$b^{(1)}_{1 }$ &$2.138$ & $2.252_{-0.21}^{+0.19}$ & $1.864$ & $2.652$ \\
$b^{(1)}_{2 }$ &$0.3709$ & $-0.1467_{-1}^{+1}$ & $-2.161$ & $1.879$ \\
$b^{(1)}_{{\G} }$ &$-0.05285$ & $-0.3078_{-0.56}^{+0.62}$ & $-1.506$ & $0.8285$ \\
$b^{(2)}_{1 }$ &$2.314$ & $2.453_{-0.23}^{+0.2}$ & $2.032$ & $2.883$ \\
$b^{(2)}_{2 }$ &$0.03068$ & $-0.02749_{-1}^{+1}$ & $-1.976$ & $1.952$ \\
$b^{(2)}_{{\G} }$ &$0.2339$ & $-0.007196_{-0.63}^{+0.65}$ & $-1.275$ & $1.261$ \\
$b^{(3)}_{1 }$ &$2.066$ & $2.173_{-0.2}^{+0.18}$ & $1.806$ & $2.555$ \\
$b^{(3)}_{2 }$ &$-0.06066$ & $0.3916_{-0.94}^{+0.94}$ & $-1.443$ & $2.258$ \\
$b^{(3)}_{{\G} }$ &$-0.5721$ & $-0.5128_{-0.5}^{+0.55}$ & $-1.563$ & $0.5285$ \\
$b^{(4)}_{1 }$ &$2.173$ & $2.266_{-0.21}^{+0.18}$ & $1.877$ & $2.666$ \\
$b^{(4)}_{2 }$ &$0.009038$ & $0.01678_{-1}^{+1}$ & $-1.948$ & $1.985$ \\
$b^{(4)}_{{\G} }$ &$-0.1313$ & $0.1145_{-0.55}^{+0.56}$ & $-0.9952$ & $1.238$ \\
\hline 
$\Omega_{m }$ &$0.2619$ & $0.273_{-0.033}^{+0.022}$ & $0.2199$ & $0.3315$ \\
$\sigma_8$ &$0.7223$ & $0.6967_{-0.055}^{+0.044}$ & $0.6002$ & $0.7966$ \\
\hline
 \end{tabular} 
 \begin{tabular}{|l|c|c|c|c|}
 \hline
  \multicolumn{5}{|c|}{BOSS $P_\ell$ with simulation-based priors} \\
\hline
Param & best-fit & mean$\pm\sigma$ & 95\% lower & 95\% upper \\ \hline
$\omega{}_{cdm }$ &$0.1184$ & $0.1225\pm 0.0071 $ & $0.1108$ & $0.1366$ \\
$H0$ &$67.73$ & $69.4^{+3.0}_{-3.5}$ & $63.35$ & $76.11$ \\
$\ln(10^{10}A_{s })$ &$2.59$ & $2.61^{+0.11}_{-0.13}$ & $2.378$ & $2.863$ \\
$b^{(1)}_{1 }$ &$2.566$ & $2.521_{-0.11}^{+0.2}$ & $2.378$ & $2.863$ \\
$b^{(1)}_{2 }$ &$0.05728$ & $0.4491_{-0.39}^{+0.27}$ & $-0.1404$ & $1.108$ \\
$b^{(1)}_{{\G} }$ &$-0.905$ & $-0.8697_{-0.32}^{+0.22}$ & $-1.412$ & $-0.2802$ \\
$b^{(2)}_{1 }$ &$2.772$ & $2.672_{-0.12}^{+0.18}$ & $2.329$ & $2.945$ \\
$b^{(2)}_{2 }$ &$0.4222$ & $0.5347_{-0.27}^{+0.21}$ & $0.09803$ & $0.9904$ \\
$b^{(2)}_{{\G} }$ &$-1.083$ & $-1.014_{-0.25}^{+0.17}$ & $-1.454$ & $-0.5176$ \\
$b^{(3)}_{1 }$ &$2.471$ & $2.409_{-0.11}^{+0.16}$ & $2.074$ & $2.650$ \\
$b^{(3)}_{2 }$ &$0.06239$ & $0.1654_{-0.31}^{+0.17}$ & $-0.248$ & $0.839$ \\
$b^{(3)}_{{\G} }$ &$-0.7845$ & $-0.7237_{-0.2}^{+0.16}$ & $-1.096$ & $-0.308$ \\
$b^{(4)}_{1 }$ &$2.498$ & $2.417_{-0.11}^{+0.16}$ & $2.102$ & $2.658$ \\
$b^{(4)}_{2 }$ &$0.07228$ & $0.2367_{-0.29}^{+0.18}$ & $-0.1902$ & $0.7457$ \\
$b^{(4)}_{{\G} }$ &$-0.8219$ & $-0.7656_{-0.19}^{+0.18}$ & $-1.131$ & $-0.3997$ \\
\hline 
$\Omega{}_{m }$ &$0.3069$ & $0.301\pm 0.021 $ & $0.2610$ & $0.3419$ \\
$\sigma_8$ &$0.651$ & $0.677^{+0.033}_{-0.039}$ & $0.6071$ & $0.7507$ \\
\hline
 \end{tabular} 
 \caption{Same as table~\ref{eq:tab3} but 
 for $\kmax=0.1~\hMpc$.}
 \label{eq:tab4}
 \end{table*}

 \begin{figure*}[ht!]
\centering
\includegraphics[width=0.7\textwidth]{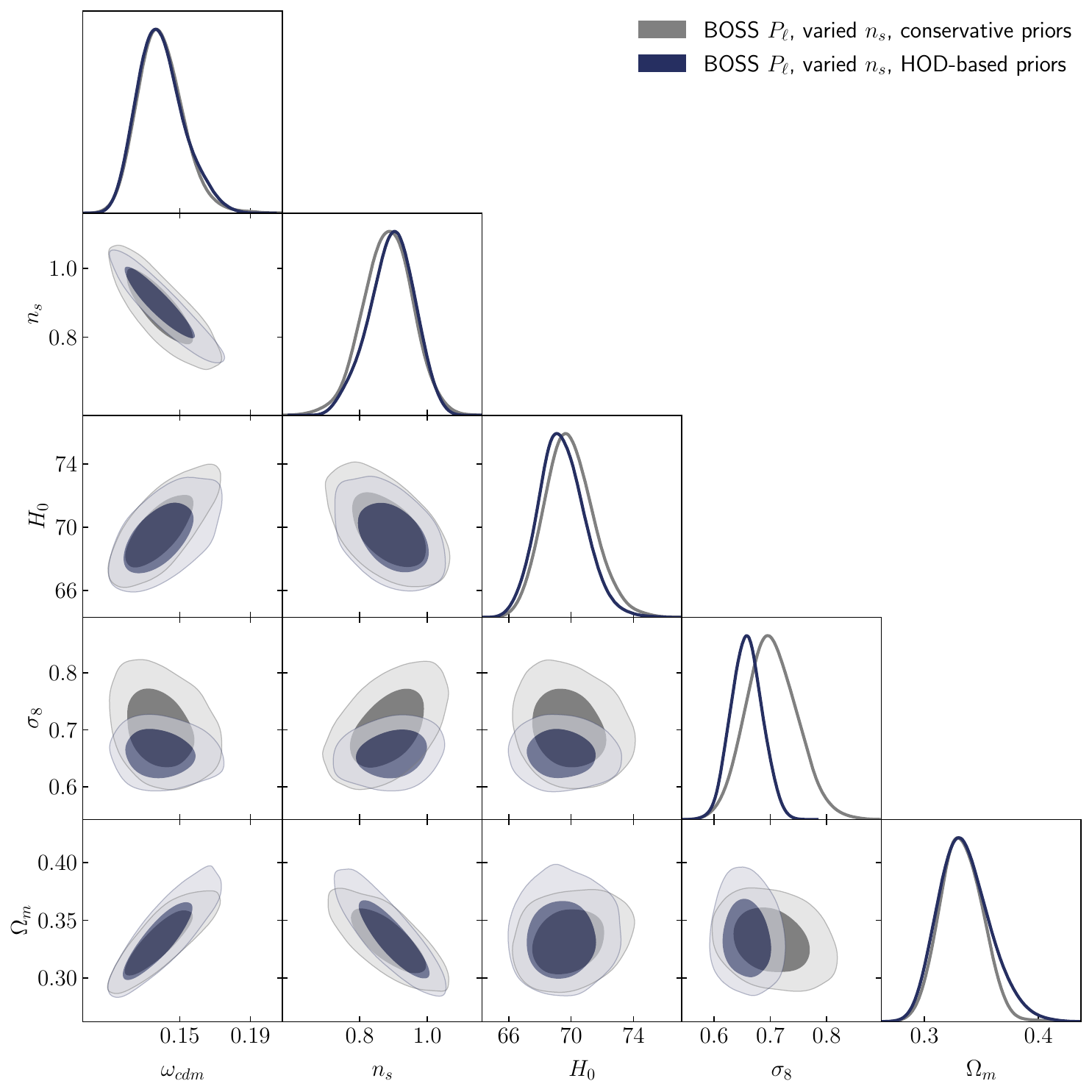}
   \caption{Same as fig.~\ref{fig:bossPk}, but with the 
   varied primordial spectral tilt $n_s$.
    } \label{fig:bossPk_ns}
\end{figure*}

\section{BOSS analysis with free $n_s$}
\label{sec:free_ns}

The results of the MCMC analysis with 
the free index of scalar primordial fluctuations $n_s$
are shown in fig.~\ref{fig:bossPk_ns}
and table~\ref{eq:tab5}.

\begin{table*}
\begin{tabular}{|l|c|c|c|c|}
 \hline
 \multicolumn{5}{|c|}{BOSS $P_\ell$ with conservative priors} \\
\hline
Param & best-fit & mean$\pm\sigma$ & 95\% lower & 95\% upper \\ \hline
$\omega_{cdm }$ &$0.13$ & $0.1393_{-0.015}^{+0.011}$ & $0.1138$ & $0.1668$ \\
$H_0$ &$68.15$ & $69.89_{-1.7}^{+1.5}$ & $66.71$ & $73.19$ \\
$\ln(10^{10}A_{s })$ &$2.788$ & $2.631_{-0.16}^{+0.15}$ & $2.312$ & $2.954$ \\
$n_{s }$ &$0.9271$ & $0.8834_{-0.072}^{+0.076}$ & $0.7337$ & $1.035$ \\
$b^{(1)}_{1 }$ &$2.234$ & $2.355_{-0.17}^{+0.16}$ & $2.02$ & $2.689$ \\
$b^{(1)}_{2 }$ &$-0.6257$ & $-1.155_{-1}^{+0.96}$ & $-3.109$ & $0.8529$ \\
$b^{(1)}_{{\G} }$ &$-0.005441$ & $-0.1835_{-0.5}^{+0.53}$ & $-1.231$ & $0.8589$ \\
$b^{(2)}_{1 }$ &$2.361$ & $2.491_{-0.17}^{+0.16}$ & $2.163$ & $2.821$ \\
$b^{(2)}_{2 }$ &$-0.154$ & $0.00792_{-0.98}^{+0.96}$ & $-1.905$ & $1.942$ \\
$b^{(2)}_{{\G} }$ &$-0.07998$ & $-0.2719_{-0.51}^{+0.54}$ & $-1.337$ & $0.77$ \\
$b^{(3)}_{1 }$ &$2.164$ & $2.247_{-0.15}^{+0.14}$ & $1.956$ & $2.545$ \\
$b^{(3)}_{2 }$ &$-0.4313$ & $-0.3574_{-0.9}^{+0.84}$ & $-2.078$ & $1.404$ \\
$b^{(3)}_{{\G} }$ &$-0.2832$ & $-0.2694_{-0.41}^{+0.43}$ & $-1.127$ & $0.5689$ \\
$b^{(4)}_{1 }$ &$2.201$ & $2.273_{-0.15}^{+0.14}$ & $1.985$ & $2.574$ \\
$b^{(4)}_{2 }$ &$0.1618$ & $-0.1207_{-0.93}^{+0.87}$ & $-1.879$ & $1.685$ \\
$b^{(4)}_{{\G} }$ &$0.1992$ & $0.2019_{-0.44}^{+0.43}$ & $-0.6637$ & $1.073$ \\
\hline 
$\Omega_{m }$ &$0.3298$ & $0.3325_{-0.02}^{+0.019}$ & $0.2953$ & $0.3704$ \\
$\sigma_8$ &$0.7403$ & $0.7036_{-0.049}^{+0.044}$ & $0.6119$ & $0.7972$ \\
\hline
 \end{tabular} 
 \begin{tabular}{|l|c|c|c|c|}
 \hline
  \multicolumn{5}{|c|}{BOSS $P_\ell$ with simulation-based priors} \\
\hline
Param & best-fit & mean$\pm\sigma$ & 95\% lower & 95\% upper \\ \hline
$\omega{}_{cdm }$ &$0.1292$ & $0.1379_{-0.014}^{+0.0094}$ & $0.116$ & $0.1642$ \\
$H_0$ &$68.49$ & $69.36_{-1.6}^{+1.4}$ & $66.44$ & $72.34$ \\
$\ln(10^{10}A_{s })$ &$2.574$ & $2.475_{-0.11}^{+0.11}$ & $2.262$ & $2.699$ \\
$n_{s }$ &$0.9641$ & $0.9029_{-0.052}^{+0.075}$ & $0.7652$ & $1.014$ \\
$b^{(1)}_{1 }$ &$2.609$ & $2.649_{-0.12}^{+0.12}$ & $2.412$ & $2.877$ \\
$b^{(1)}_{2 }$ &$0.6053$ & $0.8918_{-0.35}^{+0.22}$ & $0.3547$ & $1.51$ \\
$b^{(1)}_{{\G} }$ &$-0.8512$ & $-0.7847_{-0.39}^{+0.23}$ & $-1.363$ & $-0.0856$ \\
$b^{(2)}_{1 }$ &$2.688$ & $2.75_{-0.13}^{+0.13}$ & $2.503$ & $2.995$ \\
$b^{(2)}_{2 }$ &$0.759$ & $0.9353_{-0.34}^{+0.23}$ & $0.3907$ & $1.528$ \\
$b^{(2)}_{{\G} }$ &$-0.9797$ & $-1.044_{-0.29}^{+0.19}$ & $-1.531$ & $-0.4928$ \\
$b^{(3)}_{1 }$ &$2.39$ & $2.451_{-0.11}^{+0.11}$ & $2.237$ & $2.669$ \\
$b^{(3)}_{2 }$ &$0.4266$ & $0.5404_{-0.22}^{+0.16}$ & $0.1692$ & $0.9514$ \\
$b^{(3)}_{{\G} }$ &$-0.8925$ & $-0.8974_{-0.29}^{+0.16}$ & $-1.371$ & $-0.346$ \\
$b^{(4)}_{1 }$ &$2.423$ & $2.451_{-0.11}^{+0.11}$ & $2.236$ & $2.668$ \\
$b^{(4)}_{2 }$ &$0.4614$ & $0.5707_{-0.22}^{+0.17}$ & $0.1977$ & $0.9661$ \\
$b^{(4)}_{{\G} }$ &$-1.02$ & $-0.9687_{-0.29}^{+0.19}$ & $-1.433$ & $-0.4442$ \\
\hline 
$\Omega{}_{m }$  &$0.323$ & $0.3331_{-0.025}^{+0.017}$ & $0.2932 $ & $0.3815$ \\
$\sigma_8$ &$0.6831$ & $0.6604_{-0.03}^{+0.026}$ & $0.6084$ & $0.715$ \\
\hline
 \end{tabular} 
 \caption{Same as table~\ref{eq:tab3} but 
 with the varied tilt of the 
 primordial scalar fluctuations $n_s$.
 }
 \label{eq:tab5}
 \end{table*}

\bibliography{short.bib}

\begin{thebibliography}{152}%
\makeatletter
\providecommand \@ifxundefined [1]{%
 \@ifx{#1\undefined}
}%
\providecommand \@ifnum [1]{%
 \ifnum #1\expandafter \@firstoftwo
 \else \expandafter \@secondoftwo
 \fi
}%
\providecommand \@ifx [1]{%
 \ifx #1\expandafter \@firstoftwo
 \else \expandafter \@secondoftwo
 \fi
}%
\providecommand \natexlab [1]{#1}%
\providecommand \enquote  [1]{``#1''}%
\providecommand \bibnamefont  [1]{#1}%
\providecommand \bibfnamefont [1]{#1}%
\providecommand \citenamefont [1]{#1}%
\providecommand \href@noop [0]{\@secondoftwo}%
\providecommand \href [0]{\begingroup \@sanitize@url \@href}%
\providecommand \@href[1]{\@@startlink{#1}\@@href}%
\providecommand \@@href[1]{\endgroup#1\@@endlink}%
\providecommand \@sanitize@url [0]{\catcode `\\12\catcode `\$12\catcode
  `\&12\catcode `\#12\catcode `\^12\catcode `\_12\catcode `\%12\relax}%
\providecommand \@@startlink[1]{}%
\providecommand \@@endlink[0]{}%
\providecommand \url  [0]{\begingroup\@sanitize@url \@url }%
\providecommand \@url [1]{\endgroup\@href {#1}{\urlprefix }}%
\providecommand \urlprefix  [0]{URL }%
\providecommand \Eprint [0]{\href }%
\providecommand \doibase [0]{http://dx.doi.org/}%
\providecommand \selectlanguage [0]{\@gobble}%
\providecommand \bibinfo  [0]{\@secondoftwo}%
\providecommand \bibfield  [0]{\@secondoftwo}%
\providecommand \translation [1]{[#1]}%
\providecommand \BibitemOpen [0]{}%
\providecommand \bibitemStop [0]{}%
\providecommand \bibitemNoStop [0]{.\EOS\space}%
\providecommand \EOS [0]{\spacefactor3000\relax}%
\providecommand \BibitemShut  [1]{\csname bibitem#1\endcsname}%
\let\auto@bib@innerbib\@empty
\bibitem [{\citenamefont {Aghamousa}\ \emph {et~al.}(2016)\citenamefont
  {Aghamousa} \emph {et~al.}}]{Aghamousa:2016zmz}%
  \BibitemOpen
  \bibfield  {author} {\bibinfo {author} {\bibfnamefont {A.}~\bibnamefont
  {Aghamousa}} \emph {et~al.} (\bibinfo {collaboration} {DESI}),\ }\href@noop
  {} {\  (\bibinfo {year} {2016})},\ \Eprint {http://arxiv.org/abs/1611.00036}
  {arXiv:1611.00036 [astro-ph.IM]} \BibitemShut {NoStop}%
\bibitem [{\citenamefont {Laureijs}\ \emph {et~al.}(2011)\citenamefont
  {Laureijs} \emph {et~al.}}]{Laureijs:2011gra}%
  \BibitemOpen
  \bibfield  {author} {\bibinfo {author} {\bibfnamefont {R.}~\bibnamefont
  {Laureijs}} \emph {et~al.} (\bibinfo {collaboration} {EUCLID}),\ }\href@noop
  {} {\  (\bibinfo {year} {2011})},\ \Eprint {http://arxiv.org/abs/1110.3193}
  {arXiv:1110.3193 [astro-ph.CO]} \BibitemShut {NoStop}%
\bibitem [{\citenamefont {Ivezi\'c}\ \emph {et~al.}(2019)\citenamefont
  {Ivezi\'c} \emph {et~al.}}]{LSST:2008ijt}%
  \BibitemOpen
  \bibfield  {author} {\bibinfo {author} {\bibfnamefont {v.}~\bibnamefont
  {Ivezi\'c}} \emph {et~al.} (\bibinfo {collaboration} {LSST}),\ }\href
  {\doibase 10.3847/1538-4357/ab042c} {\bibfield  {journal} {\bibinfo
  {journal} {Astrophys. J.}\ }\textbf {\bibinfo {volume} {873}},\ \bibinfo
  {pages} {111} (\bibinfo {year} {2019})},\ \Eprint
  {http://arxiv.org/abs/0805.2366} {arXiv:0805.2366 [astro-ph]} \BibitemShut
  {NoStop}%
\bibitem [{\citenamefont {Akeson}\ \emph {et~al.}(2019)\citenamefont {Akeson}
  \emph {et~al.}}]{Akeson:2019biv}%
  \BibitemOpen
  \bibfield  {author} {\bibinfo {author} {\bibfnamefont {R.}~\bibnamefont
  {Akeson}} \emph {et~al.},\ }\href@noop {} {\  (\bibinfo {year} {2019})},\
  \Eprint {http://arxiv.org/abs/1902.05569} {arXiv:1902.05569 [astro-ph.IM]}
  \BibitemShut {NoStop}%
\bibitem [{\citenamefont {Zeldovich}(1970)}]{Zeldovich:1969sb}%
  \BibitemOpen
  \bibfield  {author} {\bibinfo {author} {\bibfnamefont {Y.~B.}\ \bibnamefont
  {Zeldovich}},\ }\href@noop {} {\bibfield  {journal} {\bibinfo  {journal}
  {Astron. Astrophys.}\ }\textbf {\bibinfo {volume} {5}},\ \bibinfo {pages}
  {84} (\bibinfo {year} {1970})}\BibitemShut {NoStop}%
\bibitem [{\citenamefont {Baumann}\ \emph {et~al.}(2012)\citenamefont
  {Baumann}, \citenamefont {Nicolis}, \citenamefont {Senatore},\ and\
  \citenamefont {Zaldarriaga}}]{Baumann:2010tm}%
  \BibitemOpen
  \bibfield  {author} {\bibinfo {author} {\bibfnamefont {D.}~\bibnamefont
  {Baumann}}, \bibinfo {author} {\bibfnamefont {A.}~\bibnamefont {Nicolis}},
  \bibinfo {author} {\bibfnamefont {L.}~\bibnamefont {Senatore}}, \ and\
  \bibinfo {author} {\bibfnamefont {M.}~\bibnamefont {Zaldarriaga}},\ }\href
  {\doibase 10.1088/1475-7516/2012/07/051} {\bibfield  {journal} {\bibinfo
  {journal} {JCAP}\ }\textbf {\bibinfo {volume} {1207}},\ \bibinfo {pages}
  {051} (\bibinfo {year} {2012})},\ \Eprint {http://arxiv.org/abs/1004.2488}
  {arXiv:1004.2488 [astro-ph.CO]} \BibitemShut {NoStop}%
\bibitem [{\citenamefont {Carrasco}\ \emph {et~al.}(2012)\citenamefont
  {Carrasco}, \citenamefont {Hertzberg},\ and\ \citenamefont
  {Senatore}}]{Carrasco:2012cv}%
  \BibitemOpen
  \bibfield  {author} {\bibinfo {author} {\bibfnamefont {J.~J.~M.}\
  \bibnamefont {Carrasco}}, \bibinfo {author} {\bibfnamefont {M.~P.}\
  \bibnamefont {Hertzberg}}, \ and\ \bibinfo {author} {\bibfnamefont
  {L.}~\bibnamefont {Senatore}},\ }\href {\doibase 10.1007/JHEP09(2012)082}
  {\bibfield  {journal} {\bibinfo  {journal} {JHEP}\ }\textbf {\bibinfo
  {volume} {09}},\ \bibinfo {pages} {082} (\bibinfo {year} {2012})},\ \Eprint
  {http://arxiv.org/abs/1206.2926} {arXiv:1206.2926 [astro-ph.CO]} \BibitemShut
  {NoStop}%
\bibitem [{\citenamefont {Ivanov}(2022)}]{Ivanov:2022mrd}%
  \BibitemOpen
  \bibfield  {author} {\bibinfo {author} {\bibfnamefont {M.~M.}\ \bibnamefont
  {Ivanov}},\ }\href@noop {} {\  (\bibinfo {year} {2022})},\ \Eprint
  {http://arxiv.org/abs/2212.08488} {arXiv:2212.08488 [astro-ph.CO]}
  \BibitemShut {NoStop}%
\bibitem [{\citenamefont {Ivanov}\ \emph
  {et~al.}(2020{\natexlab{a}})\citenamefont {Ivanov}, \citenamefont
  {Simonovi\'c},\ and\ \citenamefont {Zaldarriaga}}]{Ivanov:2019hqk}%
  \BibitemOpen
  \bibfield  {author} {\bibinfo {author} {\bibfnamefont {M.~M.}\ \bibnamefont
  {Ivanov}}, \bibinfo {author} {\bibfnamefont {M.}~\bibnamefont {Simonovi\'c}},
  \ and\ \bibinfo {author} {\bibfnamefont {M.}~\bibnamefont {Zaldarriaga}},\
  }\href {\doibase 10.1103/PhysRevD.101.083504} {\bibfield  {journal} {\bibinfo
   {journal} {Phys. Rev. D}\ }\textbf {\bibinfo {volume} {101}},\ \bibinfo
  {pages} {083504} (\bibinfo {year} {2020}{\natexlab{a}})},\ \Eprint
  {http://arxiv.org/abs/1912.08208} {arXiv:1912.08208 [astro-ph.CO]}
  \BibitemShut {NoStop}%
\bibitem [{\citenamefont {D'Amico}\ \emph {et~al.}(2019)\citenamefont
  {D'Amico}, \citenamefont {Gleyzes}, \citenamefont {Kokron}, \citenamefont
  {Markovic}, \citenamefont {Senatore}, \citenamefont {Zhang}, \citenamefont
  {Beutler},\ and\ \citenamefont {Gil-Marín}}]{DAmico:2019fhj}%
  \BibitemOpen
  \bibfield  {author} {\bibinfo {author} {\bibfnamefont {G.}~\bibnamefont
  {D'Amico}}, \bibinfo {author} {\bibfnamefont {J.}~\bibnamefont {Gleyzes}},
  \bibinfo {author} {\bibfnamefont {N.}~\bibnamefont {Kokron}}, \bibinfo
  {author} {\bibfnamefont {D.}~\bibnamefont {Markovic}}, \bibinfo {author}
  {\bibfnamefont {L.}~\bibnamefont {Senatore}}, \bibinfo {author}
  {\bibfnamefont {P.}~\bibnamefont {Zhang}}, \bibinfo {author} {\bibfnamefont
  {F.}~\bibnamefont {Beutler}}, \ and\ \bibinfo {author} {\bibfnamefont
  {H.}~\bibnamefont {Gil-Marín}},\ }\href@noop {} {\  (\bibinfo {year}
  {2019})},\ \Eprint {http://arxiv.org/abs/1909.05271} {arXiv:1909.05271
  [astro-ph.CO]} \BibitemShut {NoStop}%
\bibitem [{\citenamefont {Chen}\ \emph
  {et~al.}(2022{\natexlab{a}})\citenamefont {Chen}, \citenamefont {Vlah},\ and\
  \citenamefont {White}}]{Chen:2021wdi}%
  \BibitemOpen
  \bibfield  {author} {\bibinfo {author} {\bibfnamefont {S.-F.}\ \bibnamefont
  {Chen}}, \bibinfo {author} {\bibfnamefont {Z.}~\bibnamefont {Vlah}}, \ and\
  \bibinfo {author} {\bibfnamefont {M.}~\bibnamefont {White}},\ }\href
  {\doibase 10.1088/1475-7516/2022/02/008} {\bibfield  {journal} {\bibinfo
  {journal} {JCAP}\ }\textbf {\bibinfo {volume} {02}},\ \bibinfo {pages} {008}
  (\bibinfo {year} {2022}{\natexlab{a}})},\ \Eprint
  {http://arxiv.org/abs/2110.05530} {arXiv:2110.05530 [astro-ph.CO]}
  \BibitemShut {NoStop}%
\bibitem [{\citenamefont {Philcox}\ and\ \citenamefont
  {Ivanov}(2022)}]{Philcox:2021kcw}%
  \BibitemOpen
  \bibfield  {author} {\bibinfo {author} {\bibfnamefont {O.~H.~E.}\
  \bibnamefont {Philcox}}\ and\ \bibinfo {author} {\bibfnamefont {M.~M.}\
  \bibnamefont {Ivanov}},\ }\href {\doibase 10.1103/PhysRevD.105.043517}
  {\bibfield  {journal} {\bibinfo  {journal} {Phys. Rev. D}\ }\textbf {\bibinfo
  {volume} {105}},\ \bibinfo {pages} {043517} (\bibinfo {year} {2022})},\
  \Eprint {http://arxiv.org/abs/2112.04515} {arXiv:2112.04515 [astro-ph.CO]}
  \BibitemShut {NoStop}%
\bibitem [{\citenamefont {Chen}\ \emph
  {et~al.}(2024{\natexlab{a}})\citenamefont {Chen}, \citenamefont {Ivanov},
  \citenamefont {Philcox},\ and\ \citenamefont {Wenzl}}]{Chen:2024vuf}%
  \BibitemOpen
  \bibfield  {author} {\bibinfo {author} {\bibfnamefont {S.-F.}\ \bibnamefont
  {Chen}}, \bibinfo {author} {\bibfnamefont {M.~M.}\ \bibnamefont {Ivanov}},
  \bibinfo {author} {\bibfnamefont {O.~H.~E.}\ \bibnamefont {Philcox}}, \ and\
  \bibinfo {author} {\bibfnamefont {L.}~\bibnamefont {Wenzl}},\ }\href@noop {}
  {\  (\bibinfo {year} {2024}{\natexlab{a}})},\ \Eprint
  {http://arxiv.org/abs/2406.13388} {arXiv:2406.13388 [astro-ph.CO]}
  \BibitemShut {NoStop}%
\bibitem [{\citenamefont {Maus}\ \emph {et~al.}(2024)\citenamefont {Maus} \emph
  {et~al.}}]{Maus:2024dzi}%
  \BibitemOpen
  \bibfield  {author} {\bibinfo {author} {\bibfnamefont {M.}~\bibnamefont
  {Maus}} \emph {et~al.},\ }\href@noop {} {\  (\bibinfo {year} {2024})},\
  \Eprint {http://arxiv.org/abs/2404.07312} {arXiv:2404.07312 [astro-ph.CO]}
  \BibitemShut {NoStop}%
\bibitem [{\citenamefont {Baldauf}\ \emph
  {et~al.}(2015{\natexlab{a}})\citenamefont {Baldauf}, \citenamefont
  {Mercolli}, \citenamefont {Mirbabayi},\ and\ \citenamefont
  {Pajer}}]{Baldauf:2014qfa}%
  \BibitemOpen
  \bibfield  {author} {\bibinfo {author} {\bibfnamefont {T.}~\bibnamefont
  {Baldauf}}, \bibinfo {author} {\bibfnamefont {L.}~\bibnamefont {Mercolli}},
  \bibinfo {author} {\bibfnamefont {M.}~\bibnamefont {Mirbabayi}}, \ and\
  \bibinfo {author} {\bibfnamefont {E.}~\bibnamefont {Pajer}},\ }\href
  {\doibase 10.1088/1475-7516/2015/05/007} {\bibfield  {journal} {\bibinfo
  {journal} {JCAP}\ }\textbf {\bibinfo {volume} {1505}},\ \bibinfo {pages}
  {007} (\bibinfo {year} {2015}{\natexlab{a}})},\ \Eprint
  {http://arxiv.org/abs/1406.4135} {arXiv:1406.4135 [astro-ph.CO]} \BibitemShut
  {NoStop}%
\bibitem [{\citenamefont {Baldauf}\ \emph
  {et~al.}(2015{\natexlab{b}})\citenamefont {Baldauf}, \citenamefont
  {Mercolli},\ and\ \citenamefont {Zaldarriaga}}]{Baldauf:2015aha}%
  \BibitemOpen
  \bibfield  {author} {\bibinfo {author} {\bibfnamefont {T.}~\bibnamefont
  {Baldauf}}, \bibinfo {author} {\bibfnamefont {L.}~\bibnamefont {Mercolli}}, \
  and\ \bibinfo {author} {\bibfnamefont {M.}~\bibnamefont {Zaldarriaga}},\
  }\href {\doibase 10.1103/PhysRevD.92.123007} {\bibfield  {journal} {\bibinfo
  {journal} {Phys. Rev.}\ }\textbf {\bibinfo {volume} {D92}},\ \bibinfo {pages}
  {123007} (\bibinfo {year} {2015}{\natexlab{b}})},\ \Eprint
  {http://arxiv.org/abs/1507.02256} {arXiv:1507.02256 [astro-ph.CO]}
  \BibitemShut {NoStop}%
\bibitem [{\citenamefont {Konstandin}\ \emph {et~al.}(2019)\citenamefont
  {Konstandin}, \citenamefont {Porto},\ and\ \citenamefont
  {Rubira}}]{Konstandin:2019bay}%
  \BibitemOpen
  \bibfield  {author} {\bibinfo {author} {\bibfnamefont {T.}~\bibnamefont
  {Konstandin}}, \bibinfo {author} {\bibfnamefont {R.~A.}\ \bibnamefont
  {Porto}}, \ and\ \bibinfo {author} {\bibfnamefont {H.}~\bibnamefont
  {Rubira}},\ }\href {\doibase 10.1088/1475-7516/2019/11/027} {\bibfield
  {journal} {\bibinfo  {journal} {JCAP}\ }\textbf {\bibinfo {volume} {11}},\
  \bibinfo {pages} {027} (\bibinfo {year} {2019})},\ \Eprint
  {http://arxiv.org/abs/1906.00997} {arXiv:1906.00997 [astro-ph.CO]}
  \BibitemShut {NoStop}%
\bibitem [{\citenamefont {Cabass}\ \emph
  {et~al.}(2022{\natexlab{a}})\citenamefont {Cabass}, \citenamefont {Ivanov},
  \citenamefont {Philcox}, \citenamefont {Simonovi\'c},\ and\ \citenamefont
  {Zaldarriaga}}]{Cabass:2022wjy}%
  \BibitemOpen
  \bibfield  {author} {\bibinfo {author} {\bibfnamefont {G.}~\bibnamefont
  {Cabass}}, \bibinfo {author} {\bibfnamefont {M.~M.}\ \bibnamefont {Ivanov}},
  \bibinfo {author} {\bibfnamefont {O.~H.~E.}\ \bibnamefont {Philcox}},
  \bibinfo {author} {\bibfnamefont {M.}~\bibnamefont {Simonovi\'c}}, \ and\
  \bibinfo {author} {\bibfnamefont {M.}~\bibnamefont {Zaldarriaga}},\
  }\href@noop {} {\  (\bibinfo {year} {2022}{\natexlab{a}})},\ \Eprint
  {http://arxiv.org/abs/2201.07238} {arXiv:2201.07238 [astro-ph.CO]}
  \BibitemShut {NoStop}%
\bibitem [{\citenamefont {Cabass}\ \emph
  {et~al.}(2022{\natexlab{b}})\citenamefont {Cabass}, \citenamefont {Ivanov},
  \citenamefont {Philcox}, \citenamefont {Simonovi\'c},\ and\ \citenamefont
  {Zaldarriaga}}]{Cabass:2022ymb}%
  \BibitemOpen
  \bibfield  {author} {\bibinfo {author} {\bibfnamefont {G.}~\bibnamefont
  {Cabass}}, \bibinfo {author} {\bibfnamefont {M.~M.}\ \bibnamefont {Ivanov}},
  \bibinfo {author} {\bibfnamefont {O.~H.~E.}\ \bibnamefont {Philcox}},
  \bibinfo {author} {\bibfnamefont {M.}~\bibnamefont {Simonovi\'c}}, \ and\
  \bibinfo {author} {\bibfnamefont {M.}~\bibnamefont {Zaldarriaga}},\
  }\href@noop {} {\  (\bibinfo {year} {2022}{\natexlab{b}})},\ \Eprint
  {http://arxiv.org/abs/2204.01781} {arXiv:2204.01781 [astro-ph.CO]}
  \BibitemShut {NoStop}%
\bibitem [{\citenamefont {Cabass}\ \emph
  {et~al.}(2022{\natexlab{c}})\citenamefont {Cabass}, \citenamefont {Ivanov},
  \citenamefont {Philcox}, \citenamefont {Simonovic},\ and\ \citenamefont
  {Zaldarriaga}}]{Cabass:2022epm}%
  \BibitemOpen
  \bibfield  {author} {\bibinfo {author} {\bibfnamefont {G.}~\bibnamefont
  {Cabass}}, \bibinfo {author} {\bibfnamefont {M.~M.}\ \bibnamefont {Ivanov}},
  \bibinfo {author} {\bibfnamefont {O.~H.~E.}\ \bibnamefont {Philcox}},
  \bibinfo {author} {\bibfnamefont {M.}~\bibnamefont {Simonovic}}, \ and\
  \bibinfo {author} {\bibfnamefont {M.}~\bibnamefont {Zaldarriaga}},\
  }\href@noop {} {\  (\bibinfo {year} {2022}{\natexlab{c}})},\ \Eprint
  {http://arxiv.org/abs/2211.14899} {arXiv:2211.14899 [astro-ph.CO]}
  \BibitemShut {NoStop}%
\bibitem [{\citenamefont {McAlpine}\ \emph {et~al.}(2016)\citenamefont
  {McAlpine} \emph {et~al.}}]{McAlpine:2015tma}%
  \BibitemOpen
  \bibfield  {author} {\bibinfo {author} {\bibfnamefont {S.}~\bibnamefont
  {McAlpine}} \emph {et~al.},\ }\href {\doibase 10.1016/j.ascom.2016.02.004}
  {\bibfield  {journal} {\bibinfo  {journal} {Astron. Comput.}\ }\textbf
  {\bibinfo {volume} {15}},\ \bibinfo {pages} {72} (\bibinfo {year} {2016})},\
  \Eprint {http://arxiv.org/abs/1510.01320} {arXiv:1510.01320 [astro-ph.GA]}
  \BibitemShut {NoStop}%
\bibitem [{\citenamefont {Springel}\ \emph {et~al.}(2018)\citenamefont
  {Springel} \emph {et~al.}}]{Springel:2017tpz}%
  \BibitemOpen
  \bibfield  {author} {\bibinfo {author} {\bibfnamefont {V.}~\bibnamefont
  {Springel}} \emph {et~al.},\ }\href {\doibase 10.1093/mnras/stx3304}
  {\bibfield  {journal} {\bibinfo  {journal} {Mon. Not. Roy. Astron. Soc.}\
  }\textbf {\bibinfo {volume} {475}},\ \bibinfo {pages} {676} (\bibinfo {year}
  {2018})},\ \Eprint {http://arxiv.org/abs/1707.03397} {arXiv:1707.03397
  [astro-ph.GA]} \BibitemShut {NoStop}%
\bibitem [{\citenamefont {Hern\'andez-Aguayo}\ \emph
  {et~al.}(2023)\citenamefont {Hern\'andez-Aguayo} \emph
  {et~al.}}]{Hernandez-Aguayo:2022xcl}%
  \BibitemOpen
  \bibfield  {author} {\bibinfo {author} {\bibfnamefont {C.}~\bibnamefont
  {Hern\'andez-Aguayo}} \emph {et~al.},\ }\href {\doibase
  10.1093/mnras/stad1657} {\bibfield  {journal} {\bibinfo  {journal} {Mon. Not.
  Roy. Astron. Soc.}\ }\textbf {\bibinfo {volume} {524}},\ \bibinfo {pages}
  {2556} (\bibinfo {year} {2023})},\ \Eprint {http://arxiv.org/abs/2210.10059}
  {arXiv:2210.10059 [astro-ph.CO]} \BibitemShut {NoStop}%
\bibitem [{\citenamefont {Wechsler}\ and\ \citenamefont
  {Tinker}(2018)}]{Wechsler:2018pic}%
  \BibitemOpen
  \bibfield  {author} {\bibinfo {author} {\bibfnamefont {R.~H.}\ \bibnamefont
  {Wechsler}}\ and\ \bibinfo {author} {\bibfnamefont {J.~L.}\ \bibnamefont
  {Tinker}},\ }\href {\doibase 10.1146/annurev-astro-081817-051756} {\bibfield
  {journal} {\bibinfo  {journal} {Ann. Rev. Astron. Astrophys.}\ }\textbf
  {\bibinfo {volume} {56}},\ \bibinfo {pages} {435} (\bibinfo {year} {2018})},\
  \Eprint {http://arxiv.org/abs/1804.03097} {arXiv:1804.03097 [astro-ph.GA]}
  \BibitemShut {NoStop}%
\bibitem [{\citenamefont {Berlind}\ and\ \citenamefont
  {Weinberg}(2002)}]{Berlind:2001xk}%
  \BibitemOpen
  \bibfield  {author} {\bibinfo {author} {\bibfnamefont {A.~A.}\ \bibnamefont
  {Berlind}}\ and\ \bibinfo {author} {\bibfnamefont {D.~H.}\ \bibnamefont
  {Weinberg}},\ }\href {\doibase 10.1086/341469} {\bibfield  {journal}
  {\bibinfo  {journal} {Astrophys. J.}\ }\textbf {\bibinfo {volume} {575}},\
  \bibinfo {pages} {587} (\bibinfo {year} {2002})},\ \Eprint
  {http://arxiv.org/abs/astro-ph/0109001} {arXiv:astro-ph/0109001} \BibitemShut
  {NoStop}%
\bibitem [{\citenamefont {Kravtsov}\ \emph {et~al.}(2004)\citenamefont
  {Kravtsov}, \citenamefont {Berlind}, \citenamefont {Wechsler}, \citenamefont
  {Klypin}, \citenamefont {Gottloeber}, \citenamefont {Allgood},\ and\
  \citenamefont {Primack}}]{Kravtsov:2003sg}%
  \BibitemOpen
  \bibfield  {author} {\bibinfo {author} {\bibfnamefont {A.~V.}\ \bibnamefont
  {Kravtsov}}, \bibinfo {author} {\bibfnamefont {A.~A.}\ \bibnamefont
  {Berlind}}, \bibinfo {author} {\bibfnamefont {R.~H.}\ \bibnamefont
  {Wechsler}}, \bibinfo {author} {\bibfnamefont {A.~A.}\ \bibnamefont
  {Klypin}}, \bibinfo {author} {\bibfnamefont {S.}~\bibnamefont {Gottloeber}},
  \bibinfo {author} {\bibfnamefont {B.}~\bibnamefont {Allgood}}, \ and\
  \bibinfo {author} {\bibfnamefont {J.~R.}\ \bibnamefont {Primack}},\ }\href
  {\doibase 10.1086/420959} {\bibfield  {journal} {\bibinfo  {journal}
  {Astrophys. J.}\ }\textbf {\bibinfo {volume} {609}},\ \bibinfo {pages} {35}
  (\bibinfo {year} {2004})},\ \Eprint {http://arxiv.org/abs/astro-ph/0308519}
  {arXiv:astro-ph/0308519} \BibitemShut {NoStop}%
\bibitem [{\citenamefont {Zheng}\ \emph {et~al.}(2005)\citenamefont {Zheng},
  \citenamefont {Berlind}, \citenamefont {Weinberg}, \citenamefont {Benson},
  \citenamefont {Baugh}, \citenamefont {Cole}, \citenamefont {Dave},
  \citenamefont {Frenk}, \citenamefont {Katz},\ and\ \citenamefont
  {Lacey}}]{Zheng:2004id}%
  \BibitemOpen
  \bibfield  {author} {\bibinfo {author} {\bibfnamefont {Z.}~\bibnamefont
  {Zheng}}, \bibinfo {author} {\bibfnamefont {A.~A.}\ \bibnamefont {Berlind}},
  \bibinfo {author} {\bibfnamefont {D.~H.}\ \bibnamefont {Weinberg}}, \bibinfo
  {author} {\bibfnamefont {A.~J.}\ \bibnamefont {Benson}}, \bibinfo {author}
  {\bibfnamefont {C.~M.}\ \bibnamefont {Baugh}}, \bibinfo {author}
  {\bibfnamefont {S.}~\bibnamefont {Cole}}, \bibinfo {author} {\bibfnamefont
  {R.}~\bibnamefont {Dave}}, \bibinfo {author} {\bibfnamefont {C.~S.}\
  \bibnamefont {Frenk}}, \bibinfo {author} {\bibfnamefont {N.}~\bibnamefont
  {Katz}}, \ and\ \bibinfo {author} {\bibfnamefont {C.~G.}\ \bibnamefont
  {Lacey}},\ }\href {\doibase 10.1086/466510} {\bibfield  {journal} {\bibinfo
  {journal} {Astrophys. J.}\ }\textbf {\bibinfo {volume} {633}},\ \bibinfo
  {pages} {791} (\bibinfo {year} {2005})},\ \Eprint
  {http://arxiv.org/abs/astro-ph/0408564} {arXiv:astro-ph/0408564} \BibitemShut
  {NoStop}%
\bibitem [{\citenamefont {Hearin}\ \emph {et~al.}(2016)\citenamefont {Hearin},
  \citenamefont {Zentner}, \citenamefont {van~den Bosch}, \citenamefont
  {Campbell},\ and\ \citenamefont {Tollerud}}]{Hearin:2015jnf}%
  \BibitemOpen
  \bibfield  {author} {\bibinfo {author} {\bibfnamefont {A.~P.}\ \bibnamefont
  {Hearin}}, \bibinfo {author} {\bibfnamefont {A.~R.}\ \bibnamefont {Zentner}},
  \bibinfo {author} {\bibfnamefont {F.~C.}\ \bibnamefont {van~den Bosch}},
  \bibinfo {author} {\bibfnamefont {D.}~\bibnamefont {Campbell}}, \ and\
  \bibinfo {author} {\bibfnamefont {E.}~\bibnamefont {Tollerud}},\ }\href
  {\doibase 10.1093/mnras/stw840} {\bibfield  {journal} {\bibinfo  {journal}
  {Mon. Not. Roy. Astron. Soc.}\ }\textbf {\bibinfo {volume} {460}},\ \bibinfo
  {pages} {2552} (\bibinfo {year} {2016})},\ \Eprint
  {http://arxiv.org/abs/1512.03050} {arXiv:1512.03050 [astro-ph.CO]}
  \BibitemShut {NoStop}%
\bibitem [{\citenamefont {Krause}\ \emph {et~al.}(2024)\citenamefont {Krause}
  \emph {et~al.}}]{Beyond-2pt:2024mqz}%
  \BibitemOpen
  \bibfield  {author} {\bibinfo {author} {\bibfnamefont {E.}~\bibnamefont
  {Krause}} \emph {et~al.} (\bibinfo {collaboration} {Beyond-2pt}),\
  }\href@noop {} {\  (\bibinfo {year} {2024})},\ \Eprint
  {http://arxiv.org/abs/2405.02252} {arXiv:2405.02252 [astro-ph.CO]}
  \BibitemShut {NoStop}%
\bibitem [{\citenamefont {Kobayashi}\ \emph {et~al.}(2022)\citenamefont
  {Kobayashi}, \citenamefont {Nishimichi}, \citenamefont {Takada},\ and\
  \citenamefont {Miyatake}}]{Kobayashi:2021oud}%
  \BibitemOpen
  \bibfield  {author} {\bibinfo {author} {\bibfnamefont {Y.}~\bibnamefont
  {Kobayashi}}, \bibinfo {author} {\bibfnamefont {T.}~\bibnamefont
  {Nishimichi}}, \bibinfo {author} {\bibfnamefont {M.}~\bibnamefont {Takada}},
  \ and\ \bibinfo {author} {\bibfnamefont {H.}~\bibnamefont {Miyatake}},\
  }\href {\doibase 10.1103/PhysRevD.105.083517} {\bibfield  {journal} {\bibinfo
   {journal} {Phys. Rev. D}\ }\textbf {\bibinfo {volume} {105}},\ \bibinfo
  {pages} {083517} (\bibinfo {year} {2022})},\ \Eprint
  {http://arxiv.org/abs/2110.06969} {arXiv:2110.06969 [astro-ph.CO]}
  \BibitemShut {NoStop}%
\bibitem [{\citenamefont {Cuesta-Lazaro}\ \emph {et~al.}(2023)\citenamefont
  {Cuesta-Lazaro} \emph {et~al.}}]{Cuesta-Lazaro:2023gbv}%
  \BibitemOpen
  \bibfield  {author} {\bibinfo {author} {\bibfnamefont {C.}~\bibnamefont
  {Cuesta-Lazaro}} \emph {et~al.},\ }\href@noop {} {\  (\bibinfo {year}
  {2023})},\ \Eprint {http://arxiv.org/abs/2309.16539} {arXiv:2309.16539
  [astro-ph.CO]} \BibitemShut {NoStop}%
\bibitem [{\citenamefont {Valogiannis}\ \emph {et~al.}(2024)\citenamefont
  {Valogiannis}, \citenamefont {Yuan},\ and\ \citenamefont
  {Dvorkin}}]{Valogiannis:2023mxf}%
  \BibitemOpen
  \bibfield  {author} {\bibinfo {author} {\bibfnamefont {G.}~\bibnamefont
  {Valogiannis}}, \bibinfo {author} {\bibfnamefont {S.}~\bibnamefont {Yuan}}, \
  and\ \bibinfo {author} {\bibfnamefont {C.}~\bibnamefont {Dvorkin}},\ }\href
  {\doibase 10.1103/PhysRevD.109.103503} {\bibfield  {journal} {\bibinfo
  {journal} {Phys. Rev. D}\ }\textbf {\bibinfo {volume} {109}},\ \bibinfo
  {pages} {103503} (\bibinfo {year} {2024})},\ \Eprint
  {http://arxiv.org/abs/2310.16116} {arXiv:2310.16116 [astro-ph.CO]}
  \BibitemShut {NoStop}%
\bibitem [{\citenamefont {Hahn}\ \emph {et~al.}(2023)\citenamefont {Hahn},
  \citenamefont {Eickenberg}, \citenamefont {Ho}, \citenamefont {Hou},
  \citenamefont {Lemos}, \citenamefont {Massara}, \citenamefont {Modi},
  \citenamefont {Moradinezhad~Dizgah}, \citenamefont {Parker},\ and\
  \citenamefont {Blancard}}]{Hahn:2023kky}%
  \BibitemOpen
  \bibfield  {author} {\bibinfo {author} {\bibfnamefont {C.}~\bibnamefont
  {Hahn}}, \bibinfo {author} {\bibfnamefont {M.}~\bibnamefont {Eickenberg}},
  \bibinfo {author} {\bibfnamefont {S.}~\bibnamefont {Ho}}, \bibinfo {author}
  {\bibfnamefont {J.}~\bibnamefont {Hou}}, \bibinfo {author} {\bibfnamefont
  {P.}~\bibnamefont {Lemos}}, \bibinfo {author} {\bibfnamefont
  {E.}~\bibnamefont {Massara}}, \bibinfo {author} {\bibfnamefont
  {C.}~\bibnamefont {Modi}}, \bibinfo {author} {\bibfnamefont {A.}~\bibnamefont
  {Moradinezhad~Dizgah}}, \bibinfo {author} {\bibfnamefont {L.}~\bibnamefont
  {Parker}}, \ and\ \bibinfo {author} {\bibfnamefont {B.~R.-S.}\ \bibnamefont
  {Blancard}},\ }\href@noop {} {\  (\bibinfo {year} {2023})},\ \Eprint
  {http://arxiv.org/abs/2310.15243} {arXiv:2310.15243 [astro-ph.CO]}
  \BibitemShut {NoStop}%
\bibitem [{\citenamefont {Hou}\ \emph {et~al.}(2024)\citenamefont {Hou},
  \citenamefont {Moradinezhad~Dizgah}, \citenamefont {Hahn}, \citenamefont
  {Eickenberg}, \citenamefont {Ho}, \citenamefont {Lemos}, \citenamefont
  {Massara}, \citenamefont {Modi}, \citenamefont {Parker},\ and\ \citenamefont
  {Blancard}}]{Hou:2024blc}%
  \BibitemOpen
  \bibfield  {author} {\bibinfo {author} {\bibfnamefont {J.}~\bibnamefont
  {Hou}}, \bibinfo {author} {\bibfnamefont {A.}~\bibnamefont
  {Moradinezhad~Dizgah}}, \bibinfo {author} {\bibfnamefont {C.}~\bibnamefont
  {Hahn}}, \bibinfo {author} {\bibfnamefont {M.}~\bibnamefont {Eickenberg}},
  \bibinfo {author} {\bibfnamefont {S.}~\bibnamefont {Ho}}, \bibinfo {author}
  {\bibfnamefont {P.}~\bibnamefont {Lemos}}, \bibinfo {author} {\bibfnamefont
  {E.}~\bibnamefont {Massara}}, \bibinfo {author} {\bibfnamefont
  {C.}~\bibnamefont {Modi}}, \bibinfo {author} {\bibfnamefont {L.}~\bibnamefont
  {Parker}}, \ and\ \bibinfo {author} {\bibfnamefont {B.~R.-S.}\ \bibnamefont
  {Blancard}},\ }\href {\doibase 10.1103/PhysRevD.109.103528} {\bibfield
  {journal} {\bibinfo  {journal} {Phys. Rev. D}\ }\textbf {\bibinfo {volume}
  {109}},\ \bibinfo {pages} {103528} (\bibinfo {year} {2024})},\ \Eprint
  {http://arxiv.org/abs/2401.15074} {arXiv:2401.15074 [astro-ph.CO]}
  \BibitemShut {NoStop}%
\bibitem [{\citenamefont {Sullivan}\ \emph {et~al.}(2021)\citenamefont
  {Sullivan}, \citenamefont {Seljak},\ and\ \citenamefont
  {Singh}}]{Sullivan:2021sof}%
  \BibitemOpen
  \bibfield  {author} {\bibinfo {author} {\bibfnamefont {J.~M.}\ \bibnamefont
  {Sullivan}}, \bibinfo {author} {\bibfnamefont {U.}~\bibnamefont {Seljak}}, \
  and\ \bibinfo {author} {\bibfnamefont {S.}~\bibnamefont {Singh}},\ }\href
  {\doibase 10.1088/1475-7516/2021/11/026} {\bibfield  {journal} {\bibinfo
  {journal} {JCAP}\ }\textbf {\bibinfo {volume} {11}},\ \bibinfo {pages} {026}
  (\bibinfo {year} {2021})},\ \Eprint {http://arxiv.org/abs/2104.10676}
  {arXiv:2104.10676 [astro-ph.CO]} \BibitemShut {NoStop}%
\bibitem [{\citenamefont {Ivanov}\ \emph
  {et~al.}(2024{\natexlab{a}})\citenamefont {Ivanov}, \citenamefont
  {Cuesta-Lazaro}, \citenamefont {Mishra-Sharma}, \citenamefont {Obuljen},\
  and\ \citenamefont {Toomey}}]{Ivanov:2024hgq}%
  \BibitemOpen
  \bibfield  {author} {\bibinfo {author} {\bibfnamefont {M.~M.}\ \bibnamefont
  {Ivanov}}, \bibinfo {author} {\bibfnamefont {C.}~\bibnamefont
  {Cuesta-Lazaro}}, \bibinfo {author} {\bibfnamefont {S.}~\bibnamefont
  {Mishra-Sharma}}, \bibinfo {author} {\bibfnamefont {A.}~\bibnamefont
  {Obuljen}}, \ and\ \bibinfo {author} {\bibfnamefont {M.~W.}\ \bibnamefont
  {Toomey}},\ }\href@noop {} {\  (\bibinfo {year} {2024}{\natexlab{a}})},\
  \Eprint {http://arxiv.org/abs/2402.13310} {arXiv:2402.13310 [astro-ph.CO]}
  \BibitemShut {NoStop}%
\bibitem [{\citenamefont {Cabass}\ \emph {et~al.}(2024)\citenamefont {Cabass},
  \citenamefont {Philcox}, \citenamefont {Ivanov}, \citenamefont {Akitsu},
  \citenamefont {Chen}, \citenamefont {Simonovi\'c},\ and\ \citenamefont
  {Zaldarriaga}}]{Cabass:2024wob}%
  \BibitemOpen
  \bibfield  {author} {\bibinfo {author} {\bibfnamefont {G.}~\bibnamefont
  {Cabass}}, \bibinfo {author} {\bibfnamefont {O.~H.~E.}\ \bibnamefont
  {Philcox}}, \bibinfo {author} {\bibfnamefont {M.~M.}\ \bibnamefont {Ivanov}},
  \bibinfo {author} {\bibfnamefont {K.}~\bibnamefont {Akitsu}}, \bibinfo
  {author} {\bibfnamefont {S.-F.}\ \bibnamefont {Chen}}, \bibinfo {author}
  {\bibfnamefont {M.}~\bibnamefont {Simonovi\'c}}, \ and\ \bibinfo {author}
  {\bibfnamefont {M.}~\bibnamefont {Zaldarriaga}},\ }\href@noop {} {\
  (\bibinfo {year} {2024})},\ \Eprint {http://arxiv.org/abs/2404.01894}
  {arXiv:2404.01894 [astro-ph.CO]} \BibitemShut {NoStop}%
\bibitem [{\citenamefont {Obuljen}\ \emph {et~al.}(2023)\citenamefont
  {Obuljen}, \citenamefont {Simonovi\'c}, \citenamefont {Schneider},\ and\
  \citenamefont {Feldmann}}]{Obuljen:2022cjo}%
  \BibitemOpen
  \bibfield  {author} {\bibinfo {author} {\bibfnamefont {A.}~\bibnamefont
  {Obuljen}}, \bibinfo {author} {\bibfnamefont {M.}~\bibnamefont
  {Simonovi\'c}}, \bibinfo {author} {\bibfnamefont {A.}~\bibnamefont
  {Schneider}}, \ and\ \bibinfo {author} {\bibfnamefont {R.}~\bibnamefont
  {Feldmann}},\ }\href {\doibase 10.1103/PhysRevD.108.083528} {\bibfield
  {journal} {\bibinfo  {journal} {Phys. Rev. D}\ }\textbf {\bibinfo {volume}
  {108}},\ \bibinfo {pages} {083528} (\bibinfo {year} {2023})},\ \Eprint
  {http://arxiv.org/abs/2207.12398} {arXiv:2207.12398 [astro-ph.CO]}
  \BibitemShut {NoStop}%
\bibitem [{\citenamefont {Modi}\ and\ \citenamefont
  {Philcox}(2023)}]{Modi:2023drt}%
  \BibitemOpen
  \bibfield  {author} {\bibinfo {author} {\bibfnamefont {C.}~\bibnamefont
  {Modi}}\ and\ \bibinfo {author} {\bibfnamefont {O.~H.~E.}\ \bibnamefont
  {Philcox}},\ }\href@noop {} {\  (\bibinfo {year} {2023})},\ \Eprint
  {http://arxiv.org/abs/2309.10270} {arXiv:2309.10270 [astro-ph.CO]}
  \BibitemShut {NoStop}%
\bibitem [{\citenamefont {Modi}\ \emph {et~al.}(2020)\citenamefont {Modi},
  \citenamefont {Chen},\ and\ \citenamefont {White}}]{Modi:2019qbt}%
  \BibitemOpen
  \bibfield  {author} {\bibinfo {author} {\bibfnamefont {C.}~\bibnamefont
  {Modi}}, \bibinfo {author} {\bibfnamefont {S.-F.}\ \bibnamefont {Chen}}, \
  and\ \bibinfo {author} {\bibfnamefont {M.}~\bibnamefont {White}},\ }\href
  {\doibase 10.1093/mnras/staa251} {\bibfield  {journal} {\bibinfo  {journal}
  {Mon. Not. Roy. Astron. Soc.}\ }\textbf {\bibinfo {volume} {492}},\ \bibinfo
  {pages} {5754} (\bibinfo {year} {2020})},\ \Eprint
  {http://arxiv.org/abs/1910.07097} {arXiv:1910.07097 [astro-ph.CO]}
  \BibitemShut {NoStop}%
\bibitem [{\citenamefont {Kokron}\ \emph {et~al.}(2021)\citenamefont {Kokron},
  \citenamefont {DeRose}, \citenamefont {Chen}, \citenamefont {White},\ and\
  \citenamefont {Wechsler}}]{Kokron:2021xgh}%
  \BibitemOpen
  \bibfield  {author} {\bibinfo {author} {\bibfnamefont {N.}~\bibnamefont
  {Kokron}}, \bibinfo {author} {\bibfnamefont {J.}~\bibnamefont {DeRose}},
  \bibinfo {author} {\bibfnamefont {S.-F.}\ \bibnamefont {Chen}}, \bibinfo
  {author} {\bibfnamefont {M.}~\bibnamefont {White}}, \ and\ \bibinfo {author}
  {\bibfnamefont {R.~H.}\ \bibnamefont {Wechsler}},\ }\href {\doibase
  10.1093/mnras/stab1358} {\bibfield  {journal} {\bibinfo  {journal} {Mon. Not.
  Roy. Astron. Soc.}\ }\textbf {\bibinfo {volume} {505}},\ \bibinfo {pages}
  {1422} (\bibinfo {year} {2021})},\ \Eprint {http://arxiv.org/abs/2101.11014}
  {arXiv:2101.11014 [astro-ph.CO]} \BibitemShut {NoStop}%
\bibitem [{\citenamefont {Pellejero-Ibanez}\ \emph {et~al.}(2023)\citenamefont
  {Pellejero-Ibanez}, \citenamefont {Angulo}, \citenamefont {Zennaro},
  \citenamefont {Stuecker}, \citenamefont {Contreras}, \citenamefont {Arico},\
  and\ \citenamefont {Maion}}]{Pellejero-Ibanez:2022efv}%
  \BibitemOpen
  \bibfield  {author} {\bibinfo {author} {\bibfnamefont {M.}~\bibnamefont
  {Pellejero-Ibanez}}, \bibinfo {author} {\bibfnamefont {R.~E.}\ \bibnamefont
  {Angulo}}, \bibinfo {author} {\bibfnamefont {M.}~\bibnamefont {Zennaro}},
  \bibinfo {author} {\bibfnamefont {J.}~\bibnamefont {Stuecker}}, \bibinfo
  {author} {\bibfnamefont {S.}~\bibnamefont {Contreras}}, \bibinfo {author}
  {\bibfnamefont {G.}~\bibnamefont {Arico}}, \ and\ \bibinfo {author}
  {\bibfnamefont {F.}~\bibnamefont {Maion}},\ }\href {\doibase
  10.1093/mnras/stad368} {\bibfield  {journal} {\bibinfo  {journal} {Mon. Not.
  Roy. Astron. Soc.}\ }\textbf {\bibinfo {volume} {520}},\ \bibinfo {pages}
  {3725} (\bibinfo {year} {2023})},\ \Eprint {http://arxiv.org/abs/2207.06437}
  {arXiv:2207.06437 [astro-ph.CO]} \BibitemShut {NoStop}%
\bibitem [{\citenamefont {Baradaran}\ \emph {et~al.}(2024)\citenamefont
  {Baradaran}, \citenamefont {Hadzhiyska}, \citenamefont {White},\ and\
  \citenamefont {Sailer}}]{Baradaran:2024jlh}%
  \BibitemOpen
  \bibfield  {author} {\bibinfo {author} {\bibfnamefont {D.}~\bibnamefont
  {Baradaran}}, \bibinfo {author} {\bibfnamefont {B.}~\bibnamefont
  {Hadzhiyska}}, \bibinfo {author} {\bibfnamefont {M.~J.}\ \bibnamefont
  {White}}, \ and\ \bibinfo {author} {\bibfnamefont {N.}~\bibnamefont
  {Sailer}},\ }\href@noop {} {\  (\bibinfo {year} {2024})},\ \Eprint
  {http://arxiv.org/abs/2406.13079} {arXiv:2406.13079 [astro-ph.CO]}
  \BibitemShut {NoStop}%
\bibitem [{\citenamefont {Kokron}\ \emph {et~al.}(2022)\citenamefont {Kokron},
  \citenamefont {DeRose}, \citenamefont {Chen}, \citenamefont {White},\ and\
  \citenamefont {Wechsler}}]{Kokron:2021faa}%
  \BibitemOpen
  \bibfield  {author} {\bibinfo {author} {\bibfnamefont {N.}~\bibnamefont
  {Kokron}}, \bibinfo {author} {\bibfnamefont {J.}~\bibnamefont {DeRose}},
  \bibinfo {author} {\bibfnamefont {S.-F.}\ \bibnamefont {Chen}}, \bibinfo
  {author} {\bibfnamefont {M.}~\bibnamefont {White}}, \ and\ \bibinfo {author}
  {\bibfnamefont {R.~H.}\ \bibnamefont {Wechsler}},\ }\href {\doibase
  10.1093/mnras/stac1420} {\bibfield  {journal} {\bibinfo  {journal} {Mon. Not.
  Roy. Astron. Soc.}\ }\textbf {\bibinfo {volume} {514}},\ \bibinfo {pages}
  {2198} (\bibinfo {year} {2022})},\ \Eprint {http://arxiv.org/abs/2112.00012}
  {arXiv:2112.00012 [astro-ph.CO]} \BibitemShut {NoStop}%
\bibitem [{\citenamefont {Donoghue}\ \emph {et~al.}(2023)\citenamefont
  {Donoghue}, \citenamefont {Golowich},\ and\ \citenamefont
  {Holstein}}]{Donoghue_Golowich_Holstein_2023}%
  \BibitemOpen
  \bibfield  {author} {\bibinfo {author} {\bibfnamefont {J.~F.}\ \bibnamefont
  {Donoghue}}, \bibinfo {author} {\bibfnamefont {E.}~\bibnamefont {Golowich}},
  \ and\ \bibinfo {author} {\bibfnamefont {B.~R.}\ \bibnamefont {Holstein}},\
  }\href@noop {} {\emph {\bibinfo {title} {Dynamics of the Standard Model}}},\
  \bibinfo {edition} {2nd}\ ed.,\ Cambridge Monographs on Particle Physics,
  Nuclear Physics and Cosmology\ (\bibinfo  {publisher} {Cambridge University
  Press},\ \bibinfo {year} {2023})\BibitemShut {NoStop}%
\bibitem [{\citenamefont {Donoghue}\ \emph {et~al.}(2017)\citenamefont
  {Donoghue}, \citenamefont {Ivanov},\ and\ \citenamefont
  {Shkerin}}]{Donoghue:2017pgk}%
  \BibitemOpen
  \bibfield  {author} {\bibinfo {author} {\bibfnamefont {J.~F.}\ \bibnamefont
  {Donoghue}}, \bibinfo {author} {\bibfnamefont {M.~M.}\ \bibnamefont
  {Ivanov}}, \ and\ \bibinfo {author} {\bibfnamefont {A.}~\bibnamefont
  {Shkerin}},\ }\href@noop {} {\  (\bibinfo {year} {2017})},\ \Eprint
  {http://arxiv.org/abs/1702.00319} {arXiv:1702.00319 [hep-th]} \BibitemShut
  {NoStop}%
\bibitem [{\citenamefont {Park}\ \emph {et~al.}(2022)\citenamefont {Park},
  \citenamefont {Gupta}, \citenamefont {Yoon}, \citenamefont {Mondal},
  \citenamefont {Bhattacharya}, \citenamefont {Jang}, \citenamefont {Jo\'o},\
  and\ \citenamefont {Winter}}]{Park:2021ypf}%
  \BibitemOpen
  \bibfield  {author} {\bibinfo {author} {\bibfnamefont {S.}~\bibnamefont
  {Park}}, \bibinfo {author} {\bibfnamefont {R.}~\bibnamefont {Gupta}},
  \bibinfo {author} {\bibfnamefont {B.}~\bibnamefont {Yoon}}, \bibinfo {author}
  {\bibfnamefont {S.}~\bibnamefont {Mondal}}, \bibinfo {author} {\bibfnamefont
  {T.}~\bibnamefont {Bhattacharya}}, \bibinfo {author} {\bibfnamefont {Y.-C.}\
  \bibnamefont {Jang}}, \bibinfo {author} {\bibfnamefont {B.}~\bibnamefont
  {Jo\'o}}, \ and\ \bibinfo {author} {\bibfnamefont {F.}~\bibnamefont {Winter}}
  (\bibinfo {collaboration} {Nucleon Matrix Elements (NME)}),\ }\href {\doibase
  10.1103/PhysRevD.105.054505} {\bibfield  {journal} {\bibinfo  {journal}
  {Phys. Rev. D}\ }\textbf {\bibinfo {volume} {105}},\ \bibinfo {pages}
  {054505} (\bibinfo {year} {2022})},\ \Eprint
  {http://arxiv.org/abs/2103.05599} {arXiv:2103.05599 [hep-lat]} \BibitemShut
  {NoStop}%
\bibitem [{\citenamefont {Abbott}\ \emph {et~al.}(2024)\citenamefont {Abbott},
  \citenamefont {Detmold}, \citenamefont {Illa}, \citenamefont {Parre\~no},
  \citenamefont {Perry}, \citenamefont {Romero-L\'opez}, \citenamefont
  {Shanahan},\ and\ \citenamefont {Wagman}}]{Abbott:2024vhj}%
  \BibitemOpen
  \bibfield  {author} {\bibinfo {author} {\bibfnamefont {R.}~\bibnamefont
  {Abbott}}, \bibinfo {author} {\bibfnamefont {W.}~\bibnamefont {Detmold}},
  \bibinfo {author} {\bibfnamefont {M.}~\bibnamefont {Illa}}, \bibinfo {author}
  {\bibfnamefont {A.}~\bibnamefont {Parre\~no}}, \bibinfo {author}
  {\bibfnamefont {R.~J.}\ \bibnamefont {Perry}}, \bibinfo {author}
  {\bibfnamefont {F.}~\bibnamefont {Romero-L\'opez}}, \bibinfo {author}
  {\bibfnamefont {P.~E.}\ \bibnamefont {Shanahan}}, \ and\ \bibinfo {author}
  {\bibfnamefont {M.~L.}\ \bibnamefont {Wagman}},\ }\href@noop {} {\  (\bibinfo
  {year} {2024})},\ \Eprint {http://arxiv.org/abs/2406.09273} {arXiv:2406.09273
  [hep-lat]} \BibitemShut {NoStop}%
\bibitem [{\citenamefont {Abbott}\ \emph {et~al.}(2023)\citenamefont {Abbott},
  \citenamefont {Detmold}, \citenamefont {Romero-L\'opez}, \citenamefont
  {Davoudi}, \citenamefont {Illa}, \citenamefont {Parre\~no}, \citenamefont
  {Perry}, \citenamefont {Shanahan},\ and\ \citenamefont
  {Wagman}}]{Abbott:2023coj}%
  \BibitemOpen
  \bibfield  {author} {\bibinfo {author} {\bibfnamefont {R.}~\bibnamefont
  {Abbott}}, \bibinfo {author} {\bibfnamefont {W.}~\bibnamefont {Detmold}},
  \bibinfo {author} {\bibfnamefont {F.}~\bibnamefont {Romero-L\'opez}},
  \bibinfo {author} {\bibfnamefont {Z.}~\bibnamefont {Davoudi}}, \bibinfo
  {author} {\bibfnamefont {M.}~\bibnamefont {Illa}}, \bibinfo {author}
  {\bibfnamefont {A.}~\bibnamefont {Parre\~no}}, \bibinfo {author}
  {\bibfnamefont {R.~J.}\ \bibnamefont {Perry}}, \bibinfo {author}
  {\bibfnamefont {P.~E.}\ \bibnamefont {Shanahan}}, \ and\ \bibinfo {author}
  {\bibfnamefont {M.~L.}\ \bibnamefont {Wagman}} (\bibinfo {collaboration}
  {NPLQCD}),\ }\href {\doibase 10.1103/PhysRevD.108.114506} {\bibfield
  {journal} {\bibinfo  {journal} {Phys. Rev. D}\ }\textbf {\bibinfo {volume}
  {108}},\ \bibinfo {pages} {114506} (\bibinfo {year} {2023})},\ \Eprint
  {http://arxiv.org/abs/2307.15014} {arXiv:2307.15014 [hep-lat]} \BibitemShut
  {NoStop}%
\bibitem [{\citenamefont {He}\ and\ \citenamefont
  {Kruczenski}(2023)}]{He:2023lyy}%
  \BibitemOpen
  \bibfield  {author} {\bibinfo {author} {\bibfnamefont {Y.}~\bibnamefont
  {He}}\ and\ \bibinfo {author} {\bibfnamefont {M.}~\bibnamefont
  {Kruczenski}},\ }\href@noop {} {\  (\bibinfo {year} {2023})},\ \Eprint
  {http://arxiv.org/abs/2309.12402} {arXiv:2309.12402 [hep-th]} \BibitemShut
  {NoStop}%
\bibitem [{\citenamefont {He}\ and\ \citenamefont
  {Kruczenski}(2024)}]{He:2024nwd}%
  \BibitemOpen
  \bibfield  {author} {\bibinfo {author} {\bibfnamefont {Y.}~\bibnamefont
  {He}}\ and\ \bibinfo {author} {\bibfnamefont {M.}~\bibnamefont
  {Kruczenski}},\ }\href@noop {} {\  (\bibinfo {year} {2024})},\ \Eprint
  {http://arxiv.org/abs/2403.10772} {arXiv:2403.10772 [hep-th]} \BibitemShut
  {NoStop}%
\bibitem [{\citenamefont {Akrami}\ \emph {et~al.}(2020)\citenamefont {Akrami}
  \emph {et~al.}}]{Planck:2019kim}%
  \BibitemOpen
  \bibfield  {author} {\bibinfo {author} {\bibfnamefont {Y.}~\bibnamefont
  {Akrami}} \emph {et~al.} (\bibinfo {collaboration} {Planck}),\ }\href
  {\doibase 10.1051/0004-6361/201935891} {\bibfield  {journal} {\bibinfo
  {journal} {Astron. Astrophys.}\ }\textbf {\bibinfo {volume} {641}},\ \bibinfo
  {pages} {A9} (\bibinfo {year} {2020})},\ \Eprint
  {http://arxiv.org/abs/1905.05697} {arXiv:1905.05697 [astro-ph.CO]}
  \BibitemShut {NoStop}%
\bibitem [{\citenamefont {Moradinezhad~Dizgah}\ \emph
  {et~al.}(2021)\citenamefont {Moradinezhad~Dizgah}, \citenamefont {Biagetti},
  \citenamefont {Sefusatti}, \citenamefont {Desjacques},\ and\ \citenamefont
  {Nore\~na}}]{MoradinezhadDizgah:2020whw}%
  \BibitemOpen
  \bibfield  {author} {\bibinfo {author} {\bibfnamefont {A.}~\bibnamefont
  {Moradinezhad~Dizgah}}, \bibinfo {author} {\bibfnamefont {M.}~\bibnamefont
  {Biagetti}}, \bibinfo {author} {\bibfnamefont {E.}~\bibnamefont {Sefusatti}},
  \bibinfo {author} {\bibfnamefont {V.}~\bibnamefont {Desjacques}}, \ and\
  \bibinfo {author} {\bibfnamefont {J.}~\bibnamefont {Nore\~na}},\ }\href
  {\doibase 10.1088/1475-7516/2021/05/015} {\bibfield  {journal} {\bibinfo
  {journal} {JCAP}\ }\textbf {\bibinfo {volume} {05}},\ \bibinfo {pages} {015}
  (\bibinfo {year} {2021})},\ \Eprint {http://arxiv.org/abs/2010.14523}
  {arXiv:2010.14523 [astro-ph.CO]} \BibitemShut {NoStop}%
\bibitem [{\citenamefont {Alam}\ \emph {et~al.}(2017)\citenamefont {Alam} \emph
  {et~al.}}]{BOSS:2016wmc}%
  \BibitemOpen
  \bibfield  {author} {\bibinfo {author} {\bibfnamefont {S.}~\bibnamefont
  {Alam}} \emph {et~al.} (\bibinfo {collaboration} {BOSS}),\ }\href {\doibase
  10.1093/mnras/stx721} {\bibfield  {journal} {\bibinfo  {journal} {Mon. Not.
  Roy. Astron. Soc.}\ }\textbf {\bibinfo {volume} {470}},\ \bibinfo {pages}
  {2617} (\bibinfo {year} {2017})},\ \Eprint {http://arxiv.org/abs/1607.03155}
  {arXiv:1607.03155 [astro-ph.CO]} \BibitemShut {NoStop}%
\bibitem [{\citenamefont {Yuan}\ \emph
  {et~al.}(2022{\natexlab{a}})\citenamefont {Yuan}, \citenamefont {Garrison},
  \citenamefont {Eisenstein},\ and\ \citenamefont {Wechsler}}]{Yuan:2022jqf}%
  \BibitemOpen
  \bibfield  {author} {\bibinfo {author} {\bibfnamefont {S.}~\bibnamefont
  {Yuan}}, \bibinfo {author} {\bibfnamefont {L.~H.}\ \bibnamefont {Garrison}},
  \bibinfo {author} {\bibfnamefont {D.~J.}\ \bibnamefont {Eisenstein}}, \ and\
  \bibinfo {author} {\bibfnamefont {R.~H.}\ \bibnamefont {Wechsler}},\ }\href
  {\doibase 10.1093/mnras/stac1830} {\bibfield  {journal} {\bibinfo  {journal}
  {Mon. Not. Roy. Astron. Soc.}\ }\textbf {\bibinfo {volume} {515}},\ \bibinfo
  {pages} {871} (\bibinfo {year} {2022}{\natexlab{a}})},\ \Eprint
  {http://arxiv.org/abs/2203.11963} {arXiv:2203.11963 [astro-ph.CO]}
  \BibitemShut {NoStop}%
\bibitem [{\citenamefont {Chen}\ \emph
  {et~al.}(2022{\natexlab{b}})\citenamefont {Chen}, \citenamefont {White},
  \citenamefont {DeRose},\ and\ \citenamefont {Kokron}}]{Chen:2022jzq}%
  \BibitemOpen
  \bibfield  {author} {\bibinfo {author} {\bibfnamefont {S.-F.}\ \bibnamefont
  {Chen}}, \bibinfo {author} {\bibfnamefont {M.}~\bibnamefont {White}},
  \bibinfo {author} {\bibfnamefont {J.}~\bibnamefont {DeRose}}, \ and\ \bibinfo
  {author} {\bibfnamefont {N.}~\bibnamefont {Kokron}},\ }\href {\doibase
  10.1088/1475-7516/2022/07/041} {\bibfield  {journal} {\bibinfo  {journal}
  {JCAP}\ }\textbf {\bibinfo {volume} {07}},\ \bibinfo {pages} {041} (\bibinfo
  {year} {2022}{\natexlab{b}})},\ \Eprint {http://arxiv.org/abs/2204.10392}
  {arXiv:2204.10392 [astro-ph.CO]} \BibitemShut {NoStop}%
\bibitem [{\citenamefont {Lange}\ \emph {et~al.}(2023)\citenamefont {Lange},
  \citenamefont {Hearin}, \citenamefont {Leauthaud}, \citenamefont {van~den
  Bosch}, \citenamefont {Xhakaj}, \citenamefont {Guo}, \citenamefont
  {Wechsler},\ and\ \citenamefont {DeRose}}]{Lange:2023khv}%
  \BibitemOpen
  \bibfield  {author} {\bibinfo {author} {\bibfnamefont {J.~U.}\ \bibnamefont
  {Lange}}, \bibinfo {author} {\bibfnamefont {A.~P.}\ \bibnamefont {Hearin}},
  \bibinfo {author} {\bibfnamefont {A.}~\bibnamefont {Leauthaud}}, \bibinfo
  {author} {\bibfnamefont {F.~C.}\ \bibnamefont {van~den Bosch}}, \bibinfo
  {author} {\bibfnamefont {E.}~\bibnamefont {Xhakaj}}, \bibinfo {author}
  {\bibfnamefont {H.}~\bibnamefont {Guo}}, \bibinfo {author} {\bibfnamefont
  {R.~H.}\ \bibnamefont {Wechsler}}, \ and\ \bibinfo {author} {\bibfnamefont
  {J.}~\bibnamefont {DeRose}},\ }\href {\doibase 10.1093/mnras/stad473} {\
  (\bibinfo {year} {2023}),\ 10.1093/mnras/stad473},\ \Eprint
  {http://arxiv.org/abs/2301.08692} {arXiv:2301.08692 [astro-ph.CO]}
  \BibitemShut {NoStop}%
\bibitem [{\citenamefont {Ivanov}\ \emph {et~al.}(2023)\citenamefont {Ivanov},
  \citenamefont {Philcox}, \citenamefont {Cabass}, \citenamefont {Nishimichi},
  \citenamefont {Simonovi\'c},\ and\ \citenamefont
  {Zaldarriaga}}]{Ivanov:2023qzb}%
  \BibitemOpen
  \bibfield  {author} {\bibinfo {author} {\bibfnamefont {M.~M.}\ \bibnamefont
  {Ivanov}}, \bibinfo {author} {\bibfnamefont {O.~H.~E.}\ \bibnamefont
  {Philcox}}, \bibinfo {author} {\bibfnamefont {G.}~\bibnamefont {Cabass}},
  \bibinfo {author} {\bibfnamefont {T.}~\bibnamefont {Nishimichi}}, \bibinfo
  {author} {\bibfnamefont {M.}~\bibnamefont {Simonovi\'c}}, \ and\ \bibinfo
  {author} {\bibfnamefont {M.}~\bibnamefont {Zaldarriaga}},\ }\href {\doibase
  10.1103/PhysRevD.107.083515} {\bibfield  {journal} {\bibinfo  {journal}
  {Phys. Rev. D}\ }\textbf {\bibinfo {volume} {107}},\ \bibinfo {pages}
  {083515} (\bibinfo {year} {2023})},\ \Eprint
  {http://arxiv.org/abs/2302.04414} {arXiv:2302.04414 [astro-ph.CO]}
  \BibitemShut {NoStop}%
\bibitem [{\citenamefont {Ib\'a\~nez}\ \emph {et~al.}(2024)\citenamefont
  {Ib\'a\~nez}, \citenamefont {Angulo},\ and\ \citenamefont
  {Peacock}}]{Ibanez:2024uua}%
  \BibitemOpen
  \bibfield  {author} {\bibinfo {author} {\bibfnamefont {M.~P.}\ \bibnamefont
  {Ib\'a\~nez}}, \bibinfo {author} {\bibfnamefont {R.~E.}\ \bibnamefont
  {Angulo}}, \ and\ \bibinfo {author} {\bibfnamefont {J.~A.}\ \bibnamefont
  {Peacock}},\ }\href@noop {} {\  (\bibinfo {year} {2024})},\ \Eprint
  {http://arxiv.org/abs/2407.07949} {arXiv:2407.07949 [astro-ph.CO]}
  \BibitemShut {NoStop}%
\bibitem [{\citenamefont {Paillas}\ \emph {et~al.}(2023)\citenamefont {Paillas}
  \emph {et~al.}}]{Paillas:2023cpk}%
  \BibitemOpen
  \bibfield  {author} {\bibinfo {author} {\bibfnamefont {E.}~\bibnamefont
  {Paillas}} \emph {et~al.},\ }\href@noop {} {\  (\bibinfo {year} {2023})},\
  \Eprint {http://arxiv.org/abs/2309.16541} {arXiv:2309.16541 [astro-ph.CO]}
  \BibitemShut {NoStop}%
\bibitem [{\citenamefont {Sailer}\ \emph {et~al.}(2024)\citenamefont {Sailer}
  \emph {et~al.}}]{Sailer:2024coh}%
  \BibitemOpen
  \bibfield  {author} {\bibinfo {author} {\bibfnamefont {N.}~\bibnamefont
  {Sailer}} \emph {et~al.},\ }\href@noop {} {\  (\bibinfo {year} {2024})},\
  \Eprint {http://arxiv.org/abs/2407.04607} {arXiv:2407.04607 [astro-ph.CO]}
  \BibitemShut {NoStop}%
\bibitem [{\citenamefont {Chen}\ \emph
  {et~al.}(2024{\natexlab{b}})\citenamefont {Chen} \emph
  {et~al.}}]{Chen:2024vvk}%
  \BibitemOpen
  \bibfield  {author} {\bibinfo {author} {\bibfnamefont {S.}~\bibnamefont
  {Chen}} \emph {et~al.},\ }\href@noop {} {\  (\bibinfo {year}
  {2024}{\natexlab{b}})},\ \Eprint {http://arxiv.org/abs/2407.04795}
  {arXiv:2407.04795 [astro-ph.CO]} \BibitemShut {NoStop}%
\bibitem [{\citenamefont {Schmittfull}\ \emph {et~al.}(2019)\citenamefont
  {Schmittfull}, \citenamefont {Simonović}, \citenamefont {Assassi},\ and\
  \citenamefont {Zaldarriaga}}]{Schmittfull:2018yuk}%
  \BibitemOpen
  \bibfield  {author} {\bibinfo {author} {\bibfnamefont {M.}~\bibnamefont
  {Schmittfull}}, \bibinfo {author} {\bibfnamefont {M.}~\bibnamefont
  {Simonović}}, \bibinfo {author} {\bibfnamefont {V.}~\bibnamefont {Assassi}},
  \ and\ \bibinfo {author} {\bibfnamefont {M.}~\bibnamefont {Zaldarriaga}},\
  }\href {\doibase 10.1103/PhysRevD.100.043514} {\bibfield  {journal} {\bibinfo
   {journal} {Phys.\ Rev.\ D}\ }\textbf {\bibinfo {volume} {100}},\ \bibinfo
  {pages} {043514} (\bibinfo {year} {2019})},\ \Eprint
  {http://arxiv.org/abs/1811.10640} {arXiv:1811.10640 [astro-ph.CO]}
  \BibitemShut {NoStop}%
\bibitem [{\citenamefont {Schmittfull}\ \emph {et~al.}(2021)\citenamefont
  {Schmittfull}, \citenamefont {Simonovi\'c}, \citenamefont {Ivanov},
  \citenamefont {Philcox},\ and\ \citenamefont
  {Zaldarriaga}}]{Schmittfull:2020trd}%
  \BibitemOpen
  \bibfield  {author} {\bibinfo {author} {\bibfnamefont {M.}~\bibnamefont
  {Schmittfull}}, \bibinfo {author} {\bibfnamefont {M.}~\bibnamefont
  {Simonovi\'c}}, \bibinfo {author} {\bibfnamefont {M.~M.}\ \bibnamefont
  {Ivanov}}, \bibinfo {author} {\bibfnamefont {O.~H.~E.}\ \bibnamefont
  {Philcox}}, \ and\ \bibinfo {author} {\bibfnamefont {M.}~\bibnamefont
  {Zaldarriaga}},\ }\href {\doibase 10.1088/1475-7516/2021/05/059} {\bibfield
  {journal} {\bibinfo  {journal} {JCAP}\ }\textbf {\bibinfo {volume} {05}},\
  \bibinfo {pages} {059} (\bibinfo {year} {2021})},\ \Eprint
  {http://arxiv.org/abs/2012.03334} {arXiv:2012.03334 [astro-ph.CO]}
  \BibitemShut {NoStop}%
\bibitem [{\citenamefont {Foreman}\ \emph {et~al.}(2024)\citenamefont
  {Foreman}, \citenamefont {Obuljen},\ and\ \citenamefont
  {Simonovi\'c}}]{Foreman:2024kzw}%
  \BibitemOpen
  \bibfield  {author} {\bibinfo {author} {\bibfnamefont {S.}~\bibnamefont
  {Foreman}}, \bibinfo {author} {\bibfnamefont {A.}~\bibnamefont {Obuljen}}, \
  and\ \bibinfo {author} {\bibfnamefont {M.}~\bibnamefont {Simonovi\'c}},\
  }\href@noop {} {\  (\bibinfo {year} {2024})},\ \Eprint
  {http://arxiv.org/abs/2405.18559} {arXiv:2405.18559 [astro-ph.CO]}
  \BibitemShut {NoStop}%
\bibitem [{\citenamefont {Schmittfull}\ \emph {et~al.}(2015)\citenamefont
  {Schmittfull}, \citenamefont {Baldauf},\ and\ \citenamefont
  {Seljak}}]{Schmittfull:2014tca}%
  \BibitemOpen
  \bibfield  {author} {\bibinfo {author} {\bibfnamefont {M.}~\bibnamefont
  {Schmittfull}}, \bibinfo {author} {\bibfnamefont {T.}~\bibnamefont
  {Baldauf}}, \ and\ \bibinfo {author} {\bibfnamefont {U.}~\bibnamefont
  {Seljak}},\ }\href {\doibase 10.1103/PhysRevD.91.043530} {\bibfield
  {journal} {\bibinfo  {journal} {Phys. Rev.}\ }\textbf {\bibinfo {volume}
  {D91}},\ \bibinfo {pages} {043530} (\bibinfo {year} {2015})},\ \Eprint
  {http://arxiv.org/abs/1411.6595} {arXiv:1411.6595 [astro-ph.CO]} \BibitemShut
  {NoStop}%
\bibitem [{\citenamefont {Lazeyras}\ and\ \citenamefont
  {Schmidt}(2018)}]{Lazeyras:2017hxw}%
  \BibitemOpen
  \bibfield  {author} {\bibinfo {author} {\bibfnamefont {T.}~\bibnamefont
  {Lazeyras}}\ and\ \bibinfo {author} {\bibfnamefont {F.}~\bibnamefont
  {Schmidt}},\ }\href {\doibase 10.1088/1475-7516/2018/09/008} {\bibfield
  {journal} {\bibinfo  {journal} {JCAP}\ }\textbf {\bibinfo {volume} {1809}},\
  \bibinfo {pages} {008} (\bibinfo {year} {2018})},\ \Eprint
  {http://arxiv.org/abs/1712.07531} {arXiv:1712.07531 [astro-ph.CO]}
  \BibitemShut {NoStop}%
\bibitem [{\citenamefont {Abidi}\ and\ \citenamefont
  {Baldauf}(2018)}]{Abidi:2018eyd}%
  \BibitemOpen
  \bibfield  {author} {\bibinfo {author} {\bibfnamefont {M.~M.}\ \bibnamefont
  {Abidi}}\ and\ \bibinfo {author} {\bibfnamefont {T.}~\bibnamefont
  {Baldauf}},\ }\href {\doibase 10.1088/1475-7516/2018/07/029} {\bibfield
  {journal} {\bibinfo  {journal} {JCAP}\ }\textbf {\bibinfo {volume} {1807}},\
  \bibinfo {pages} {029} (\bibinfo {year} {2018})},\ \Eprint
  {http://arxiv.org/abs/1802.07622} {arXiv:1802.07622 [astro-ph.CO]}
  \BibitemShut {NoStop}%
\bibitem [{\citenamefont {Schmidt}\ \emph {et~al.}(2019)\citenamefont
  {Schmidt}, \citenamefont {Elsner}, \citenamefont {Jasche}, \citenamefont
  {Nguyen},\ and\ \citenamefont {Lavaux}}]{Schmidt:2018bkr}%
  \BibitemOpen
  \bibfield  {author} {\bibinfo {author} {\bibfnamefont {F.}~\bibnamefont
  {Schmidt}}, \bibinfo {author} {\bibfnamefont {F.}~\bibnamefont {Elsner}},
  \bibinfo {author} {\bibfnamefont {J.}~\bibnamefont {Jasche}}, \bibinfo
  {author} {\bibfnamefont {N.~M.}\ \bibnamefont {Nguyen}}, \ and\ \bibinfo
  {author} {\bibfnamefont {G.}~\bibnamefont {Lavaux}},\ }\href {\doibase
  10.1088/1475-7516/2019/01/042} {\bibfield  {journal} {\bibinfo  {journal}
  {JCAP}\ }\textbf {\bibinfo {volume} {01}},\ \bibinfo {pages} {042} (\bibinfo
  {year} {2019})},\ \Eprint {http://arxiv.org/abs/1808.02002} {arXiv:1808.02002
  [astro-ph.CO]} \BibitemShut {NoStop}%
\bibitem [{\citenamefont {Elsner}\ \emph {et~al.}(2020)\citenamefont {Elsner},
  \citenamefont {Schmidt}, \citenamefont {Jasche}, \citenamefont {Lavaux},\
  and\ \citenamefont {Nguyen}}]{Elsner:2019rql}%
  \BibitemOpen
  \bibfield  {author} {\bibinfo {author} {\bibfnamefont {F.}~\bibnamefont
  {Elsner}}, \bibinfo {author} {\bibfnamefont {F.}~\bibnamefont {Schmidt}},
  \bibinfo {author} {\bibfnamefont {J.}~\bibnamefont {Jasche}}, \bibinfo
  {author} {\bibfnamefont {G.}~\bibnamefont {Lavaux}}, \ and\ \bibinfo {author}
  {\bibfnamefont {N.-M.}\ \bibnamefont {Nguyen}},\ }\href {\doibase
  10.1088/1475-7516/2020/01/029} {\bibfield  {journal} {\bibinfo  {journal}
  {JCAP}\ }\textbf {\bibinfo {volume} {01}},\ \bibinfo {pages} {029} (\bibinfo
  {year} {2020})},\ \Eprint {http://arxiv.org/abs/1906.07143} {arXiv:1906.07143
  [astro-ph.CO]} \BibitemShut {NoStop}%
\bibitem [{\citenamefont {Cabass}\ and\ \citenamefont
  {Schmidt}(2020)}]{Cabass:2019lqx}%
  \BibitemOpen
  \bibfield  {author} {\bibinfo {author} {\bibfnamefont {G.}~\bibnamefont
  {Cabass}}\ and\ \bibinfo {author} {\bibfnamefont {F.}~\bibnamefont
  {Schmidt}},\ }\href {\doibase 10.1088/1475-7516/2020/04/042} {\bibfield
  {journal} {\bibinfo  {journal} {JCAP}\ }\textbf {\bibinfo {volume} {04}},\
  \bibinfo {pages} {042} (\bibinfo {year} {2020})},\ \Eprint
  {http://arxiv.org/abs/1909.04022} {arXiv:1909.04022 [astro-ph.CO]}
  \BibitemShut {NoStop}%
\bibitem [{\citenamefont {Schmidt}(2020)}]{Schmidt:2020tao}%
  \BibitemOpen
  \bibfield  {author} {\bibinfo {author} {\bibfnamefont {F.}~\bibnamefont
  {Schmidt}},\ }\href@noop {} {\  (\bibinfo {year} {2020})},\ \Eprint
  {http://arxiv.org/abs/2009.14176} {arXiv:2009.14176 [astro-ph.CO]}
  \BibitemShut {NoStop}%
\bibitem [{\citenamefont {Schmidt}\ \emph {et~al.}(2020)\citenamefont
  {Schmidt}, \citenamefont {Cabass}, \citenamefont {Jasche},\ and\
  \citenamefont {Lavaux}}]{Schmidt:2020viy}%
  \BibitemOpen
  \bibfield  {author} {\bibinfo {author} {\bibfnamefont {F.}~\bibnamefont
  {Schmidt}}, \bibinfo {author} {\bibfnamefont {G.}~\bibnamefont {Cabass}},
  \bibinfo {author} {\bibfnamefont {J.}~\bibnamefont {Jasche}}, \ and\ \bibinfo
  {author} {\bibfnamefont {G.}~\bibnamefont {Lavaux}},\ }\href {\doibase
  10.1088/1475-7516/2020/11/008} {\bibfield  {journal} {\bibinfo  {journal}
  {JCAP}\ }\textbf {\bibinfo {volume} {11}},\ \bibinfo {pages} {008} (\bibinfo
  {year} {2020})},\ \Eprint {http://arxiv.org/abs/2004.06707} {arXiv:2004.06707
  [astro-ph.CO]} \BibitemShut {NoStop}%
\bibitem [{\citenamefont {Lazeyras}\ \emph {et~al.}(2021)\citenamefont
  {Lazeyras}, \citenamefont {Barreira},\ and\ \citenamefont
  {Schmidt}}]{Lazeyras:2021dar}%
  \BibitemOpen
  \bibfield  {author} {\bibinfo {author} {\bibfnamefont {T.}~\bibnamefont
  {Lazeyras}}, \bibinfo {author} {\bibfnamefont {A.}~\bibnamefont {Barreira}},
  \ and\ \bibinfo {author} {\bibfnamefont {F.}~\bibnamefont {Schmidt}},\ }\href
  {\doibase 10.1088/1475-7516/2021/10/063} {\bibfield  {journal} {\bibinfo
  {journal} {JCAP}\ }\textbf {\bibinfo {volume} {10}},\ \bibinfo {pages} {063}
  (\bibinfo {year} {2021})},\ \Eprint {http://arxiv.org/abs/2106.14713}
  {arXiv:2106.14713 [astro-ph.CO]} \BibitemShut {NoStop}%
\bibitem [{\citenamefont {Stadler}\ \emph {et~al.}(2023)\citenamefont
  {Stadler}, \citenamefont {Schmidt},\ and\ \citenamefont
  {Reinecke}}]{Stadler:2023hea}%
  \BibitemOpen
  \bibfield  {author} {\bibinfo {author} {\bibfnamefont {J.}~\bibnamefont
  {Stadler}}, \bibinfo {author} {\bibfnamefont {F.}~\bibnamefont {Schmidt}}, \
  and\ \bibinfo {author} {\bibfnamefont {M.}~\bibnamefont {Reinecke}},\ }\href
  {\doibase 10.1088/1475-7516/2023/10/069} {\bibfield  {journal} {\bibinfo
  {journal} {JCAP}\ }\textbf {\bibinfo {volume} {10}},\ \bibinfo {pages} {069}
  (\bibinfo {year} {2023})},\ \Eprint {http://arxiv.org/abs/2303.09876}
  {arXiv:2303.09876 [astro-ph.CO]} \BibitemShut {NoStop}%
\bibitem [{\citenamefont {Nguyen}\ \emph {et~al.}(2024)\citenamefont {Nguyen},
  \citenamefont {Schmidt}, \citenamefont {Tucci}, \citenamefont {Reinecke},\
  and\ \citenamefont {Kosti\'c}}]{Nguyen:2024yth}%
  \BibitemOpen
  \bibfield  {author} {\bibinfo {author} {\bibfnamefont {N.-M.}\ \bibnamefont
  {Nguyen}}, \bibinfo {author} {\bibfnamefont {F.}~\bibnamefont {Schmidt}},
  \bibinfo {author} {\bibfnamefont {B.}~\bibnamefont {Tucci}}, \bibinfo
  {author} {\bibfnamefont {M.}~\bibnamefont {Reinecke}}, \ and\ \bibinfo
  {author} {\bibfnamefont {A.}~\bibnamefont {Kosti\'c}},\ }\href@noop {} {\
  (\bibinfo {year} {2024})},\ \Eprint {http://arxiv.org/abs/2403.03220}
  {arXiv:2403.03220 [astro-ph.CO]} \BibitemShut {NoStop}%
\bibitem [{\citenamefont {Chudaykin}\ \emph {et~al.}(2020)\citenamefont
  {Chudaykin}, \citenamefont {Ivanov}, \citenamefont {Philcox},\ and\
  \citenamefont {Simonovi\'c}}]{Chudaykin:2020aoj}%
  \BibitemOpen
  \bibfield  {author} {\bibinfo {author} {\bibfnamefont {A.}~\bibnamefont
  {Chudaykin}}, \bibinfo {author} {\bibfnamefont {M.~M.}\ \bibnamefont
  {Ivanov}}, \bibinfo {author} {\bibfnamefont {O.~H.~E.}\ \bibnamefont
  {Philcox}}, \ and\ \bibinfo {author} {\bibfnamefont {M.}~\bibnamefont
  {Simonovi\'c}},\ }\href {\doibase 10.1103/PhysRevD.102.063533} {\bibfield
  {journal} {\bibinfo  {journal} {Phys. Rev. D}\ }\textbf {\bibinfo {volume}
  {102}},\ \bibinfo {pages} {063533} (\bibinfo {year} {2020})},\ \Eprint
  {http://arxiv.org/abs/2004.10607} {arXiv:2004.10607 [astro-ph.CO]}
  \BibitemShut {NoStop}%
\bibitem [{\citenamefont {Blas}\ \emph
  {et~al.}(2016{\natexlab{a}})\citenamefont {Blas}, \citenamefont {Garny},
  \citenamefont {Ivanov},\ and\ \citenamefont {Sibiryakov}}]{Blas:2015qsi}%
  \BibitemOpen
  \bibfield  {author} {\bibinfo {author} {\bibfnamefont {D.}~\bibnamefont
  {Blas}}, \bibinfo {author} {\bibfnamefont {M.}~\bibnamefont {Garny}},
  \bibinfo {author} {\bibfnamefont {M.~M.}\ \bibnamefont {Ivanov}}, \ and\
  \bibinfo {author} {\bibfnamefont {S.}~\bibnamefont {Sibiryakov}},\ }\href
  {\doibase 10.1088/1475-7516/2016/07/052} {\bibfield  {journal} {\bibinfo
  {journal} {JCAP}\ }\textbf {\bibinfo {volume} {1607}},\ \bibinfo {pages}
  {052} (\bibinfo {year} {2016}{\natexlab{a}})},\ \Eprint
  {http://arxiv.org/abs/1512.05807} {arXiv:1512.05807 [astro-ph.CO]}
  \BibitemShut {NoStop}%
\bibitem [{\citenamefont {Blas}\ \emph
  {et~al.}(2016{\natexlab{b}})\citenamefont {Blas}, \citenamefont {Garny},
  \citenamefont {Ivanov},\ and\ \citenamefont {Sibiryakov}}]{Blas:2016sfa}%
  \BibitemOpen
  \bibfield  {author} {\bibinfo {author} {\bibfnamefont {D.}~\bibnamefont
  {Blas}}, \bibinfo {author} {\bibfnamefont {M.}~\bibnamefont {Garny}},
  \bibinfo {author} {\bibfnamefont {M.~M.}\ \bibnamefont {Ivanov}}, \ and\
  \bibinfo {author} {\bibfnamefont {S.}~\bibnamefont {Sibiryakov}},\ }\href
  {\doibase 10.1088/1475-7516/2016/07/028} {\bibfield  {journal} {\bibinfo
  {journal} {JCAP}\ }\textbf {\bibinfo {volume} {1607}},\ \bibinfo {pages}
  {028} (\bibinfo {year} {2016}{\natexlab{b}})},\ \Eprint
  {http://arxiv.org/abs/1605.02149} {arXiv:1605.02149 [astro-ph.CO]}
  \BibitemShut {NoStop}%
\bibitem [{\citenamefont {Ivanov}\ and\ \citenamefont
  {Sibiryakov}(2018)}]{Ivanov:2018gjr}%
  \BibitemOpen
  \bibfield  {author} {\bibinfo {author} {\bibfnamefont {M.~M.}\ \bibnamefont
  {Ivanov}}\ and\ \bibinfo {author} {\bibfnamefont {S.}~\bibnamefont
  {Sibiryakov}},\ }\href {\doibase 10.1088/1475-7516/2018/07/053} {\bibfield
  {journal} {\bibinfo  {journal} {JCAP}\ }\textbf {\bibinfo {volume} {1807}},\
  \bibinfo {pages} {053} (\bibinfo {year} {2018})},\ \Eprint
  {http://arxiv.org/abs/1804.05080} {arXiv:1804.05080 [astro-ph.CO]}
  \BibitemShut {NoStop}%
\bibitem [{\citenamefont {Vasudevan}\ \emph {et~al.}(2019)\citenamefont
  {Vasudevan}, \citenamefont {Ivanov}, \citenamefont {Sibiryakov},\ and\
  \citenamefont {Lesgourgues}}]{Vasudevan:2019ewf}%
  \BibitemOpen
  \bibfield  {author} {\bibinfo {author} {\bibfnamefont {A.}~\bibnamefont
  {Vasudevan}}, \bibinfo {author} {\bibfnamefont {M.~M.}\ \bibnamefont
  {Ivanov}}, \bibinfo {author} {\bibfnamefont {S.}~\bibnamefont {Sibiryakov}},
  \ and\ \bibinfo {author} {\bibfnamefont {J.}~\bibnamefont {Lesgourgues}},\
  }\href {\doibase 10.1088/1475-7516/2019/09/037} {\bibfield  {journal}
  {\bibinfo  {journal} {JCAP}\ }\textbf {\bibinfo {volume} {09}},\ \bibinfo
  {pages} {037} (\bibinfo {year} {2019})},\ \Eprint
  {http://arxiv.org/abs/1906.08697} {arXiv:1906.08697 [astro-ph.CO]}
  \BibitemShut {NoStop}%
\bibitem [{\citenamefont {Crocce}\ and\ \citenamefont
  {Scoccimarro}(2008)}]{Crocce:2007dt}%
  \BibitemOpen
  \bibfield  {author} {\bibinfo {author} {\bibfnamefont {M.}~\bibnamefont
  {Crocce}}\ and\ \bibinfo {author} {\bibfnamefont {R.}~\bibnamefont
  {Scoccimarro}},\ }\href {\doibase 10.1103/PhysRevD.77.023533} {\bibfield
  {journal} {\bibinfo  {journal} {Phys. Rev.}\ }\textbf {\bibinfo {volume}
  {D77}},\ \bibinfo {pages} {023533} (\bibinfo {year} {2008})},\ \Eprint
  {http://arxiv.org/abs/0704.2783} {arXiv:0704.2783 [astro-ph]} \BibitemShut
  {NoStop}%
\bibitem [{\citenamefont {Senatore}\ and\ \citenamefont
  {Zaldarriaga}(2015)}]{Senatore:2014via}%
  \BibitemOpen
  \bibfield  {author} {\bibinfo {author} {\bibfnamefont {L.}~\bibnamefont
  {Senatore}}\ and\ \bibinfo {author} {\bibfnamefont {M.}~\bibnamefont
  {Zaldarriaga}},\ }\href {\doibase 10.1088/1475-7516/2015/02/013} {\bibfield
  {journal} {\bibinfo  {journal} {JCAP}\ }\textbf {\bibinfo {volume} {1502}},\
  \bibinfo {pages} {013} (\bibinfo {year} {2015})},\ \Eprint
  {http://arxiv.org/abs/1404.5954} {arXiv:1404.5954 [astro-ph.CO]} \BibitemShut
  {NoStop}%
\bibitem [{\citenamefont {Baldauf}\ \emph
  {et~al.}(2015{\natexlab{c}})\citenamefont {Baldauf}, \citenamefont
  {Mirbabayi}, \citenamefont {Simonović},\ and\ \citenamefont
  {Zaldarriaga}}]{Baldauf:2015xfa}%
  \BibitemOpen
  \bibfield  {author} {\bibinfo {author} {\bibfnamefont {T.}~\bibnamefont
  {Baldauf}}, \bibinfo {author} {\bibfnamefont {M.}~\bibnamefont {Mirbabayi}},
  \bibinfo {author} {\bibfnamefont {M.}~\bibnamefont {Simonović}}, \ and\
  \bibinfo {author} {\bibfnamefont {M.}~\bibnamefont {Zaldarriaga}},\ }\href
  {\doibase 10.1103/PhysRevD.92.043514} {\bibfield  {journal} {\bibinfo
  {journal} {Phys. Rev.}\ }\textbf {\bibinfo {volume} {D92}},\ \bibinfo {pages}
  {043514} (\bibinfo {year} {2015}{\natexlab{c}})},\ \Eprint
  {http://arxiv.org/abs/1504.04366} {arXiv:1504.04366 [astro-ph.CO]}
  \BibitemShut {NoStop}%
\bibitem [{\citenamefont {Simonovic}\ \emph {et~al.}(2018)\citenamefont
  {Simonovic}, \citenamefont {Baldauf}, \citenamefont {Zaldarriaga},
  \citenamefont {Carrasco},\ and\ \citenamefont
  {Kollmeier}}]{Simonovic:2017mhp}%
  \BibitemOpen
  \bibfield  {author} {\bibinfo {author} {\bibfnamefont {M.}~\bibnamefont
  {Simonovic}}, \bibinfo {author} {\bibfnamefont {T.}~\bibnamefont {Baldauf}},
  \bibinfo {author} {\bibfnamefont {M.}~\bibnamefont {Zaldarriaga}}, \bibinfo
  {author} {\bibfnamefont {J.~J.}\ \bibnamefont {Carrasco}}, \ and\ \bibinfo
  {author} {\bibfnamefont {J.~A.}\ \bibnamefont {Kollmeier}},\ }\href {\doibase
  10.1088/1475-7516/2018/04/030} {\bibfield  {journal} {\bibinfo  {journal}
  {JCAP}\ }\textbf {\bibinfo {volume} {1804}},\ \bibinfo {pages} {030}
  (\bibinfo {year} {2018})},\ \Eprint {http://arxiv.org/abs/1708.08130}
  {arXiv:1708.08130 [astro-ph.CO]} \BibitemShut {NoStop}%
\bibitem [{\citenamefont {Villaescusa-Navarro}\ \emph
  {et~al.}(2020)\citenamefont {Villaescusa-Navarro} \emph
  {et~al.}}]{Villaescusa-Navarro:2019bje}%
  \BibitemOpen
  \bibfield  {author} {\bibinfo {author} {\bibfnamefont {F.}~\bibnamefont
  {Villaescusa-Navarro}} \emph {et~al.},\ }\href {\doibase
  10.3847/1538-4365/ab9d82} {\bibfield  {journal} {\bibinfo  {journal}
  {Astrophys. J. Suppl.}\ }\textbf {\bibinfo {volume} {250}},\ \bibinfo {pages}
  {2} (\bibinfo {year} {2020})},\ \Eprint {http://arxiv.org/abs/1909.05273}
  {arXiv:1909.05273 [astro-ph.CO]} \BibitemShut {NoStop}%
\bibitem [{\citenamefont {Seljak}(2000)}]{Seljak:2000gq}%
  \BibitemOpen
  \bibfield  {author} {\bibinfo {author} {\bibfnamefont {U.}~\bibnamefont
  {Seljak}},\ }\href {\doibase 10.1046/j.1365-8711.2000.03715.x} {\bibfield
  {journal} {\bibinfo  {journal} {Mon. Not. Roy. Astron. Soc.}\ }\textbf
  {\bibinfo {volume} {318}},\ \bibinfo {pages} {203} (\bibinfo {year}
  {2000})},\ \Eprint {http://arxiv.org/abs/astro-ph/0001493}
  {arXiv:astro-ph/0001493} \BibitemShut {NoStop}%
\bibitem [{\citenamefont {Cooray}\ and\ \citenamefont
  {Sheth}(2002)}]{Cooray:2002dia}%
  \BibitemOpen
  \bibfield  {author} {\bibinfo {author} {\bibfnamefont {A.}~\bibnamefont
  {Cooray}}\ and\ \bibinfo {author} {\bibfnamefont {R.~K.}\ \bibnamefont
  {Sheth}},\ }\href {\doibase 10.1016/S0370-1573(02)00276-4} {\bibfield
  {journal} {\bibinfo  {journal} {Phys. Rept.}\ }\textbf {\bibinfo {volume}
  {372}},\ \bibinfo {pages} {1} (\bibinfo {year} {2002})},\ \Eprint
  {http://arxiv.org/abs/astro-ph/0206508} {arXiv:astro-ph/0206508} \BibitemShut
  {NoStop}%
\bibitem [{\citenamefont {Nishimichi}\ \emph {et~al.}(2020)\citenamefont
  {Nishimichi}, \citenamefont {D'Amico}, \citenamefont {Ivanov}, \citenamefont
  {Senatore}, \citenamefont {Simonovi\'c}, \citenamefont {Takada},
  \citenamefont {Zaldarriaga},\ and\ \citenamefont
  {Zhang}}]{Nishimichi:2020tvu}%
  \BibitemOpen
  \bibfield  {author} {\bibinfo {author} {\bibfnamefont {T.}~\bibnamefont
  {Nishimichi}}, \bibinfo {author} {\bibfnamefont {G.}~\bibnamefont {D'Amico}},
  \bibinfo {author} {\bibfnamefont {M.~M.}\ \bibnamefont {Ivanov}}, \bibinfo
  {author} {\bibfnamefont {L.}~\bibnamefont {Senatore}}, \bibinfo {author}
  {\bibfnamefont {M.}~\bibnamefont {Simonovi\'c}}, \bibinfo {author}
  {\bibfnamefont {M.}~\bibnamefont {Takada}}, \bibinfo {author} {\bibfnamefont
  {M.}~\bibnamefont {Zaldarriaga}}, \ and\ \bibinfo {author} {\bibfnamefont
  {P.}~\bibnamefont {Zhang}},\ }\href {\doibase 10.1103/PhysRevD.102.123541}
  {\bibfield  {journal} {\bibinfo  {journal} {Phys. Rev. D}\ }\textbf {\bibinfo
  {volume} {102}},\ \bibinfo {pages} {123541} (\bibinfo {year} {2020})},\
  \Eprint {http://arxiv.org/abs/2003.08277} {arXiv:2003.08277 [astro-ph.CO]}
  \BibitemShut {NoStop}%
\bibitem [{\citenamefont {Chudaykin}\ \emph
  {et~al.}(2021{\natexlab{a}})\citenamefont {Chudaykin}, \citenamefont
  {Dolgikh},\ and\ \citenamefont {Ivanov}}]{Chudaykin:2020ghx}%
  \BibitemOpen
  \bibfield  {author} {\bibinfo {author} {\bibfnamefont {A.}~\bibnamefont
  {Chudaykin}}, \bibinfo {author} {\bibfnamefont {K.}~\bibnamefont {Dolgikh}},
  \ and\ \bibinfo {author} {\bibfnamefont {M.~M.}\ \bibnamefont {Ivanov}},\
  }\href {\doibase 10.1103/PhysRevD.103.023507} {\bibfield  {journal} {\bibinfo
   {journal} {Phys. Rev. D}\ }\textbf {\bibinfo {volume} {103}},\ \bibinfo
  {pages} {023507} (\bibinfo {year} {2021}{\natexlab{a}})},\ \Eprint
  {http://arxiv.org/abs/2009.10106} {arXiv:2009.10106 [astro-ph.CO]}
  \BibitemShut {NoStop}%
\bibitem [{\citenamefont {Ivanov}\ \emph
  {et~al.}(2020{\natexlab{b}})\citenamefont {Ivanov}, \citenamefont
  {McDonough}, \citenamefont {Hill}, \citenamefont {Simonovi\'c}, \citenamefont
  {Toomey}, \citenamefont {Alexander},\ and\ \citenamefont
  {Zaldarriaga}}]{Ivanov:2020ril}%
  \BibitemOpen
  \bibfield  {author} {\bibinfo {author} {\bibfnamefont {M.~M.}\ \bibnamefont
  {Ivanov}}, \bibinfo {author} {\bibfnamefont {E.}~\bibnamefont {McDonough}},
  \bibinfo {author} {\bibfnamefont {J.~C.}\ \bibnamefont {Hill}}, \bibinfo
  {author} {\bibfnamefont {M.}~\bibnamefont {Simonovi\'c}}, \bibinfo {author}
  {\bibfnamefont {M.~W.}\ \bibnamefont {Toomey}}, \bibinfo {author}
  {\bibfnamefont {S.}~\bibnamefont {Alexander}}, \ and\ \bibinfo {author}
  {\bibfnamefont {M.}~\bibnamefont {Zaldarriaga}},\ }\href {\doibase
  10.1103/PhysRevD.102.103502} {\bibfield  {journal} {\bibinfo  {journal}
  {Phys. Rev. D}\ }\textbf {\bibinfo {volume} {102}},\ \bibinfo {pages}
  {103502} (\bibinfo {year} {2020}{\natexlab{b}})},\ \Eprint
  {http://arxiv.org/abs/2006.11235} {arXiv:2006.11235 [astro-ph.CO]}
  \BibitemShut {NoStop}%
\bibitem [{\citenamefont {He}\ \emph {et~al.}(2023)\citenamefont {He},
  \citenamefont {Ivanov}, \citenamefont {An},\ and\ \citenamefont
  {Gluscevic}}]{He:2023dbn}%
  \BibitemOpen
  \bibfield  {author} {\bibinfo {author} {\bibfnamefont {A.}~\bibnamefont
  {He}}, \bibinfo {author} {\bibfnamefont {M.~M.}\ \bibnamefont {Ivanov}},
  \bibinfo {author} {\bibfnamefont {R.}~\bibnamefont {An}}, \ and\ \bibinfo
  {author} {\bibfnamefont {V.}~\bibnamefont {Gluscevic}},\ }\href {\doibase
  10.3847/2041-8213/acdb63} {\bibfield  {journal} {\bibinfo  {journal}
  {Astrophys. J. Lett.}\ }\textbf {\bibinfo {volume} {954}},\ \bibinfo {pages}
  {L8} (\bibinfo {year} {2023})},\ \Eprint {http://arxiv.org/abs/2301.08260}
  {arXiv:2301.08260 [astro-ph.CO]} \BibitemShut {NoStop}%
\bibitem [{\citenamefont {Camarena}\ \emph {et~al.}(2023)\citenamefont
  {Camarena}, \citenamefont {Cyr-Racine},\ and\ \citenamefont
  {Houghteling}}]{Camarena:2023cku}%
  \BibitemOpen
  \bibfield  {author} {\bibinfo {author} {\bibfnamefont {D.}~\bibnamefont
  {Camarena}}, \bibinfo {author} {\bibfnamefont {F.-Y.}\ \bibnamefont
  {Cyr-Racine}}, \ and\ \bibinfo {author} {\bibfnamefont {J.}~\bibnamefont
  {Houghteling}},\ }\href {\doibase 10.1103/PhysRevD.108.103535} {\bibfield
  {journal} {\bibinfo  {journal} {Phys. Rev. D}\ }\textbf {\bibinfo {volume}
  {108}},\ \bibinfo {pages} {103535} (\bibinfo {year} {2023})},\ \Eprint
  {http://arxiv.org/abs/2309.03941} {arXiv:2309.03941 [astro-ph.CO]}
  \BibitemShut {NoStop}%
\bibitem [{\citenamefont {He}\ \emph {et~al.}(2024)\citenamefont {He},
  \citenamefont {An}, \citenamefont {Ivanov},\ and\ \citenamefont
  {Gluscevic}}]{He:2023oke}%
  \BibitemOpen
  \bibfield  {author} {\bibinfo {author} {\bibfnamefont {A.}~\bibnamefont
  {He}}, \bibinfo {author} {\bibfnamefont {R.}~\bibnamefont {An}}, \bibinfo
  {author} {\bibfnamefont {M.~M.}\ \bibnamefont {Ivanov}}, \ and\ \bibinfo
  {author} {\bibfnamefont {V.}~\bibnamefont {Gluscevic}},\ }\href {\doibase
  10.1103/PhysRevD.109.103527} {\bibfield  {journal} {\bibinfo  {journal}
  {Phys. Rev. D}\ }\textbf {\bibinfo {volume} {109}},\ \bibinfo {pages}
  {103527} (\bibinfo {year} {2024})},\ \Eprint
  {http://arxiv.org/abs/2309.03956} {arXiv:2309.03956 [astro-ph.CO]}
  \BibitemShut {NoStop}%
\bibitem [{\citenamefont {Chudaykin}\ and\ \citenamefont
  {Ivanov}(2019)}]{Chudaykin:2019ock}%
  \BibitemOpen
  \bibfield  {author} {\bibinfo {author} {\bibfnamefont {A.}~\bibnamefont
  {Chudaykin}}\ and\ \bibinfo {author} {\bibfnamefont {M.~M.}\ \bibnamefont
  {Ivanov}},\ }\href {\doibase 10.1088/1475-7516/2019/11/034} {\bibfield
  {journal} {\bibinfo  {journal} {JCAP}\ }\textbf {\bibinfo {volume} {11}},\
  \bibinfo {pages} {034} (\bibinfo {year} {2019})},\ \Eprint
  {http://arxiv.org/abs/1907.06666} {arXiv:1907.06666 [astro-ph.CO]}
  \BibitemShut {NoStop}%
\bibitem [{\citenamefont {Rogers}\ \emph {et~al.}(2023)\citenamefont {Rogers},
  \citenamefont {Hlo\v{z}ek}, \citenamefont {Lagu\"e}, \citenamefont {Ivanov},
  \citenamefont {Philcox}, \citenamefont {Cabass}, \citenamefont {Akitsu},\
  and\ \citenamefont {Marsh}}]{Rogers:2023ezo}%
  \BibitemOpen
  \bibfield  {author} {\bibinfo {author} {\bibfnamefont {K.~K.}\ \bibnamefont
  {Rogers}}, \bibinfo {author} {\bibfnamefont {R.}~\bibnamefont {Hlo\v{z}ek}},
  \bibinfo {author} {\bibfnamefont {A.}~\bibnamefont {Lagu\"e}}, \bibinfo
  {author} {\bibfnamefont {M.~M.}\ \bibnamefont {Ivanov}}, \bibinfo {author}
  {\bibfnamefont {O.~H.~E.}\ \bibnamefont {Philcox}}, \bibinfo {author}
  {\bibfnamefont {G.}~\bibnamefont {Cabass}}, \bibinfo {author} {\bibfnamefont
  {K.}~\bibnamefont {Akitsu}}, \ and\ \bibinfo {author} {\bibfnamefont
  {D.~J.~E.}\ \bibnamefont {Marsh}},\ }\href {\doibase
  10.1088/1475-7516/2023/06/023} {\bibfield  {journal} {\bibinfo  {journal}
  {JCAP}\ }\textbf {\bibinfo {volume} {06}},\ \bibinfo {pages} {023} (\bibinfo
  {year} {2023})},\ \Eprint {http://arxiv.org/abs/2301.08361} {arXiv:2301.08361
  [astro-ph.CO]} \BibitemShut {NoStop}%
\bibitem [{\citenamefont {Xu}\ \emph {et~al.}(2022)\citenamefont {Xu},
  \citenamefont {Mu\~noz},\ and\ \citenamefont {Dvorkin}}]{Xu:2021rwg}%
  \BibitemOpen
  \bibfield  {author} {\bibinfo {author} {\bibfnamefont {W.~L.}\ \bibnamefont
  {Xu}}, \bibinfo {author} {\bibfnamefont {J.~B.}\ \bibnamefont {Mu\~noz}}, \
  and\ \bibinfo {author} {\bibfnamefont {C.}~\bibnamefont {Dvorkin}},\ }\href
  {\doibase 10.1103/PhysRevD.105.095029} {\bibfield  {journal} {\bibinfo
  {journal} {Phys. Rev. D}\ }\textbf {\bibinfo {volume} {105}},\ \bibinfo
  {pages} {095029} (\bibinfo {year} {2022})},\ \Eprint
  {http://arxiv.org/abs/2107.09664} {arXiv:2107.09664 [astro-ph.CO]}
  \BibitemShut {NoStop}%
\bibitem [{\citenamefont {Ivanov}\ \emph
  {et~al.}(2020{\natexlab{c}})\citenamefont {Ivanov}, \citenamefont
  {Simonovi\'c},\ and\ \citenamefont {Zaldarriaga}}]{Ivanov:2019pdj}%
  \BibitemOpen
  \bibfield  {author} {\bibinfo {author} {\bibfnamefont {M.~M.}\ \bibnamefont
  {Ivanov}}, \bibinfo {author} {\bibfnamefont {M.}~\bibnamefont {Simonovi\'c}},
  \ and\ \bibinfo {author} {\bibfnamefont {M.}~\bibnamefont {Zaldarriaga}},\
  }\href {\doibase 10.1088/1475-7516/2020/05/042} {\bibfield  {journal}
  {\bibinfo  {journal} {JCAP}\ }\textbf {\bibinfo {volume} {05}},\ \bibinfo
  {pages} {042} (\bibinfo {year} {2020}{\natexlab{c}})},\ \Eprint
  {http://arxiv.org/abs/1909.05277} {arXiv:1909.05277 [astro-ph.CO]}
  \BibitemShut {NoStop}%
\bibitem [{\citenamefont {Chudaykin}\ \emph
  {et~al.}(2021{\natexlab{b}})\citenamefont {Chudaykin}, \citenamefont
  {Ivanov},\ and\ \citenamefont {Simonovi\'c}}]{Chudaykin:2020hbf}%
  \BibitemOpen
  \bibfield  {author} {\bibinfo {author} {\bibfnamefont {A.}~\bibnamefont
  {Chudaykin}}, \bibinfo {author} {\bibfnamefont {M.~M.}\ \bibnamefont
  {Ivanov}}, \ and\ \bibinfo {author} {\bibfnamefont {M.}~\bibnamefont
  {Simonovi\'c}},\ }\href {\doibase 10.1103/PhysRevD.103.043525} {\bibfield
  {journal} {\bibinfo  {journal} {Phys. Rev. D}\ }\textbf {\bibinfo {volume}
  {103}},\ \bibinfo {pages} {043525} (\bibinfo {year} {2021}{\natexlab{b}})},\
  \Eprint {http://arxiv.org/abs/2009.10724} {arXiv:2009.10724 [astro-ph.CO]}
  \BibitemShut {NoStop}%
\bibitem [{\citenamefont {Aghanim}\ \emph {et~al.}(2017)\citenamefont {Aghanim}
  \emph {et~al.}}]{Aghanim:2016sns}%
  \BibitemOpen
  \bibfield  {author} {\bibinfo {author} {\bibfnamefont {N.}~\bibnamefont
  {Aghanim}} \emph {et~al.} (\bibinfo {collaboration} {Planck}),\ }\href
  {\doibase 10.1051/0004-6361/201629504} {\bibfield  {journal} {\bibinfo
  {journal} {Astron. Astrophys.}\ }\textbf {\bibinfo {volume} {607}},\ \bibinfo
  {pages} {A95} (\bibinfo {year} {2017})},\ \Eprint
  {http://arxiv.org/abs/1608.02487} {arXiv:1608.02487 [astro-ph.CO]}
  \BibitemShut {NoStop}%
\bibitem [{\citenamefont {Chen}\ \emph {et~al.}(2020)\citenamefont {Chen},
  \citenamefont {Vlah},\ and\ \citenamefont {White}}]{Chen:2020fxs}%
  \BibitemOpen
  \bibfield  {author} {\bibinfo {author} {\bibfnamefont {S.-F.}\ \bibnamefont
  {Chen}}, \bibinfo {author} {\bibfnamefont {Z.}~\bibnamefont {Vlah}}, \ and\
  \bibinfo {author} {\bibfnamefont {M.}~\bibnamefont {White}},\ }\href
  {\doibase 10.1088/1475-7516/2020/07/062} {\bibfield  {journal} {\bibinfo
  {journal} {JCAP}\ }\textbf {\bibinfo {volume} {07}},\ \bibinfo {pages} {062}
  (\bibinfo {year} {2020})},\ \Eprint {http://arxiv.org/abs/2005.00523}
  {arXiv:2005.00523 [astro-ph.CO]} \BibitemShut {NoStop}%
\bibitem [{\citenamefont {Chen}\ \emph {et~al.}(2021)\citenamefont {Chen},
  \citenamefont {Vlah}, \citenamefont {Castorina},\ and\ \citenamefont
  {White}}]{Chen:2020zjt}%
  \BibitemOpen
  \bibfield  {author} {\bibinfo {author} {\bibfnamefont {S.-F.}\ \bibnamefont
  {Chen}}, \bibinfo {author} {\bibfnamefont {Z.}~\bibnamefont {Vlah}}, \bibinfo
  {author} {\bibfnamefont {E.}~\bibnamefont {Castorina}}, \ and\ \bibinfo
  {author} {\bibfnamefont {M.}~\bibnamefont {White}},\ }\href {\doibase
  10.1088/1475-7516/2021/03/100} {\bibfield  {journal} {\bibinfo  {journal}
  {JCAP}\ }\textbf {\bibinfo {volume} {03}},\ \bibinfo {pages} {100} (\bibinfo
  {year} {2021})},\ \Eprint {http://arxiv.org/abs/2012.04636} {arXiv:2012.04636
  [astro-ph.CO]} \BibitemShut {NoStop}%
\bibitem [{\citenamefont {D'Amico}\ \emph {et~al.}(2021)\citenamefont
  {D'Amico}, \citenamefont {Senatore},\ and\ \citenamefont
  {Zhang}}]{DAmico:2020kxu}%
  \BibitemOpen
  \bibfield  {author} {\bibinfo {author} {\bibfnamefont {G.}~\bibnamefont
  {D'Amico}}, \bibinfo {author} {\bibfnamefont {L.}~\bibnamefont {Senatore}}, \
  and\ \bibinfo {author} {\bibfnamefont {P.}~\bibnamefont {Zhang}},\ }\href
  {\doibase 10.1088/1475-7516/2021/01/006} {\bibfield  {journal} {\bibinfo
  {journal} {JCAP}\ }\textbf {\bibinfo {volume} {01}},\ \bibinfo {pages} {006}
  (\bibinfo {year} {2021})},\ \Eprint {http://arxiv.org/abs/2003.07956}
  {arXiv:2003.07956 [astro-ph.CO]} \BibitemShut {NoStop}%
\bibitem [{\citenamefont {Ade}\ \emph {et~al.}(2016)\citenamefont {Ade} \emph
  {et~al.}}]{Ade:2015xua}%
  \BibitemOpen
  \bibfield  {author} {\bibinfo {author} {\bibfnamefont {P.~A.~R.}\
  \bibnamefont {Ade}} \emph {et~al.} (\bibinfo {collaboration} {Planck}),\
  }\href {\doibase 10.1051/0004-6361/201525830} {\bibfield  {journal} {\bibinfo
   {journal} {Astron. Astrophys.}\ }\textbf {\bibinfo {volume} {594}},\
  \bibinfo {pages} {A13} (\bibinfo {year} {2016})},\ \Eprint
  {http://arxiv.org/abs/1502.01589} {arXiv:1502.01589 [astro-ph.CO]}
  \BibitemShut {NoStop}%
\bibitem [{\citenamefont {Maksimova}\ \emph {et~al.}(2021)\citenamefont
  {Maksimova}, \citenamefont {Garrison}, \citenamefont {Eisenstein},
  \citenamefont {Hadzhiyska}, \citenamefont {Bose},\ and\ \citenamefont
  {Satterthwaite}}]{Maksimova:2021ynf}%
  \BibitemOpen
  \bibfield  {author} {\bibinfo {author} {\bibfnamefont {N.~A.}\ \bibnamefont
  {Maksimova}}, \bibinfo {author} {\bibfnamefont {L.~H.}\ \bibnamefont
  {Garrison}}, \bibinfo {author} {\bibfnamefont {D.~J.}\ \bibnamefont
  {Eisenstein}}, \bibinfo {author} {\bibfnamefont {B.}~\bibnamefont
  {Hadzhiyska}}, \bibinfo {author} {\bibfnamefont {S.}~\bibnamefont {Bose}}, \
  and\ \bibinfo {author} {\bibfnamefont {T.~P.}\ \bibnamefont
  {Satterthwaite}},\ }\href {\doibase 10.1093/mnras/stab2484} {\bibfield
  {journal} {\bibinfo  {journal} {Mon. Not. Roy. Astron. Soc.}\ }\textbf
  {\bibinfo {volume} {508}},\ \bibinfo {pages} {4017} (\bibinfo {year}
  {2021})},\ \Eprint {http://arxiv.org/abs/2110.11398} {arXiv:2110.11398
  [astro-ph.CO]} \BibitemShut {NoStop}%
\bibitem [{\citenamefont {Aghanim}\ \emph {et~al.}(2018)\citenamefont {Aghanim}
  \emph {et~al.}}]{Aghanim:2018eyx}%
  \BibitemOpen
  \bibfield  {author} {\bibinfo {author} {\bibfnamefont {N.}~\bibnamefont
  {Aghanim}} \emph {et~al.} (\bibinfo {collaboration} {Planck}),\ }\href@noop
  {} {\  (\bibinfo {year} {2018})},\ \Eprint {http://arxiv.org/abs/1807.06209}
  {arXiv:1807.06209 [astro-ph.CO]} \BibitemShut {NoStop}%
\bibitem [{\citenamefont {Hadzhiyska}\ \emph {et~al.}(2021)\citenamefont
  {Hadzhiyska}, \citenamefont {Eisenstein}, \citenamefont {Bose}, \citenamefont
  {Garrison},\ and\ \citenamefont {Maksimova}}]{Hadzhiyska:2021zbd}%
  \BibitemOpen
  \bibfield  {author} {\bibinfo {author} {\bibfnamefont {B.}~\bibnamefont
  {Hadzhiyska}}, \bibinfo {author} {\bibfnamefont {D.}~\bibnamefont
  {Eisenstein}}, \bibinfo {author} {\bibfnamefont {S.}~\bibnamefont {Bose}},
  \bibinfo {author} {\bibfnamefont {L.~H.}\ \bibnamefont {Garrison}}, \ and\
  \bibinfo {author} {\bibfnamefont {N.}~\bibnamefont {Maksimova}},\ }\href
  {\doibase 10.1093/mnras/stab2980} {\bibfield  {journal} {\bibinfo  {journal}
  {Mon. Not. Roy. Astron. Soc.}\ }\textbf {\bibinfo {volume} {509}},\ \bibinfo
  {pages} {501} (\bibinfo {year} {2021})},\ \Eprint
  {http://arxiv.org/abs/2110.11408} {arXiv:2110.11408 [astro-ph.CO]}
  \BibitemShut {NoStop}%
\bibitem [{\citenamefont {Assassi}\ \emph {et~al.}(2014)\citenamefont
  {Assassi}, \citenamefont {Baumann}, \citenamefont {Green},\ and\
  \citenamefont {Zaldarriaga}}]{Assassi:2014fva}%
  \BibitemOpen
  \bibfield  {author} {\bibinfo {author} {\bibfnamefont {V.}~\bibnamefont
  {Assassi}}, \bibinfo {author} {\bibfnamefont {D.}~\bibnamefont {Baumann}},
  \bibinfo {author} {\bibfnamefont {D.}~\bibnamefont {Green}}, \ and\ \bibinfo
  {author} {\bibfnamefont {M.}~\bibnamefont {Zaldarriaga}},\ }\href {\doibase
  10.1088/1475-7516/2014/08/056} {\bibfield  {journal} {\bibinfo  {journal}
  {JCAP}\ }\textbf {\bibinfo {volume} {1408}},\ \bibinfo {pages} {056}
  (\bibinfo {year} {2014})},\ \Eprint {http://arxiv.org/abs/1402.5916}
  {arXiv:1402.5916 [astro-ph.CO]} \BibitemShut {NoStop}%
\bibitem [{\citenamefont {Desjacques}\ \emph
  {et~al.}(2018{\natexlab{a}})\citenamefont {Desjacques}, \citenamefont
  {Jeong},\ and\ \citenamefont {Schmidt}}]{Desjacques:2016bnm}%
  \BibitemOpen
  \bibfield  {author} {\bibinfo {author} {\bibfnamefont {V.}~\bibnamefont
  {Desjacques}}, \bibinfo {author} {\bibfnamefont {D.}~\bibnamefont {Jeong}}, \
  and\ \bibinfo {author} {\bibfnamefont {F.}~\bibnamefont {Schmidt}},\ }\href
  {\doibase 10.1016/j.physrep.2017.12.002} {\bibfield  {journal} {\bibinfo
  {journal} {Phys. Rept.}\ }\textbf {\bibinfo {volume} {733}},\ \bibinfo
  {pages} {1} (\bibinfo {year} {2018}{\natexlab{a}})},\ \Eprint
  {http://arxiv.org/abs/1611.09787} {arXiv:1611.09787 [astro-ph.CO]}
  \BibitemShut {NoStop}%
\bibitem [{\citenamefont {Perko}\ \emph {et~al.}(2016)\citenamefont {Perko},
  \citenamefont {Senatore}, \citenamefont {Jennings},\ and\ \citenamefont
  {Wechsler}}]{Perko:2016puo}%
  \BibitemOpen
  \bibfield  {author} {\bibinfo {author} {\bibfnamefont {A.}~\bibnamefont
  {Perko}}, \bibinfo {author} {\bibfnamefont {L.}~\bibnamefont {Senatore}},
  \bibinfo {author} {\bibfnamefont {E.}~\bibnamefont {Jennings}}, \ and\
  \bibinfo {author} {\bibfnamefont {R.~H.}\ \bibnamefont {Wechsler}},\
  }\href@noop {} {\  (\bibinfo {year} {2016})},\ \Eprint
  {http://arxiv.org/abs/1610.09321} {arXiv:1610.09321 [astro-ph.CO]}
  \BibitemShut {NoStop}%
\bibitem [{\citenamefont {Bernardeau}\ \emph {et~al.}(2002)\citenamefont
  {Bernardeau}, \citenamefont {Colombi}, \citenamefont {Gaztanaga},\ and\
  \citenamefont {Scoccimarro}}]{Bernardeau:2001qr}%
  \BibitemOpen
  \bibfield  {author} {\bibinfo {author} {\bibfnamefont {F.}~\bibnamefont
  {Bernardeau}}, \bibinfo {author} {\bibfnamefont {S.}~\bibnamefont {Colombi}},
  \bibinfo {author} {\bibfnamefont {E.}~\bibnamefont {Gaztanaga}}, \ and\
  \bibinfo {author} {\bibfnamefont {R.}~\bibnamefont {Scoccimarro}},\ }\href
  {\doibase 10.1016/S0370-1573(02)00135-7} {\bibfield  {journal} {\bibinfo
  {journal} {Phys. Rept.}\ }\textbf {\bibinfo {volume} {367}},\ \bibinfo
  {pages} {1} (\bibinfo {year} {2002})},\ \Eprint
  {http://arxiv.org/abs/astro-ph/0112551} {arXiv:astro-ph/0112551 [astro-ph]}
  \BibitemShut {NoStop}%
\bibitem [{\citenamefont {Blas}\ \emph {et~al.}(2013)\citenamefont {Blas},
  \citenamefont {Garny},\ and\ \citenamefont {Konstandin}}]{Blas:2013bpa}%
  \BibitemOpen
  \bibfield  {author} {\bibinfo {author} {\bibfnamefont {D.}~\bibnamefont
  {Blas}}, \bibinfo {author} {\bibfnamefont {M.}~\bibnamefont {Garny}}, \ and\
  \bibinfo {author} {\bibfnamefont {T.}~\bibnamefont {Konstandin}},\ }\href
  {\doibase 10.1088/1475-7516/2013/09/024} {\bibfield  {journal} {\bibinfo
  {journal} {JCAP}\ }\textbf {\bibinfo {volume} {09}},\ \bibinfo {pages} {024}
  (\bibinfo {year} {2013})},\ \Eprint {http://arxiv.org/abs/1304.1546}
  {arXiv:1304.1546 [astro-ph.CO]} \BibitemShut {NoStop}%
\bibitem [{\citenamefont {Scoccimarro}\ and\ \citenamefont
  {Frieman}(1996{\natexlab{a}})}]{Scoccimarro:1995if}%
  \BibitemOpen
  \bibfield  {author} {\bibinfo {author} {\bibfnamefont {R.}~\bibnamefont
  {Scoccimarro}}\ and\ \bibinfo {author} {\bibfnamefont {J.}~\bibnamefont
  {Frieman}},\ }\href {\doibase 10.1086/192306} {\bibfield  {journal} {\bibinfo
   {journal} {Astrophys.\ J.\ Suppl.}\ }\textbf {\bibinfo {volume} {105}},\
  \bibinfo {pages} {37} (\bibinfo {year} {1996}{\natexlab{a}})},\ \Eprint
  {http://arxiv.org/abs/astro-ph/9509047} {arXiv:astro-ph/9509047} \BibitemShut
  {NoStop}%
\bibitem [{\citenamefont {Scoccimarro}\ and\ \citenamefont
  {Frieman}(1996{\natexlab{b}})}]{Scoccimarro:1996se}%
  \BibitemOpen
  \bibfield  {author} {\bibinfo {author} {\bibfnamefont {R.}~\bibnamefont
  {Scoccimarro}}\ and\ \bibinfo {author} {\bibfnamefont {J.}~\bibnamefont
  {Frieman}},\ }\href {\doibase 10.1086/178177} {\bibfield  {journal} {\bibinfo
   {journal} {Astrophys.\ J.}\ }\textbf {\bibinfo {volume} {473}},\ \bibinfo
  {pages} {620} (\bibinfo {year} {1996}{\natexlab{b}})},\ \Eprint
  {http://arxiv.org/abs/astro-ph/9602070} {arXiv:astro-ph/9602070} \BibitemShut
  {NoStop}%
\bibitem [{\citenamefont {Senatore}\ and\ \citenamefont
  {Zaldarriaga}(2014)}]{Senatore:2014vja}%
  \BibitemOpen
  \bibfield  {author} {\bibinfo {author} {\bibfnamefont {L.}~\bibnamefont
  {Senatore}}\ and\ \bibinfo {author} {\bibfnamefont {M.}~\bibnamefont
  {Zaldarriaga}},\ }\href@noop {} {\  (\bibinfo {year} {2014})},\ \Eprint
  {http://arxiv.org/abs/1409.1225} {arXiv:1409.1225 [astro-ph.CO]} \BibitemShut
  {NoStop}%
\bibitem [{\citenamefont {Ivanov}\ \emph
  {et~al.}(2022{\natexlab{a}})\citenamefont {Ivanov}, \citenamefont {Philcox},
  \citenamefont {Simonovi\'c}, \citenamefont {Zaldarriaga}, \citenamefont
  {Nischimichi},\ and\ \citenamefont {Takada}}]{Ivanov:2021fbu}%
  \BibitemOpen
  \bibfield  {author} {\bibinfo {author} {\bibfnamefont {M.~M.}\ \bibnamefont
  {Ivanov}}, \bibinfo {author} {\bibfnamefont {O.~H.~E.}\ \bibnamefont
  {Philcox}}, \bibinfo {author} {\bibfnamefont {M.}~\bibnamefont
  {Simonovi\'c}}, \bibinfo {author} {\bibfnamefont {M.}~\bibnamefont
  {Zaldarriaga}}, \bibinfo {author} {\bibfnamefont {T.}~\bibnamefont
  {Nischimichi}}, \ and\ \bibinfo {author} {\bibfnamefont {M.}~\bibnamefont
  {Takada}},\ }\href {\doibase 10.1103/PhysRevD.105.043531} {\bibfield
  {journal} {\bibinfo  {journal} {Phys. Rev. D}\ }\textbf {\bibinfo {volume}
  {105}},\ \bibinfo {pages} {043531} (\bibinfo {year} {2022}{\natexlab{a}})},\
  \Eprint {http://arxiv.org/abs/2110.00006} {arXiv:2110.00006 [astro-ph.CO]}
  \BibitemShut {NoStop}%
\bibitem [{\citenamefont {Taule}\ and\ \citenamefont
  {Garny}(2023)}]{Taule:2023izt}%
  \BibitemOpen
  \bibfield  {author} {\bibinfo {author} {\bibfnamefont {P.}~\bibnamefont
  {Taule}}\ and\ \bibinfo {author} {\bibfnamefont {M.}~\bibnamefont {Garny}},\
  }\href {\doibase 10.1088/1475-7516/2023/11/078} {\bibfield  {journal}
  {\bibinfo  {journal} {JCAP}\ }\textbf {\bibinfo {volume} {11}},\ \bibinfo
  {pages} {078} (\bibinfo {year} {2023})},\ \Eprint
  {http://arxiv.org/abs/2308.07379} {arXiv:2308.07379 [astro-ph.CO]}
  \BibitemShut {NoStop}%
\bibitem [{\citenamefont {Hand}\ \emph {et~al.}(2018)\citenamefont {Hand},
  \citenamefont {Feng}, \citenamefont {Beutler}, \citenamefont {Li},
  \citenamefont {Modi}, \citenamefont {Seljak},\ and\ \citenamefont
  {Slepian}}]{Hand:2017pqn}%
  \BibitemOpen
  \bibfield  {author} {\bibinfo {author} {\bibfnamefont {N.}~\bibnamefont
  {Hand}}, \bibinfo {author} {\bibfnamefont {Y.}~\bibnamefont {Feng}}, \bibinfo
  {author} {\bibfnamefont {F.}~\bibnamefont {Beutler}}, \bibinfo {author}
  {\bibfnamefont {Y.}~\bibnamefont {Li}}, \bibinfo {author} {\bibfnamefont
  {C.}~\bibnamefont {Modi}}, \bibinfo {author} {\bibfnamefont {U.}~\bibnamefont
  {Seljak}}, \ and\ \bibinfo {author} {\bibfnamefont {Z.}~\bibnamefont
  {Slepian}},\ }\href {\doibase 10.3847/1538-3881/aadae0} {\bibfield  {journal}
  {\bibinfo  {journal} {Astron. J.}\ }\textbf {\bibinfo {volume} {156}},\
  \bibinfo {pages} {160} (\bibinfo {year} {2018})},\ \Eprint
  {http://arxiv.org/abs/1712.05834} {arXiv:1712.05834 [astro-ph.IM]}
  \BibitemShut {NoStop}%
\bibitem [{\citenamefont {Jackson}(1972)}]{Jackson:2008yv}%
  \BibitemOpen
  \bibfield  {author} {\bibinfo {author} {\bibfnamefont {J.~C.}\ \bibnamefont
  {Jackson}},\ }\href {\doibase 10.1093/mnras/156.1.1P} {\bibfield  {journal}
  {\bibinfo  {journal} {Mon. Not. Roy. Astron. Soc.}\ }\textbf {\bibinfo
  {volume} {156}},\ \bibinfo {pages} {1P} (\bibinfo {year} {1972})},\ \Eprint
  {http://arxiv.org/abs/0810.3908} {arXiv:0810.3908 [astro-ph]} \BibitemShut
  {NoStop}%
\bibitem [{\citenamefont {Benson}\ \emph {et~al.}(2000)\citenamefont {Benson},
  \citenamefont {Cole}, \citenamefont {Frenk}, \citenamefont {Baugh},\ and\
  \citenamefont {Lacey}}]{Benson:1999mva}%
  \BibitemOpen
  \bibfield  {author} {\bibinfo {author} {\bibfnamefont {A.~J.}\ \bibnamefont
  {Benson}}, \bibinfo {author} {\bibfnamefont {S.}~\bibnamefont {Cole}},
  \bibinfo {author} {\bibfnamefont {C.~S.}\ \bibnamefont {Frenk}}, \bibinfo
  {author} {\bibfnamefont {C.~M.}\ \bibnamefont {Baugh}}, \ and\ \bibinfo
  {author} {\bibfnamefont {C.~G.}\ \bibnamefont {Lacey}},\ }\href {\doibase
  10.1046/j.1365-8711.2000.03101.x} {\bibfield  {journal} {\bibinfo  {journal}
  {Mon. Not. Roy. Astron. Soc.}\ }\textbf {\bibinfo {volume} {311}},\ \bibinfo
  {pages} {793} (\bibinfo {year} {2000})},\ \Eprint
  {http://arxiv.org/abs/astro-ph/9903343} {arXiv:astro-ph/9903343} \BibitemShut
  {NoStop}%
\bibitem [{\citenamefont {Press}\ and\ \citenamefont
  {Schechter}(1974)}]{Press:1973iz}%
  \BibitemOpen
  \bibfield  {author} {\bibinfo {author} {\bibfnamefont {W.~H.}\ \bibnamefont
  {Press}}\ and\ \bibinfo {author} {\bibfnamefont {P.}~\bibnamefont
  {Schechter}},\ }\href {\doibase 10.1086/152650} {\bibfield  {journal}
  {\bibinfo  {journal} {Astrophys. J.}\ }\textbf {\bibinfo {volume} {187}},\
  \bibinfo {pages} {425} (\bibinfo {year} {1974})}\BibitemShut {NoStop}%
\bibitem [{\citenamefont {Ivanov}\ \emph {et~al.}(2019)\citenamefont {Ivanov},
  \citenamefont {Kaurov},\ and\ \citenamefont {Sibiryakov}}]{Ivanov:2018lcg}%
  \BibitemOpen
  \bibfield  {author} {\bibinfo {author} {\bibfnamefont {M.~M.}\ \bibnamefont
  {Ivanov}}, \bibinfo {author} {\bibfnamefont {A.~A.}\ \bibnamefont {Kaurov}},
  \ and\ \bibinfo {author} {\bibfnamefont {S.}~\bibnamefont {Sibiryakov}},\
  }\href {\doibase 10.1088/1475-7516/2019/03/009} {\bibfield  {journal}
  {\bibinfo  {journal} {JCAP}\ }\textbf {\bibinfo {volume} {1903}},\ \bibinfo
  {pages} {009} (\bibinfo {year} {2019})},\ \Eprint
  {http://arxiv.org/abs/1811.07913} {arXiv:1811.07913 [astro-ph.CO]}
  \BibitemShut {NoStop}%
\bibitem [{\citenamefont {Sheth}\ and\ \citenamefont
  {Tormen}(1999)}]{Sheth:1999mn}%
  \BibitemOpen
  \bibfield  {author} {\bibinfo {author} {\bibfnamefont {R.~K.}\ \bibnamefont
  {Sheth}}\ and\ \bibinfo {author} {\bibfnamefont {G.}~\bibnamefont {Tormen}},\
  }\href {\doibase 10.1046/j.1365-8711.1999.02692.x} {\bibfield  {journal}
  {\bibinfo  {journal} {Mon. Not. Roy. Astron. Soc.}\ }\textbf {\bibinfo
  {volume} {308}},\ \bibinfo {pages} {119} (\bibinfo {year} {1999})},\ \Eprint
  {http://arxiv.org/abs/astro-ph/9901122} {arXiv:astro-ph/9901122} \BibitemShut
  {NoStop}%
\bibitem [{\citenamefont {Tinker}\ \emph {et~al.}(2008)\citenamefont {Tinker},
  \citenamefont {Kravtsov}, \citenamefont {Klypin}, \citenamefont {Abazajian},
  \citenamefont {Warren}, \citenamefont {Yepes}, \citenamefont {Gottlober},\
  and\ \citenamefont {Holz}}]{Tinker:2008ff}%
  \BibitemOpen
  \bibfield  {author} {\bibinfo {author} {\bibfnamefont {J.~L.}\ \bibnamefont
  {Tinker}}, \bibinfo {author} {\bibfnamefont {A.~V.}\ \bibnamefont
  {Kravtsov}}, \bibinfo {author} {\bibfnamefont {A.}~\bibnamefont {Klypin}},
  \bibinfo {author} {\bibfnamefont {K.}~\bibnamefont {Abazajian}}, \bibinfo
  {author} {\bibfnamefont {M.~S.}\ \bibnamefont {Warren}}, \bibinfo {author}
  {\bibfnamefont {G.}~\bibnamefont {Yepes}}, \bibinfo {author} {\bibfnamefont
  {S.}~\bibnamefont {Gottlober}}, \ and\ \bibinfo {author} {\bibfnamefont
  {D.~E.}\ \bibnamefont {Holz}},\ }\href {\doibase 10.1086/591439} {\bibfield
  {journal} {\bibinfo  {journal} {Astrophys. J.}\ }\textbf {\bibinfo {volume}
  {688}},\ \bibinfo {pages} {709} (\bibinfo {year} {2008})},\ \Eprint
  {http://arxiv.org/abs/0803.2706} {arXiv:0803.2706 [astro-ph]} \BibitemShut
  {NoStop}%
\bibitem [{\citenamefont {Lazeyras}\ \emph {et~al.}(2016)\citenamefont
  {Lazeyras}, \citenamefont {Wagner}, \citenamefont {Baldauf},\ and\
  \citenamefont {Schmidt}}]{Lazeyras:2015lgp}%
  \BibitemOpen
  \bibfield  {author} {\bibinfo {author} {\bibfnamefont {T.}~\bibnamefont
  {Lazeyras}}, \bibinfo {author} {\bibfnamefont {C.}~\bibnamefont {Wagner}},
  \bibinfo {author} {\bibfnamefont {T.}~\bibnamefont {Baldauf}}, \ and\
  \bibinfo {author} {\bibfnamefont {F.}~\bibnamefont {Schmidt}},\ }\href
  {\doibase 10.1088/1475-7516/2016/02/018} {\bibfield  {journal} {\bibinfo
  {journal} {JCAP}\ }\textbf {\bibinfo {volume} {1602}},\ \bibinfo {pages}
  {018} (\bibinfo {year} {2016})},\ \Eprint {http://arxiv.org/abs/1511.01096}
  {arXiv:1511.01096 [astro-ph.CO]} \BibitemShut {NoStop}%
\bibitem [{\citenamefont {Baldauf}\ \emph {et~al.}(2013)\citenamefont
  {Baldauf}, \citenamefont {Seljak}, \citenamefont {Smith}, \citenamefont
  {Hamaus},\ and\ \citenamefont {Desjacques}}]{Baldauf:2013hka}%
  \BibitemOpen
  \bibfield  {author} {\bibinfo {author} {\bibfnamefont {T.}~\bibnamefont
  {Baldauf}}, \bibinfo {author} {\bibfnamefont {U.}~\bibnamefont {Seljak}},
  \bibinfo {author} {\bibfnamefont {R.~E.}\ \bibnamefont {Smith}}, \bibinfo
  {author} {\bibfnamefont {N.}~\bibnamefont {Hamaus}}, \ and\ \bibinfo {author}
  {\bibfnamefont {V.}~\bibnamefont {Desjacques}},\ }\href {\doibase
  10.1103/PhysRevD.88.083507} {\bibfield  {journal} {\bibinfo  {journal} {Phys.
  Rev. D}\ }\textbf {\bibinfo {volume} {88}},\ \bibinfo {pages} {083507}
  (\bibinfo {year} {2013})},\ \Eprint {http://arxiv.org/abs/1305.2917}
  {arXiv:1305.2917 [astro-ph.CO]} \BibitemShut {NoStop}%
\bibitem [{\citenamefont {Li}\ and\ \citenamefont {Smith}(2024)}]{Li:2024wco}%
  \BibitemOpen
  \bibfield  {author} {\bibinfo {author} {\bibfnamefont {Y.}~\bibnamefont
  {Li}}\ and\ \bibinfo {author} {\bibfnamefont {R.~E.}\ \bibnamefont {Smith}},\
  }\href@noop {} {\  (\bibinfo {year} {2024})},\ \Eprint
  {http://arxiv.org/abs/2411.18722} {arXiv:2411.18722 [astro-ph.CO]}
  \BibitemShut {NoStop}%
\bibitem [{\citenamefont {Yuan}\ \emph
  {et~al.}(2022{\natexlab{b}})\citenamefont {Yuan}, \citenamefont {Garrison},
  \citenamefont {Hadzhiyska}, \citenamefont {Bose},\ and\ \citenamefont
  {Eisenstein}}]{Yuan:2021izi}%
  \BibitemOpen
  \bibfield  {author} {\bibinfo {author} {\bibfnamefont {S.}~\bibnamefont
  {Yuan}}, \bibinfo {author} {\bibfnamefont {L.~H.}\ \bibnamefont {Garrison}},
  \bibinfo {author} {\bibfnamefont {B.}~\bibnamefont {Hadzhiyska}}, \bibinfo
  {author} {\bibfnamefont {S.}~\bibnamefont {Bose}}, \ and\ \bibinfo {author}
  {\bibfnamefont {D.~J.}\ \bibnamefont {Eisenstein}},\ }\href {\doibase
  10.1093/mnras/stab3355} {\bibfield  {journal} {\bibinfo  {journal} {Mon. Not.
  Roy. Astron. Soc.}\ }\textbf {\bibinfo {volume} {510}},\ \bibinfo {pages}
  {3301} (\bibinfo {year} {2022}{\natexlab{b}})},\ \Eprint
  {http://arxiv.org/abs/2110.11412} {arXiv:2110.11412 [astro-ph.CO]}
  \BibitemShut {NoStop}%
\bibitem [{\citenamefont {Ivanov}\ \emph
  {et~al.}(2024{\natexlab{b}})\citenamefont {Ivanov} \emph
  {et~al.}}]{Ivanov:2024dgv}%
  \BibitemOpen
  \bibfield  {author} {\bibinfo {author} {\bibfnamefont {M.~M.}\ \bibnamefont
  {Ivanov}} \emph {et~al.},\ }\href@noop {} {\  (\bibinfo {year}
  {2024}{\natexlab{b}})},\ \Eprint {http://arxiv.org/abs/2412.01888}
  {arXiv:2412.01888 [astro-ph.CO]} \BibitemShut {NoStop}%
\bibitem [{\citenamefont {Reid}\ \emph {et~al.}(2016)\citenamefont {Reid} \emph
  {et~al.}}]{Reid:2015gra}%
  \BibitemOpen
  \bibfield  {author} {\bibinfo {author} {\bibfnamefont {B.}~\bibnamefont
  {Reid}} \emph {et~al.},\ }\href {\doibase 10.1093/mnras/stv2382} {\bibfield
  {journal} {\bibinfo  {journal} {Mon. Not. Roy. Astron. Soc.}\ }\textbf
  {\bibinfo {volume} {455}},\ \bibinfo {pages} {1553} (\bibinfo {year}
  {2016})},\ \Eprint {http://arxiv.org/abs/1509.06529} {arXiv:1509.06529
  [astro-ph.CO]} \BibitemShut {NoStop}%
\bibitem [{\citenamefont {Adame}\ \emph {et~al.}(2025)\citenamefont {Adame}
  \emph {et~al.}}]{DESI:2024mwx}%
  \BibitemOpen
  \bibfield  {author} {\bibinfo {author} {\bibfnamefont {A.~G.}\ \bibnamefont
  {Adame}} \emph {et~al.} (\bibinfo {collaboration} {DESI}),\ }\href {\doibase
  10.1088/1475-7516/2025/02/021} {\bibfield  {journal} {\bibinfo  {journal}
  {JCAP}\ }\textbf {\bibinfo {volume} {02}},\ \bibinfo {pages} {021} (\bibinfo
  {year} {2025})},\ \Eprint {http://arxiv.org/abs/2404.03002} {arXiv:2404.03002
  [astro-ph.CO]} \BibitemShut {NoStop}%
\bibitem [{\citenamefont {Casas-Miranda}\ \emph {et~al.}(2002)\citenamefont
  {Casas-Miranda}, \citenamefont {Mo}, \citenamefont {Sheth},\ and\
  \citenamefont {Boerner}}]{Casas-Miranda:2001dwz}%
  \BibitemOpen
  \bibfield  {author} {\bibinfo {author} {\bibfnamefont {R.}~\bibnamefont
  {Casas-Miranda}}, \bibinfo {author} {\bibfnamefont {H.~J.}\ \bibnamefont
  {Mo}}, \bibinfo {author} {\bibfnamefont {R.~K.}\ \bibnamefont {Sheth}}, \
  and\ \bibinfo {author} {\bibfnamefont {G.}~\bibnamefont {Boerner}},\ }\href
  {\doibase 10.1046/j.1365-8711.2002.05378.x} {\bibfield  {journal} {\bibinfo
  {journal} {Mon. Not. Roy. Astron. Soc.}\ }\textbf {\bibinfo {volume} {333}},\
  \bibinfo {pages} {730} (\bibinfo {year} {2002})},\ \Eprint
  {http://arxiv.org/abs/astro-ph/0105008} {arXiv:astro-ph/0105008} \BibitemShut
  {NoStop}%
\bibitem [{\citenamefont {Papamakarios}\ \emph {et~al.}(2018)\citenamefont
  {Papamakarios}, \citenamefont {Pavlakou},\ and\ \citenamefont
  {Murray}}]{papamakarios2018maskedautoregressiveflowdensity}%
  \BibitemOpen
  \bibfield  {author} {\bibinfo {author} {\bibfnamefont {G.}~\bibnamefont
  {Papamakarios}}, \bibinfo {author} {\bibfnamefont {T.}~\bibnamefont
  {Pavlakou}}, \ and\ \bibinfo {author} {\bibfnamefont {I.}~\bibnamefont
  {Murray}},\ }\href {https://arxiv.org/abs/1705.07057} {\enquote {\bibinfo
  {title} {Masked autoregressive flow for density estimation},}\ } (\bibinfo
  {year} {2018}),\ \Eprint {http://arxiv.org/abs/1705.07057} {arXiv:1705.07057
  [stat.ML]} \BibitemShut {NoStop}%
\bibitem [{\citenamefont {Paszke}\ \emph {et~al.}(2019)\citenamefont {Paszke},
  \citenamefont {Gross}, \citenamefont {Massa}, \citenamefont {Lerer},
  \citenamefont {Bradbury}, \citenamefont {Chanan}, \citenamefont {Killeen},
  \citenamefont {Lin}, \citenamefont {Gimelshein}, \citenamefont {Antiga} \emph
  {et~al.}}]{paszke2019pytorch}%
  \BibitemOpen
  \bibfield  {author} {\bibinfo {author} {\bibfnamefont {A.}~\bibnamefont
  {Paszke}}, \bibinfo {author} {\bibfnamefont {S.}~\bibnamefont {Gross}},
  \bibinfo {author} {\bibfnamefont {F.}~\bibnamefont {Massa}}, \bibinfo
  {author} {\bibfnamefont {A.}~\bibnamefont {Lerer}}, \bibinfo {author}
  {\bibfnamefont {J.}~\bibnamefont {Bradbury}}, \bibinfo {author}
  {\bibfnamefont {G.}~\bibnamefont {Chanan}}, \bibinfo {author} {\bibfnamefont
  {T.}~\bibnamefont {Killeen}}, \bibinfo {author} {\bibfnamefont
  {Z.}~\bibnamefont {Lin}}, \bibinfo {author} {\bibfnamefont {N.}~\bibnamefont
  {Gimelshein}}, \bibinfo {author} {\bibfnamefont {L.}~\bibnamefont {Antiga}},
  \emph {et~al.},\ }\href@noop {} {\bibfield  {journal} {\bibinfo  {journal}
  {Advances in neural information processing systems}\ }\textbf {\bibinfo
  {volume} {32}} (\bibinfo {year} {2019})}\BibitemShut {NoStop}%
\bibitem [{\citenamefont {Hahn}\ \emph {et~al.}(2017)\citenamefont {Hahn},
  \citenamefont {Scoccimarro}, \citenamefont {Blanton}, \citenamefont
  {Tinker},\ and\ \citenamefont {Rodr\'\i{}guez-Torres}}]{Hahn:2016kiy}%
  \BibitemOpen
  \bibfield  {author} {\bibinfo {author} {\bibfnamefont {C.}~\bibnamefont
  {Hahn}}, \bibinfo {author} {\bibfnamefont {R.}~\bibnamefont {Scoccimarro}},
  \bibinfo {author} {\bibfnamefont {M.~R.}\ \bibnamefont {Blanton}}, \bibinfo
  {author} {\bibfnamefont {J.~L.}\ \bibnamefont {Tinker}}, \ and\ \bibinfo
  {author} {\bibfnamefont {S.~A.}\ \bibnamefont {Rodr\'\i{}guez-Torres}},\
  }\href {\doibase 10.1093/mnras/stx185} {\bibfield  {journal} {\bibinfo
  {journal} {Mon. Not. Roy. Astron. Soc.}\ }\textbf {\bibinfo {volume} {467}},\
  \bibinfo {pages} {1940} (\bibinfo {year} {2017})},\ \Eprint
  {http://arxiv.org/abs/1609.01714} {arXiv:1609.01714 [astro-ph.CO]}
  \BibitemShut {NoStop}%
\bibitem [{\citenamefont {Ivanov}(2021)}]{Ivanov:2021zmi}%
  \BibitemOpen
  \bibfield  {author} {\bibinfo {author} {\bibfnamefont {M.~M.}\ \bibnamefont
  {Ivanov}},\ }\href@noop {} {\  (\bibinfo {year} {2021})},\ \Eprint
  {http://arxiv.org/abs/2106.12580} {arXiv:2106.12580 [astro-ph.CO]}
  \BibitemShut {NoStop}%
\bibitem [{\citenamefont {Chudaykin}\ and\ \citenamefont
  {Ivanov}(2022)}]{Chudaykin:2022nru}%
  \BibitemOpen
  \bibfield  {author} {\bibinfo {author} {\bibfnamefont {A.}~\bibnamefont
  {Chudaykin}}\ and\ \bibinfo {author} {\bibfnamefont {M.~M.}\ \bibnamefont
  {Ivanov}},\ }\href@noop {} {\  (\bibinfo {year} {2022})},\ \Eprint
  {http://arxiv.org/abs/2210.17044} {arXiv:2210.17044 [astro-ph.CO]}
  \BibitemShut {NoStop}%
\bibitem [{\citenamefont {Ivanov}\ \emph
  {et~al.}(2022{\natexlab{b}})\citenamefont {Ivanov}, \citenamefont {Philcox},
  \citenamefont {Nishimichi}, \citenamefont {Simonovi\'c}, \citenamefont
  {Takada},\ and\ \citenamefont {Zaldarriaga}}]{Ivanov:2021kcd}%
  \BibitemOpen
  \bibfield  {author} {\bibinfo {author} {\bibfnamefont {M.~M.}\ \bibnamefont
  {Ivanov}}, \bibinfo {author} {\bibfnamefont {O.~H.~E.}\ \bibnamefont
  {Philcox}}, \bibinfo {author} {\bibfnamefont {T.}~\bibnamefont {Nishimichi}},
  \bibinfo {author} {\bibfnamefont {M.}~\bibnamefont {Simonovi\'c}}, \bibinfo
  {author} {\bibfnamefont {M.}~\bibnamefont {Takada}}, \ and\ \bibinfo {author}
  {\bibfnamefont {M.}~\bibnamefont {Zaldarriaga}},\ }\href {\doibase
  10.1103/PhysRevD.105.063512} {\bibfield  {journal} {\bibinfo  {journal}
  {Phys. Rev. D}\ }\textbf {\bibinfo {volume} {105}},\ \bibinfo {pages}
  {063512} (\bibinfo {year} {2022}{\natexlab{b}})},\ \Eprint
  {http://arxiv.org/abs/2110.10161} {arXiv:2110.10161 [astro-ph.CO]}
  \BibitemShut {NoStop}%
\bibitem [{\citenamefont {Philcox}\ \emph {et~al.}(2022)\citenamefont
  {Philcox}, \citenamefont {Ivanov}, \citenamefont {Cabass}, \citenamefont
  {Simonovi\'c}, \citenamefont {Zaldarriaga},\ and\ \citenamefont
  {Nishimichi}}]{Philcox:2022frc}%
  \BibitemOpen
  \bibfield  {author} {\bibinfo {author} {\bibfnamefont {O.~H.~E.}\
  \bibnamefont {Philcox}}, \bibinfo {author} {\bibfnamefont {M.~M.}\
  \bibnamefont {Ivanov}}, \bibinfo {author} {\bibfnamefont {G.}~\bibnamefont
  {Cabass}}, \bibinfo {author} {\bibfnamefont {M.}~\bibnamefont {Simonovi\'c}},
  \bibinfo {author} {\bibfnamefont {M.}~\bibnamefont {Zaldarriaga}}, \ and\
  \bibinfo {author} {\bibfnamefont {T.}~\bibnamefont {Nishimichi}},\ }\href
  {\doibase 10.1103/PhysRevD.106.043530} {\bibfield  {journal} {\bibinfo
  {journal} {Phys. Rev. D}\ }\textbf {\bibinfo {volume} {106}},\ \bibinfo
  {pages} {043530} (\bibinfo {year} {2022})},\ \Eprint
  {http://arxiv.org/abs/2206.02800} {arXiv:2206.02800 [astro-ph.CO]}
  \BibitemShut {NoStop}%
\bibitem [{\citenamefont {Beutler}\ \emph {et~al.}(2017)\citenamefont {Beutler}
  \emph {et~al.}}]{Beutler:2016arn}%
  \BibitemOpen
  \bibfield  {author} {\bibinfo {author} {\bibfnamefont {F.}~\bibnamefont
  {Beutler}} \emph {et~al.} (\bibinfo {collaboration} {BOSS}),\ }\href
  {\doibase 10.1093/mnras/stw3298} {\bibfield  {journal} {\bibinfo  {journal}
  {Mon. Not. Roy. Astron. Soc.}\ }\textbf {\bibinfo {volume} {466}},\ \bibinfo
  {pages} {2242} (\bibinfo {year} {2017})},\ \Eprint
  {http://arxiv.org/abs/1607.03150} {arXiv:1607.03150 [astro-ph.CO]}
  \BibitemShut {NoStop}%
\bibitem [{\citenamefont {Philcox}(2021{\natexlab{a}})}]{Philcox:2020vbm}%
  \BibitemOpen
  \bibfield  {author} {\bibinfo {author} {\bibfnamefont {O.~H.~E.}\
  \bibnamefont {Philcox}},\ }\href {\doibase 10.1103/PhysRevD.103.103504}
  {\bibfield  {journal} {\bibinfo  {journal} {Phys. Rev. D}\ }\textbf {\bibinfo
  {volume} {103}},\ \bibinfo {pages} {103504} (\bibinfo {year}
  {2021}{\natexlab{a}})},\ \Eprint {http://arxiv.org/abs/2012.09389}
  {arXiv:2012.09389 [astro-ph.CO]} \BibitemShut {NoStop}%
\bibitem [{\citenamefont {Philcox}(2021{\natexlab{b}})}]{Philcox:2021ukg}%
  \BibitemOpen
  \bibfield  {author} {\bibinfo {author} {\bibfnamefont {O.~H.~E.}\
  \bibnamefont {Philcox}},\ }\href@noop {} {\  (\bibinfo {year}
  {2021}{\natexlab{b}})},\ \Eprint {http://arxiv.org/abs/2107.06287}
  {arXiv:2107.06287 [astro-ph.CO]} \BibitemShut {NoStop}%
\bibitem [{\citenamefont {Philcox}\ \emph {et~al.}(2020)\citenamefont
  {Philcox}, \citenamefont {Ivanov}, \citenamefont {Simonovi\'c},\ and\
  \citenamefont {Zaldarriaga}}]{Philcox:2020vvt}%
  \BibitemOpen
  \bibfield  {author} {\bibinfo {author} {\bibfnamefont {O.~H.~E.}\
  \bibnamefont {Philcox}}, \bibinfo {author} {\bibfnamefont {M.~M.}\
  \bibnamefont {Ivanov}}, \bibinfo {author} {\bibfnamefont {M.}~\bibnamefont
  {Simonovi\'c}}, \ and\ \bibinfo {author} {\bibfnamefont {M.}~\bibnamefont
  {Zaldarriaga}},\ }\href {\doibase 10.1088/1475-7516/2020/05/032} {\bibfield
  {journal} {\bibinfo  {journal} {JCAP}\ }\textbf {\bibinfo {volume} {05}},\
  \bibinfo {pages} {032} (\bibinfo {year} {2020})},\ \Eprint
  {http://arxiv.org/abs/2002.04035} {arXiv:2002.04035 [astro-ph.CO]}
  \BibitemShut {NoStop}%
\bibitem [{\citenamefont {Wadekar}\ \emph {et~al.}(2020)\citenamefont
  {Wadekar}, \citenamefont {Ivanov},\ and\ \citenamefont
  {Scoccimarro}}]{Wadekar:2020hax}%
  \BibitemOpen
  \bibfield  {author} {\bibinfo {author} {\bibfnamefont {D.}~\bibnamefont
  {Wadekar}}, \bibinfo {author} {\bibfnamefont {M.~M.}\ \bibnamefont {Ivanov}},
  \ and\ \bibinfo {author} {\bibfnamefont {R.}~\bibnamefont {Scoccimarro}},\
  }\href {\doibase 10.1103/PhysRevD.102.123521} {\bibfield  {journal} {\bibinfo
   {journal} {Phys. Rev. D}\ }\textbf {\bibinfo {volume} {102}},\ \bibinfo
  {pages} {123521} (\bibinfo {year} {2020})},\ \Eprint
  {http://arxiv.org/abs/2009.00622} {arXiv:2009.00622 [astro-ph.CO]}
  \BibitemShut {NoStop}%
\bibitem [{\citenamefont {Lange}\ \emph {et~al.}(2021)\citenamefont {Lange},
  \citenamefont {Hearin}, \citenamefont {Leauthaud}, \citenamefont {van~den
  Bosch}, \citenamefont {Guo},\ and\ \citenamefont {DeRose}}]{Lange:2021zre}%
  \BibitemOpen
  \bibfield  {author} {\bibinfo {author} {\bibfnamefont {J.~U.}\ \bibnamefont
  {Lange}}, \bibinfo {author} {\bibfnamefont {A.~P.}\ \bibnamefont {Hearin}},
  \bibinfo {author} {\bibfnamefont {A.}~\bibnamefont {Leauthaud}}, \bibinfo
  {author} {\bibfnamefont {F.~C.}\ \bibnamefont {van~den Bosch}}, \bibinfo
  {author} {\bibfnamefont {H.}~\bibnamefont {Guo}}, \ and\ \bibinfo {author}
  {\bibfnamefont {J.}~\bibnamefont {DeRose}},\ }\href {\doibase
  10.1093/mnras/stab3111} {\bibfield  {journal} {\bibinfo  {journal} {Mon. Not.
  Roy. Astron. Soc.}\ }\textbf {\bibinfo {volume} {509}},\ \bibinfo {pages}
  {1779} (\bibinfo {year} {2021})},\ \Eprint {http://arxiv.org/abs/2101.12261}
  {arXiv:2101.12261 [astro-ph.CO]} \BibitemShut {NoStop}%
\bibitem [{\citenamefont {de~Mattia}\ \emph {et~al.}(2021)\citenamefont
  {de~Mattia} \emph {et~al.}}]{deMattia:2020fkb}%
  \BibitemOpen
  \bibfield  {author} {\bibinfo {author} {\bibfnamefont {A.}~\bibnamefont
  {de~Mattia}} \emph {et~al.},\ }\href {\doibase 10.1093/mnras/staa3891}
  {\bibfield  {journal} {\bibinfo  {journal} {Mon. Not. Roy. Astron. Soc.}\
  }\textbf {\bibinfo {volume} {501}},\ \bibinfo {pages} {5616} (\bibinfo {year}
  {2021})},\ \Eprint {http://arxiv.org/abs/2007.09008} {arXiv:2007.09008
  [astro-ph.CO]} \BibitemShut {NoStop}%
\bibitem [{\citenamefont {Neveux}\ \emph {et~al.}(2020)\citenamefont {Neveux}
  \emph {et~al.}}]{eBOSS:2020uxp}%
  \BibitemOpen
  \bibfield  {author} {\bibinfo {author} {\bibfnamefont {R.}~\bibnamefont
  {Neveux}} \emph {et~al.} (\bibinfo {collaboration} {eBOSS}),\ }\href
  {\doibase 10.1093/mnras/staa2780} {\bibfield  {journal} {\bibinfo  {journal}
  {Mon. Not. Roy. Astron. Soc.}\ }\textbf {\bibinfo {volume} {499}},\ \bibinfo
  {pages} {210} (\bibinfo {year} {2020})},\ \Eprint
  {http://arxiv.org/abs/2007.08999} {arXiv:2007.08999 [astro-ph.CO]}
  \BibitemShut {NoStop}%
\bibitem [{\citenamefont {Desjacques}\ \emph
  {et~al.}(2018{\natexlab{b}})\citenamefont {Desjacques}, \citenamefont
  {Jeong},\ and\ \citenamefont {Schmidt}}]{Desjacques:2018pfv}%
  \BibitemOpen
  \bibfield  {author} {\bibinfo {author} {\bibfnamefont {V.}~\bibnamefont
  {Desjacques}}, \bibinfo {author} {\bibfnamefont {D.}~\bibnamefont {Jeong}}, \
  and\ \bibinfo {author} {\bibfnamefont {F.}~\bibnamefont {Schmidt}},\ }\href
  {\doibase 10.1088/1475-7516/2018/12/035} {\bibfield  {journal} {\bibinfo
  {journal} {JCAP}\ }\textbf {\bibinfo {volume} {1812}},\ \bibinfo {pages}
  {035} (\bibinfo {year} {2018}{\natexlab{b}})},\ \Eprint
  {http://arxiv.org/abs/1806.04015} {arXiv:1806.04015 [astro-ph.CO]}
  \BibitemShut {NoStop}%
\bibitem [{\citenamefont {Ivanov}(2024)}]{Ivanov:2023yla}%
  \BibitemOpen
  \bibfield  {author} {\bibinfo {author} {\bibfnamefont {M.~M.}\ \bibnamefont
  {Ivanov}},\ }\href {\doibase 10.1103/PhysRevD.109.023507} {\bibfield
  {journal} {\bibinfo  {journal} {Phys. Rev. D}\ }\textbf {\bibinfo {volume}
  {109}},\ \bibinfo {pages} {023507} (\bibinfo {year} {2024})},\ \Eprint
  {http://arxiv.org/abs/2309.10133} {arXiv:2309.10133 [astro-ph.CO]}
  \BibitemShut {NoStop}%
\bibitem [{\citenamefont {Ivanov}\ \emph
  {et~al.}(2024{\natexlab{c}})\citenamefont {Ivanov}, \citenamefont {Toomey},\
  and\ \citenamefont {Kara\c{c}ayl\i{}}}]{Ivanov:2024jtl}%
  \BibitemOpen
  \bibfield  {author} {\bibinfo {author} {\bibfnamefont {M.~M.}\ \bibnamefont
  {Ivanov}}, \bibinfo {author} {\bibfnamefont {M.~W.}\ \bibnamefont {Toomey}},
  \ and\ \bibinfo {author} {\bibfnamefont {N.~G.}\ \bibnamefont
  {Kara\c{c}ayl\i{}}},\ }\href@noop {} {\  (\bibinfo {year}
  {2024}{\natexlab{c}})},\ \Eprint {http://arxiv.org/abs/2405.13208}
  {arXiv:2405.13208 [astro-ph.CO]} \BibitemShut {NoStop}%
\bibitem [{\citenamefont {Baldauf}\ \emph {et~al.}(2016)\citenamefont
  {Baldauf}, \citenamefont {Schaan},\ and\ \citenamefont
  {Zaldarriaga}}]{Baldauf:2015zga}%
  \BibitemOpen
  \bibfield  {author} {\bibinfo {author} {\bibfnamefont {T.}~\bibnamefont
  {Baldauf}}, \bibinfo {author} {\bibfnamefont {E.}~\bibnamefont {Schaan}}, \
  and\ \bibinfo {author} {\bibfnamefont {M.}~\bibnamefont {Zaldarriaga}},\
  }\href {\doibase 10.1088/1475-7516/2016/03/007} {\bibfield  {journal}
  {\bibinfo  {journal} {JCAP}\ }\textbf {\bibinfo {volume} {03}},\ \bibinfo
  {pages} {007} (\bibinfo {year} {2016})},\ \Eprint
  {http://arxiv.org/abs/1507.02255} {arXiv:1507.02255 [astro-ph.CO]}
  \BibitemShut {NoStop}%
\bibitem [{\citenamefont {Philcox}\ \emph {et~al.}(2021)\citenamefont
  {Philcox}, \citenamefont {Sherwin}, \citenamefont {Farren},\ and\
  \citenamefont {Baxter}}]{Philcox:2020xbv}%
  \BibitemOpen
  \bibfield  {author} {\bibinfo {author} {\bibfnamefont {O.~H.~E.}\
  \bibnamefont {Philcox}}, \bibinfo {author} {\bibfnamefont {B.~D.}\
  \bibnamefont {Sherwin}}, \bibinfo {author} {\bibfnamefont {G.~S.}\
  \bibnamefont {Farren}}, \ and\ \bibinfo {author} {\bibfnamefont {E.~J.}\
  \bibnamefont {Baxter}},\ }\href {\doibase 10.1103/PhysRevD.103.023538}
  {\bibfield  {journal} {\bibinfo  {journal} {Phys. Rev. D}\ }\textbf {\bibinfo
  {volume} {103}},\ \bibinfo {pages} {023538} (\bibinfo {year} {2021})},\
  \Eprint {http://arxiv.org/abs/2008.08084} {arXiv:2008.08084 [astro-ph.CO]}
  \BibitemShut {NoStop}%
\end{thebibliography}%

\end{document}